\documentclass[twocolumn,prb,eqsecnum,superscriptaddress]{revtex4}%showpacs

\usepackage{graphicx}
\usepackage{color}\usepackage{enumerate}
\usepackage{amsmath}\usepackage{amssymb}\usepackage{centernot}
\usepackage{multirow}
\usepackage{float}
\usepackage{longtable,booktabs}
\usepackage{hyperref}\usepackage{url}%Turn on when output to pdf file

\begin{document}

\title{Unconventional Fusion and Braiding of Topological Defects in a Lattice Model}

\author{Jeffrey C.Y. Teo}\email{cteo@illinois.edu}
\affiliation{Department of Physics, Institute for Condensed Matter Theory, University of Illinois at Urbana-Champaign, IL 61801, USA}
\author{Abhishek Roy}
\affiliation{Department of Physics, Institute for Condensed Matter Theory, University of Illinois at Urbana-Champaign, IL 61801, USA}
\affiliation{Institute of Theoretical Physics, University of Cologne}
\author{Xiao Chen}
\affiliation{Department of Physics, Institute for Condensed Matter Theory, University of Illinois at Urbana-Champaign, IL 61801, USA}

\begin{abstract}
We examine non-abelian topological defects in an abelian lattice model in two dimensions. We first construct an exact solvable lattice model that exhibits coexisting and intertwined topological and classical order. The anyon types of quasiparticle excitations are permuted by lattice symmetry operations like translations, rotations, and reflections. The global anyon permutation symmetry has a group structure of $S_3$, the permutation group of three elements. Topological crystalline defects -- dislocations and disclinations -- change the anyon type of an orbiting quasiparticle. They exhibits multichannel order dependent fusion rules and projective braiding operations. Their braiding and exchange statistics breaks modular invariance and violates conventional spin-statistics theorem. We develop a framework to characterize these unconventional properties that originate from the semiclassical nature of defects.

\end{abstract}

%\pacs{}
\maketitle
\tableofcontents
	
\section{Introduction}\label{sec:introduction}

The search for Majorana fermions~\cite{Majorana37} has attracted tremendous attention in recent years~\cite{Wilczek09, HasanKane10, QiZhangreview11, Beenakker11, Alicea12} due to their potential application in topological quantum computation~\cite{Kitaev97, OgburnPreskill99, Preskilllecturenotes, FreedmanKitaevLarsenWang01, ChetanSimonSternFreedmanDasSarma, Wangbook}. With the discovery of topological insulators~\cite{KaneMele2D1, KaneMele2D2, Molenkamp07, MooreBalents07, Roy07, FuKaneMele3D, QiHughesZhang08, Hasan08}, the quest has shifted from $p$-wave superconductors~\cite{Kitaevchain, Volovik99, ReadGreen, Ivanov} and quantum Hall states~\cite{MooreRead, GreiterWenWilczek91, NayakWilczek96} to superconductor-ferromagnet (SC-FM) heterostructures with quantum spin Hall insulators~\cite{FuKane08, AkhmerovNilssonBeenakker09, FuKanechargetransport09, LawLeeNg09, GoldhaberGordon12} and strong spin-orbit coupled semiconductors~\cite{SauLutchynTewariDasSarma, OregRefaelvonOppen10, Kouwenhoven12}. The more exotic fractional Majorana fermions that carry richer fusion and braiding characteristics are predicted at the SC-FM edge~\cite{LindnerBergRefaelStern, ClarkeAliceaKirill, MChen, Vaezi} of fractional topological insulators~\cite{LevinStern09, LevinStern12, MaciejkoQiKarchZhang10, SwingleBarkeshliMcGreevySenthil11, LevinBurnellKochStern11} and helical 1D Luttinger liquids~\cite{OregSelaStern13}.

Fractional Majorana fermions can be conceptually studied by {\em twist defects} in exact solvable lattice models. These include Ising-type dislocations in the Kitaev toric code~\cite{Kitaev06, Bombin, KitaevKong12}, fractional Majorana-type dislocations in the Wen plaquette $\mathbb{Z}_k$-rotor model~\cite{YouWen, YouJianWen}, and colored Majorana defects in a string net model~\cite{Bombin11}. Twist defects even appear at dislocation line defects in 3D topological phases~\cite{BiRasmussenXu13, MesarosKimRan13}. Similar non-abelian defects can be constructed as dislocations in abelian {\em topological nematic states}~\cite{BarkeshliQi, BarkeshliQi13} such as a multiple Chern band with symmetry~\cite{WangRan11, LuRan12}, described by {\em genons} in effective field theory~\cite{BarkeshliJianQi} and classified by Wilson structures of non-chiral gapped edges~\cite{LevinGu12, Levin13, BarkeshliJianQi13, BarkeshliJianQi13long}. 

The underlying topological state of all systems mentioned above carries an ungauged symmetry, e.g. charge-flux duality in the toric code~\cite{Kitaev97} and Wen plaquette model~\cite{Wenplaquettemodel}, color permutation symmetry in the color code~\cite{BombinMartin06}, and bilayer symmetry in a topological nematic state~\cite{BarkeshliQi}. Symmetries also appear in many strongly correlated systems such as electronic liquid-crystal phases of a doped Mott insulator~\cite{KivelsonFradkinEmery98} and spin liquid~\cite{Wenspinliquid02}. They intertwine with topology and offer a finer classification of topological phases~\cite{ChenGuLiuWen11, GuWen12, LevinGu12, MesarosRan12, ChenGuLiuWen12, VishwanathSenthil12, BurnellChenFidkowskiVishwanath13}. Even if the symmetry is not broken spontaneously by a Landau order parameter, it may still be {\em weakly broken} in the Kitaev sense~\cite{Kitaev06} by anyon labeling. Extrinsic twist defects further break the symmetry locally by winding the anyon labels. In this article, we demonstrate, using twist defects in an exact solvable lattice model, some fundamental distinctions in fusion and braiding~\cite{Kitaev06, Turaevbook, BakalovKirillovlecturenotes, Wangbook} that separate semiclassical symmetry defects from quantum deconfined anyons~\cite{ArovasSchriefferWilczek84, Wilczekbook} in true topological phases~\cite{Wentopologicalorder90, Wenbook, Fradkinbook} such as the Kitaev honeycomb Ising phase~\cite{Kitaev06} or quasi-topological phases~\cite{BondersonNayak12} such as the physical Pfaffian quantum Hall state~\cite{MooreRead, GreiterWenWilczek91, NayakWilczek96}.

Topological defects involve windings of certain non-dynamical extensive order parameters~\cite{TeoKane} such as pairing phase, Dirac mass, spin polarization, etc. The semiclassical order parameter forbids quantum superposition of different defect configurations. Although defect excitations in SC-FM heterostructures have exponentially localized wavefunctions and twist defects in integrable models only violate symmetry locally at a point, they should be treated as quasi-extensive objects because of the bulk order parameter associated with each of them. This extensiveness provides the means to circumvent locality restrictions~\cite{DoplicherHaagRoberts71, DoplicherHaagRoberts74, Read03} and gives rise to non-abelian statistics in $(3+1)$-dimensions~\cite{TeoKane09, FreedmanHastingsNayakQiWalkerWang, FreedmanHastingsNayakQi}.

In this article, we address three major consequences of the quasi-extensive nature of twist defects in an exact solvable topologically ordered abelian system with symmetry. 
\begin{enumerate} 
\item {\bf Non-commutative fusion} Contrary to a non-abelian discrete gauge theory~\cite{PropitiusBais96, Mochon04, Preskilllecturenotes} where fluxons are labeled by conjugacy classes, twist defects are instead labeled by symmetry group elements and the fusion of defects  depends on their order.
\item {\bf Modified spin statistics} Topological spin for a defect $\lambda$ can only be robustly defined through a $2\pi\times\mbox{ord}(\lambda)$ rotation, where $\mbox{ord}(\lambda)$ is the order of the defect, so that the initial and rotated systems are classically indistinguishable. The exchange phase is then identical to this new definition of spin.\cite{spinstatisicsnotes}
\item {\bf Modified modular invariance} Unlike modular functors in conformal field theory~\cite{Segal88, Verlinde88, WittenJonespolynomials}, fractional quantum Hall states~\cite{Wentopologicalorder90, KeskiWen93, Wenmodulartransformation12} or topological quantum field theories~\cite{Walkernotes91, Turaevbook, FreedmanLarsenWang00, BakalovKirillovlecturenotes}, defect exchange and braiding does not obey the full modular group $SL(2;\mathbb{Z})$, but is restricted to a congruent subgroup. 

%It is evident for instance along a 2D toric interface between topological insulator and superconductor in 3D where the longitudinal and meridian cycles are distinguished.
\end{enumerate}

\subsection{Outline and Summary}
We begin in section~\ref{sec:honeycombmodel} by presenting an abelian $\mathbb{Z}_k$ rotor lattice model on a honeycomb lattice or in general bipartite trivalent planar graph. Its $\mathbb{Z}_2$ version has been studied by Bombin and Martin-Delgado~\cite{BombinMartin06} and is called the color code. The model possesses a hidden non-abelian symmetry $S_3=\mathbb{Z}_2\ltimes\mathbb{Z}_3$, the permutation of group on three elements. There is a $k^4$ ground state degeneracy on a torus. The $k^4$ abelian anyon excitations can be labelled and bipartitioned into ${\bf a}=({\bf a}_\bullet,{\bf a}_\circ)$, where each spinless component ${\bf a}_{\bullet/\circ}$ lives on a two dimensional triangular $\mathbb{Z}_k$-lattice (see figure~\ref{fig:abeliananyonlattice}). Threefold {\em cyclic color permutations} $\Lambda_3,\Lambda_3^{-1}$ in $S_3$ act as rotations on the anyon lattice, leaving the bipartite $\bullet,\circ$-label untouched. Twofold {\em color sublattice transpositions} $\Lambda_Y,\Lambda_R,\Lambda_B$ in $S_3$ interchange $\bullet\leftrightarrow\circ$ and act as three mirror planes on the anyon lattice. This model has fusion and braiding properties identical to two copies of the $\mathbb{Z}_k$ toric code. However the symmetry group is non-trivially extended from charge-flux $\mathbb{Z}_2$ duality to $S_3$ by a threefold cyclic color permutation.

We construct non-abelian twist defects in section~\ref{sec:twistdefect}. A defect is classified by a group element $\Lambda$ in $S_3$ that characterizes the change of the label ${\bf a}\to\Lambda\cdot{\bf a}$ of an encirling abelian anyon (see figure~\ref{fig:anyontwist}). There are two threefold defects $[1/3]$ and its anti-particle $[\overline{1/3}]$, and three twofold defects $[1/2]_\chi$ labeled by color $\chi=Y,R,B$. Primitive threefold defects are constructed in the lattice level by $\pm120^\circ$ disclinations at tetravalent/bivalent vertex, and twofold defects are associated with $\pm60^\circ$ disclinations at pentagons and heptagons (see figure~\ref{fig:twistdefects}). Quantum dimensions can be deduced either by counting plaquette and vertex degrees of freedom (section~\ref{sec:latticedefecthamiltonian}) or evaluated by ground state degeneracy corresponding to the non-local Wilson loop algebra (section~\ref{sec:nonlocalWilsonalgebra}). They are given by \begin{align}d_{[1/3]}=d_{[\overline{1/3}]}=\left\{\begin{array}{*{20}c}k^2,&\mbox{if $3\centernot\mid k$}\\k^2/3,&\mbox{if $3\mid k$}\end{array}\right.,\quad d_{[1/2]}=k\end{align} Note that the dimension for the twofold defect matches that of two copies of $\mathbb{Z}_k$-fractional Majorana fermion. A defect contains a phase parameter that determines the value of a local Wilson observable (see figure~\ref{fig:defectlocalloop}). It subdivides twofold defects into $k^2$ species ${\bf l}$ and threefold defects into 9 species ${\bf s}$ when $k$ is divisible by 3. Species labels can mutate by absorbing or releasing abelian anyons, a process driven by continuously tuning the local defect phase parameter. This novel species characterization of defect-anyon composites is essential in a complete description of fusion and braiding. The Wilson loop algebra of an arbitrary multi-defect system is studied using word presentation consisting of open Wilson paths in section~\ref{sec:AlphabeticpresentationofWilsonalgebra}, where the $S_3$-transformation of defects and symmetry structure of the Wilson algebra are discussed.

In section~\ref{sec:defectfusion}, we investigate the non-commutative defect fusion category. The objects consist of abelian anyons and twist defects labeled with species. Due to the semi-classical nature of the non-dynamical $S_3$-symmetry, defects are not grouped into conjugacy classes of $S_3$-fluxes and anyons are not projected into $S_3$-orbifold superselection sectors. Loosely speaking the fusion rules originate from the non-abelian group structure of $S_3$ and take the following multi-channel form \begin{align}{\sum_{\bf a}}'[{\bf a}]&\simeq[1/2]_\chi\times[1/2]_\chi\nonumber\\&\simeq[1/3]\times[1/3]\times[1/3]\nonumber\\&\simeq[1/2]_\chi\times[1/2]_{\chi+1}\times[\overline{1/3}]\nonumber\\&\simeq[1/2]_\chi\times[1/2]_{\chi-1}\times[1/3]\label{fusionrulesummary}\end{align} where $\chi=0,1,2$ mod 3 represents the three colors $Y,R,B$, the equation is unaffected by cyclic permutation of defect order on the right, and the sum of abelian anyons ${\bf a}$ on the left is restricted by defect species so that the particle-antiparticle duality requires a change of species label $[1/2]_{\chi,{\bf l}_1}\times[{\bf a}]=[1/2]_{\chi,{\bf l}_2}$ when absorbing or releasing an abelian anyon in general. The second equality requires fusion degeneracy $[1/3]\times[1/3]=d_{[1/3]}[\overline{1/3}]$, and the third and forth equalities exhibit fusion non-commutativity. A more precise set of fusion rules can be found in section~\ref{sec:fusionrules}. The choice of a particular set of splitting states (or Wilson string configurations) shown in figure~\ref{fig:splittingspaces} in section~\ref{sec:splittingspaces} fixes the gauge for a consistent set of basis transformations between different maximally commuting sets of observables, called $F$-symbols, that characterizes the fusion category. A few calculated examples are illustrated in section~\ref{sec:Fmoves} and a complete list of $F$-matrices (up to $S_3$-symmetry) is included in table~\ref{tab:Fsymbols} in appendix~\ref{sec:Fsymbols}.

We describe exchange and braiding between defects of the same type in section~\ref{sec:defectexchangebraiding}. Although the non-commutative fusion category cannot be fully braided, a subset of $180^\circ$ rotation operations or $R$-symbols between commuting objects can be defined and are evaluated in section~\ref{sec:statisticsZ3} and \ref{sec:statisticsZ2} by rotating Wilson strings. While threefold defects do not carry spin, twofold defects have non-trivial species dependent statistics as shown in section~\ref{sec:statisticsZ2} and they are identified with the spin phase of $2\times(360^\circ)$ rotation. The braiding $S$-matrices are defined among defects of the same $S_3$-type with entries labeled by species. The $S$ and $T$ matrices for twofold defects are identified in section~\ref{sec:defectmodulartransformation} with Dehn twist $t_y$ and double Dehn twist $t_x^2$ respectively on the ground states on a torus decorated with a branch cut along the $y$-cycle. In general, they form a unitary group structure of the congruent subgroup $\Gamma_0(2)$ rather than the full modular group $SL(2;\mathbb{Z})$. Physical unitary braiding operations or $B$-matrices between defects of the same type are computed in section~\ref{sec:Bmatrices} and they demonstrate the non-abelian nature of twist defects. A certain compactification braiding identity of the sphere braid group that is expected to hold in a closed anyon system is now only {\em projectively} satisfied for defects.

\section{The \texorpdfstring{$\mathbb{Z}_k$}{Zk} rotor lattice model}\label{sec:honeycombmodel}
\begin{figure}[ht] \centering
  \includegraphics[width=3.5in]{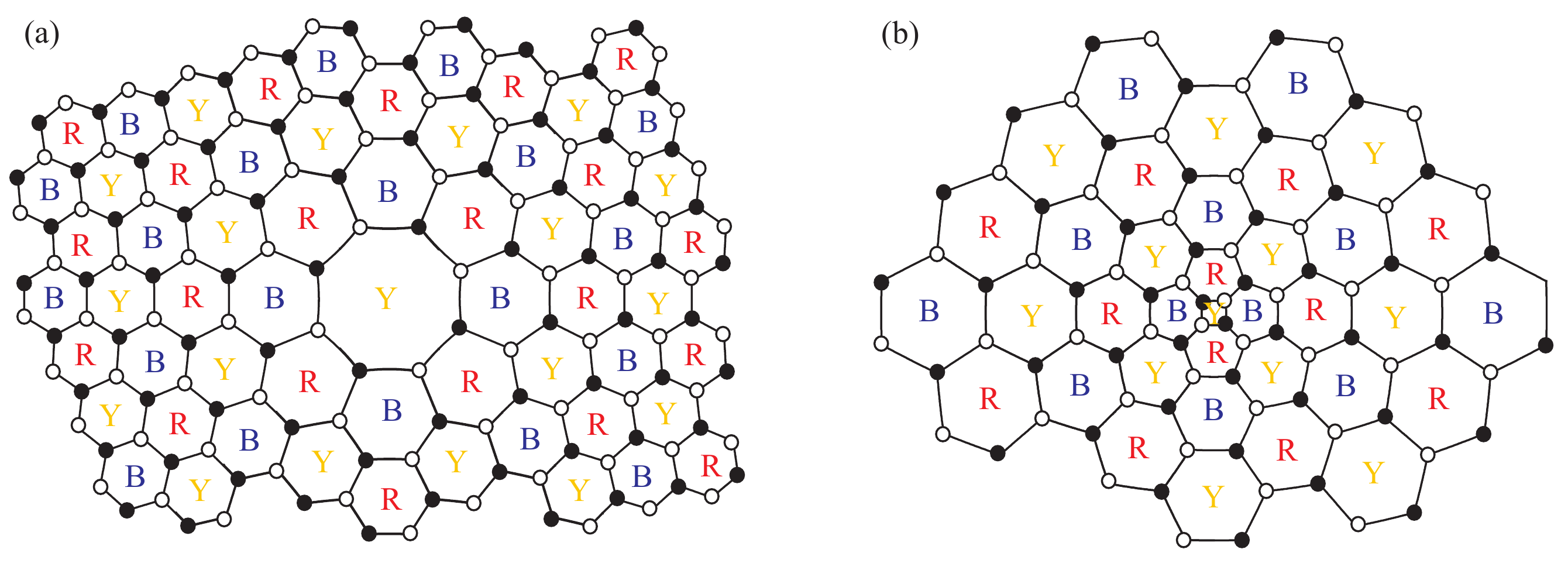} \caption{Trivial {\em twistless} defects that preserve local tri-coloring (YRB) bipartite ($\bullet,\circ$) order. (a) Negatively curved disclination with Frank angle $-120^\circ$; (b) Positively curved disclination with Frank angle $+120^\circ$.}\label{fig:squareoctagon} 
\end{figure}

We consider a $\mathbb{Z}_k$ rotor model (or spin-$1/2$ model for $k=2$) on vertices of a honeycomb lattice, or in general a bipartite trivalent planar graph with the following properties.
\begin{enumerate}
	\item {\bf Tricoloring} Each plaquette may be colored with one of three colors, say yellow (Y), red (R) and blue (B), such that adjacent plaquettes never carry the same color. 
  \item {\bf Bipartite} There is a sublattice structure so that vertices can be labeled by black $(\bullet)$ or white $(\circ)$ and adjacent vertices are always of differing types.  
\end{enumerate}
In addition, in the absence of twist defects and branch cuts which will be explained later, we require the graph to satisfy the above constraints globally. On a torus, they will be globally preserved with compatible periodic boundary conditions, when the length of the two primitive cycles are multiples of three. The model can be put on a closed surface with arbitrary genus by adding an appropriate number of defect squares or octagons (figure~\ref{fig:squareoctagon}) on the regular honeycomb lattice. These are trivial {\em twistless} defects in the sense that they do not violate the tricoloring and bipartite order. 

\subsection{The Hamiltonian} 
The degrees of freedom (the ``spins'' for $k=2$) of the $\mathbb{Z}_k$-lattice model live on vertices. $\mathbb{Z}_k$ rotors are operators $\sigma$ and $\tau$ that take eigenvalues in $\{1,e^{2\pi i/k},\ldots,e^{2\pi i(k-1)/k}\}$ and commute up to a $\mathbb{Z}_k$ phase \begin{equation}\tau\sigma=w\sigma\tau,\quad\sigma^k=\tau^k=1\label{sigmataucomm}\end{equation} where $w$ is a $k^{th}$ root of unity. We will assume $w=e^{2\pi i/k}$ hereafter. They can be represented by $k$-dimensional matrices \begin{equation}\tau=\left(\begin{array}{*{20}c}0&1&\cdots&0\\\vdots&\vdots&\ddots&\vdots\\0&0&\cdots&1\\1&0&\cdots&0\end{array}\right),\quad\sigma=\left(\begin{array}{*{20}c}1&0&\cdots&0\\0&w&\cdots&0\\\vdots&\vdots&\ddots&\vdots\\0&0&\cdots&w^{k-1}\end{array}\right)\label{sigmataurep}\end{equation} Each vertex carries a $k$-dimensional Hilbert space and a set of rotors. 
The total Hilbert space is a tensor product over vertices, and rotors at different vertices commute. 

\begin{figure}[ht]
	\centering
	\includegraphics[width=1.3in]{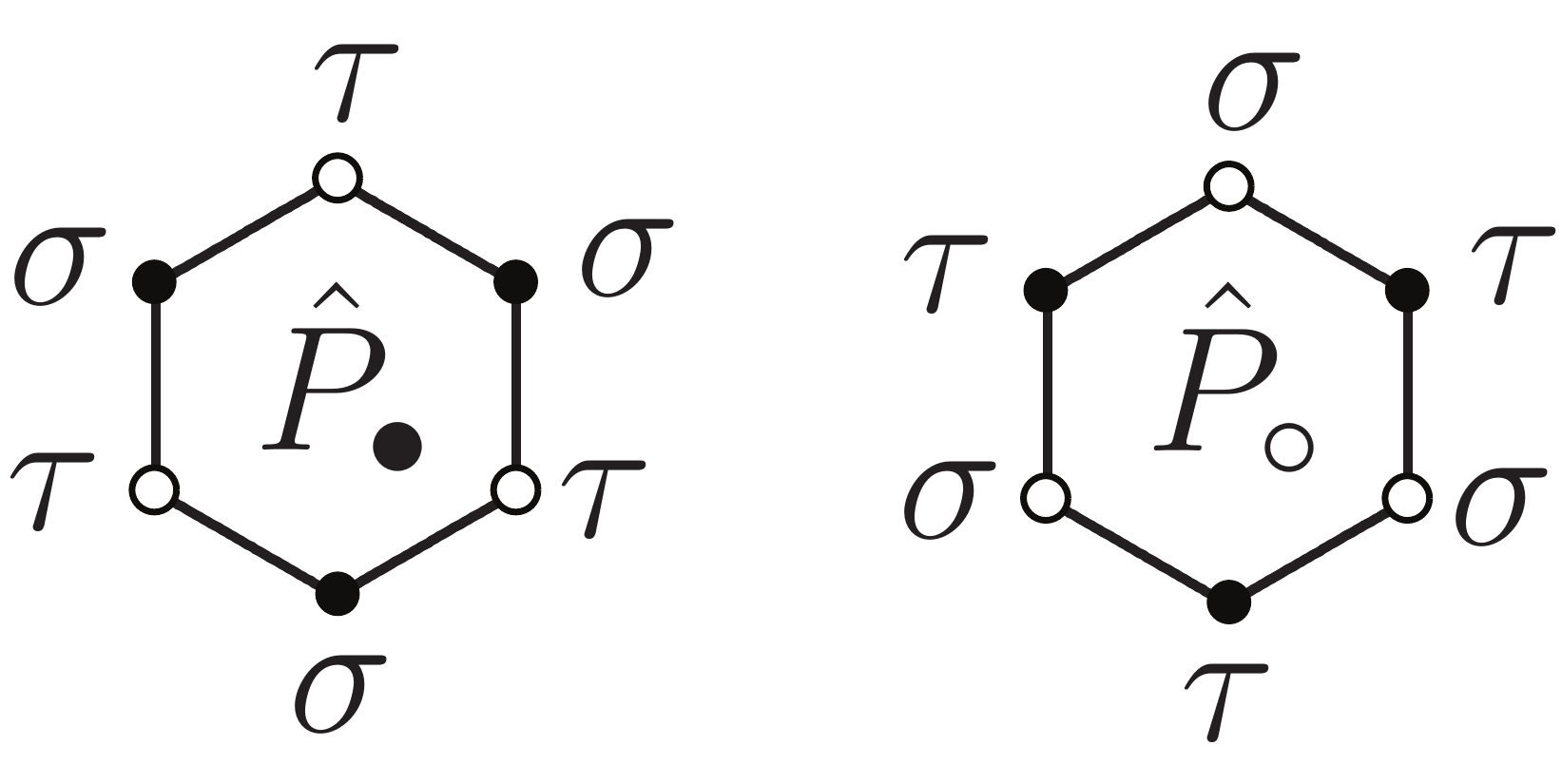}
	\caption{Plaquete stabilizers $\hat{P}_\bullet$ and $\hat{P}_\circ$ as tensor products of rotors $\sigma,\tau$.}\label{fig:stabilzers}
\end{figure}

Given a bipartite $\bullet,\circ$ assignment of vertices, each plaquette $P$ carries two stablilizer operators \begin{equation}\hat{P}_\bullet=\prod_{v_\bullet\in P}\sigma_{v_\bullet}\prod_{v_\circ\in P}\tau_{v_\circ},\quad\hat{P}_\circ=\prod_{v_\bullet\in P}\tau_{v_\bullet}\prod_{v_\circ\in P}\sigma_{v_\circ}\label{stabilizers1}\end{equation} where $\sigma_v,\tau_v$ are rotors at vertices $v$ around the plaquette and tensor products have been suppressed. The bipartite structure ensures all plaquettes have the same number of $\bullet$ and $\circ$ vertices and two neighboring plaquettes share exactly one $\bullet$ and one $\circ$ vertex. Then \eqref{sigmataucomm} ensures mutual commutativity of the plaquettes operators, which form a set of good quantum numbers referred to as $\mathbb{Z}_k$ stabilizers or fluxes. The Hamiltonian is defined by the sum of stablilizers, \begin{equation}H=-J\sum_P(\hat{P}_\bullet+\hat{P}_\bullet^\dagger)-J\sum_P(\hat{P}_\circ+\hat{P}_\circ^\dagger)\label{ham1}\end{equation}Ground states are trivial flux configurations where $\hat{P}_\bullet=\hat{P}_\circ=1$ for all stabilizers. 

It is indicative to notice that the model \eqref{ham1} describes the topological phase of two copies of $\mathbb{Z}_k$ version of Kitaev toric code. However, Hamiltonian \eqref{ham1} realizes a much richer $S_3$ {\em geometric symmetry} that extends the original charge-flux $\mathbb{Z}_2$ duality in the quantum double model. The $S_3$ symmetry can be understood in the microscopic geometric level by the action of space group on the tricoloring and bipartite pattern of the honeycomb lattice. These correspond the threefold and twofold generators of the permutation group $S_3$ and will be discussed in detail in section~\ref{sec:compositelatticedefects}. This generalizes of the $e$-$m$ duality in the Kitaev's toric code that involves geometrically switching vertices and plaquettes.

The ground state degeneracy of the Hamiltonian \eqref{ham1} on a trivalent graph on any closed orientable genus $g$ surface can be evaluated by counting vertices and independent plaquette stabilizers. Denote the total number of vertices by $\#V$ and the number of plaquettes by $\#P$. For a regular honeycomb lattice on a torus, $\#V=2\times\#P$, which is commensurate with the two stabilizer operators eq.\eqref{stabilizers1} located at each plaquette. However there is an overcounting since certain products of stabilizers are identical to the identity. The number of these relations is independent of system size and therefore gives rise to a {\em topological} ground state degeneracy ($k^4$ on the torus if the honeycomb is globally tricolorable).

As mentioned earlier, the model can be put on any closed orientable surface without violating {\em local} tricolorability and the $\bullet,\circ$-sublattice structure, by adding square or octagon defects as showen in figure~\ref{fig:squareoctagon}. The Gauss-Bonnet theorem (or Euler characteristic) requires \begin{equation}\#\mbox{octagons}-\#\mbox{squares}=6(g-1).\end{equation} Since an octagon/square carries two greater/fewer vertices than a hexagon, \begin{equation}\#V=2\times\#P+4(g-1)\label{VPcount}.\end{equation} on a genus $g$ surface.

We will first investigate this overcounting in the case when the trivalent graph is {\em globally} tricolorable. This is an additional topological constraint that ensures that the tricoloration remains unchanged around a non-trivial cycle. In particular, the total number of plaquettes must be a multiple of three and appropriate twisted periodic boundary conditions must be applied if individual cycles have lengths not divisible by three.

Given a global $YRB$-coloration of the plaquettes $P^Y$, $P^R$ and $P^B$, stabilizer operators \eqref{stabilizers1} are overcounted by the following four cocycle relations, \begin{align}1&=\prod_{P^Y}\hat{P}^Y_\bullet\left(\prod_{P^R}\hat{P}_\bullet^R\right)^\dagger=\prod_{P^R}\hat{P}^R_\bullet\left(\prod_{P^B}\hat{P}_\bullet^B\right)^\dagger\nonumber\\&=\prod_{P^Y}\hat{P}^Y_\circ\left(\prod_{P^R}\hat{P}_\circ^R\right)^\dagger=\prod_{P^R}\hat{P}^R_\circ\left(\prod_{P^B}\hat{P}_\circ^B\right)^\dagger\label{stabilizerovercounting}\end{align} These can be understood by observing that the product of all vertex rotors can be given by product of plaquettes of a particular color, i.e. $\prod_{v_\bullet}\sigma_{v_\bullet}\prod_{v_\circ}\tau_{v_\circ}=\prod_{P^Y}\hat{P}^Y_{\bullet}=\prod_{P^R}\hat{P}^R_{\bullet}=\prod_{P^B}\hat{P}^B_{\bullet}$ and similarly for the other set, $\hat{P}_\circ$. Thus the number of indepedent stabilizers is, \begin{equation}\#\mbox{stabilizers}=2\times\#P-4\end{equation} and the ground state degeneracy (G.S.D.) of a globally tricolorable graph on a genus $g$ closed surface is \begin{equation}G.S.D.=k^{\#V-\#\mbox{\small stabilizers}}=k^{4g}\end{equation}

\subsection{Wilson algebra and ground state degeneracy}
\begin{figure}[ht]
	\centering
	\includegraphics[width=3.4in]{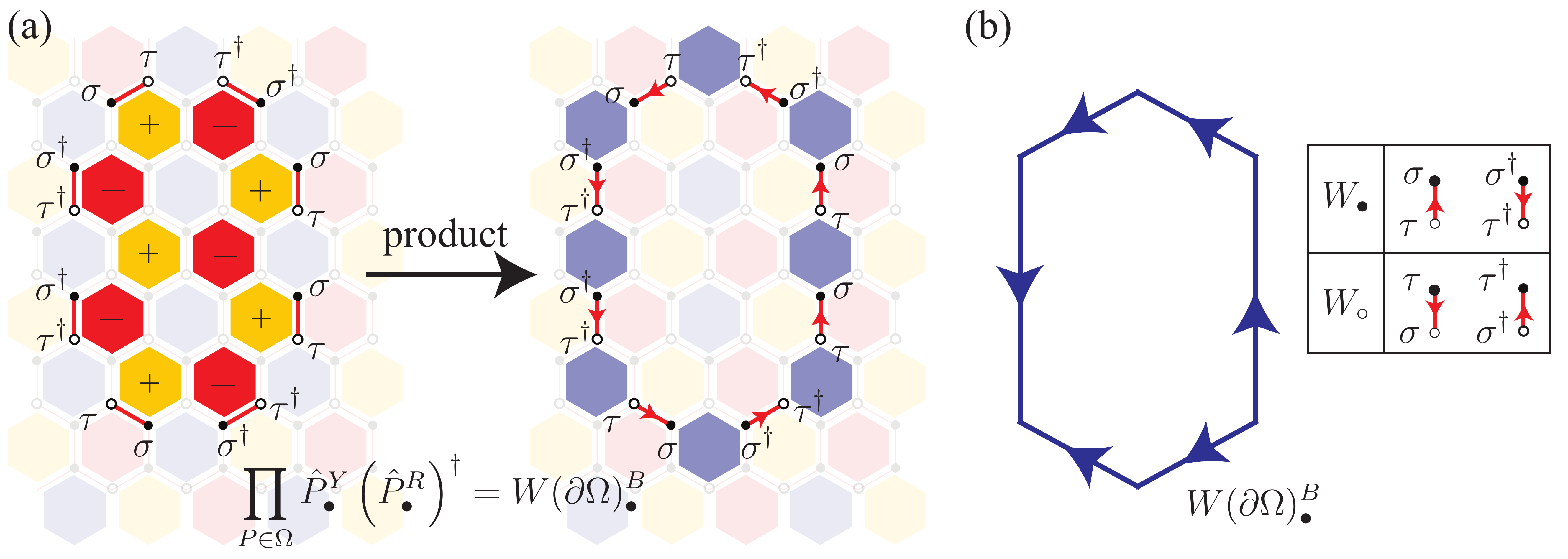}
\caption{Contractible Wilson loop on a honeycomb lattice. (a) Product of yellow plaquette operators $\hat{P}^Y_{\bullet}$ and inverses of red plaquette operators $\left(\hat{P}^R_{\bullet}\right)^{-1}$ inside a domain $\Omega$ leaving a string of $\sigma$ and $\tau$ rotor operators $W(\partial\Omega)_\bullet^B$ (on highlighted $\bullet$ and $\circ$ sites) along the boundary $\partial\Omega$ that connects a series of blue plaquettes. (b) Schematics of the boundaries $W(\partial\Omega)^B_{\bullet}$, where the arrows denote the rotors $\sigma,\tau$ or $\sigma^{-1},\tau^{-1}$ along the strings in the lattice.}\label{fig:trivialloop}
\end{figure}
\begin{figure}[ht]
	\centering
	\includegraphics[width=1.7in]{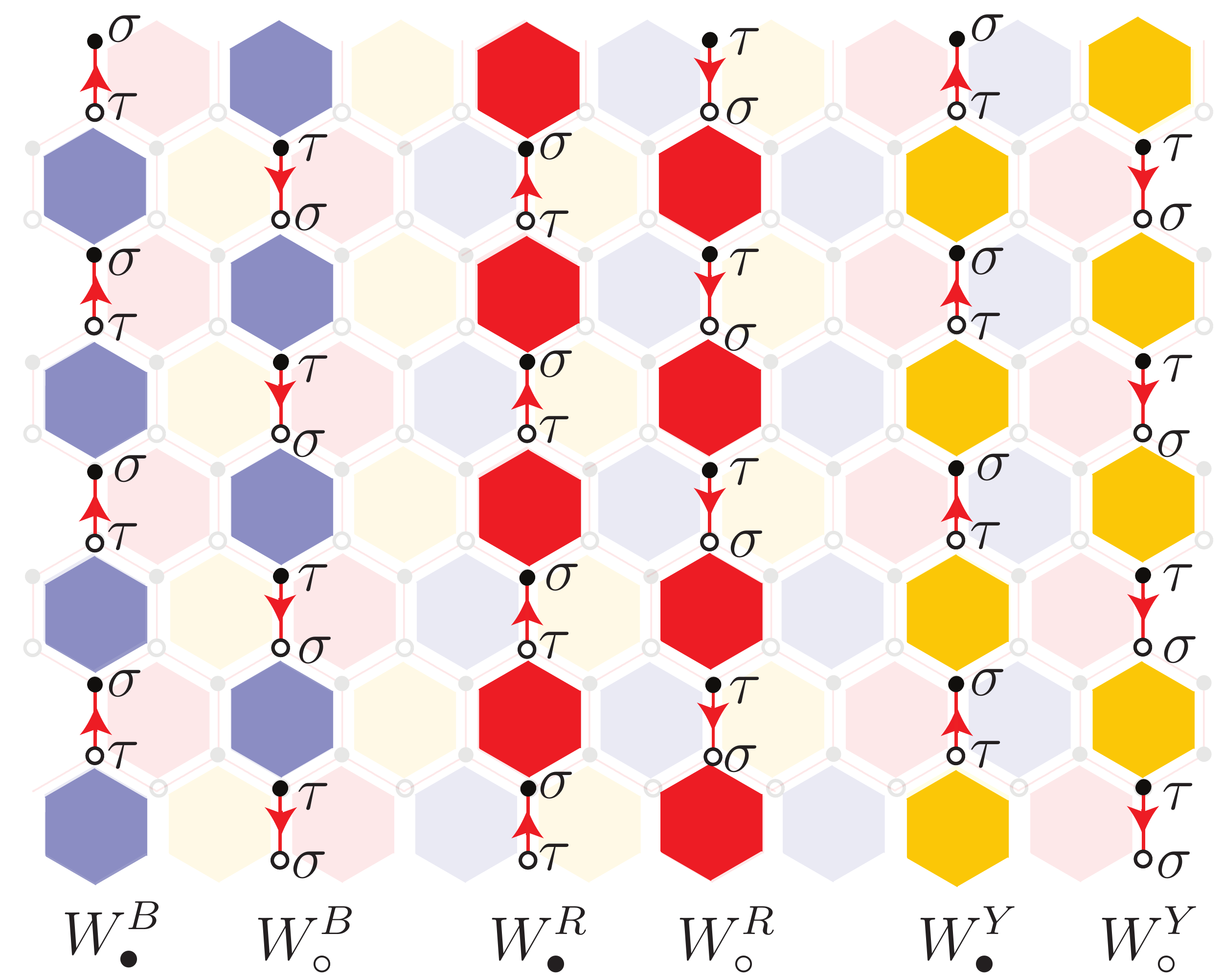}
\caption{Product of rotor operators along a Wilson string $W^\chi_\bullet=\prod\sigma_{v_\bullet}\tau_{v_\circ}$ and $W^\chi_\circ=\prod\tau_{v_\bullet}\sigma_{v_\circ}$ for $\chi=Y,R,B$. $(\sigma,\tau)\to(\sigma^{-1},\tau^{-1})$ upon arrow reversal.}\label{fig:Wilsonstring}
\end{figure}

Ground states can be written down as an equally weighted sum of plaquette operators acting on a suitably chosen state. One example is, \begin{equation}|0\rangle_\bullet=\frac{1}{\sqrt{\mathcal{N}}}\prod_{P}\left(\sum_{r=0}^{k-1}\hat{P}_\circ^r\right)|\sigma_{v_\bullet}=\tau_{v_\circ}=+1\rangle\label{GS0}\end{equation} where $|\sigma_{v_\bullet}=\tau_{v_\circ}=+1\rangle$ is the tensor product eigenstate state of $\sigma$ for each $\bullet$-vertex and $\tau$ for $\circ$-vertex, and $\mathcal{N}=k^{\#P-2}$ is a normalization factor. It is a simultaneous eigenstate for all plaquette operators $\hat{P}_\bullet|0\rangle=\hat{P}_\circ|0\rangle=|0\rangle$. The ground state can be interpreted as a condensate of of trivial \emph{Wilson loops}. A Wilson loop is a string of rotors that commutes with each stabilizer, and trivial if it can be written as a product of stabilizers. They are labeled by the $\bullet,\circ$-sublattice type and colors $YRB$, so that a Wilson loop along the boundary of a domain $\Omega$ is of the form \begin{align}W(\partial\Omega)^\chi_{\bullet}&=\prod_{P\subseteq\Omega}\hat{P}^{\chi+1}_{\bullet}(\hat{P}^{\chi-1}_{\bullet})^{-1}\\W(\partial\Omega)^\chi_{\circ}&=\prod_{P\subseteq\Omega}\hat{P}^{\chi-1}_{\circ}(\hat{P}^{\chi+1}_{\circ})^{-1}\end{align} for $\chi\in\{0,1,2\}$ labels the colors $\{Y,R,B\}$ mod 3, and the product is taken over either $\bullet$- or $\circ$-plaquettes of two complementing colors $\chi\pm1$ inside the open domain $\Omega$. Operators in the interior of $\Omega$ cancel each other and leave a string of rotors along the boundary $\partial\Omega$ connecting a strand of plaquettes of the same color $\chi$. This is illustrated by figure~\ref{fig:trivialloop} for $\chi=B$.

\begin{figure}[ht]
	\centering
	\includegraphics[width=3.5in]{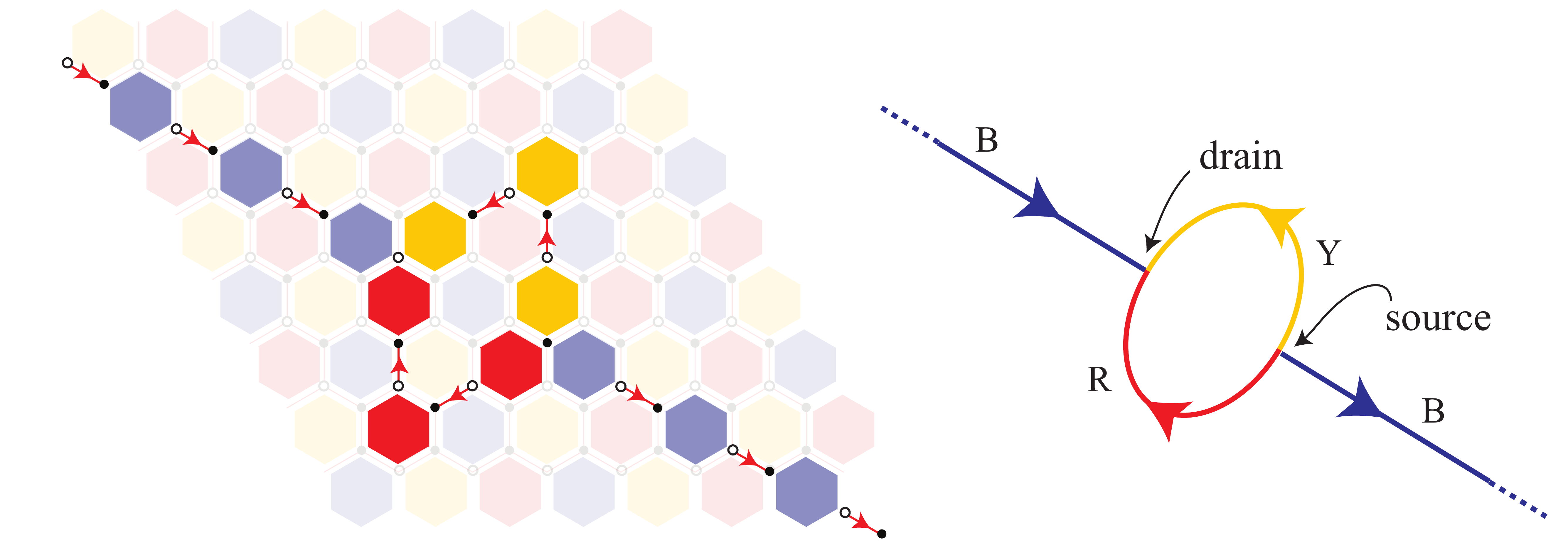}
 \caption{Color splitting of Wilson string through tricolor, trivalent, source and drain. Product of rotor operators are taken over highlighted $\bullet$ and $\circ$ vertices.}\label{fig:colorspliting}
\end{figure}

There is an important redundancy in the color labels. A blue string can split into a yellow and red pair, both propagating in the opposite direction. This can be done by multiplying plaquettes on the string, and results in a pair of tricolor, trivalent, sources and drains as shown in figure~\ref{fig:colorspliting}. Similarly, a parallel triplet of co-propagating Wilson strings of three different colors can be locally cancelled by plaquette operators. The color redundancy can be summarized by the fusion \begin{equation}Y_\bullet\times R_\bullet\times B_\bullet=Y_\circ\times R_\circ\times B_\circ=1\label{3colorfusion}\end{equation} for 1 being the vacuum.

In general a closed Wilson loop is built by joining colored, directed paths emanating from sources and drains. Each colored path is a string of $\sigma^{\pm1}$ and $\tau^{\pm1}$ rotors that connects plaquettes of that color as illustrated in figure~\ref{fig:Wilsonstring}. \begin{equation}W^\chi_\bullet=\prod_{v_\bullet}\sigma_{v_\bullet}^{\pm1}\prod_{v_\circ}\tau_{v_\circ}^{\pm1},\quad W^\chi_\circ=\prod_{v_\bullet}\tau_{v_\bullet}^{\pm1}\prod_{v_\circ}\sigma_{v_\circ}^{\pm1}\end{equation} The signs of the rotors are dictated by the arrow directions in figure~\ref{fig:trivialloop}(b) and \ref{fig:Wilsonstring}. For instance, rotors along a string switch signs about a $\pm60^\circ$ or $180^\circ$ corner as shown in figure~\ref{fig:trivialloop}. It is straightforward to check that closed Wilson strings commute with all stabilizers through phase cancellation. Evidently, any product of plaquette operators is a closed Wilson string, which can be pulled out of the condensate ground state \eqref{GS0} without an energy cost. However, the converse is not true --- a closed Wilson string is not necessarily the product of plaquettes. There are non-trivial cycles on the genus $g\geq1$ surface that do not bound an open domain.

\begin{figure}[ht]
	\centering
	\includegraphics[width=2in]{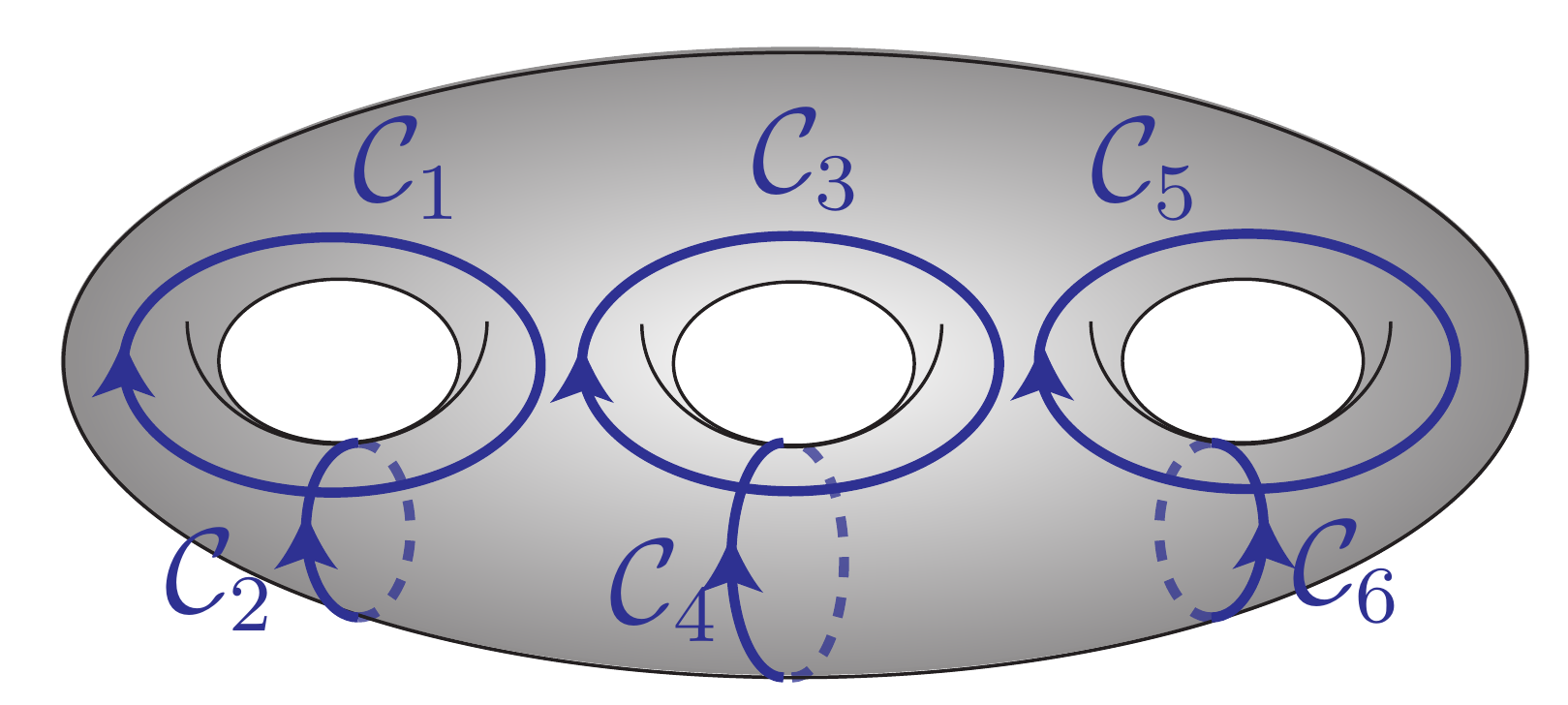}
 \caption{Primitive cycles on a genus $g=3$ surface.}\label{fig:genusgcycles}
\end{figure}
\begin{figure}[ht]
	\centering
	\includegraphics[width=3.5in]{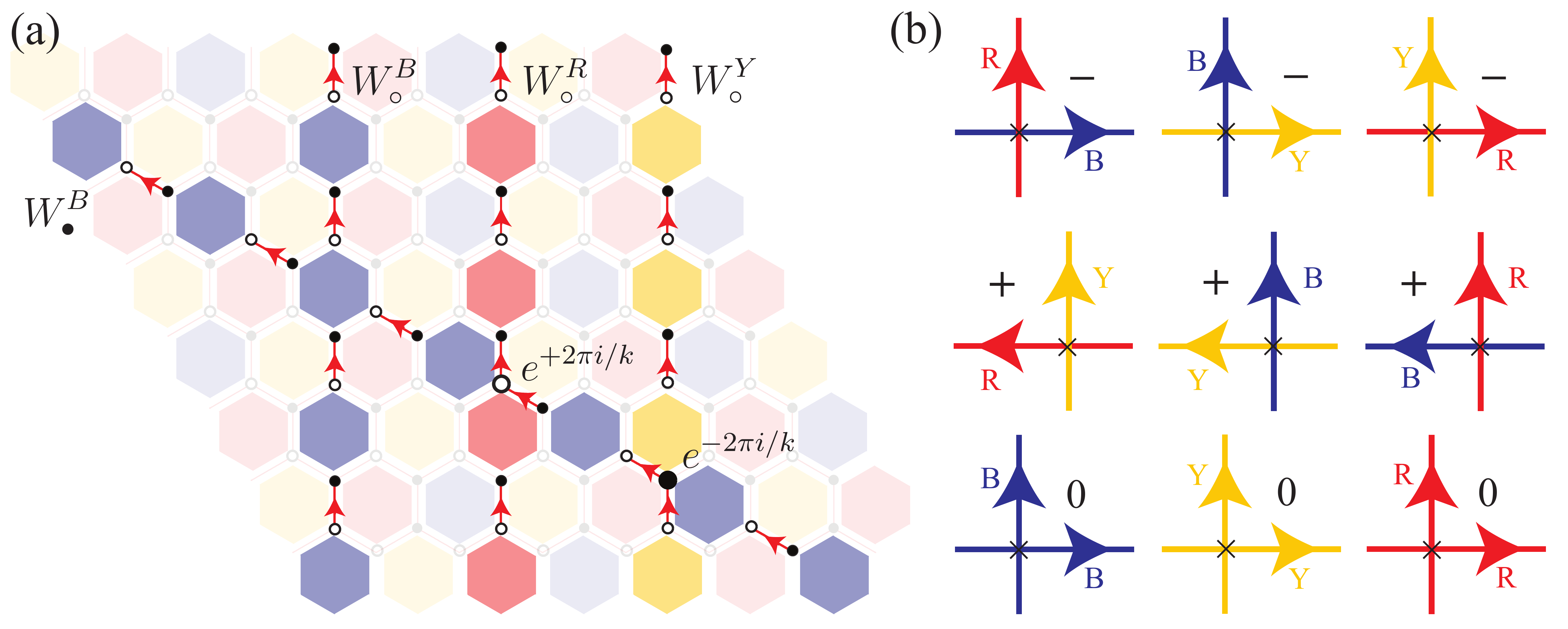}
 \caption{Intersection of Wilson strings of opposite $\bullet,\circ$-types. (a) Abelian $\mathbb{Z}_k$-phases $e^{\pm2\pi i/k}$ accumulated by passing a $\circ$-Wilson string in front of an intersecting $\bullet$-one of different color. Intersections are highlighted at bold vertices. (b) Schematics showing intersection number $\langle\ast,\ast\rangle=-1,0,+1$ between different directed Wilson strings.}\label{fig:intersection}
\end{figure}
On a genus $g$ surface, there are $2g$ non-trivial primitive cycles $\mathcal{C}_i$, where $\mathcal{C}_{2l-1},\mathcal{C}_{2l}$ are the two cycles associated with the $l^{th}$ handle (see figure~\ref{fig:genusgcycles}). Each one can be labeled by sublattice type $\bullet,\circ$ and color $YRB$. Together, the Wilson loops form a non-commutative algebra with the algebraic relations \begin{align}\left(W(\mathcal{C}_i)^{\chi}_{\bullet/\circ}\right)^k&=1\label{wilsoncomm1}\\\left[W(\mathcal{C}_{i})^{\chi_1}_\bullet,W(\mathcal{C}_{j})^{\chi_2}_\bullet\right]&=\left[W(\mathcal{C}_{j})^{\chi_1}_\circ,W(\mathcal{C}_{j})^{\chi_2}_\circ\right]=0\label{wilsoncomm2}\\ W(\mathcal{C}_i)^{\chi_1}_\bullet W(\mathcal{C}_{j})^{\chi_2}_\circ&=e^{i\frac{2\pi}{k}\langle\mathcal{C}_i^{\chi_1},\mathcal{C}_j^{\chi_2}\rangle}W(\mathcal{C}_{j})^{\chi_2}_\circ W(\mathcal{C}_{i})^{\chi_1}_\bullet\label{wilsoncomm3}\end{align} where the $\chi$'s run over the colors $YRB$ and the pairing $\langle\mathcal{C}_i^{\chi_1},\mathcal{C}_j^{\chi_2}\rangle$ is determined by intersection number between Wilson strings, summarized in figure~\ref{fig:intersection}. The intersection form $\langle\ast,\ast\rangle$ is bilinear and symmetric. It counts the total $\mathbb{Z}_k$-phase accumulated by interchanging rotor operators $\sigma,\tau$ at overlapping vertices according to eq.\eqref{sigmataucomm}. Intersection is invariant under cyclic permutation of colors, and changes sign if the direction of one of the Wilson string is reversed. The color fusion rule eq.\eqref{3colorfusion} forces null intersection between Wilson cycles of the same color. Since primitive cycles only intersect when they correspond to the same handle, the intersection $\langle\mathcal{C}_i^{\chi_1},\mathcal{C}_j^{\chi_2}\rangle$ is zero unless $i=2l-1$, $j=2l$ or vice versa, in which the number according to figure~\ref{fig:genusgcycles} and \ref{fig:intersection}(b) would be \begin{equation}\langle\mathcal{C}_{2l-1}^{\chi_1},\mathcal{C}_{2l}^{\chi_2}\rangle=\left\{\begin{array}{*{20}c}0&\mbox{for $\chi_1=\chi_2$}\\-1&\mbox{for $\chi_1=\chi_2-1$}\\1&\mbox{for $\chi_1=\chi_2+1$}\end{array}\right.\label{wilsoncomm4}\end{equation} for $\chi=Y,R,B=0,1,2$ modulo 3.

The three colors are not independent due to color fusion \eqref{3colorfusion}. For example, the products of the parallel triplets
\begin{equation}W(\mathcal{C}_i)^Y_\bullet W(\mathcal{C}_i)^R_\bullet W(\mathcal{C}_i)^B_\bullet=W(\mathcal{C}_i)^Y_\circ W(\mathcal{C}_i)^R_\circ W(\mathcal{C}_i)^B_\circ=1\label{colorfusion}\end{equation} since these can be written as products of plaquette operators and act as the identity on the ground state. Therefore, there are $k^{8g}$ independent Wilson loops generated by the 
primitive cycles $W(\mathcal{C}_i)^Y_\bullet$, $W(\mathcal{C}_i)^R_\bullet$, $W(\mathcal{C}_i)^Y_\circ$ and $W(\mathcal{C}_i)^R_\circ$, for $i=1,\ldots,2g$, 

An orthonormal basis for the ground state Hilbert space can be written down using the Wilson algebra starting with the particular ground state $|0\rangle_\bullet$ in eq.\eqref{GS0}. Define the normalized ground state \begin{equation}|{\bf m},{\bf n}\rangle_\bullet=\left[\prod_{i=1,\ldots,2g}\left(W(\mathcal{C}_i)^Y_\circ\right)^{m_i}\left(W(\mathcal{C}_i)^R_\circ\right)^{n_i}\right]|0\rangle_\bullet\label{GSbasis}\end{equation} where ${\bf m}=(m_i)$, ${\bf n}=(n_i)$ have integers modulo $k$ entries, for $i=1,\ldots,2g$. The $\bullet$-Wilson loops $W(\mathcal{C}_i)^\chi_\bullet$ form a maximal set of commuting generators and share simultaneous eigenstates. The eigenvalues can be evaluated using the intersection relation \eqref{wilsoncomm3} and \eqref{wilsoncomm4}.
\begin{equation}W(\mathcal{C}_j)^\chi_\bullet|{\bf m},{\bf n}\rangle_\bullet=e^{i\frac{2\pi}{k}\varphi_j^\chi({\bf m},{\bf n})}|{\bf m},{\bf n}\rangle_\bullet\label{GSeigenvalues}\end{equation} where the $\mathbb{Z}_k$ phase is given by \begin{align}\varphi_j^\chi({\bf m},{\bf n})&=(-1)^j\left[n_{j-(-1)^j}(\delta^\chi_0-\delta^\chi_2)\right.\nonumber\\&\quad\quad\left.+m_{j-(-1)^j}(\delta^\chi_2-\delta^\chi_1)\right]\end{align} modulo $k$, where $\chi=0,1,2$ index the colors $Y,R,B$. As different ground states in eq.\eqref{GSbasis} are distinguished by their eigenvalues with respect to the $\bullet$-Wilson loops, they must be mutually orthogonal. The $k^{4g}$ dimensional space of degenerate ground states forms an irreducible unitary representation of the Wilson algebra.

\subsubsection{Obstruction to global tricolorability}\label{sec:obstructiontoglobaltricolorability}
\begin{figure}[ht]
	\centering
	\includegraphics[width=3in]{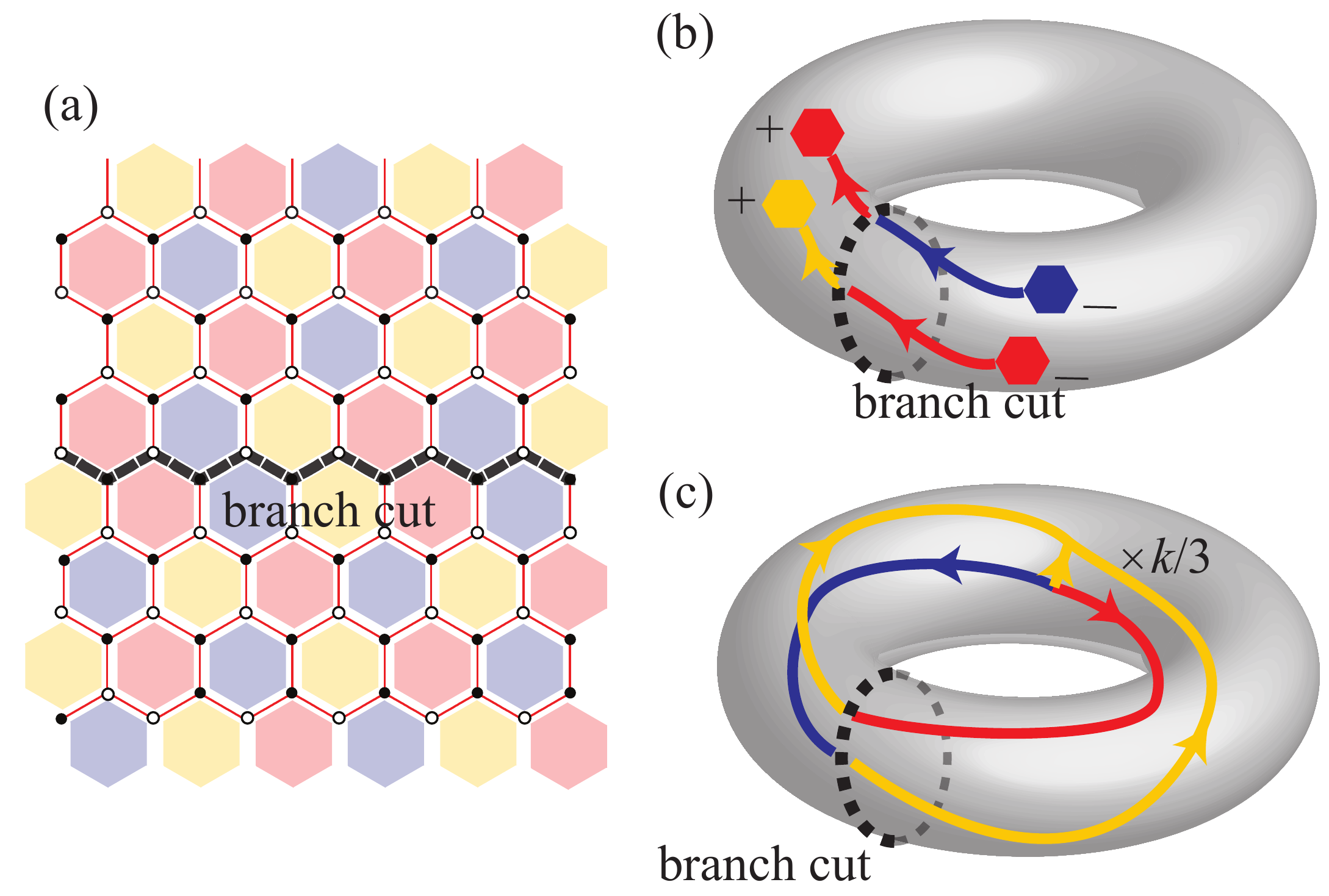}
 \caption{(a) Color branch cut along a zig-zag edge where same color hexagons share edges and vertices. (b) Wilson strings and abelian anyon changes color across a threefold branch cut. (c) A closed Wilson string $\Sigma_1$ composed of $k/3$ copies of the open path that closes at the uni-color drain.}\label{fig:torusbranchcut}
\end{figure}

We will spend the remaining of the subsection on the Wilson algebra and ground state degeneracy when the trivalent graph is {\em not} globally tricolorable. The topological obstruction is characterized by closed branch cuts where same color plaquettes share edges and vertices (see figure~\ref{fig:torusbranchcut}(a)). Branch cuts are not physical domain walls as the Hamiltonian \eqref{ham1} does not depend on an explicit plaquette color definition. A closed branch cut that runs along a trivial loop can be removed by cyclic permuting the colors inside the area bounded by the loop. A branch cut going along a non-contractible cycle is however irremovable (unless canceled by another branch cut). This topological color inconsistency has a reducing effect on the Wilson algebra and consequently ground state degeneracy. Similar issue of branch cuts also arise for $\mathbb{Z}_k$ toric code over a checkerboard lattice on a torus~\cite{YouWen}, where the charge-flux duality is realized as the bicolor structure of checkerboard plaquettes. The situation in the tricolored model \eqref{ham1} is qualitative different as the Wilson algebra and ground state degeneracy in the presence of branch cuts depend on the divisibility of $k$ by 3.

For simplicity, we will only demonstrate the case when the model is put on a torus and there is a single branch cut along a non-trivial cycle, say the meridian direction. This can be achieved by adapting a twisted boundary condition on a regular honeycomb lattice through introducing a lattice displacement along an zig-zag edge illustrated in figure~\ref{fig:torusbranchcut}(a). We will see in the section~\ref{sec:twistdefect} that an open color branch cut ends at a conjugate pair of non-abelian threefold {\em twist defects}. And therefore the twisted boundary condition can be constructed by draging a threefold defect around a cycle. Note that this is fundamentally different from threading a non-abelian quantum flux in a true topological phase as the underlying semiclassical configuration explicitly breaks $SL(2;\mathbb{Z})$ modular invariance.  The branch cut picks out a particular non-trivial cycle on the torus and the Wilson algebra does not close under the $SL(2;\mathbb{Z})$ action. Modular transformations will be discussed in more detail in section~\ref{sec:defectmodulartransformation}.

A Wilson string will change color through cyclic permutation across the branch cut as shown in figure~\ref{fig:torusbranchcut}(b). As a result the longitudinal cycle no longer corresponds a Wilson loop as the string will not close back onto itself after passing across the branch cut. Wilson loops along the meridian direction on the other hand are still closed as they do not intersect the parallel branch cut. However, they will change color if the entire loop is dragged around the torus. This gives rise to a color ambiguity since meridian Wilson loops of different colors are now interchangable through plaquette stabilizers, and they are indistinguishable on the ground states. The color fusion rule $Y\times R\times B=1$ then implies all meridian Wilson loops to be of order 3, $W(\mathcal{C}_2)^3=1$. Being built by $\mathbb{Z}_k$ rotors, a Wilson loop is automatically of order $k$ as seen in \eqref{wilsoncomm1}. And therefore unless $k$ is divisible by 3, the meridian Wilson loop is trivial and is a product of plaquette operators \begin{equation}W(\mathcal{C}_2)_{\bullet/\circ}=\prod_P\left(\hat{P}^Y_{\bullet/\circ}\right)^{2s}\left(\hat{P}^R_{\bullet/\circ}\right)^{-s}\left(\hat{P}^B_{\bullet/\circ}\right)^{-s}\end{equation} where $s$ is an integer so that $3s\equiv1$ mod $k$.

When $k$ is a multiple of 3, the meridian Wilson loop $W(\mathcal{C}_2)_{\bullet/\circ}$ is not a contractible boundary and it intersects with a closed Wilson string $W(\Sigma_1)_{\bullet/\circ}$ in the longitudinal direction consists of $k/3$ tricolor trivalent sources and a unicolor $k$-valent drain depicted in figure~\ref{fig:torusbranchcut}(c). It commutes with all plaquette stabilizers because of the color fusion $Y\times R\times B=1$ at the tricolor sources and $\mathbb{Z}_k$ fusion $Y^k=1$ at the unicolor drain. $W(\Sigma_1)_{\bullet/\circ}$ is equivalent to dragging the abelian anyon $\boldsymbol\kappa_{\bullet/\circ}$ (figure~\ref{fig:abeliananyonlattice}), a non-trivial anyon invariant under cyclic color permutation only when $k$ divisible by 3, around the longitudinal cycle. The Wilson algebra is then generated by $W(\Sigma_1)_\bullet$, $W(\mathcal{C}_2)_\bullet$, $W(\Sigma_1)_\circ$ and $W(\mathcal{C}_2)_\circ$ that satisfy the following algebraic relations. \begin{align}\left(W(\Sigma_1)_{\bullet/\circ}\right)^3&=\left(W(\mathcal{C}_2)_{\bullet/\circ}\right)^3=1\\\left[W(\Sigma_1)_\bullet,W(\mathcal{C}_2)_\bullet\right]&=\left[W(\Sigma_1)_\circ,W(\mathcal{C}_2)_\circ\right]=0\\W(\Sigma_1)_\bullet W(\mathcal{C}_2)_\circ&=e^{2\pi i/3}W(\mathcal{C}_2)_\circ W(\Sigma_2)_\bullet\\W(\mathcal{C}_2)_\bullet W(\Sigma_1)_\circ&=e^{2\pi i/3}W(\Sigma_1)_\circ W(\mathcal{C}_2)_\bullet\end{align} This gives rise to a ground state degeneracy of $9=3^2$ on a torus.

The ground state degeneracy on a torus is identical to the total number of deconfined anyon types~\cite{Wentopologicalorder89, WenNiu90}. On a globally tricolorable graph, abelian anyons can be uniquely labeled by the particle numbers mod $k$ of fundamental constituents $Y_\bullet,R_\bullet,Y_\circ,R_\circ$, and thus the ground state degeneracy is $k^4$. When there is a color ambiguity from non-contractible branch cut, there will be less particle types which are now referred as species because $Y_\bullet=R_\bullet=B_\bullet$ and $Y_\circ=R_\circ=B_\circ$. The three colors and $\mathbb{Z}_k$ fusion implies $B_{\bullet/\circ}^3=B_{\bullet/\circ}^k=1$. And hence there will not be non-trivial species unless $k$ is a multiple of 3, in which case they will be labeled by the particle number of the two fundamental generators $B_{\bullet},B_{\circ}$ modulo 3. This gives rise to a $3^2$-fold degeneracy on a torus and corresponds $3^2$ species of threefold defects distinguishable by $\Sigma_{\bullet/\circ}$ discussed in more detail in section~\ref{sec:twistdefect}. 

\subsection{Abelian anyon excitations, effective field theory and \texorpdfstring{$S_3$}{S3}-symmetry}
\begin{figure}[ht]
	\centering
	\includegraphics[width=3.5in]{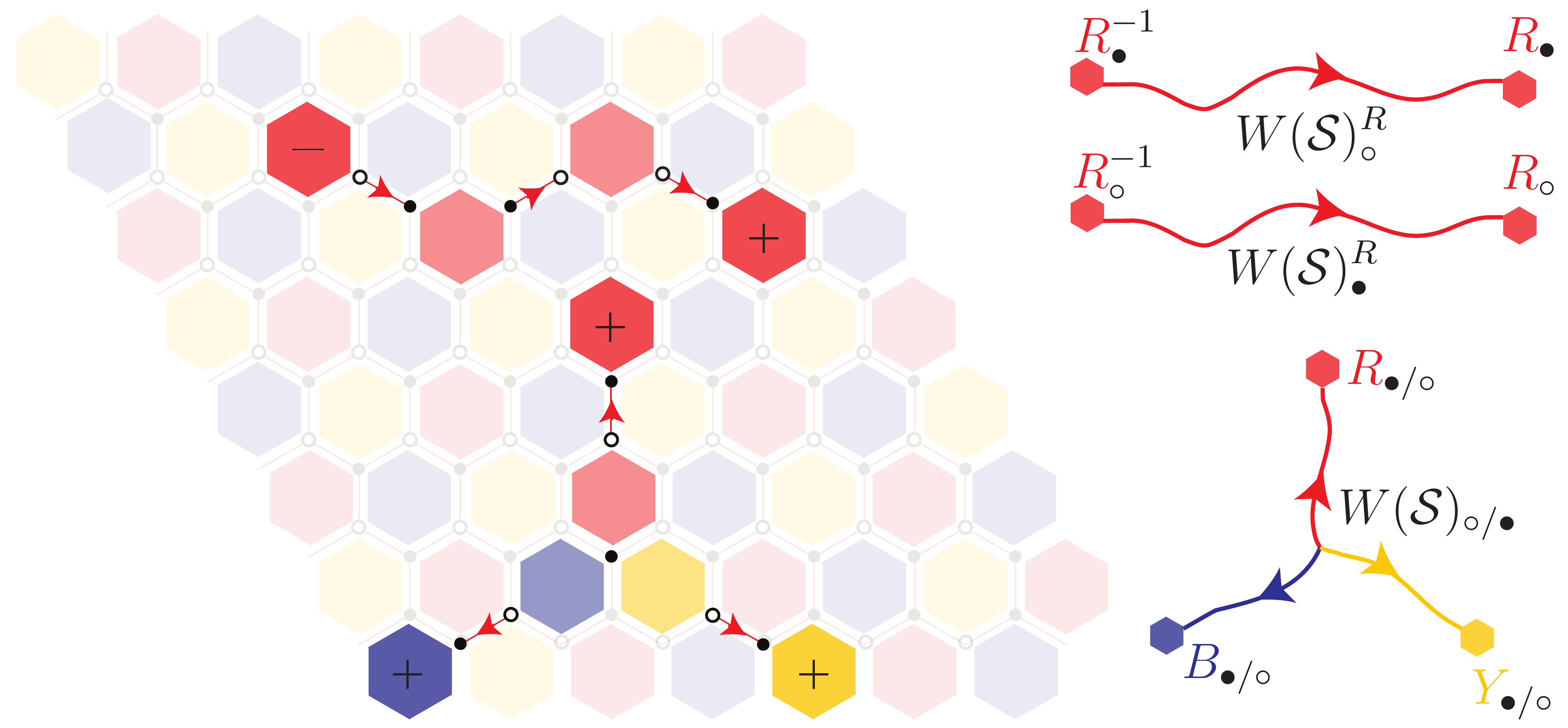}
 \caption{Abelian anyon excitations (highlighted hexagons with signs) connected by open Wilson strings. Product of rotors are taken over highlighted vertices along strings with appropriate signs according to the arrow rule in figure~\ref{fig:trivialloop}(b).}\label{fig:excitations}
\end{figure}

\begin{figure}[ht]
	\centering
	\includegraphics[width=2.5in]{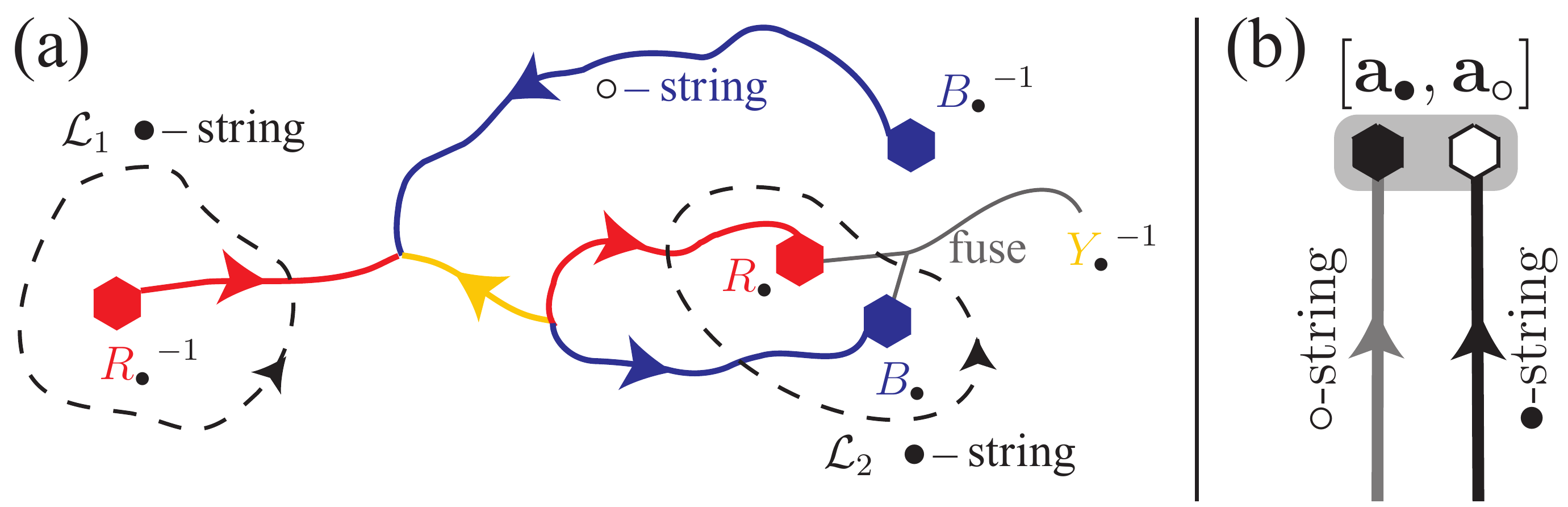}
 \caption{(a) Abelian $\bullet$-anyons (hexagons) connected by a Jordan-Wigner $\circ$-string. Anyons are probed by Wilson loops $\mathcal{L}_1$ and $\mathcal{L}_2$, which detect a ${R_\bullet}^{-1}$ and a ${Y_\bullet}^{-1}=R_\bullet\times B_\bullet$ inside respectively. (b) Diagramatic representation of a composite anyon so that a $\bullet$-anyon is attached with a $\circ$-string and vice versa}\label{fig:coloranyons}
\end{figure}

Excitations of the Hamiltonian \eqref{ham1} are eigenstates of plaquette stabilizers \eqref{stabilizers1} with non-unit eigenvalues. They can be constructed by letting open Wilson (also called Jordan-Wigner) string operators $W(\mathcal{S})_{\circ/\bullet}$ act on a ground state $|GS\rangle$. \begin{equation}|\partial\mathcal{S}\rangle_{\bullet/\circ}=W(\mathcal{S})_{\circ/\bullet}|GS\rangle\label{excitation}\end{equation} Open Wilson strings ($\bullet$ or $\circ$) do not commute with local Wilson loops ($W(\mathcal{L})_\circ$ and $W(\mathcal{L})_\bullet$ resp.) surrounding the end points in $\partial\mathcal{S}$ (see figure~\ref{fig:excitations} and \ref{fig:coloranyons}). Since trivial closed Wilson loops condense in the ground state, the excitation state $|\partial\mathcal{S}\rangle$ in \eqref{excitation} depends only on its plaquette eigenvalues at the end points of the string $\mathcal{S}$ rather than the path itself as long as it does not wrap an extra non-trivial cycle. The excited state (or in general a collection of states due to ground state degeneracy) can therefore be labeled by local {\em abelian anyon} configurations \eqref{abeliananyonlabel1}, \eqref{abeliananyonlabel2}, measured by the eigenvalues of plaquette stabilizers \begin{equation}\hat{P}_{\bullet/\circ}|\partial\mathcal{S}\rangle_{\bullet/\circ}^\pm=e^{\pm2\pi i/k}|\partial\mathcal{S}\rangle_{\bullet/\circ}\end{equation} for $\partial\mathcal{S}=\sum P^+-\sum P^-$ are directed plaquettes where the open string $\mathcal{S}$ ends (see figure~\ref{fig:excitations} and \ref{fig:coloranyons} for sign definition).  

\begin{figure}[ht]
	\centering
	\includegraphics[width=2.7in]{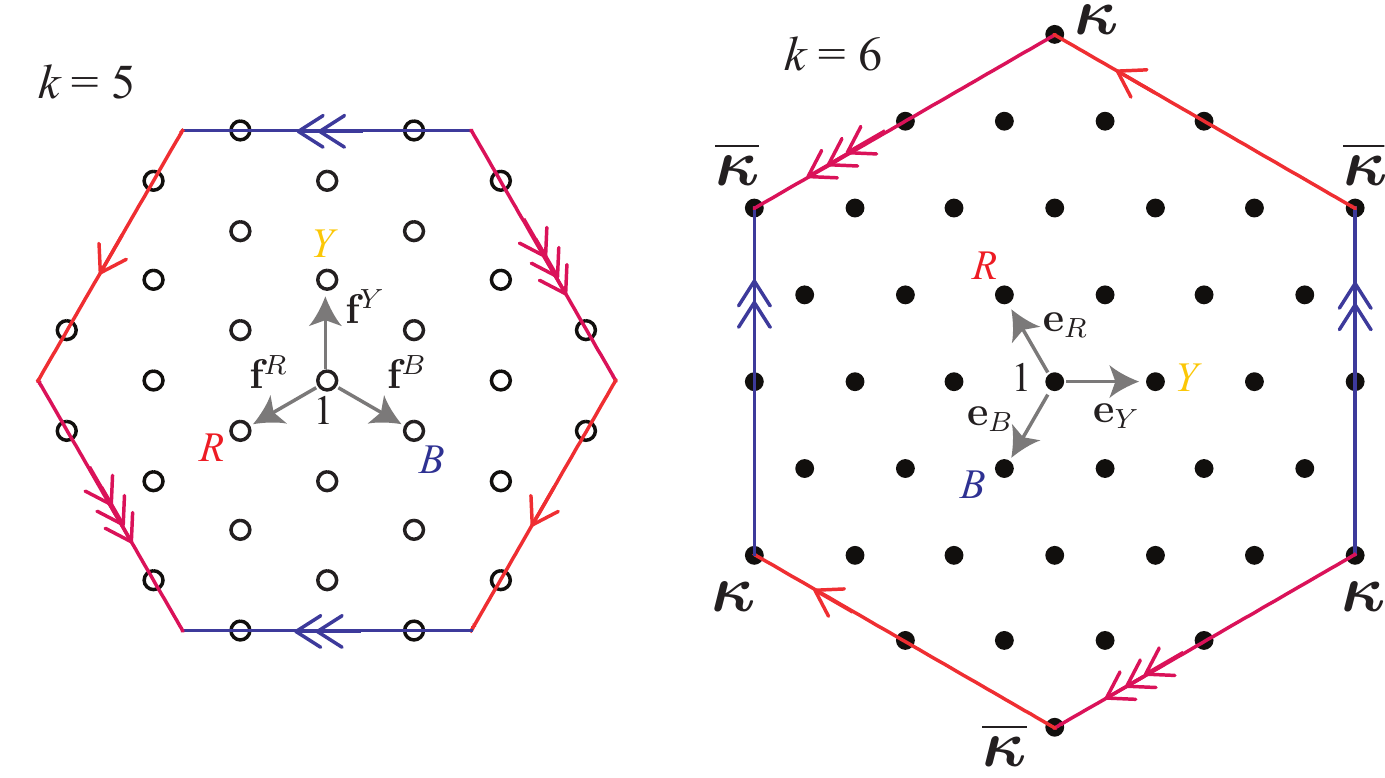}
	\caption{$\bullet$- or $\circ$-type abelian anyons (black of white dots) on an integer mod $k$ triangular lattice. Boundary of the hexagon is identified so that opposite dots along boundary lines represent the same anyon. The {\em vacuum} is represented by the origin and the three fundamental anyons $Y$, $R$ and $B=Y^{-1}R^{-1}$ are represented by the primitive lattice vectors ${\bf e}_Y$, ${\bf e}_R$, ${\bf e}_B=-{\bf e}_Y-{\bf e}_R$ for $\bullet$-type or ${\bf f}^Y$, ${\bf f}^R$, ${\bf f}^B=-{\bf f}^Y-{\bf f}^R$ for $\circ$-type.}\label{fig:abeliananyonlattice}
\end{figure}

Anyons are in general detected by local Wilson loops that encircle the quasi-particle excitation (see figure~\ref{fig:coloranyons}) with eigenvalues given by Wilson strings intersection (see figure~\ref{fig:intersection}). Primitive anyons are labelled by color, $Y,R,B$, and sublattice type, $\bullet,\circ$ with fusion relation \begin{equation}\underbrace{\chi_{\bullet/\circ}\times\ldots\times\chi_{\bullet/\circ}}_{k}=Y_{\bullet/\circ}\times R_{\bullet/\circ}\times B_{\bullet/\circ}=1\label{orderkcolorfusion}\end{equation} for $\chi=Y,R,B$, so that no Wilson loop measurement can tell apart these combination from the trivial vacuum $1$ without enclosing a proper subset. Composite anyons $[{\bf a}]$ are labeled by particle numbers (modulo $k$) of independent primitive anyons. \begin{align}[{\bf a}]&=(Y_\bullet)^{y_1}(R_\bullet)^{r_1}(Y_\circ)^{y_2}(R_\circ)^{r_2}\label{abeliananyonlabel1}\\{\bf	a}&=({\bf a}_\bullet,{\bf a}_\circ),\quad\left\{\begin{array}{*{20}c}{\bf a}_\bullet=y_1{\bf e}_Y+r_1{\bf e}_R\\{\bf a}_\circ=y_2{\bf f}^Y+r_2{\bf f}^R\hfill\end{array}\right.\label{abeliananyonlabel2}\end{align} for $y_1,y_2,r_1,r_2$ in $\mathbb{Z}_k$. The $k^4$ anyon are mutually distinguishable by Wilson loops. They are represented as a pair of $\mathbb{Z}_k$-valued two dimensional vectors, ${\bf a}_\bullet$ and ${\bf a}_\circ$, on two triangular lattices (see figure~\ref{fig:abeliananyonlattice}), one for $\bullet$-anyons, another for $\circ$-anyons.

\begin{figure}[ht]
	\centering
	\includegraphics[width=3.3in]{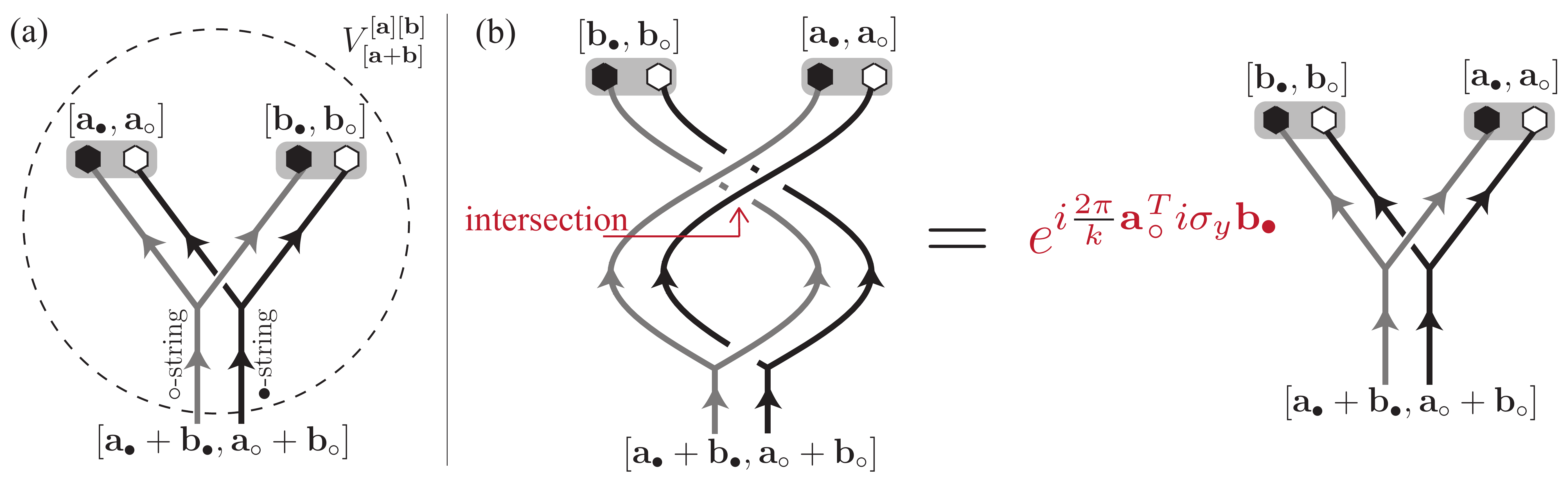}
	\caption{(a) Basis choice of splitting space $V^{[{\bf a}][{\bf b}]}_{[{\bf a}+{\bf b}]}$. The black and grey lines are Jordan-Wigner $\bullet$- and $\circ$-strings respectively. (b) The $R$-symbol of exchanging abelian anyons $[{\bf a}]$ and $[{\bf b}]$ under the basis choice for splitting space $V^{[{\bf a}][{\bf b}]}_{[{\bf a}+{\bf b}]}$ in (a). The phase is obtained by passing the grey line in front of the black one at the intersection.}\label{fig:abeliananyonsplittingspace}
\end{figure}

Abelian anyons support single channel fusion \begin{equation}[{\bf a}_\bullet,{\bf a}_\circ]\times[{\bf b}_\bullet,{\bf b}_\circ]=[{\bf a}_\bullet+{\bf b}_\bullet,{\bf a}_\circ+{\bf b}_\circ]\label{abeliananyonfusion}\end{equation} and carry unit quantum dimension $d_{[\bf a]}=1$. A basis of the one dimensional splitting space $V^{[{\bf a}][{\bf b}]}_{[{\bf a}+{\bf b}]}$ is given by letting the Jordan-Wigner string in figure~\ref{fig:abeliananyonsplittingspace}(a) act on the ground state \eqref{GS0} (projected locally inside the dotted line with fixed boundary condition). We adopt the time-ordering convention that a $\bullet$-string always acts on the ground state before a $\circ$-one. \begin{align}\left|V^{[{\bf a}][{\bf b}]}_{[{\bf a}+{\bf b}]}\right\rangle=W\left(\vcenter{\hbox{\includegraphics[width=0.2in]{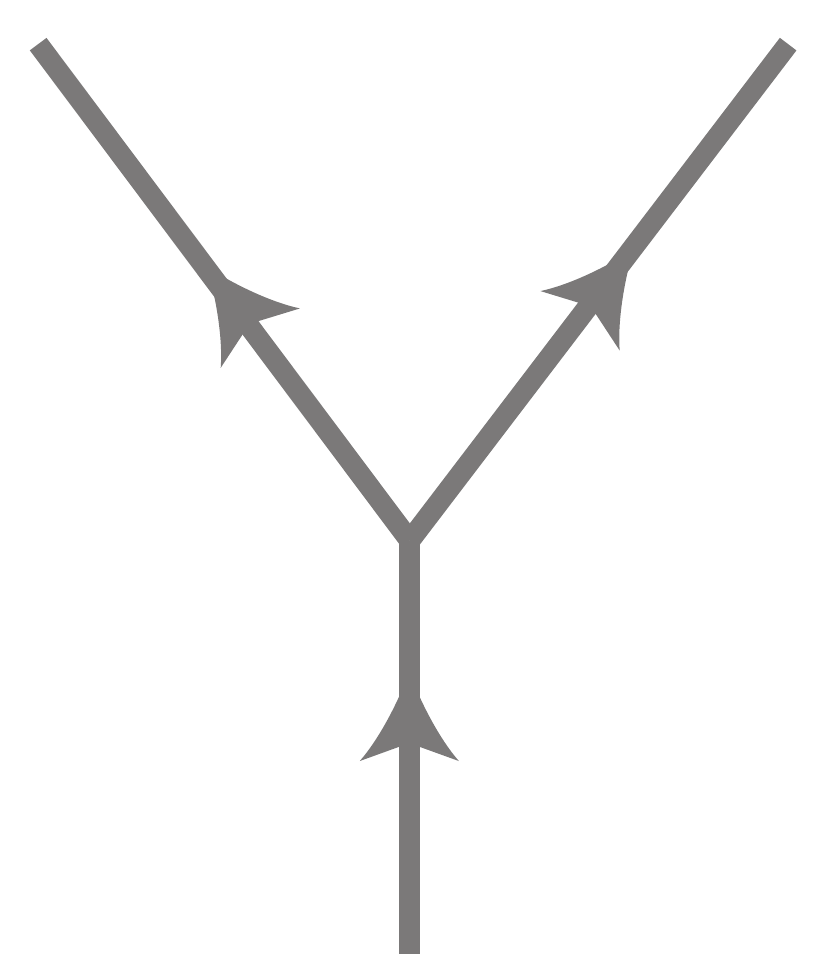}}}\right)_\circ W\left(\vcenter{\hbox{\includegraphics[width=0.2in]{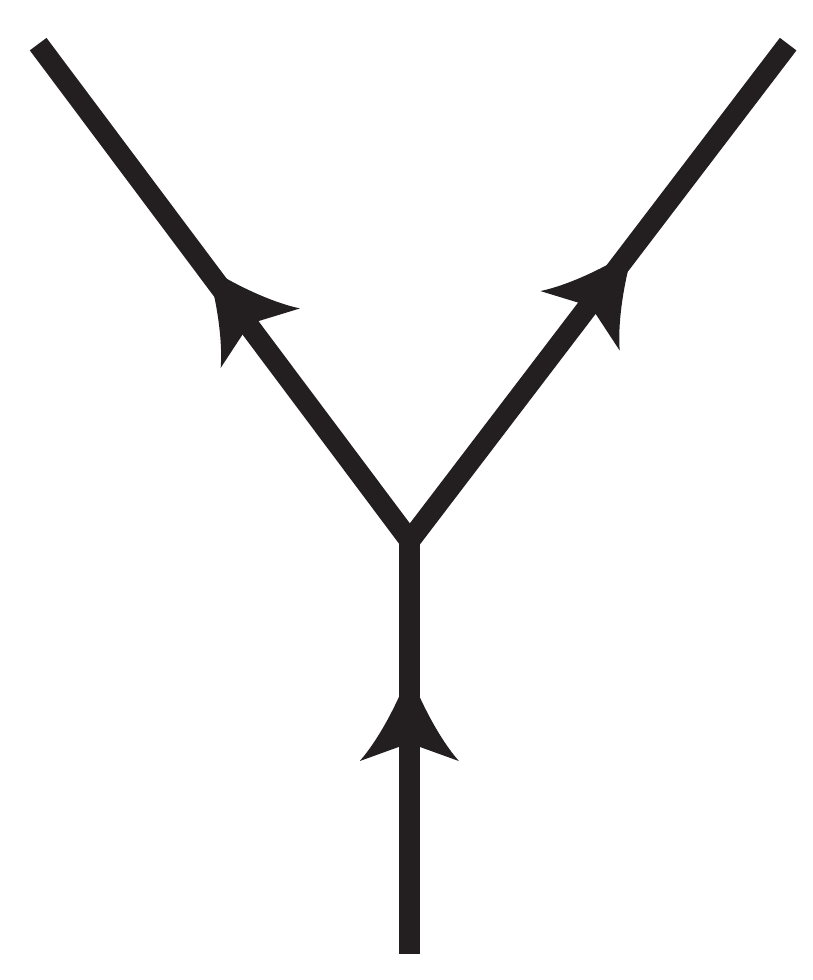}}}\right)_\bullet|GS\rangle\label{anyonsplittingbasis}\end{align} The string ordering is a gauge choice for splitting state. This will be generalized to twist defects in figure~\ref{fig:splittingspaces} in section~\ref{sec:defectfusion}.

\begin{figure}[ht]
	\includegraphics[width=3.3in]{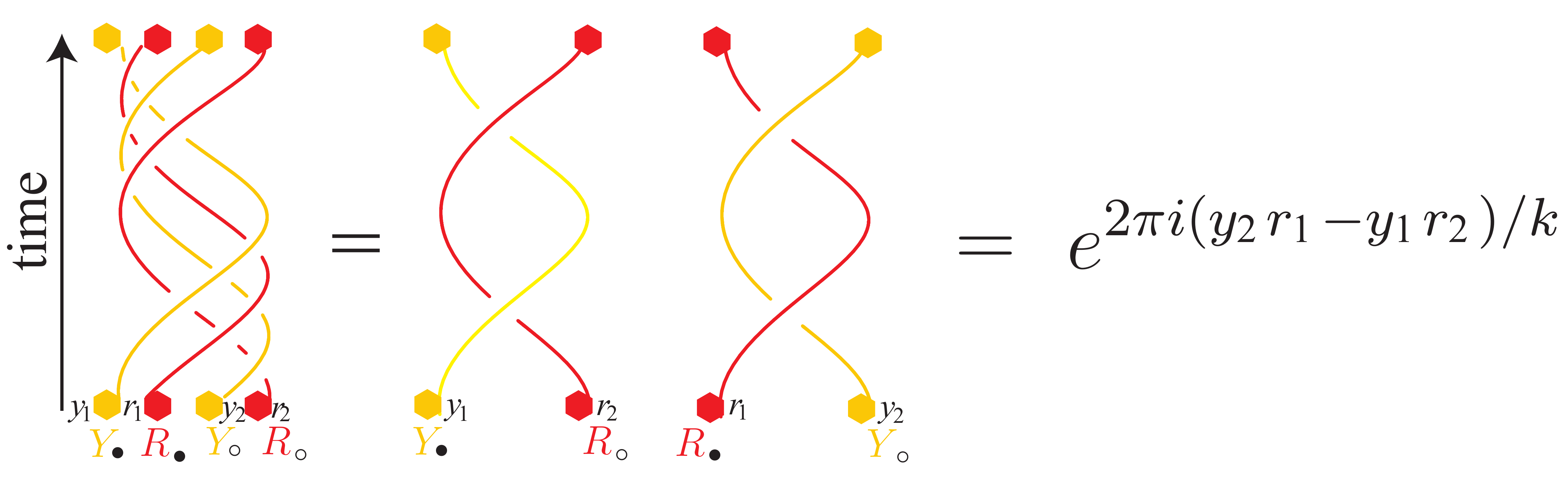}
 \caption{Topological spin of the abelian anyon $(Y_\bullet)^{y_1}(R_\bullet)^{r_1}(Y_\circ)^{y_2}(R_\circ)^{r_2}$ evaluated by $360^\circ$ rotation.}\label{fig:abelianspin}
\end{figure}

Exchange and braiding operations are represented by abelian $\mathbb{Z}_k$ phases. They stem from intersection between the anyon worldlines and Jordan-Wigner strings, which can be shown to be identical to the linking number between worldlines of anyons~\cite{Kitaev97,WilczekZee,HorowitzSrednicki} such as those shown in figure~\ref{fig:abelianspin}. The $R$-symbol of exchanging abelian anyon $[{\bf a}]$ and $[{\bf b}]$ under the basis choice of splitting space $V^{[{\bf a}][{\bf b}]}_{[{\bf a}+{\bf b}]}$ in eq.\eqref{anyonsplittingbasis} and figure~\ref{fig:abeliananyonsplittingspace}(a) is given by \begin{equation}R^{[{\bf a}][{\bf b}]}_{[{\bf a}+{\bf b}]}=e^{i\frac{2\pi}{k}{\bf a}_\circ^Ti\sigma_y{\bf b}_\bullet}=e^{i\frac{2\pi}{k}(y_2r'_1-r_2y'_1)}\label{Rabeliananyon}\end{equation} and is illustrated in figure~\ref{fig:abeliananyonsplittingspace}(b), for ${\bf a}_\circ=y_2{\bf f}^Y+r_2{\bf f}^R$ and ${\bf b}_\bullet=y'_1{\bf e}_Y+r'_1{\bf e}_R$. The topological spin of an anyon can be evaluated by a $360^\circ$ rotation (figure~\ref{fig:abelianspin}) or exchange, giving the spin-statisticcal phase \begin{equation}\theta_{[{\bf a}]}=R^{[{\bf a}][{\bf a}]}_{[2{\bf a}]}=\vcenter{\hbox{\includegraphics[width=0.7in]{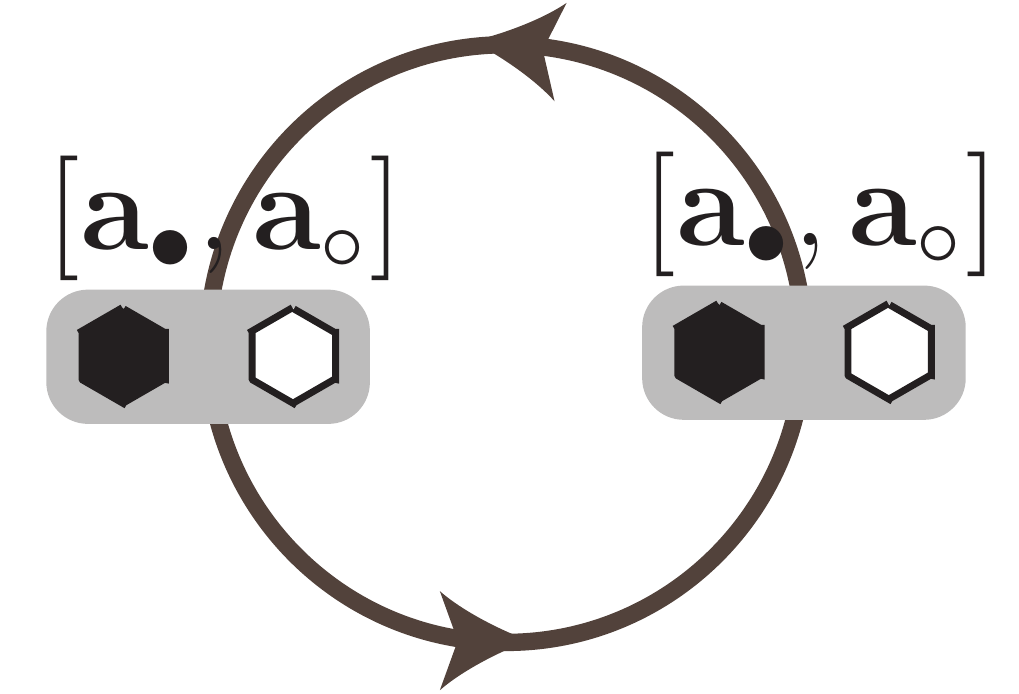}}}=e^{i\frac{2\pi}{k}{\bf a}_\circ^Ti\sigma_y{\bf a}_\bullet}\label{anyonexchangespin}\end{equation} The full braiding between anyon $[{\bf a}]$ and $[{\bf b}]$ is given by the symmetric
\begin{equation}S_{[{\bf a}][{\bf b}]}=\frac{1}{k^2}\vcenter{\hbox{\includegraphics[width=0.5in]{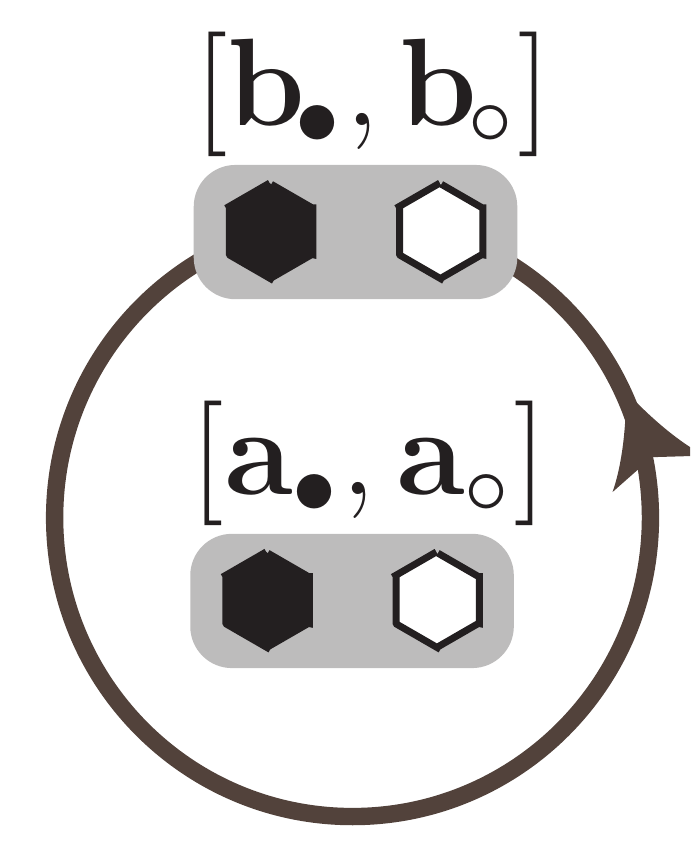}}}=\frac{1}{k^2}e^{i\frac{2\pi}{k}({\bf a}_\circ^Ti\sigma_y{\bf b}_\bullet+{\bf b}_\circ^Ti\sigma_y{\bf a}_\bullet)}\label{anyonfullbraiding}\end{equation} where the normalization $\mathcal{D}=\sqrt{\sum_{\bf a}d^2_{\bf a}}=k^2$ equals the total quantum dimension of the abelian topological phase that is responsible for its topological entanglement entropy~\cite{KitaevPreskill06}, and is added so that the $S$-matrix is unitary. Both spin $\theta$ and braiding $S$ are gauge invariant.

\subsubsection{Low energy effective field theory}\label{sec:lowenergyeffectivefieldtheory}
The model Hamiltonian \eqref{ham1} can be described by the low energy effective Chern-Simons theory~\cite{ZhangHanssonKivelson89, BlokWen90} \begin{equation}\mathcal{L}=\frac{1}{4\pi}\int K_{IJ}a_I\wedge da_J\label{CS}\end{equation} with the $4\times4$ $K$-matrix \begin{equation}K=k\left(\begin{array}{*{20}c}0&0&0&-1\\0&0&1&0\\0&1&0&0\\-1&0&0&0\end{array}\right)\label{Kmatrix}\end{equation} where $Y_\bullet,R_\bullet,Y_\circ,R_\circ$ are anyon charges for the $U(1)$-gauge fields $a_1,a_2,a_3,a_4$ respectively. This Chern-Simons theory has the same $k^4$ anyon types with identical fusion and braiding. It also supports an identical Wilson algebra eq.(\ref{wilsoncomm1}-\ref{wilsoncomm4}) on a genus $g$ surface. The Hamiltonian \eqref{ham1} therefore has the same low energy description as two copies of the $\mathbb{Z}_k$ version of Kitaev's toric code~\cite{Kitaev97,Bullock,Schulz}. The novelty of \eqref{ham1} is the apparent $S_3$-symmetry relating the tricoloring and bipartite structure of the lattice, which enriches the charge-flux (or plaquette-vertex) $\mathbb{Z}_2$-duality in Kitaev's toric code. Dislocations or twist defects are topological defects that violate certain duality or symmetry and carry non-abelian signature. The lattice structure in Hamiltonian \eqref{ham1} facilitates $S_3$-twist defects naturally through lattice dislocations and disclinations. Similar topological defects are more obscure and less motivated in the low energy Chern-Simons theory \eqref{CS} or double $\mathbb{Z}_k$-plaquette model. Their field theoretical constructions rely on an explicit branch cut in real space where the gauge fields are discontinuous. The cut can be {\em gauged away} only when the $K$-matrix is symmetric under symmetry transformation. In a lattice description of twist defects, we will see that branch cuts are absent completely.

Although the Chern-Simons theory \eqref{CS} is not our fundamental tool in this article, we expect a {\em genon} description of the $S_3$-twist defects similar to ref.[\onlinecite{BarkeshliQi,BarkeshliJianQi}]. Here we identify the $S_3$-symmetry action on the multicomponent $U(1)$-gauge fields. The permutation group $S_3$ of three elements is generated by a non-commutative threefold and twofold symmetries that act respectively as cyclic color permutation \begin{equation}\Lambda_3:\left(\begin{array}{*{20}c}Y_\bullet&R_\bullet&B_\bullet\\Y_\circ&R_\circ&B_\circ\end{array}\right)\to\left(\begin{array}{*{20}c}R_\bullet&B_\bullet&Y_\bullet\\R_\circ&B_\circ&Y_\circ\end{array}\right)\label{ST}\end{equation} and transposition for color and rotor types \begin{equation}\Lambda_{B}:\left(\begin{array}{*{20}c}Y_\bullet&R_\bullet&B_\bullet\\Y_\circ&R_\circ&B_\circ\end{array}\right)\to\left(\begin{array}{*{20}c}R_\circ&Y_\circ&B_\circ\\R_\bullet&Y_\bullet&B_\bullet\end{array}\right)\label{S}\end{equation} $\Lambda_{B}$ represents the charge-flux duality within each copy of $\mathbb{Z}_k$-Kitaev toric code pair, and $\Lambda_3$ corresponds an extra $\mathbb{Z}_3$ symmetry that intertwines the two copies. These will arise naturally as space group operators of the lattice model described in section~\ref{sec:compositelatticedefects}. They are represented by the $4\times4$ matrices \begin{equation}\Lambda_3=\left(\begin{array}{*{20}c}0&-1&0&0\\1&-1&0&0\\0&0&0&-1\\0&0&1&-1\end{array}\right),\quad\Lambda_{B}=\left(\begin{array}{*{20}c}0&0&0&1\\0&0&1&0\\0&1&0&0\\1&0&0&0\end{array}\right)\label{STCS}\end{equation} acting on the gauge fields $(a_1,a_2,a_3,a_4)$. The $K$-matrix \eqref{Kmatrix} is invariant under symmetric transformation, $K=\Lambda^TK\Lambda$ for $\Lambda\in S_3$, and therefore the theory is symmetric under the $S_3$-transformation $a_i\to\Lambda_{ij}a_j$.

Consequently fusion and braiding are also invariant under the symmetry transformation $[{\bf a}_\bullet,{\bf a}_\circ]\to\Lambda\cdot[{\bf a}_\bullet,{\bf a}_\circ]$ according to \eqref{ST} and \eqref{S}. These transformations rename the color and sublattice labels for abelian anyons. Cyclic color permutation $\Lambda_3$ corresponds a threefold rotation of the triangular anyon lattice in figure~\ref{fig:abeliananyonlattice} while keeping the sublattice label $\bullet,\circ$ fixes, and transposition $\Lambda_B$ corresponds a mirror operation while flipping between a $\bullet\leftrightarrow\circ$. Notice that for $k$ divisible by 3, there are non-trivial anyons $\boldsymbol\kappa_\bullet=\overline{\boldsymbol\kappa}_\bullet^\dagger=(Y_\bullet)^{k/3}(R_\bullet)^{-k/3}$ and $\boldsymbol\kappa_\circ=\overline{\boldsymbol\kappa}_\circ^\dagger=(Y_\circ)^{k/3}(R_\circ)^{-k/3}$ that are invariant under cyclic color permutation $(ST)\cdot\boldsymbol\kappa_{\bullet/\circ}=\boldsymbol\kappa_{\bullet/\circ}$. Furthermore, unlike over complex coefficients where the finite group $S_3$ has only 2-dimenional faithful irreducible representation, the two 4-dimensional matrices \eqref{STCS} cannot be simultaneously further block diagonalized with discrete coefficients. This means that the anyon Hilbert space cannot be decomposed into tensor product without violating $S_3$-symmetry.

In mathematical terms, the symmetry group $S_3$ is a subgroup of $\Gamma_1$, the group of relabeling of anyons, or precisely the group of invertible functors $\mathcal{A}\to\mathcal{A}$ of the unitary braided fusion category $\mathcal{A}$.\cite{Kitaev06} In the lattice rotor model with $K$-matrix \eqref{Kmatrix}, $\Gamma_1$ is given by the group of outer automorphisms \begin{equation}\Gamma_1=\mbox{Outer}(K)\equiv\frac{\mbox{Aut}(K)}{\mbox{Inner}(K)}\end{equation} where the automorphism group \begin{align}\mbox{Aut}(K)=O(K;\mathbb{Z})\equiv\left\{g\in GL(4;\mathbb{Z}):g^TKg=K\right\}\end{align} is given by orthogonal transformations that leave the $K$-matrix invariant, and the subgroup of inner automorphism \begin{align}\mbox{Inner}(K)\equiv\left\{g\in\mbox{Aut}(K):g\cdot[{\bf a}_\bullet,{\bf a}_\circ]\equiv[{\bf a}_\bullet,{\bf a}_\circ]\;\mbox{mod $K$}\right\}\end{align} contain orthogonal transformations that fix the anyon lattice $\mathbb{Z}^4/K\mathbb{Z}^4$. As shown later in eq.\eqref{spacegroupquotient}, the symmetry subgroup $S_3$ is inherited from and identical to the symmetry of the underlying trivalent bipartite planar graph. For example the color permutation $\Lambda_3$ is induced by a lattice translation on the honeycomb and color sublattice transposition $\Lambda_B$ is induced by a lattice inversion. The correspondence between symmetry of the microscopic Hamiltonian and anyon relabeling symmetry \begin{equation}\omega_1:S_3\to\Gamma_1=\mbox{Outer}(K)\label{omega1}\end{equation} is a {\em first level weak symmetry breaking} according to Kitaev.\cite{Kitaev06} Twist defects are explicit local violations of the underlying symmetry, and because of \eqref{omega1} they are also symmetry defects that alter anyon sectors.

\section{Non-Abelian \texorpdfstring{$S_3$}{S3} Twist Defects}\label{sec:twistdefect}

\begin{figure}[ht]
	\centering
	\includegraphics[width=3.5in]{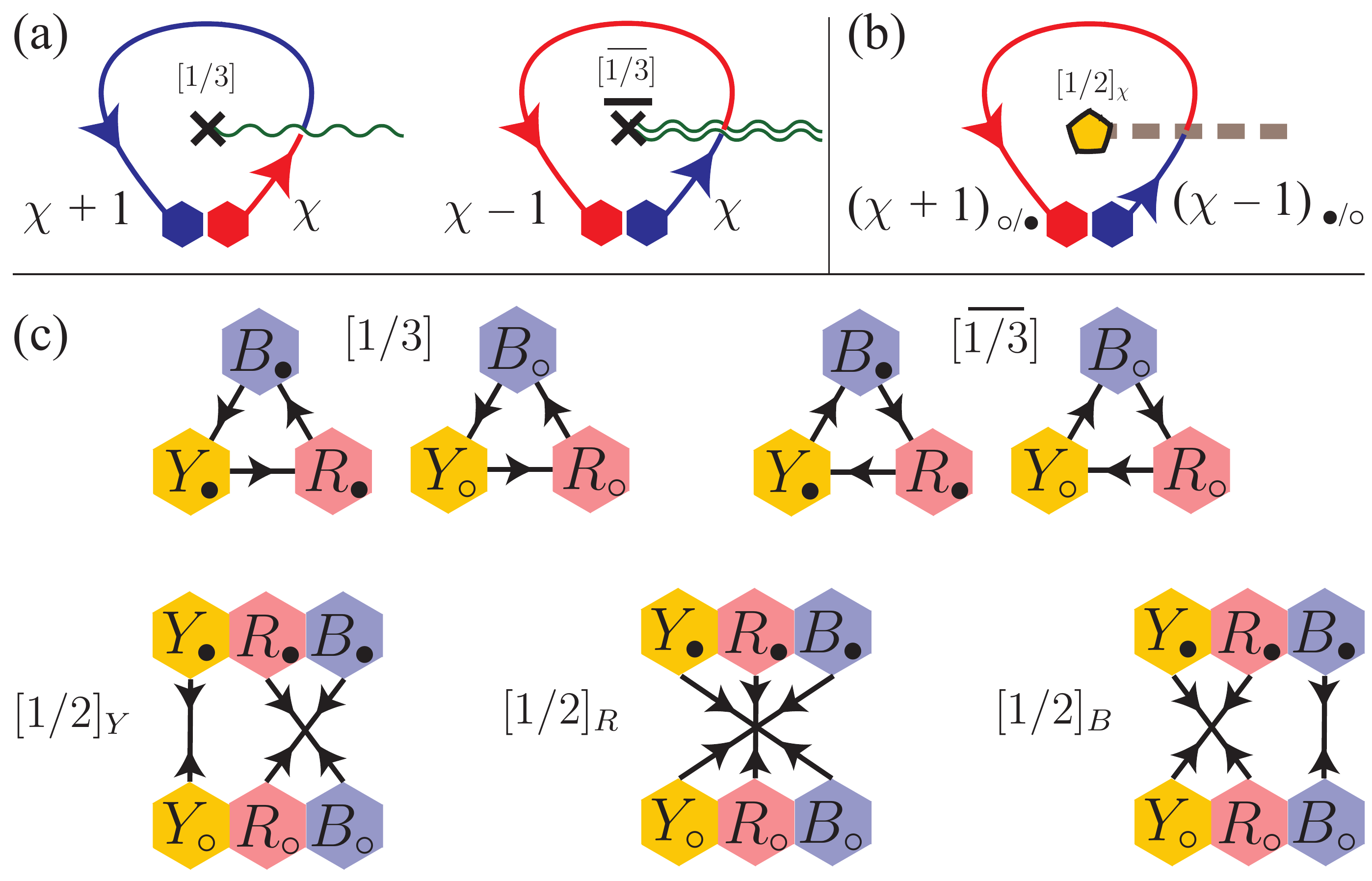}
 \caption{Twisting abelian anyons (colored hexagons) around twist defects (black crosses and colored pentagon). (a) Cyclic color permutation, $\chi\in\{Y,R,B\}=\{0,1,2\}$, of a threefold twist defect $[1/3]$ and its anti-partner $[\overline{1/3}]$. (b) Color and sublattice transposition of a twofold twist defect $[1/2]_\chi$ with color $\chi$. (c) Color and sublattice twisting of the two threefold and three twofold twist defects.}\label{fig:anyontwist}
\end{figure}

The $YRB$-plaquette coloration and $\bullet,\circ$-sublattice types give rise to the four fundamental abelian anyon excitations $Y_\bullet,Y_\circ,R_\bullet,R_\circ$ in the lattice rotor model \eqref{ham1} (recall the color redundancy $Y\times R\times B=1$). The arbitrariness of color and sublattice labeling of abelian anyons is summarized by the $S_3$-symmetry generated by cyclic color permutation $\Lambda_3$ and transposition $\Lambda_B$ in eq.\eqref{ST} and \eqref{S}. A twist defect is a topological defect that locally violates the symmetry by altering, or {\em twisting}, the color and rotor label of an anyon that goes around it (see figure~\ref{fig:anyontwist}). In other words, a Wilson string that circles around a twist defect does not close back to itself, and therefore unlike abelian anyons which can be locally detected by small Wilson loops such as plaquette stabilizers, there are no local observables measuring a twist defect state. This non-locality is a central theme of many non-abelian anyons, such as vortex-bound Majorana fermions in chiral $p+ip$ superconductors~\cite{Ivanov,ReadGreen}, Ising anyon in the Kitaev's honeycomb model~\cite{Kitaev06} and Pfaffian fractional quantum Hall state~\cite{MooreRead}. The non-abelian anyons associated with twist defects considered in this article, however, are not fundamental deconfined excitations of a true topological phase. They are qualitatively more similar to (fractional) Majorana excitations at SC-FM heterostructures with (fractional) topological insulators~\cite{FuKane08, LindnerBergRefaelStern, ClarkeAliceaKirill, MChen, Vaezi} or strongly spin-orbit coupled quantum wires~\cite{SauLutchynTewariDasSarma, OregSelaStern13}. Their existence rely on the topological winding of certain classical non-dynamical {\em order parameter field}, such as pairing and spin/charge gap~\cite{TeoKane}. The tricoloring and bipartite structure of the lattice Hamiltonian \eqref{ham1} can be regarded as a discrete {\em order parameter} of the condensate, and its winding around a point defect supports the color and sublattice twisting.

A twist defect in our lattice model \eqref{ham1} is labeled by an element $\Lambda$ in the symmetry group $S_3$ according to its action on the anyon label so that when an anyon $[{\bf a}]=Y_\bullet^{y_1}R_\bullet^{r_1}Y_\circ^{y_2}R_\circ^{r_2}$ passes counter-clockwise around the twist defect, it changes into $\Lambda\cdot[{\bf a}]$, where $\Lambda$ is some product combinations of $\Lambda_B$ and $\Lambda_3$ in \eqref{ST} and \eqref{S}. Threefold cyclic permutation and twofold transposition are the two conjugacy classes of $S_3$ and correspond to two threefold twist defects $[1/3],[\overline{1/3}]$ and three twofold ones $[1/2]_Y,[1/2]_R,[1/2]_B$ respectively. Their twisting actions on abelian anyons going around them are summarized in figure~\ref{fig:anyontwist}. The fraction label is chosen to match with the {\em fractionalization} of abelian anyons so that the denominator shows the minimal number of identical defect copies required to fuse into an abelian channel (see \eqref{fusionrulesummary}). 

\begin{figure}[ht]
	\centering
	\includegraphics[width=3.5in]{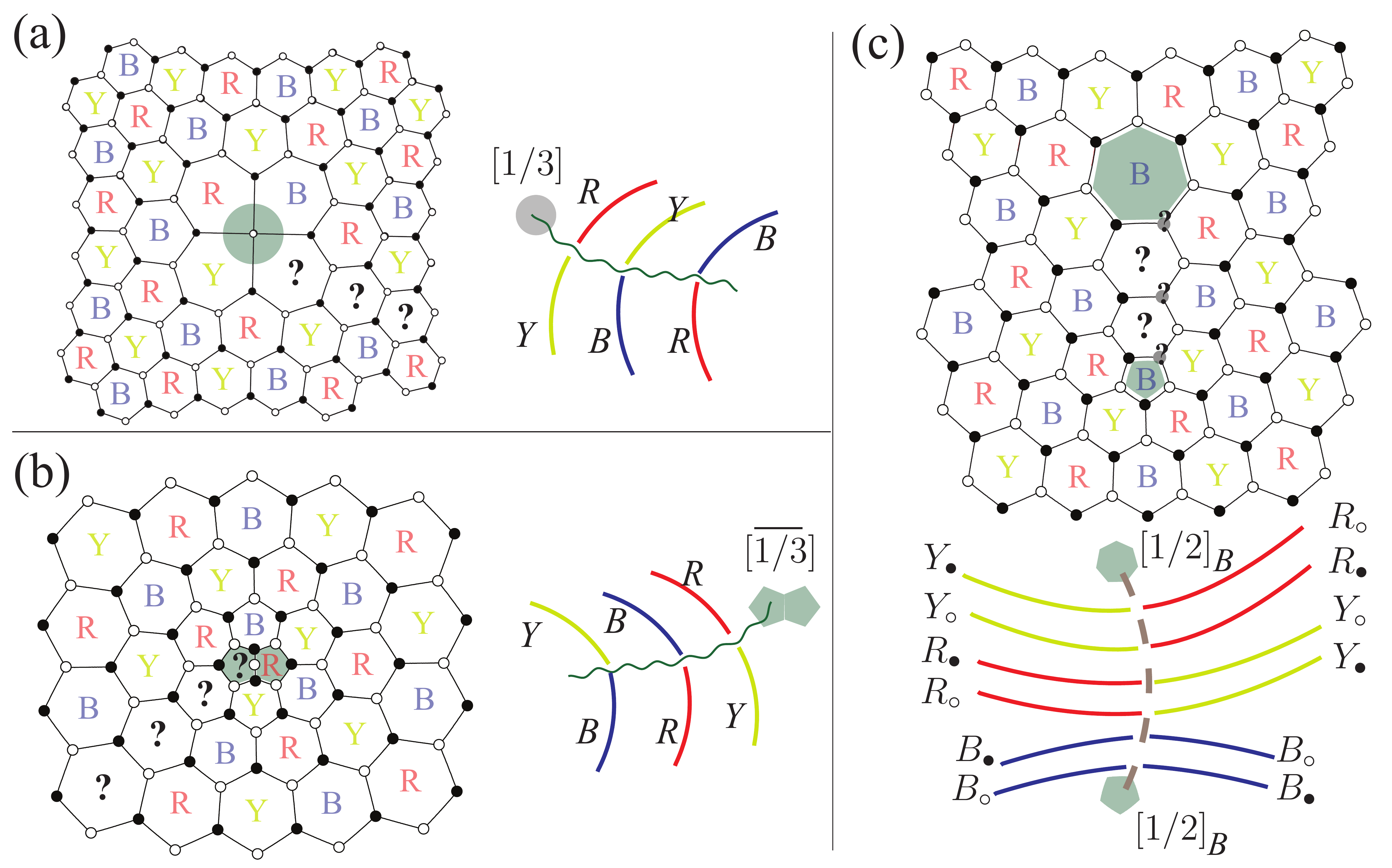}
 \caption{Disclination twist defects (shaded plaquettes and vertices). Color or sublattice frustrations (question marks) give rise to branch cuts (curly or dashed lines) that alter anyon types (colored lines) across. (a) A threefold twist defect generated by a $+120^\circ$-disclination centered at a tetravalent vertex. Sublattice types $\bullet,\circ$ are not affected by the defect. (b) An anti-threefold twist defect generated by a $-120^\circ$-disclination centered at a bivalent vertex and permutes colors in the opposite direction around. (c) A pair of twofold twist defects generated by a $\pm60^\circ$-disclination dipole. Each violates both tricoloring and bipartite structure.}\label{fig:twistdefects}
\end{figure}
\begin{figure}[ht]
	\centering
	\includegraphics[width=3in]{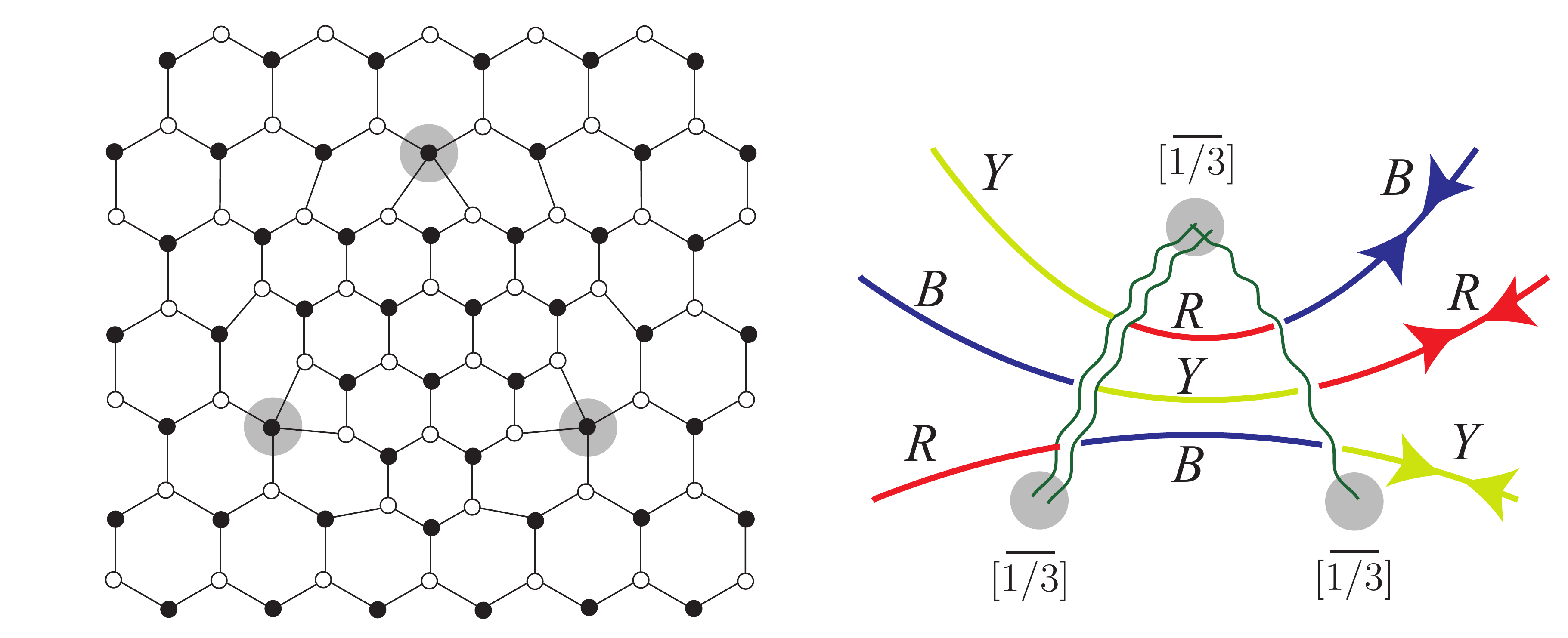}
 \caption{A triplet of dislocation threefold twist defects connected by two color permutation branch cuts (curly lines).}\label{fig:Z3dislocation}
\end{figure}
Crystalline defects are predicted to carry topologically protected excitations in topological insulators\cite{RanZhangVishwanath, TeoKane} and superconductors~\cite{TeoHughes}. They are expected to hold fractional quantum vortices in FFLO states~\cite{RadzihovskyVishwanath08, AgterbergTsunetsugu08, BergFradkinKivelson09, GopalakrishnanTeoHughes13}. Here we realize twist defects in the lattice model \eqref{ham1} as crystalline defects, such as the disclinations and dislocations illustrated in figure~\ref{fig:twistdefects} and~\ref{fig:Z3dislocation}, that center at non trivalent vertices for threefold twist defects, or odd sided plaquettes for twofold twist defects. These are topological lattice defects that carry curvature or torsion singularities and locally breaks lattice rotation and translation symmetry. Through local violation of tricolorability and/or bipartite structure of the Hamiltonian \eqref{ham1}, they change the color and/or sublattice type of abelian anyons that go around them. This gives rise to additional non-contractible Wilson loops, ground state degeneracies and non-trivial quantum dimensions ($d>1$) associate with the twist defects. 

We begin this section by writing down the local lattice Hamiltonian for the two kinds of twist defects. Their quantum dimensions (or ground state degeneracies) can be deduced by counting plaquette stabilizers and vertices. Next we show the topological degeneracies are inherited from non-trivial Wilson string operators, each surrounds multiple defects. These form a set of non-commuting physical observables with $\mathbb{Z}_k$-valued measurements. As a quantum state is labeled by its simultaneous eigenvalues of a maximal set of commuting Wilson operators, it cannot be {\em accidentally} measured by local observation since the Wilson strings are non-local operators passing through spatially separated twist defects. This non-local storage of quantum information between non-abelian anyons provides topological protection against decoherence and forms the basis for fault tolerant topological quantum computation~\cite{Kitaev97, OgburnPreskill99, Preskilllecturenotes, FreedmanKitaevLarsenWang01, ChetanSimonSternFreedmanDasSarma, Wangbook}. Similar to the algebraic relation eq.\eqref{wilsoncomm3}, the  non-commutativity of Wilson operators are characterized by an intersection form $\langle\mathcal{C}^{\chi_1}_i,\mathcal{C}^{\chi_2}_j\rangle$ between Wilson strings. We compute these pairings explicitly in this section and show their covariant behavior under $S_3$-transformation that differs from the invariant one from the previous section. The basis of Wilson strings and their intersection properties will be useful for characterizing defect fusion and braiding in the following section.

\subsection{Lattice defect Hamiltonian}\label{sec:latticedefecthamiltonian}
We describe the lattice model modifications at primitive $\pm120^\circ$-disclinations centered at tetravalent or bivalent vertices and $\pm60^\circ$-disclinations at heptagon or pentagon plaquettes (see figure~\ref{fig:twistdefects}) corresponding to threefold and twofold twist defects respectively. Unlike the square or octagon disclinations in figure~\ref{fig:squareoctagon}, these non-trivial lattice defects require additional sets of vertex rotors or allow less plaquette operators in order for stabilizers to remain mutually commutative. The extra rotor degree of freedom increases the ground state degeneracy and associates non-trivial quantum dimensions to twist defects. The $S_3$-classification of primitive disclinations are summarized in table~\ref{tab:twistdefects}.
\begin{table}[ht]
\centering
\begin{tabular}{cl}
\hline\hline
$S_3$-label & \multicolumn{1}{c}{lattice disclinations}\\\hline
\multirow{2}{*}{$[1/3]$} & $+120^\circ$ disclination at a tetravalent $\circ$-vertex\\
& $-120^\circ$ disclination at a bivalent $\bullet$-vertex\\\noalign{\smallskip}
\multirow{2}{*}{$[\overline{1/3}]$} & $+120^\circ$ disclination at a tetravalent $\bullet$-vertex\\
& $-120^\circ$ disclination at a bivalent $\circ$-vertex\\\noalign{\smallskip}
\multirow{2}{*}{$[1/2]_\chi$} & $+60^\circ$ disclination at a $\chi$-colored heptagon\\
& $-60^\circ$ disclination at a $\chi$-colored pentagon\\\hline\hline
\end{tabular}
\caption{Types of primitive twist defects at disclinations, $\chi=Y,R,B$.}\label{tab:twistdefects}
\end{table}

\subsubsection{Threefold twist defects} 
\begin{figure}[ht]
	\centering
	\includegraphics[width=2.5in]{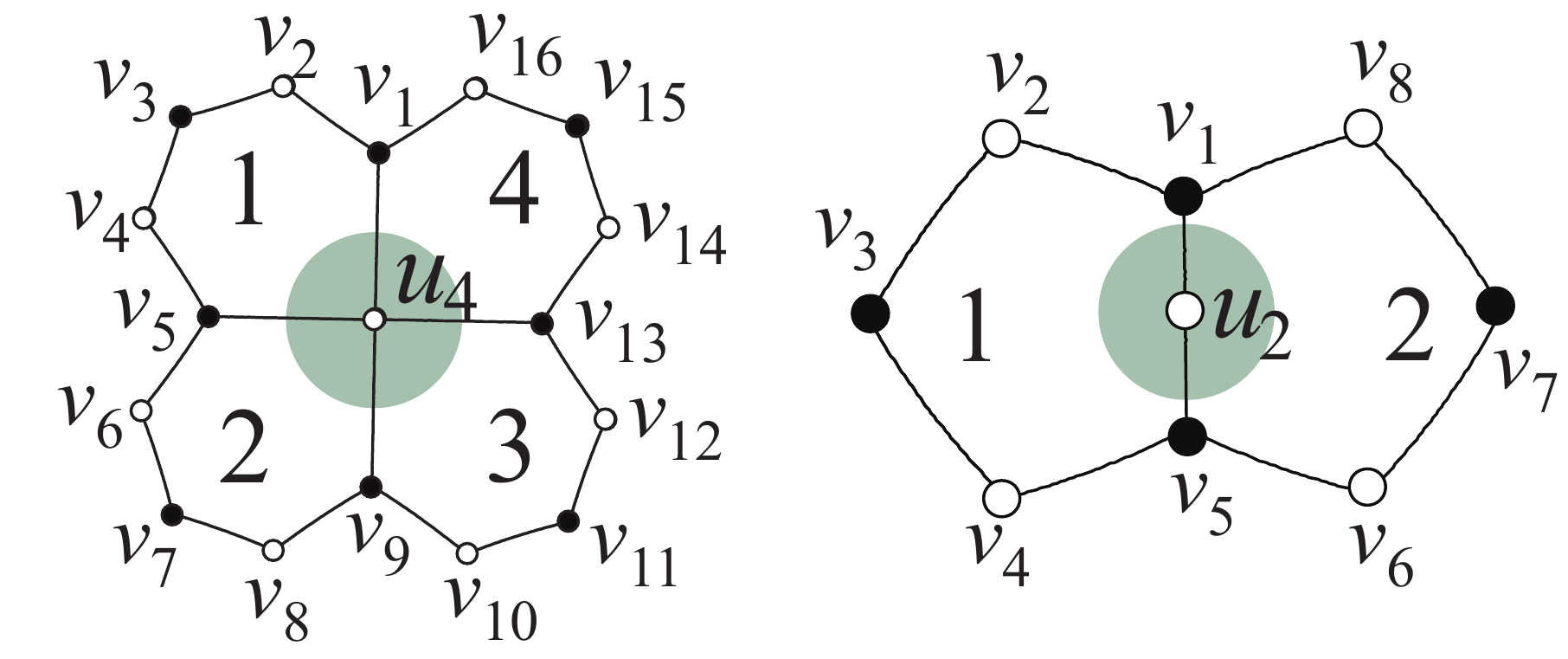}
 \caption{Tetravalent and bivalent vertices $u_4,u_2$ and their nearest plaquettes.}\label{fig:4valentvertex}
\end{figure}
Instead of accommodating a single set of rotors $\sigma,\tau$ and a $k$-dimensional Hilbert space like an ordinary trivalent vertex, a tetravalent vertex hosts four sets of rotors $\sigma(i),\tau(i)$ and a $k^4$-dimensional tensor product space. We define the rotor operators at a tetravalent vertex by the following tensor products \begin{equation}\begin{array}{*{20}c}\sigma(1)=\sigma\otimes\openone_k\otimes\openone_k\otimes\openone_k\\\sigma(2)=\openone_k\otimes\sigma\otimes\openone_k\otimes\openone_k\\\sigma(3)=\openone_k\otimes\openone_k\otimes\sigma\otimes\openone_k\\\sigma(4)=\openone_k\otimes\openone_k\otimes\openone_k\otimes\sigma\end{array}\quad\begin{array}{*{20}c}\tau(1)=\tau\otimes\tau\otimes\openone_k\otimes\tau\\\tau(2)=\tau\otimes\tau\otimes\tau\otimes\openone_k\\\tau(3)=\openone_k\otimes\tau\otimes\tau\otimes\tau\\\tau(4)=\tau\otimes\openone_k\otimes\tau\otimes\tau\end{array}\label{4valentrotorrep}\end{equation} where $\openone_k$ is the $k\times k$ identity matrix and $\sigma,\tau$ are the usual rotors with matrix representations \eqref{sigmataurep}. They satisfy a modified algebraic relation \begin{align}\tau(i)\sigma(j)\tau(i)^{-1}\sigma(j)^{-1}=\left\{\begin{array}{*{20}c}w,&\mbox{if $|i-j|\neq2$}\\1,&\mbox{if $|i-j|=2$}\end{array}\right.\label{4valentrotorcomm1}\\\sigma(i)^k=\tau(i)^k=1,\quad[\sigma(i),\sigma(j)]=[\tau(i),\tau(j)]=0\label{4valentrotorcomm2}\end{align} where $w=e^{2\pi i/k}$. The four adjacent plaquette stabilizers around the tetravalent vertex $u_4$ in figure~\ref{fig:4valentvertex} are defined with the new rotor operators (suppressing tensor products) \begin{align}\begin{array}{*{20}c}\hat{P}_\bullet(1)=\tau_{u_4}(1)\sigma_{v_1}\tau_{v_2}\sigma_{v_3}\tau_{v_4}\sigma_{v_5}\hfill\\\hat{P}_\bullet(2)=\tau_{u_4}(2)\sigma_{v_5}\tau_{v_6}\sigma_{v_7}\tau_{v_8}\sigma_{v_9}\hfill\\\hat{P}_\bullet(3)=\tau_{u_4}(3)\sigma_{v_9}\tau_{v_{10}}\sigma_{v_{11}}\tau_{v_{12}}\sigma_{v_{13}}\\\hat{P}_\bullet(4)=\tau_{u_4}(4)\sigma_{v_{13}}\tau_{v_{14}}\sigma_{v_{15}}\tau_{v_{16}}\sigma_{v_1}\end{array}\end{align} and similarly the $\circ$-plaquette operators $\hat{P}_\circ(i)$ are defined by interchanging $\sigma\leftrightarrow\tau$.

A bivalent vertex on the other hand is hosting two sets of rotor operators and a $k^2$-dimensional tensor product space. The rotors are defined by \begin{equation}\begin{array}{*{20}c}\sigma(1)=\sigma\otimes\openone_k\\\sigma(2)=\openone_k\otimes\sigma\end{array}\quad\begin{array}{*{20}c}\tau(1)=\tau\otimes\tau^2\\\tau(2)=\tau^2\otimes\tau\end{array}\label{2valentrotorrep}\end{equation} and satisfy the algebraic relations \eqref{4valentrotorcomm2} as well as \begin{equation}\tau(i)\sigma(j)\tau(i)^{-1}\sigma(j)^{-1}=\left\{\begin{array}{*{20}c}w,&\mbox{if $i=j$}\\w^2,&\mbox{if $i\neq j$}\end{array}\right.\label{2valentrotorcomm1}\end{equation} The two adjacent plaquette stabilizers of the bivalent vertex $u_2$ in figure~\ref{fig:4valentvertex} are defined by \begin{equation}\begin{array}{*{20}c}\hat{P}_\bullet(1)=\tau_{u_2}(1)\sigma_{v_1}\tau_{v_2}\sigma_{v_3}\tau_{v_4}\sigma_{v_5}\\\hat{P}_\bullet(2)=\tau_{u_2}(2)\sigma_{v_5}\tau_{v_6}\sigma_{v_7}\tau_{v_8}\sigma_{v_1}\end{array}\end{equation} and similarly for the $\circ$-plaquette operators. 

Thanks to the modified algebraic relations \eqref{4valentrotorcomm1} and \eqref{2valentrotorcomm1}, all plaquette operators around a tetravalent or bivalent vertex mutually commute. The Hamiltonian is then defined just as in eq.\eqref{ham1} by summing over all plaquette stabilizers. In fact \eqref{4valentrotorcomm1} and \eqref{2valentrotorcomm1} are not just sufficient but also necessary for the stabilizers to form good quantum numbers. The $k^4$ or $k^2$ dimensional tensor product rotor representations \eqref{4valentrotorrep} or \eqref{2valentrotorrep} are not accidental. Their dimensionality are {\em topologically protected} by non-contractible Wilson strings around them and are directly related to the quantum dimension of threefold twist defect. These will be explained in detail in the following subsection.

There is however a caveat when $k$ is a multiple of 3. By examining the rotor representation at a tetravalent vertex \eqref{4valentrotorrep} or that at a bivalent one \eqref{2valentrotorrep}, we have the additional torsion relation \begin{equation}\left(\prod_{i=1}^4\tau_{u_4}(i)\right)^{k/3}=\left(\prod_{i=1}^2\tau_{u_2}(i)\right)^{k/3}=1\end{equation} and the order three center operators that commute with all rotors \begin{equation}\Sigma_{u_4}=\left(\prod_{i=1}^4\sigma_{u_4}(i)\right)^{k/3},\quad\Sigma_{u_2}=\left(\prod_{i=1}^2\sigma_{u_2}(i)\right)^{k/3}\label{24Sigma}\end{equation} This means that the $k^4$ (or $k^2$) dimensional tensor product space at the tetravalent (resp. bivalent) vertex is no longer irreducible for the rotor algebra eq.\eqref{4valentrotorcomm1} (or eq.\eqref{2valentrotorcomm1}) as it can be decomposed according to the $\mathbb{Z}_3$-valued good quantum number according to $\Sigma_{u_4}$ (resp.~$\Sigma_{u_2}$). In this case, we restrict the tensor product space at any tetravalent (or bivalent) vertices to one of the $k^4/3$ (resp.~$k^2/3$) dimensional sector by specifying the $\mathbb{Z}_3$-phase for the central observables $\Sigma_{u_4}$ (resp.~$\Sigma_{u_2}$). Or equivalently the $k^4$ (resp.~$k^2$) dimensional rotor space can be broken down by the local defect Hamiltonian \begin{equation}H_{u_4/u_2}=-J_\ast\left(e^{-i\phi}\Sigma_{u_4/u_2}+e^{i\phi}\Sigma_{u_4/u_2}^\dagger\right)\label{Z33defectham}\end{equation} where the phase variable $\phi$ controls the $\mathbb{Z}_3$-value of $\Sigma_{u_4/u_2}$ in the ground state except at $\phi=\pi,\pm\pi/3$ where level crossings occur. Consequently, when $k$ is divisible by 3, threefold twist defects are further subdivided into nine different species distinguished by the eigenvalues of the $\Sigma$'s, which are local measurements described by certain order 3 Wilson strings described in the upcoming subsection.

Using the Euler characteristics $2-2g=\#P-\#E+\#V$, eq.\eqref{VPcount} is modified into \begin{equation}\#V=2\times\#P+4(g-1)-\#u_4+\#u_2\label{Z3VPcount}\end{equation} where $\#V$, $\#P$ are the total numbers of vertices, plaquettes and $\#u_4$, $\#u_2$ are the numbers of tetravalent, bivalent vertices respectively. We are interested in how the ground state degeneracy (G.S.D.) scales with $\#u_4$ and $\#u_2$ in the thermodynamic limit. For this, we will ignore the genus $g$ and the overcounting of plaquettes such as \eqref{stabilizerovercounting} that will contribute to the G.S.D. by a proportionality constant independent from the twist defect number. 

The total Hilbert space is a tensor product of rotor spaces of dimension $k^{\#V+3\#u_4+\#u_2}$ for $k$ not divisible by 3, or $k^{\#V+3\#u_4+\#u_2}/3^{\#u_4+\#u_2}$ otherwise. Stabilized by the two operators $\hat{P}_\bullet,\hat{P}_\circ$ per plaquette, the ground state degeneracy scales as \begin{equation}G.S.D\sim\left\{\begin{array}{*{20}c}(k^2)^{\#u_4+\#u_2},&\mbox{if $3\centernot\mid k$}\\(k^2/3)^{\#u_4+\#u_2},&\mbox{if $3\mid k$}\end{array}\right.\end{equation} where $\#u_4+\#u_2$ is the total number of defects. Or equivalently the quantum	dimensions of a threefold twist defect is given by \begin{equation}d_{[1/3]}=d_{[\overline{1/3}]}=\left\{\begin{array}{*{20}c}k^2,&\mbox{if $3\centernot\mid k$}\\k^2/3,&\mbox{if $3\mid k$}\end{array}\right.\label{Z3dimension}\end{equation}

\subsubsection{Twofold twist defects}
\begin{figure}[ht]
	\centering
	\includegraphics[width=2.5in]{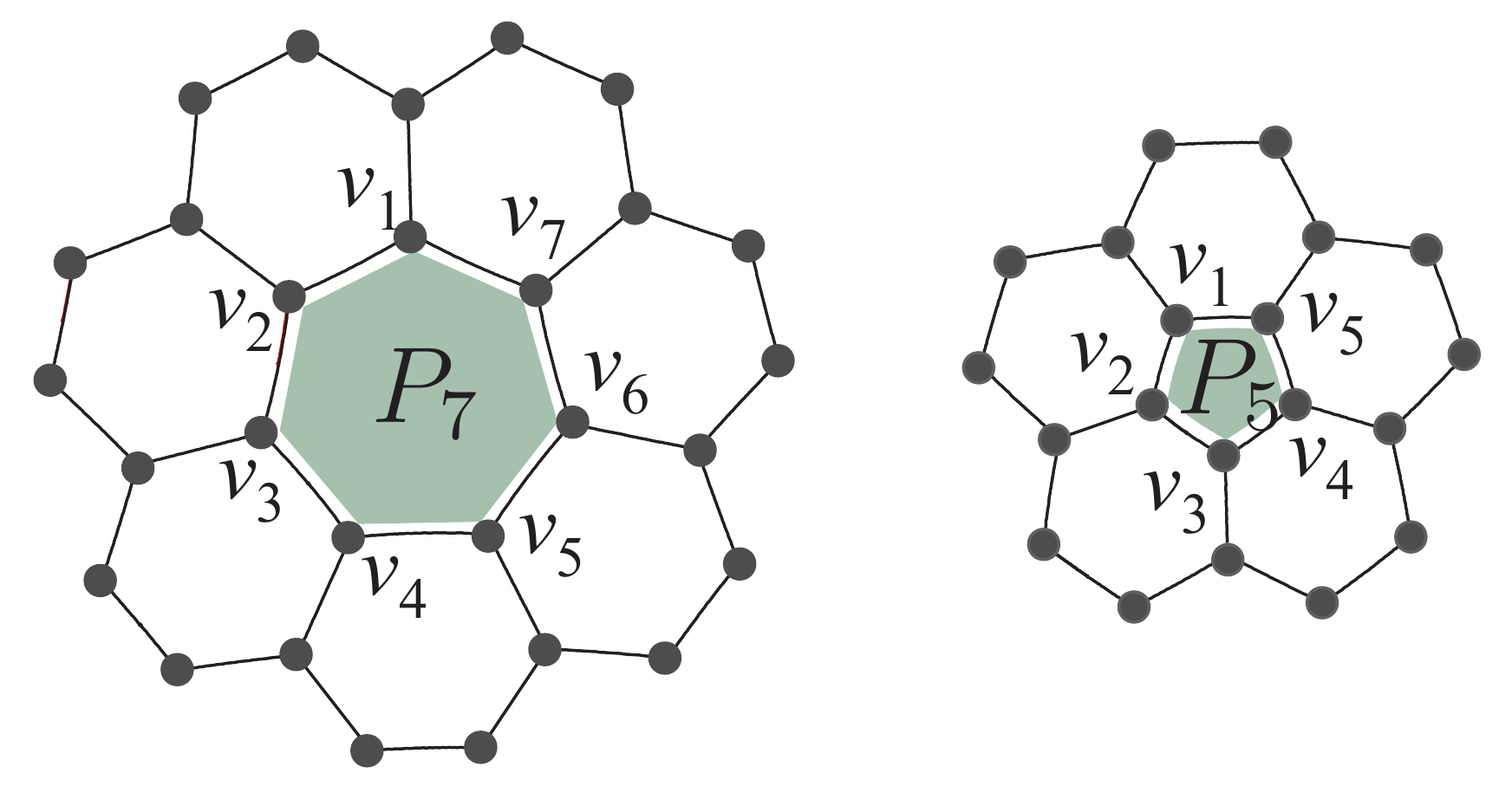}
 \caption{Heptagon and pentagon plaquettes}\label{fig:pentagonheptagon}
\end{figure}
\begin{figure}[ht]
	\centering
	\includegraphics[width=3.2in]{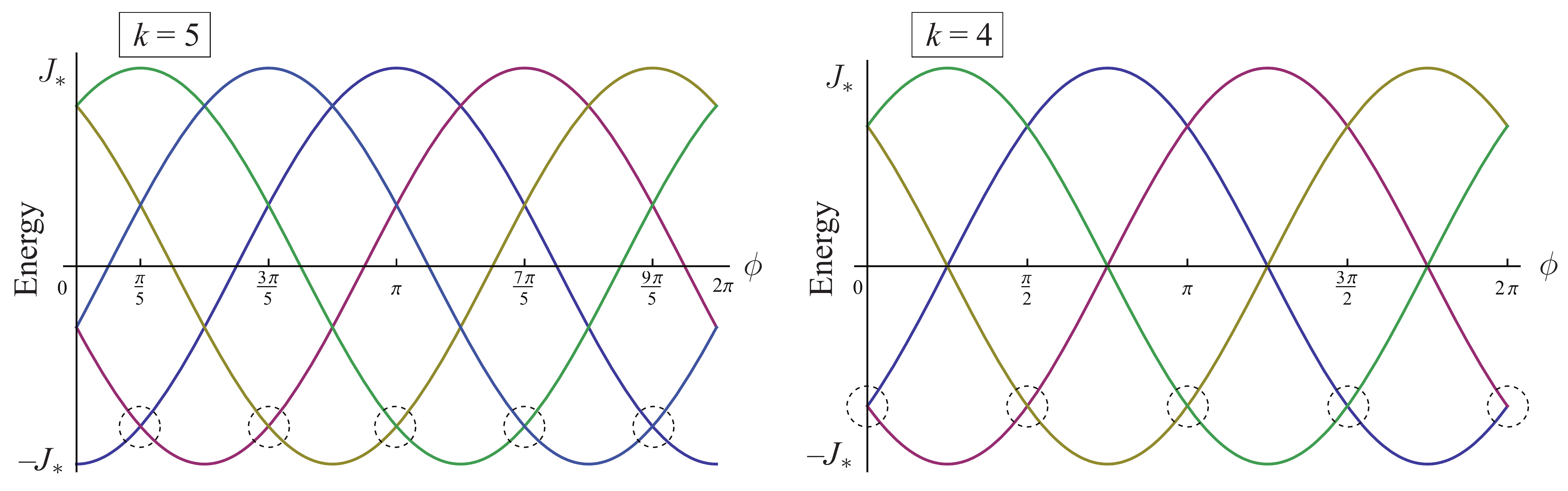}
	\caption{Energy levels of $H_{5,7}$ as a function of phase variable $\phi$. Level crossings of ground states are shown in dashed circles. Colors label $k$ different discrete values for $p(\phi)$.}\label{fig:levelcrossing}
\end{figure}

Heptagon and pentagon are odd-sided plaquettes and therefore the two operators \eqref{stabilizers1} will not commute with one another. There are instead only one plaquette stabilizer at a heptagon and pentagon (or any odd-sided plaquettes) \begin{equation}\hat{P}_7=\prod_{p=i}^7\sigma_{v_i}\tau_{v_i},\quad\hat{P}_5=\prod_{p=i}^5\sigma_{v_i}\tau_{v_i}\label{pentagonheptagonoperator}\end{equation} All faces around the heptagon and pentagon are ordinary even-sided plaquettes. Each carries two usual stabilizers \eqref{stabilizers1} and commute with the heptagon and pentagon. Similar to dislocation operators in Kitaev's toric code~\cite{Bombin} or $\mathbb{Z}_k$-Wen plaquette model~\cite{YouWen}, the problem with $\hat{P}_7$ and $\hat{P}_5$ is that they are not necessarily $k^{th}$ roots of unity since $(\sigma\tau)^k=w^{k(k-1)/2}=(-1)^{k-1}$. And in general the defect Hamiltonian can depend on a phase \begin{equation}H_{5,7}(\phi)=-J_\ast\left(e^{-i\phi}\hat{P}_{5,7}+e^{i\phi}\hat{P}_{5,7}^\dagger\right)\label{H57}\end{equation} Its ground state is labeled by the stabilizer eigenvalues $\hat{P}_{5,7}=e^{2\pi ip(\phi)/k}$ for $p(\phi)$ being the integer for $k$ odd (or half-integer for $k$ even) between \begin{equation}\frac{k\phi}{2\pi}-\frac{1}{2}<p(\phi)<\frac{k\phi}{2\pi}+\frac{1}{2}\end{equation} There are level crossings when $k\phi/2\pi$ is a half-integer (or integer) when $k$ is odd (resp~even) where two eigenstates for $\hat{P}_{5,7}$ have the same energy and the system becomes gapless (see figure~\ref{fig:levelcrossing}).

We treat this phase variable $\phi$ as a non-dynamic classical parameter associate to a twofold twist defect. This is a fundamental feature differing the twist defect from usual deconfined non-abelian anyons in topologically ordered system. $p(\phi)$ modulo $k$ is a locally measurable quantity by the pentagon or heptagon plaquette operator $\hat{P}_{5,7}$. In general twofold twist defects are labeled by small Wilson loops circling them (to be discussed in the following subsection) and are divided into $k^2$ species apart from their colors $\chi$. Each species has distinct fusion and braiding characteristics such as pair measurement restrictions and topological spin as will be seen in section~\ref{sec:defectfusion} and ~\ref{sec:defectexchangebraiding}. Slow evolution of the phase variable $\phi\to\phi+2\pi/k$ drives a ground state adiabatically to an excited state due to a level crossing shown in figure~\ref{fig:levelcrossing}. The excited state can relax to a different ground state by absorbing or emitting an abelian anyon and thereby driving a {\em species mutation} of the twofold twist defect. A successive phase winding of a series of twofold twist defects can lead to non-local transportation of abelian anyons without actually moving the defects. This pumping process is a $\mathbb{Z}_k$ analogue of the $U(1)$ Thouless charge pump~\cite{Thouless,NiuThouless} or the $\mathbb{Z}_2$ fermion parity pump~\cite{Kitaevchain,TeoKane}.

The ground state degeneracy of multiple twofold twist defects can be estimated in the thermodynamic limit by counting degree of freedom. Assume the number heptagons and pentagons are identical so that they do not contribute to the net curvature in the Gauss Bonnet theorem. Since there are only one stabilizer per pentagon or heptagon, the ground state degeneracy scales as $k^{N}$ where $N$ is the total number of twofold twist defects. And therefore its quantum dimension is given by \begin{equation}d_{[1/2]}=k\label{Z2quantumdimension}\end{equation} This matches the $\sqrt{k}$ quantum dimension for dislocation twist defects in $\mathbb{Z}_k$-rotor model~\cite{YouWen} or $\sqrt{2}$ for Kitaev's toric code~\cite{Bombin} since our Hamiltonian \eqref{ham1} is identical to two copies of quantum double $\mathbb{Z}_k$-model. The difference here is that there are three types of twofold twist defects $[1/2]_\chi$ labeled by colors $\chi$, which is the plaquette color for primitive pentagon or heptagon defects. Two identical twofold defects fuse to $k^2$ measurable abelian channels while a pair of twofold defects of different colors fuse into a threefold one, $[1/2]_\chi\times[1/2]_{\chi\pm1}=[1/3]$ or $[\overline{1/3}]$ as will be discussed further in section~\ref{sec:defectfusion}.

\subsubsection{Composite lattice defects}\label{sec:compositelatticedefects}
Classical lattice defects are topologically classified by the {holonomy} around it~\cite{Mermin79, ChaikinLubensky, Nelsonbook, KlemanFriedel08} such as the Burgers' vector ${\bf B}$ of a dislocation or the Frank angle $\Omega$ of a disclination. The holonomy is an element in the space group counting the net amount of discrete rotations $r(\Omega)$ and translations ${\bf t}$ along a loop around a point defect or a combination of them on a lattice. The space group in our model on a honeycomb lattice is the semi-direct product $P6=C_6\ltimes\mathcal{L}$, where $C_6$ is the 6-fold rotation group and $\mathcal{L}=\mathbb{Z}^2$ is the translation group of a triangular lattice. The holonomy $(r(\Omega),{\bf t})$ is path independent as long as the loop encloses the same defects but it may change according to conjugacy transformation into $(r(\Omega),{\bf t}+r(\Omega)\cdot{\bf d}-{\bf d})$ if the starting and ending point of the loop is displaced by ${\bf d}$. Thus lattice defects are precisely characterized by the conjugacy classes of holonomy $(r(\Omega),[{\bf t}]_\Omega)$, where $r(\Omega)=e^{i\Omega\sigma_y}$ is rotation with Frank angle $\Omega$, a multiple of $\pi/3$, and $[{\bf t}]_\Omega$ is the conjugacy class of translation in the quotient group \begin{equation}\frac{\mathcal{L}}{(r(\Omega)-1)\mathcal{L}}=\left\{\begin{array}{*{20}c}\mathcal{L},\hfill&\mbox{for $\Omega=0$}\hfill\\0,\hfill&\mbox{for $\Omega=\pm60^\circ$}\hfill\\\mathbb{Z}_3,\hfill&\mbox{for $\Omega=\pm120^\circ$}\hfill\\\mathbb{Z}_2\oplus\mathbb{Z}_2,&\mbox{for $\Omega=180^\circ$}\hfill\end{array}\right.\label{disclinationclassification}\end{equation} In particular $[{\bf t}]_\Omega$ differentiates disclinations with the same Frank angle and curvature singularity. The $\mathbb{Z}_3$ classification distinguishes $120^\circ$ disclinations centered at octagons such as figure~\ref{fig:squareoctagon}(a), tetravalent $\circ$-vertices such as figure~\ref{fig:twistdefects}(a) and a tetravalent $\bullet$-vertices. And therefore tricolorability is violated when $[{\bf t}]_{120^\circ}$ is non-trivial in $\mathbb{Z}_3$. This remains true for any composite defect with an overall $\pm120^\circ$ Frank angle. Since the classification for $\pm60^\circ$-disclinations is trivial, all such composite defect is holonomically equivalent to a pentagon or heptagon which breaks both tricolorability and bipartite structure. %$\mathbb{Z}_2\oplus\mathbb{Z}_2$ classification for $180^\circ$ disclinations shows a nonagon defect and a vertex surrounded by three heptagons are fundamentally distinct.

Dislocations are composite defects consist of a collection of disclinations with canceling Frank angles and curvature. The torsional singularity of a dislocation, characterized holonomically by a Burgers' vector ${\bf B}$ in $\mathcal{L}$, originates from the spacial separation of its constituent disclinations. For example, each dislocation in figure~\ref{fig:Z3dislocation} is a disclination dipole with a tetravalent vertex ($\Omega=+120^\circ$) separated from a square ($\Omega=-120^\circ$) by half a lattice spacing. And figure~\ref{fig:twistdefects}(c) shows an overall dislocation from the $\pm60$-disclination pair. All dislocation preserves the $\bullet,\circ$ bipartite structure since it always consists of the same number of pentagons and heptagons. It violates tricolorability when its Burgers' vector ${\bf B}$ sends a hexagon plaquette to a different color one. Examples of dislocation twist defects include $\pm120^\circ$-disclination dipole with unequal translation pieces $[{\bf t}]_{+120^\circ}\neq[{\bf t}]_{+120^\circ}\in\mathbb{Z}_3$ (i.e. inequivalent defect centers, $\bullet,\circ$-vertex or plaquette) such as those in figure~\ref{fig:Z3dislocation}, and $\pm60^\circ$ disclination dipole with different color pentagon and heptagon. 

Let $\mathcal{L}'$ be the translation subgroup of $\mathcal{L}$ that preserves tricoloration and that $\mathcal{L}/\mathcal{L}'=\mathbb{Z}_3$ classified $\pm120^\circ$-disclinations in \eqref{disclinationclassification}. Any dislocation with Burgers' vector in $\mathcal{L}'$ and any $\pm120^\circ$-disclination with trivial $\mathbb{Z}_3$ translation piece $[{\bf t}]_{120^\circ}$ are {\em twistless} defects that do not violate tricolorability and bipartite strucutre such as those in figure~\ref{fig:squareoctagon}. The $P6$ honeycomb space group symmetry collapses when there is are twistless dislocations and $\pm120^\circ$-disclinations in the lattice that break discrete $\mathcal{L}'$ translation and $C_3$ rotation symmetry respectively. 
%The trivalent lattice structure collapses into a continuum upon
%proliferation of twistless disolcations and $\pm120^\circ$-disclinations that
%break discrete $\mathcal{L}'$ translation and $C_3$ rotation symmetry
%respectively, and thereby restoring continuous translation and rotation
%symmetry in long length scale. 
Certain {\em discreteness} in the space group that represents tricoloration and bipartite structure is however left over. The presence of twistless lattice defects breaks the space group symmetry into the following residue \begin{equation}\frac{C_6\ltimes\mathcal{L}}{C_3\ltimes\mathcal{L}'}=\mathbb{Z}_2\ltimes\mathbb{Z}_3=S_3\label{spacegroupquotient}\end{equation} which is not surprisingly identical to the group of symmetry between abelian anyons \eqref{ST} and \eqref{S} or gauge fields \eqref{STCS} in \eqref{CS}. In a trivalent bipartite graph, defects are holonomically classified by the {\em residue} group $S_3$. For instance, a pentagon defect is indistinguishable from a heptagon one as $\pm60^\circ$-disclinations are interchangeable by absorbing or releasing a square or octagon disclination. A primitive dislocation is equivalent to a tetravalent vertex by releasing a square disclination and in the long length scale indistinguishable from a pentagon-heptagon dipole separated by one lattice spacing. 

\subsection{Non-local Wilson algebra}\label{sec:nonlocalWilsonalgebra}
Figure~\ref{fig:anyontwist} shows how an abelian anyon changes type around a twist defect and its Wilson string does not close back to itself after one cycle. A closed Wilson string is either non-local so that it encloses multiple twist defects or wraps around a twist defect multiple times. The former contributes to a ground state degeneracy as it intersects and therefore does not commute with other non-local Wilson strings. The latter can be represented by an abelian phase as it can be shrunk to the point defect and will not intersect with other Wilson strings. 

\begin{figure}[ht]
	\centering
	\includegraphics[width=3.2in]{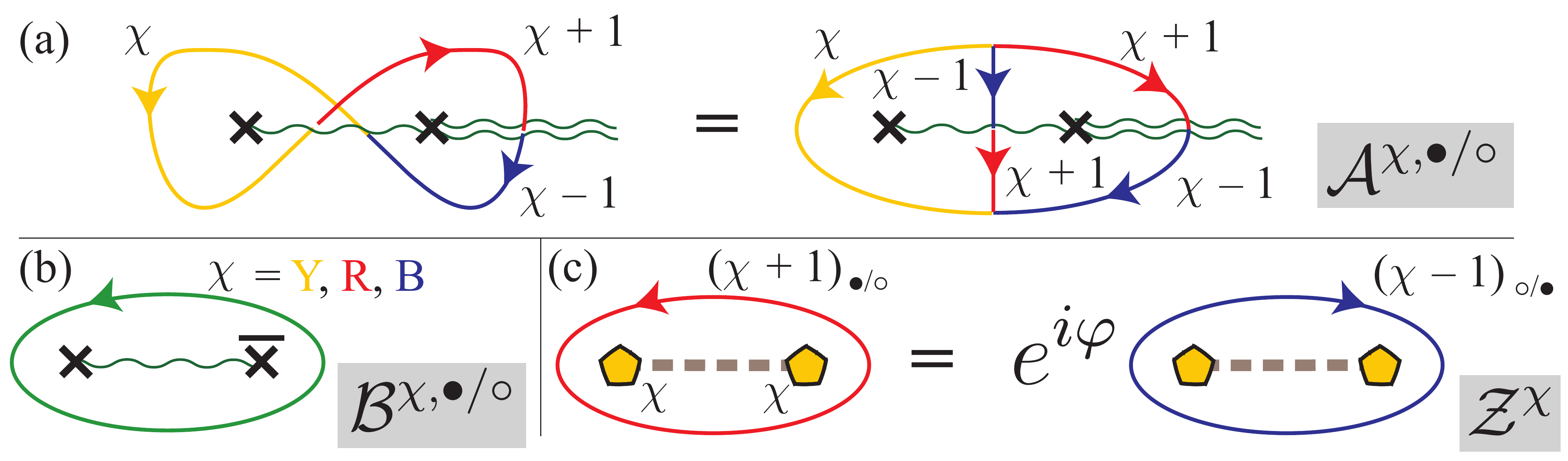}
	\caption{Prototypes of non-local closed Wilson loops around twist defects. (a) A closed string $\mathcal{A}^{\chi,\bullet/\circ}$, either entirely $\bullet$-type or $\circ$-type, enclosing two threefold twist defects $[1/3]$ (black crosses). It can be equivalently represented by a loop (left) or a closed branched path with a tricolored source and drain (right).\cite{notesZ3Wilsonloop} Wilson strings change colors $\chi\in\{Y,R,B\}=\{0,1,2\}$ across threefold branch cuts (curly lines). (b) A Wilson loop $\mathcal{B}^{\chi,\bullet/\circ}$ surrounding a threefold twist defect $[1/3]$ and its anti-particle $[\overline{1/3}]$. (c) A Wilson loop $\mathcal{Z}^{\chi}$ containing two twofold twist defects of the same color $[1/2]_\chi$ (colored pentagons). Up to an abelian phase, the loop can be flipped inside out and changes color and sublattice types according to figure~\ref{fig:anyontwist}(c).}\label{fig:defectWilsonloop}
\end{figure}
\begin{figure}[ht]
	\centering
	\includegraphics[width=3in]{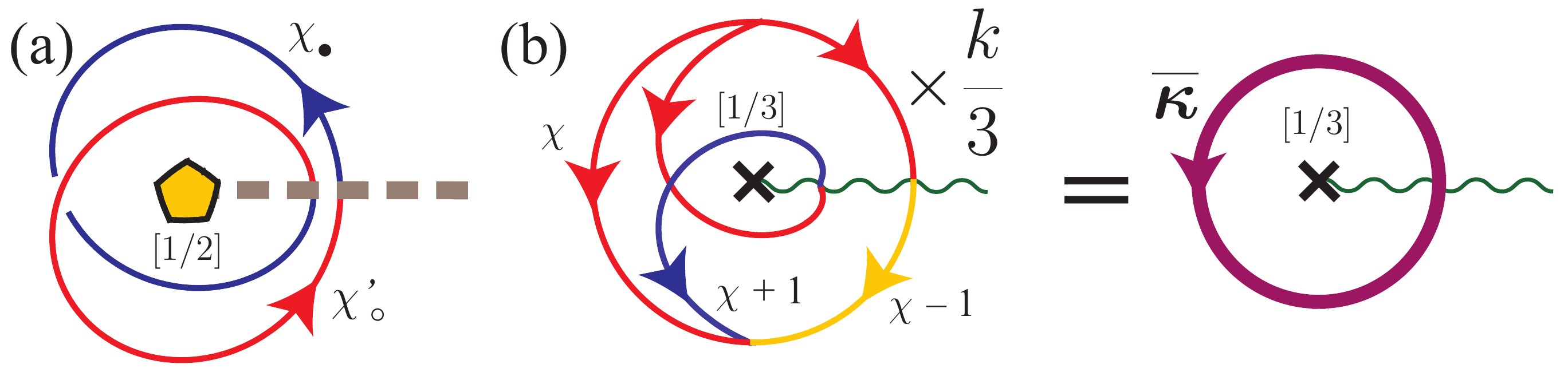}
	\caption{Local Wilson loops that circle a single defect. (a) Species identification of a twofold twist defect by small Wilson loops $\Theta_\chi$ that circle twice. The $\bullet$-string is chosen by convention to sit below a $\circ$-one at the intersection. (b) A local Wilson loop, $\Sigma_\bullet$ or $\Sigma_\circ$, surrounding a single threefold twist defect $k/3$ times. It consists of a $k$-valent uni-color source and $k/3$ tricolored drains, or equivalently constructed by dragging the color permutation invariant abelian anyon $\overline{\boldsymbol\kappa}_{\bullet/\circ}=(Y_{\bullet/\circ})^{-k/3}(R_{\bullet/\circ})^{k/3}$ (see figure~\ref{fig:abeliananyonlattice}) around the defect. It exists only when $k$ is divisible by 3.}\label{fig:defectlocalloop}
\end{figure}
All non-local Wilson loops are generated by primitive ones shown in figure~\ref{fig:defectWilsonloop} (and another one in figure~\ref{fig:splittingspaces}(g) which will not be used in this section), each encircles two twist defects. A
particular arbitrary choice of branch cut is assigned in the figure to keep track of color and sublattice transformation according to figure~\ref{fig:anyontwist} and \ref{fig:twistdefects}. Twist defects are further subdivided into species according to the abelian phase eigenvalues of local Wilson loops, $\Theta_\chi$ around twofold twist defect and $\Sigma_\bullet,\Sigma_\circ$ around threefold twist defects when $k$ is divisible by 3. Each circles a single defect multiple times as shown in figure~\ref{fig:defectlocalloop}. $\Theta_\chi$ are continuum versions of defect pentagon and heptagon operators \eqref{H57}, and $\Sigma_\bullet,\Sigma_\circ$ are related to the good quantum numbers associate with each tetravalent and bivalent vertex \eqref{24Sigma}.

\subsubsection{Threefold twist defects}\label{sec:Z3twistdefects}
A threefold twist defect is characterized by its 3-fold anyon color twisting described in figure~\ref{fig:anyontwist}. It can be generated by primitive dislocation, tetravalent or bivalent lattice disclination, or any composite defect with the same overall color permutation character. 
\begin{figure}[ht]
	\centering
	\includegraphics[width=2in]{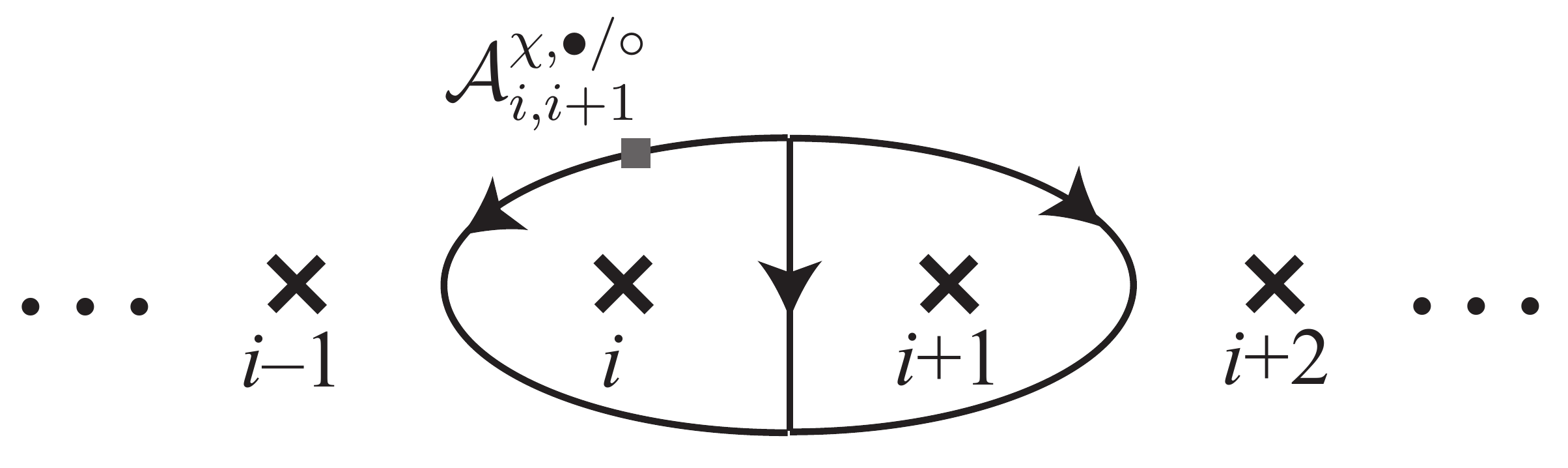}
	\caption{Wilson loop $\mathcal{A}_{i,i+1}^{\chi,\bullet}$ and $\mathcal{A}_{i,i+1}^{\chi,\circ}$ between the $i^{th}$ and $(i+1)^{th}$ threefold twist defect. The sublattice type $\bullet,\circ$ and color $\chi=Y,R$ is view at the base point (grey square) on the loop.}\label{fig:Z3loopdefinition}
\end{figure}
We consider a system of $N$ threefold twist defects $[1/3]$ for $N$ is some multiple of three so that the system can be closed on a sphere, i.e. the $N$ defects fuse to the trivial abelian anyon channel. All non-local Wilson loops are combinations of the one in figure~\ref{fig:defectWilsonloop}(a) between neighboring defects and we label the one enclosing the $i^{th}$ and $(i+1)^{th}$ defect by $\mathcal{A}_{i,i+1}^{\chi,\bullet}$ and $\mathcal{A}_{i,i+1}^{\chi,\circ}$ (see figure~\ref{fig:Z3loopdefinition}), where $\chi$ is the color of the string at the fixed view point (grey square). Due to color redundancy $Y\times R\times B=1$, it is enough to take $\chi=Y,R$. And the final loop $\mathcal{A}_{N-1,N}$ can be expressed in terms of the previous ones because the system is closed and the Wilson loop enclosing all $N$ defects is trivial. \begin{align}\prod_{i=1}^{N/3}\mathcal{A}_{3i-2,3i-1}^{\chi,\bullet/\circ}\left(\mathcal{A}_{3i-1,3i}^{\chi-1,\bullet/\circ}\right)^\dagger=\vcenter{\hbox{\includegraphics[width=0.8in]{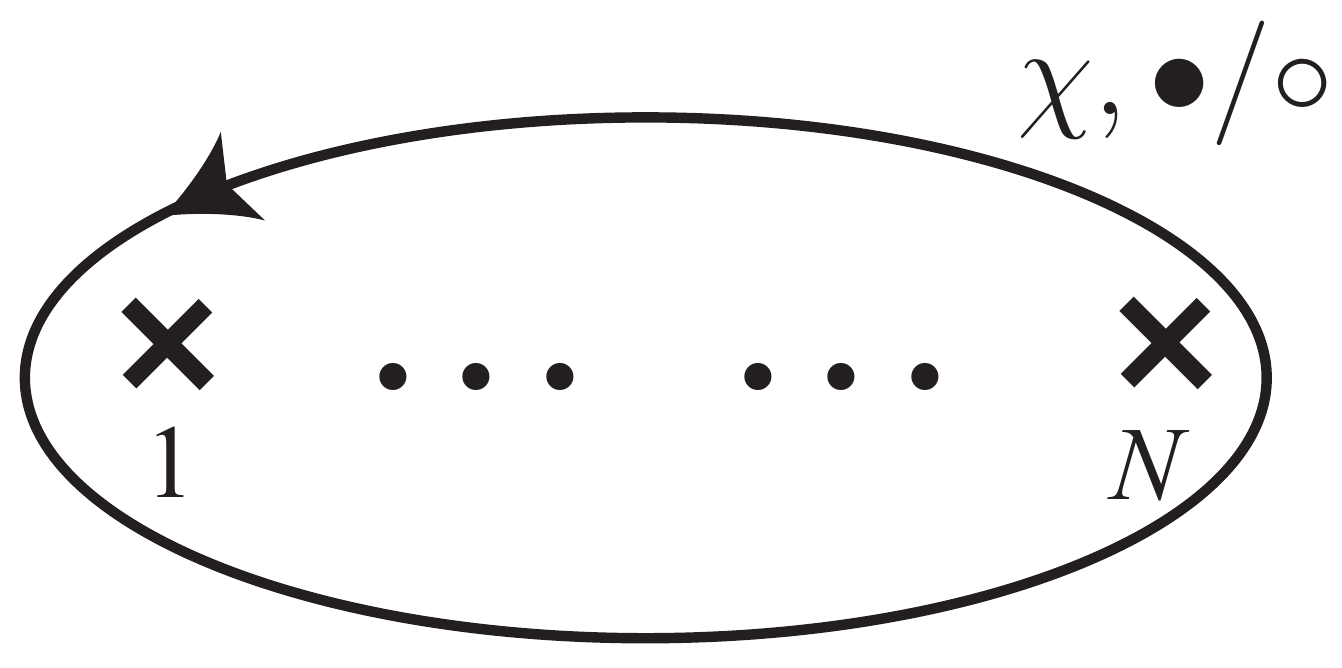}}}=1\label{Z3closeness}\end{align} for $[i]_3\equiv i$ modulo 3 cyclically permutes the color $\chi$. 

\begin{figure}[ht]
	\centering
	\includegraphics[width=3.5in]{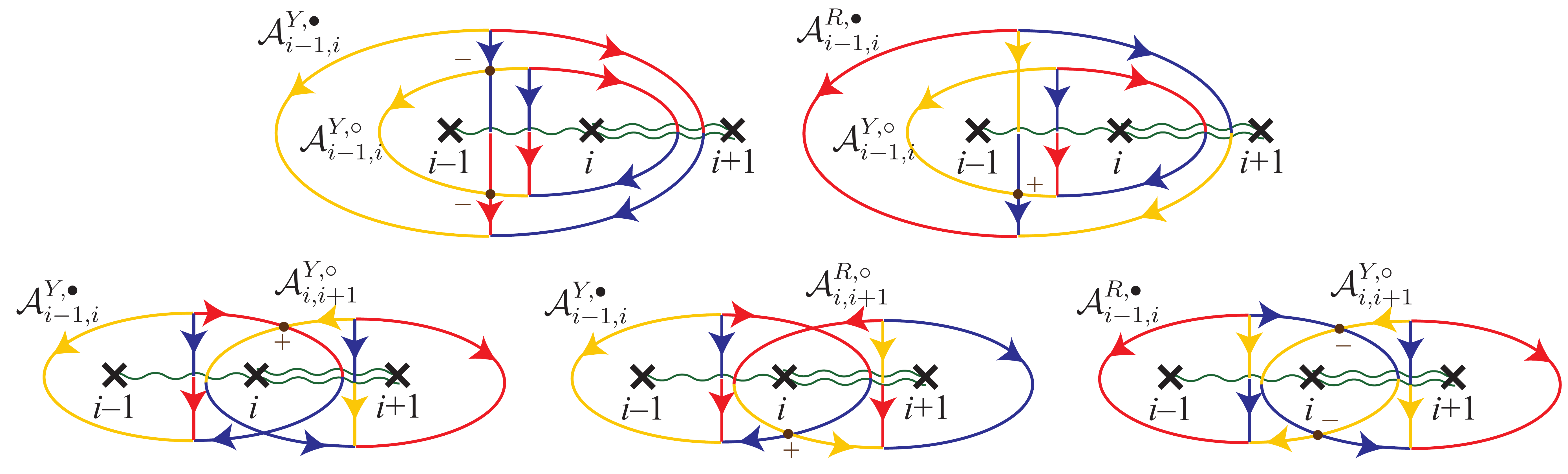}
	\caption{Intersection (brown dots) between nearest and next nearest $\bullet$- and $\circ$-type Wilson loops.}\label{fig:Z3intersection}
\end{figure}
Since $\bullet$-Wilson loops mutually commute, the $2(N-2)$ $\bullet$-loops form a maximal commuting set of Wilson operators. The Wilson algebra \begin{equation}\mathcal{A}_{i,i+1}^{\chi,\bullet}\mathcal{A}_{j,j+1}^{\chi',\circ}=e^{i\frac{2\pi}{k}\langle\mathcal{A}_{i,i+1}^\chi,\mathcal{A}_{j,j+1}^{\chi'}\rangle}\mathcal{A}_{j,j+1}^{\chi',\circ}\mathcal{A}_{i,i+1}^{\chi,\bullet}\label{Z3Wilsonalgebra}\end{equation} is determined by the symmetric pairing matrix $\mathbb{I}=\langle\ast,\ast\rangle$ between colored strings. It is non-zero only when $|i-j|\leq1$ and can be evaluated by local intersection number according to figure~\ref{fig:intersection}. All non-trivial intersection are shown in figure~\ref{fig:Z3intersection}. Note that the result is independent from the arbitrary choice of branch cut. The intersection form for $N$ threefold twist defects is given by the $2(N-2)\times2(N-2)$ symmetric matrix $\mathbb{I}$ \begin{equation}\langle\mathcal{A}_{i,i+1},\mathcal{A}_{j,j+1}\rangle=\left(\begin{array}{*{20}c}I_0&I_1&0&0&\ldots&0\\I_1^T&I_0&I_1&0&\ldots&0\\0&I_1^T&I_0&I_1&\ldots&0\\0&0&I_1^T&I_0&\ldots&0\\\vdots&\vdots&\vdots&\ddots&\ddots&\vdots\\0&0&0&0&\ldots&I_0\end{array}\right)\label{Z3intersectionmatrix}\end{equation} \begin{equation}I_0=\left(\begin{array}{*{20}c}-2&1\\1&-2\end{array}\right),\quad I_1=I_0\Lambda_3=\left(\begin{array}{*{20}c}1&1\\-2&1\end{array}\right)\label{I0I1}\end{equation} where the $(N-2)$ rows and columns of the intersection matrix \eqref{Z3intersectionmatrix} is enumerated according to the loops $\{\mathcal{A}_{i,i+1}:i=1,\ldots,N-2\}$, its $2\times2$ matrix entries \eqref{I0I1} act on color degree of freedom $\chi=Y,R$, and $\Lambda_3=\left(\begin{array}{*{20}c}0&-1\\1&-1\end{array}\right)$ is cyclic color permutation. $I_0$ comes from the intersection between Wilson loops about the same defect pair (first two diagrams in figure~\ref{fig:Z3intersection}) while $I_1$ comes from the intersection between Wilson loops about adjacent defect pairs (last three diagrams in figure~\ref{fig:Z3intersection}). 

We notice that the Wilson algebra \eqref{Z3Wilsonalgebra} is symmetric under cyclic color permutation $\Lambda_3^{\pm1}:\chi\to\chi\pm1$ but is not invariant under color and sublattice transposition $\Lambda_B:Y_{\bullet/\circ}\leftrightarrow R_{\circ/\bullet}$. This can be understood by observing that the Wilson loop in figure~\ref{fig:defectWilsonloop}(a) violates transposition explicitly. Color label $\chi$ of the Wilson operator $\mathcal{A}_{i,i+1}^{\chi,\bullet/\circ}$ in figure~\ref{fig:Z3loopdefinition} is defined with respect to a base point, and a change of base point does not commute, hence inconsistent, with all transpositions $\Lambda_Y,\Lambda_R,\Lambda_B$.  In fact color and sublattice transposition switches a threefold defect into its anti-partner \begin{equation}\Lambda_B:[1/3]\leftrightarrow[\overline{1/3}]\end{equation} and the intersection form is {\em covariant} under $\Lambda_B$ \begin{equation}\Lambda_B^T\cdot\mathbb{I}_{[1/3]}\cdot\Lambda_B=-\mathbb{I}_{[\overline{1/3}]}\end{equation} where $\Lambda_B=\sigma_x$ acts on the colors $\chi=Y,R$ and $\mathbb{I}_{[\overline{1/3}]}$ is the intersection matrix for $N$ anti-threefold defects. In the microscopic lattice level, transposition $\Lambda_B$ originates from inversion which interchanges the $\bullet,\circ$ sublattice type and switches $[1/3]\leftrightarrow[\overline{1/3}]$ according to table~\ref{tab:twistdefects} for primitive disclination twist defects. This can also be understood in the continuum by seeing transposition interchanges the abelian anyon twisting between $[1/3]$ and $[\overline{1/3}]$ in figure~\ref{fig:anyontwist}(a) and (c). 

For later convenience, we adapt the multi-exponents notation ${\bf m}=(m_1,m_2,\ldots,m_{2N-5},m_{2N-4})$ for Wilson operator product \begin{equation}\left(\mathcal{A}^\bullet\right)^{\bf m}=\prod_{i=1}^{N-2}\left(\mathcal{A}_{i,i+1}^{Y,\bullet}\right)^{m_{2i-1}}\left(\mathcal{A}_{i,i+1}^{R,\bullet}\right)^{m_{2i}}\end{equation} and similarly for the dual ones $\left(\mathcal{A}^\circ\right)^{\bf m}$, where $m_j$ are integers modulo $k$.

When $k$ is {\em not} divisible by 3, starting with the particular ground state $|0\rangle_\bullet$ in eq.\eqref{GS0} fixed by the $\bullet$-Wilson operators $\left(\mathcal{A}^\bullet\right)^{\bf m}|0\rangle_\bullet=|0\rangle_\bullet$, the dual $\circ$-Wilson operators generate all ground states by product combinations \begin{equation}|{\bf m}\rangle_\bullet=\left(\mathcal{A}^\circ\right)^{\bf m}|0\rangle_\bullet\label{GSZ3}\end{equation} These are simultaneous eigenstates for the maximally commuting set of $\bullet$-Wilson operators. The matrix elements of $\mathcal{A}^\bullet$ and $\mathcal{A}^\circ$ with respect to this basis are given by \begin{align}\langle{\bf m}'|\left(\mathcal{A}^\bullet\right)^{\bf n}|{\bf m}\rangle&=e^{i\frac{2\pi}{k}{\bf n}^T\mathbb{I}{\bf m}}\delta_{{\bf m}',{\bf m}}\label{Z3Wilsoneigenvalues}\\\langle{\bf m}'|\left(\mathcal{A}^\circ\right)^{\bf n}|{\bf m}\rangle&=\delta_{{\bf m}',{\bf m}+{\bf n}}\end{align} where $\mathbb{I}$ is the intersection form in \eqref{Z3intersectionmatrix}.

\begin{figure}[ht]
	\centering
	\includegraphics[width=2.5in]{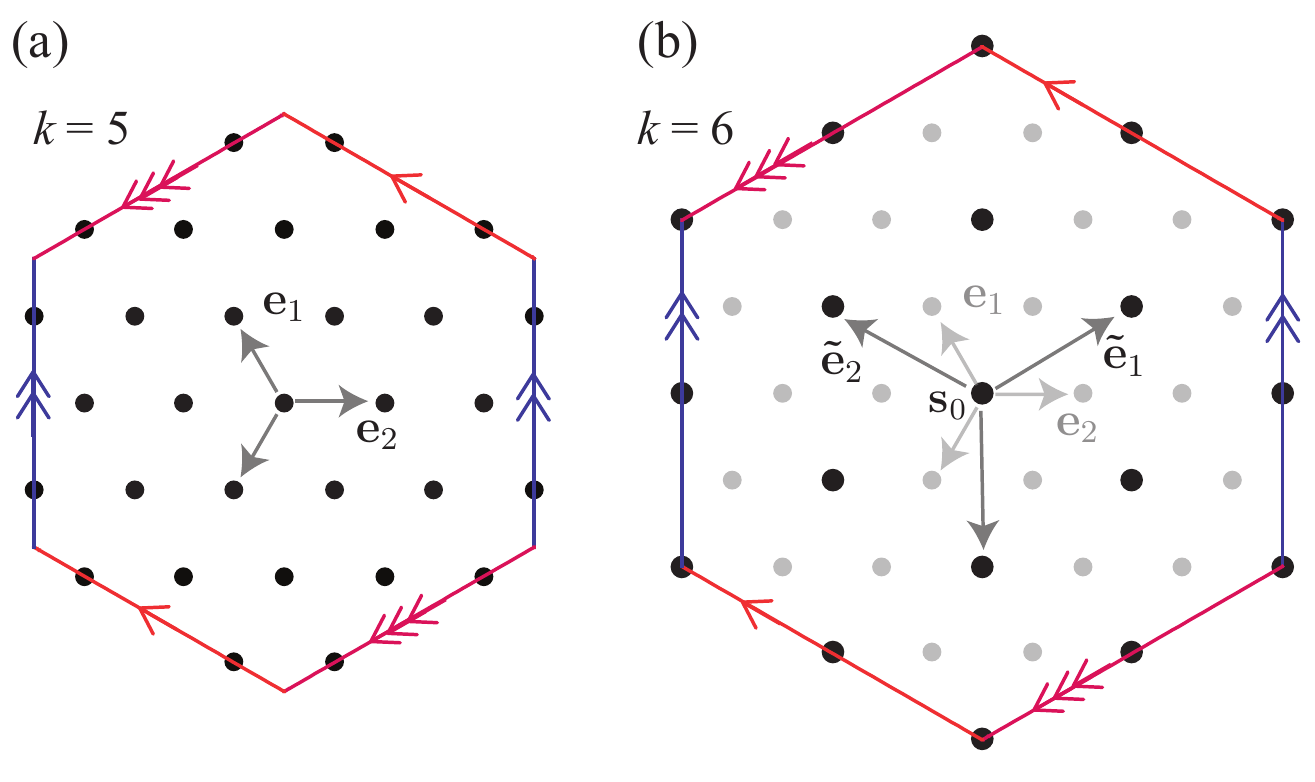}
	\caption{Eigenvalue lattice for the $\bullet$-Wilson operators for each $i=1,2,\ldots,N-2$. $(\mathcal{A}_{i,i+1}^{Y,\bullet},\mathcal{A}_{i,i+1}^{R,\bullet})=(e^{2\pi iy_i/k},e^{2\pi ir_i/k})$ at the lattice point ${\bf a}_i=y_i{\bf e}_1+r_i{\bf e}_2$. (a) Lattice for $k$ not divisible by 3. (b) Lattice for $k$ divisible by 3 with shifted origin ${\bf s}_0$ that depends on the $\mathbb{Z}_3$-value of \eqref{Sigmaii+1}. The thickened sublattice contains the eigenvalues allowed by \eqref{Sigmaii+1}.}\label{fig:3foldsublattice}
\end{figure}
We notice that $3\mathbb{I}^{-1}$ has integer entries, and therefore the intersection matrix $\mathbb{I}$ is invertible (with $\mathbb{Z}_k$ entries) when there is an integer $s$ with $3s\equiv1$ mod $k$, in other words $k$ is not divisible by 3. This means different ground states $|{\bf m}\rangle_\bullet$ in \eqref{GSZ3} can be distinguished by their set of eigenvalues for the $\bullet$-Wilson operators, and thus \eqref{GSZ3} forms a complete orthonormal basis for the ground state. Alternatively one can label the ground states using $\bullet$-Wilson operator eigenvalues ${\boldsymbol\alpha}=(y_1,r_1,\ldots,y_{N-2},r_{N-2})$ by \begin{equation}|{\boldsymbol\alpha}\rangle_\bullet=\left(\mathcal{A}^\circ\right)^{\bf m}|0\rangle_\bullet,\quad{\bf m}=\mathbb{I}^{-1}{\boldsymbol\alpha}\label{Z3groundstatealpha}\end{equation} where $y_i,r_i$ are integers modulo $k$, so that the matrix elements of Wilson operators are \begin{align}\langle{\boldsymbol\alpha}'|\left(\mathcal{A}^{\bullet}\right)^{\bf n}|{\boldsymbol\alpha}\rangle&=e^{i\frac{2\pi}{k}{\bf n}^T{\boldsymbol\alpha}}\delta_{{\boldsymbol\alpha}',{\boldsymbol\alpha}}\label{Z3Wilsonmatix1}\\\langle{\boldsymbol\alpha}'|\left(\mathcal{A}^{\circ}\right)^{\bf n}|{\boldsymbol\alpha}\rangle&=\delta_{{\boldsymbol\alpha}',\mathbb{I}{\bf n}+{\boldsymbol\alpha}}\label{Z3Wilsonmatix2}\end{align} One can put the ground states $|{\boldsymbol\alpha}\rangle_\bullet$ on the Cartesian product of $N-2$ periodic 2D lattices shown in figure~\ref{fig:3foldsublattice}(a). Each lattice point ${\boldsymbol\alpha}_i=y_i{\bf e}_1+r_i{\bf e}_2$ on the integer mod $k$ triangular lattice represents the eigenvalues $e^{2\pi iy_i/k},e^{2\pi ir_i/k}$ for $\mathcal{A}_{i,i+1}^{Y,\bullet},\mathcal{A}_{i,i+1}^{R,\bullet}$. The ground state degeneracy (G.S.D.) for a closed spherical system with $N$ threefold twist defects is given by \begin{equation}G.S.D.=k^{2(N-2)}\label{Z3dimensionGSD}\end{equation} which matches the quantum dimension $d_{[1/3]}=k^2$ predicted by counting lattice degree of freedom in eq.\eqref{Z3dimension}.

When $k$ is divisible by 3, there is a non-trivial center that commute with all Wilson operators. It is generated by \begin{align}\Sigma_{i,i+1}^{\bullet/\circ}&=\left(\mathcal{A}_{i,i+1}^{Y,\bullet/\circ}\right)^{-k/3}\left(\mathcal{A}_{i,i+1}^{R,\bullet/\circ}\right)^{k/3}\nonumber\\&=\Sigma_i^{\bullet/\circ}\left(\Sigma_{i+1}^{\bullet/\circ}\right)^{-1}=e^{i\frac{2\pi}{3}\left(s^{\bullet/\circ}_i-s^{\bullet/\circ}_{i+1}\right)}\in\mathbb{Z}_3\label{Sigmaii+1}\end{align} where the eigenvalues of the local Wilson operator $\Sigma^\bullet=e^{2\pi is_\bullet/3}$ and $\Sigma^\circ=e^{2\pi is_\circ/3}$ in figure~\ref{fig:defectlocalloop}(b) distinguishes the nine species ${\bf s}=(s_\bullet,s_\circ)\in\mathbb{Z}_3\oplus\mathbb{Z}_3$ for each threefold defects $[1/3]_{\bf s}$. It can be shown, up to plaquette stabilizers, that the local Wilson loop $\Sigma^{\bullet/\circ}_i$ around a tetravalent (or bivalent) $\bullet/\circ$-vertex is $\Sigma_{u_4}^{-1}$ (resp.~$\Sigma_{u_2}$) in eq.\eqref{24Sigma} or 1 around a $\circ/\bullet$-vertex. And therefore the $\mathbb{Z}_3$-phases for the small loops $\Sigma^{\bullet/\circ}_i$ are determined by local defect Hamiltonian \eqref{Z33defectham}. The species are interchangeable by tuning the defect phase $\phi\to\phi\pm2\pi/3$ in \eqref{Z33defectham} that drive the system into an excited state and locally relax back to a ground state by emitting or absorbing an abelian anyon. This process changes the $\mathbb{Z}_3$-values for $\Sigma^\bullet$ and $\Sigma^\circ$ and thus switches the species. This extra phase degree of freedom provides a possibility for non-local transport of abelian anyon along a series of coupled $[1/3]$ defects that is a unique feature only when $k$ is a multiple of 3. 

The ground states $|{\bf m}\rangle_\bullet$ in \eqref{GSZ3} do not form an orthonormal set as they overcount the ground state degeneracy. In fact the allowed eigenvalues $e^{i\frac{2\pi}{k}{\bf n}^T\boldsymbol\alpha}$ for $(\mathcal{A}^\bullet)^{\bf n}$ are restricted by the species $s^{\bullet/\circ}$ in \eqref{Sigmaii+1}. The ground states $|\boldsymbol\alpha\rangle$ now form a sublattice in figure~\ref{fig:3foldsublattice}(b), which contains a third of lattice points in the origin triangular lattices. The ground state degeneracy (G.S.D.) is thus reduced by 3 for each defects. \begin{equation}G.S.D.=\left(\frac{k^2}{3}\right)^{N-2}\label{Z33dimensionGSD}\end{equation} This matches the quantum dimension $d_{[1/3]}=k^2/3$ in \eqref{Z3dimension} predicted by microscopic lattice derivation. A complete description for the Wilson structure of threefold defects when 3 divides $k$ can be found in appendix~\ref{sec:Z33Wilsonalgebraappendix}.

\subsubsection{Twofold twist defects}
A twofold twist defect is characterized by its twofold twisting of abelian anyons circling around as shown in figure~\ref{fig:anyontwist}. It can be realized as pentagon or heptagon defects in the lattice, or any composite defects that violate tricolorability and bipartite structure. Since there are three transpositions $\Lambda_Y,\Lambda_R,\Lambda_B$ in the permutation group $S_3$, twofold defects are classified into three types according to colors, $[1/2]_Y$, $[1/2]_R$ and $[1/2]_B$. We consider a closed collection of $N$ twofold twist defects for some even $N$. We assume for simplicity that all defects are of the same color type, say blue ($B$), and they fuse to the overall vacuum channel and the system is compactified on a sphere. A more general discussion on coexisting multi-type defects will be given in the upcomming subsection.

\begin{figure}[ht]
	\centering
	\includegraphics[width=2in]{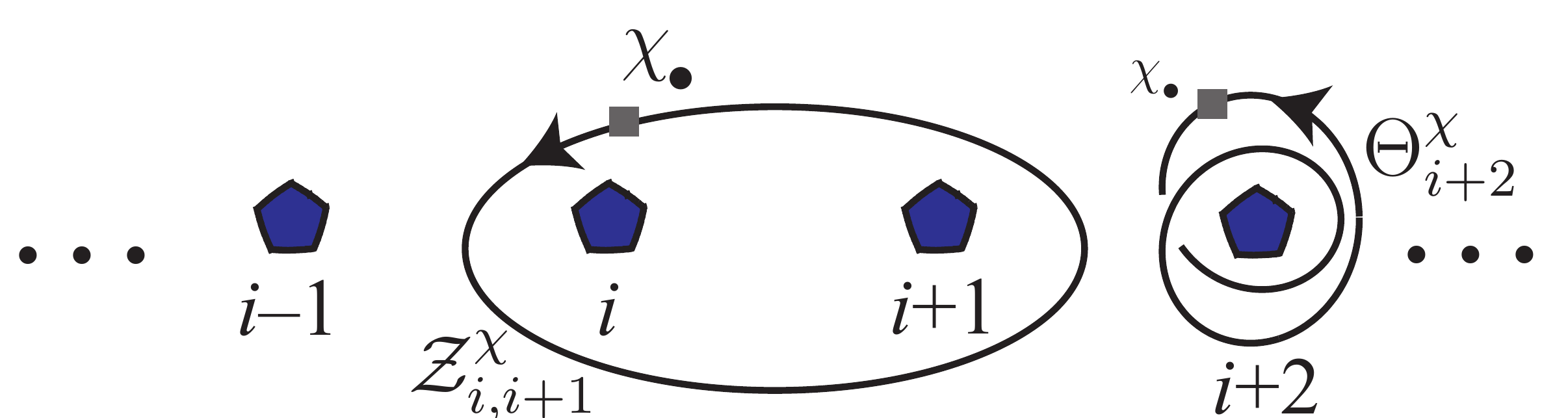}
	\caption{Wilson operator $\mathcal{Z}^\chi_{i,i+1}$ with color $\chi=Y,R,B$ and $\bullet$-sublattice type at the based point chosen at the grey square. Local Wilson loop $\Theta_i^{\chi}=w^{l_i^\chi}$, $w=e^{2\pi i/k}$, about the $i^{th}$ defect with $\chi$-colored and $\bullet$-sublattice starting point.}\label{fig:Z2loopdefinition}
\end{figure}

The twofold defect $[1/2]_B$ is further subdivided into $k^2$ species according to eigenvalues of two {\em local} measurements at the defect, the small Wilson loops $\Theta_\chi$ (see figure~\ref{fig:defectlocalloop}(a) and \ref{fig:Z2loopdefinition}) that circles the defect twice for $\chi=Y,R,B$. For primitive pentagon or heptagon defects, they are identical (up to a $\mathbb{Z}_k$ phase) to the defect plaquette operator $\hat{P}_{5,7}$ in eq.\eqref{pentagonheptagonoperator}, and therefore their eigenvalues $w^{l_\chi}$, $w=e^{2\pi i/k}$, is fixed by the local defect Hamiltonian \eqref{H57}. Similar to $\hat{P}_{5,7}$, the self-intersection of the small Wilson loop $\Theta_\chi$ causes a minus sign for even $k$ so that $(\Theta_\chi)^k=(-1)^{k-1}$ for $\chi$ not equal $B$, the color of the twofold defect. This means that $l_\chi$ is an integer modulo $k$ for $k$ odd but a half-integer modulo $k$ when $k$ is even and $\chi\neq B$. Color redundancy $Y\times R\times B=1$ requires $l_Y+l_R+l_B=0$. The two independent phases ${\bf l}=(l_Y,l_R)$ in $\mathbb{Z}_k\oplus\mathbb{Z}_k$ for $k$ odd or in $(\mathbb{Z}+1/2)_k\oplus(\mathbb{Z}+1/2)_k$ for $k$ even form the species label for $[1/2]_{B,{\bf l}}$.

\begin{figure}[ht]
	\centering
	\includegraphics[width=2.5in]{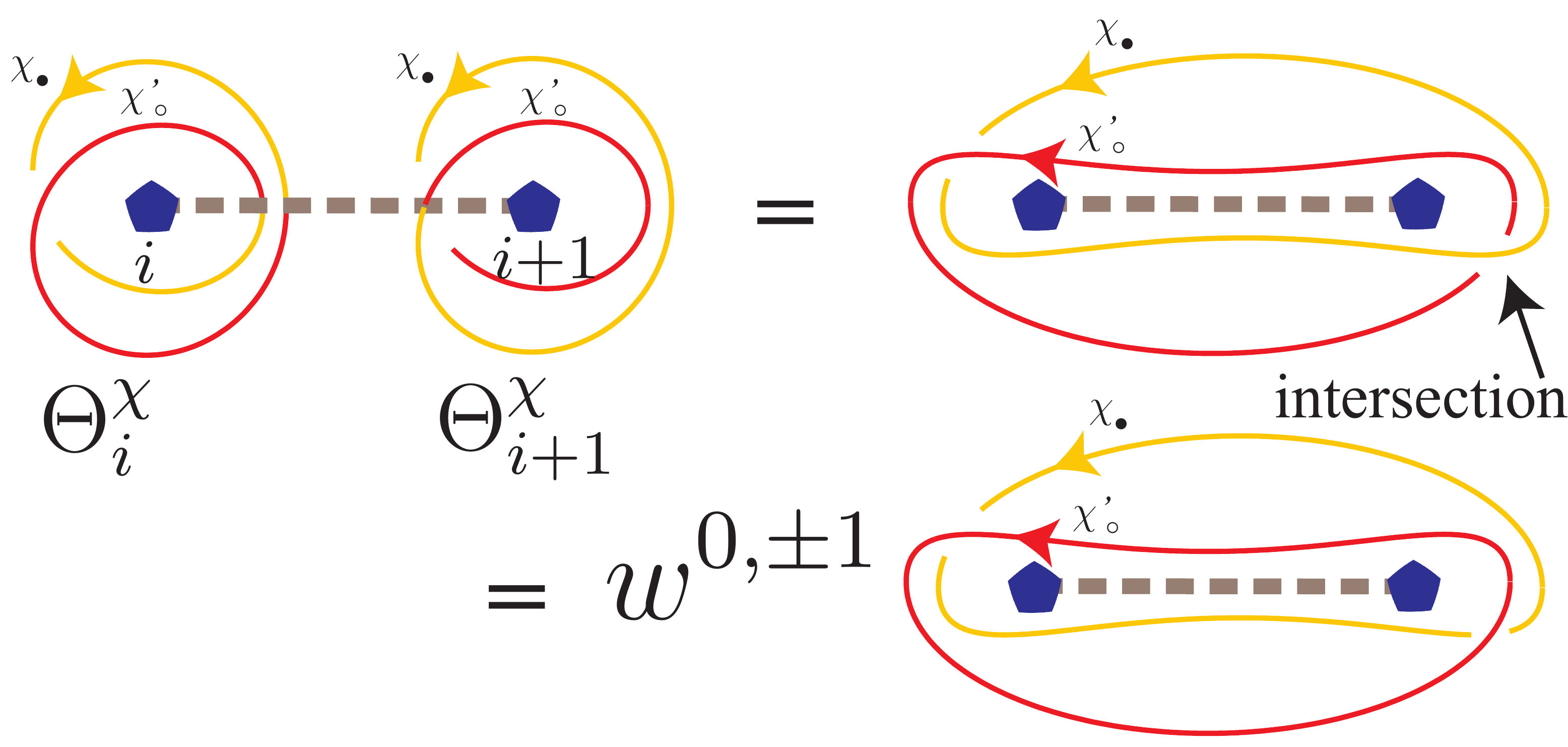}
	\caption{Two local Wilson operators $\Theta_i^\chi$ and $\Theta_{i+1}^\chi$ can be deformed, joined and unlinked into a pair of non-local Wilson loops of type $\chi_\bullet$ and $\chi'_\circ$ related by twofold symmetry transformation induced by defects.}\label{fig:doublelooppair}
\end{figure}
The prototype of a Wilson loop is depicted in figure~\ref{fig:defectWilsonloop}(c) and \ref{fig:Z2loopdefinition}. We denote the Wilson loop enclosing the $i^{th}$ and $(i+1)^{th}$ defect by $\mathcal{Z}^\chi_{i,i+1}$ according to the string color $\chi=Y,R,B$ observed at the base point (grey square). We can assume the loop always originates as a $\bullet$-sublattice type string from the base point because different sublattice types are interchangeable up to a phase as shown in figure~\ref{fig:defectWilsonloop}(c). The phase is fixed precisely by the local Wilson loops $\Theta_i^\chi=w^{l^\chi_i}$ and $\Theta_{i+1}^\chi=w^{l^\chi_{i+1}}$ and an unlinking procedure illustrated in figure~\ref{fig:doublelooppair} and is shown in the right side of the equation below. \begin{equation}\mathcal{Z}_{i,i+1}^{\chi',\circ}\mathcal{Z}_{i,i+1}^{\chi,\bullet}=e^{i\frac{2\pi}{k}\left(\delta^\chi_{\chi'+1}-\delta^\chi_{\chi'-1}+l^\chi_i+l^\chi_{i+1}\right)}\label{Zloopinsideout}\end{equation} where the color $\chi,\chi'=Y,R,B$ are indexed by $0,1,2$ mod 3, and $(\chi',\circ)$ and $(\chi,\bullet)$ are related by the transposition that characterizes the twisting of twofold defects defined in figure~\ref{fig:anyontwist}.

\begin{figure}[ht]
	\centering
	\includegraphics[width=3in]{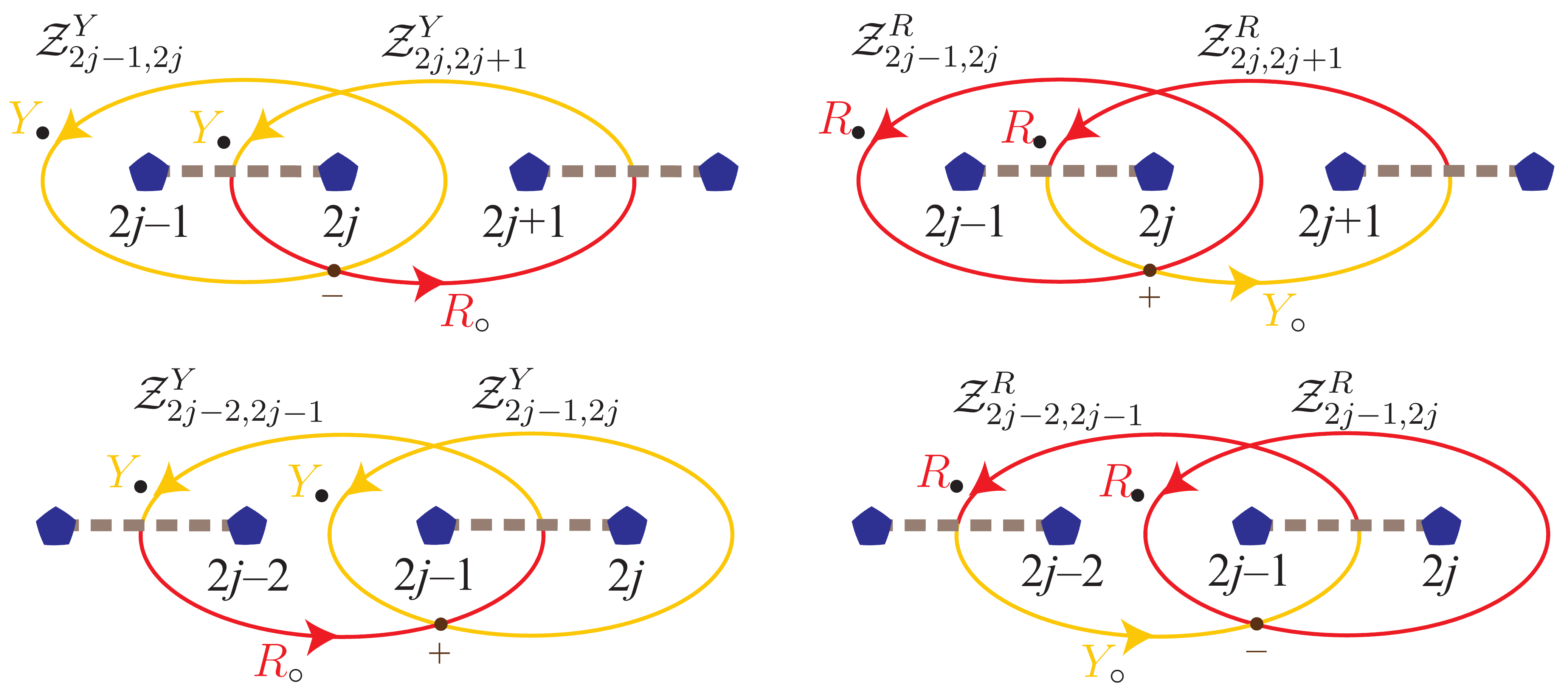}
	\caption{Intersection (brown dots) between adjacent Wilson operators $\mathcal{Z}_{2j-1,2j}^\chi$ and $\mathcal{Z}_{2j,2j+1}^{\chi'}$ about blue twofold defects $[1/2]_B$. Only different colors and sublattice types strings intersect.}\label{fig:Z2intersection}
\end{figure}
All Wilson operators can be generated by the prototype $\mathcal{Z}_{i,i+1}^\chi$ for $\chi=Y,R$ (recall $B=Y^{-1}R^{-1}$) and $i=1,\ldots,N-2$ since $\mathcal{Z}_{N-1,N}^\chi$ can be generated by the compactification relation \begin{align}\prod_{j=1}^{N/2}\mathcal{Z}_{2j-1,2j}^\chi=\vcenter{\hbox{\includegraphics[width=1in]{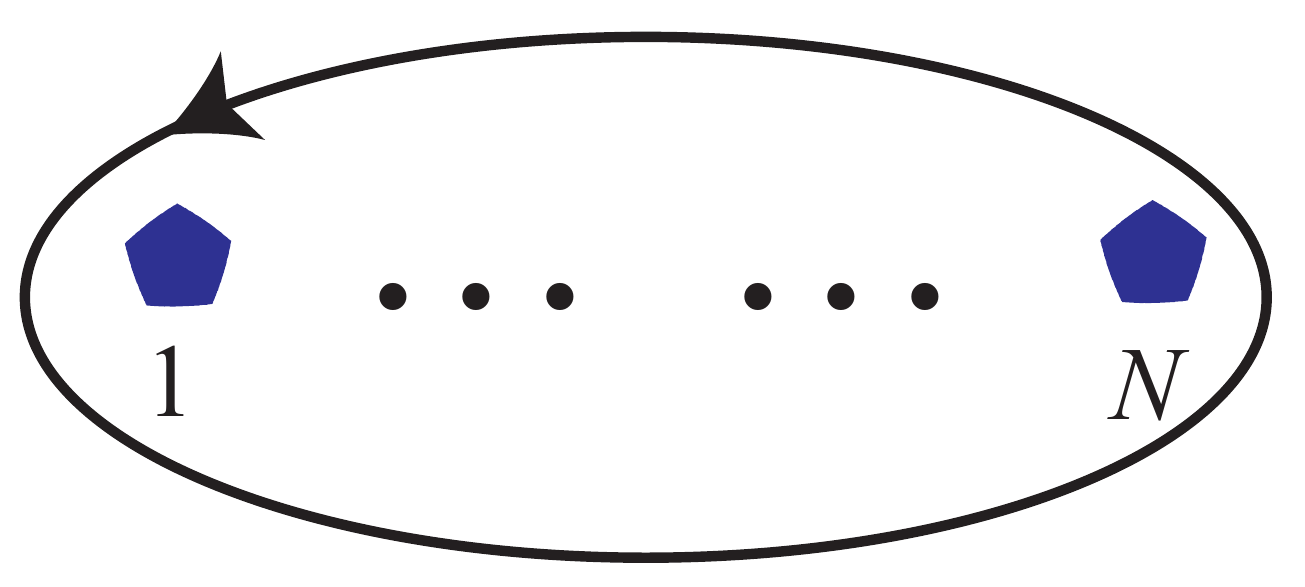}}}=1\label{Z2closeness}\end{align} $\{\mathcal{Z}_{2j-1,2j}^\chi\}_{j=1,\ldots,N/2-1}$ and $\{\mathcal{Z}_{2j,2j+1}^\chi\}_{j=1,\ldots,N/2-1}$ forms two maximal commuting set of Wilson operators because loops in the same set do not intersect. The Wilson algebra is characterized by the intersection relation between the two sets. \begin{equation}\mathcal{Z}_{2i-1,2i}^\chi\mathcal{Z}_{2j,2j+1}^{\chi'}=e^{i\frac{2\pi}{k}\langle\mathcal{Z}_{2i-1,2i}^\chi,\mathcal{Z}_{2j,2j+1}^{\chi'}\rangle}\mathcal{Z}_{2j,2j+1}^{\chi'}\mathcal{Z}_{2i-1,2i}^\chi\end{equation} where the pairing $\mathbb{I}=\langle\ast,\ast\rangle$ can be deduced by counting intersections between adjacent loops in figure~\ref{fig:Z2intersection} according to the rules set by figure~\ref{fig:intersection}, and is given by the following $(N-2)\times(N-2)$ invertible matrix $\mathbb{I}$. \begin{equation}\mathbb{I}=\langle\mathcal{Z}_{2i-1,2i}^\chi,\mathcal{Z}_{2j,2j+1}^{\chi'}\rangle=J\otimes\mathbb{J}_0\label{Z2intersectionmatrix}\end{equation} \begin{equation}J_Y=\left(\begin{array}{*{20}c}0&1\\1&-1\end{array}\right),\quad J_R=\left(\begin{array}{*{20}c}1&-1\\-1&0\end{array}\right),\quad J_B=\left(\begin{array}{*{20}c}-1&0\\0&1\end{array}\right)\label{Z2JYRB}\end{equation} \begin{equation}\mathbb{J}_0=\left(\begin{array}{*{20}c}1&0&0&\ldots&0\\-1&1&0&\ldots&0\\0&-1&1&\ldots&0\\\vdots&\vdots&\vdots&\ddots&\vdots\\0&0&0&\ldots&1\\\end{array}\right)\label{Z2J0}\end{equation} where the $(N-2)/2$ rows and columns of $\mathbb{J}_0$ in \eqref{Z2J0} run over the maximal commuting sets of Wilson loops $\{\mathcal{Z}_{2i-1,2i}\}_{i=1,\ldots,N/2-1}$ and $\{\mathcal{Z}_{2j,2j+1}\}_{i=1,\ldots,N/2-1}$ respectively, the $2\times2$ entries $J=J_Y,J_R,J_B$ are the corresponding matrices for the three types of twofold defects $[1/2]_Y,[1/2]_R,[1/2]_B$ that act on $\chi$-degree of freedom of $\mathcal{Z}^\chi$, for $\chi=Y,R$. Contrary to threefold defects, the Wilson algebra specified by the intersection form \eqref{Z2intersectionmatrix} preserves transposition but not symmetric under color permutation. 
%This is because even Wilson loops $\mathcal{Z}_{2j,2j+1}$ that crosses branch
%cuts in figure~\ref{fig:Z2intersection} explicitly violate color permutation,
%and the change of base point of the Wilson loop (grey square in
%figure~\ref{fig:Z2loopdefinition}) does not commute with color permutation.
It is evident from table~\ref{tab:twistdefects} and figure~\ref{fig:anyontwist} that twofold defects change type upon color permutation \begin{equation}\Lambda_3:[1/2]_\chi\to[1/2]_{\chi+1}\end{equation} where $\chi=0,1,2$ mod 3 index the colors $Y,R,B$, and the intersection matrix is $\Lambda_3$-{\em covariant} \begin{equation}\Lambda_3^T\cdot\mathbb{I}_{[1/2]_\chi}\cdot\Lambda_3=\mathbb{I}_{[1/2]_{\chi-1}}\end{equation} where $\Lambda_3=\left(\begin{array}{*{20}c}0&-1\\1&-1\end{array}\right)$ acts on the color degree of freedom $\chi=Y,R$.

Given an arbitrary choice of branch cut that specified the sublattice type $\bullet,\circ$ of all vertices, one can write down a particular ground state (up to a normalization constant) \begin{equation}|0\rangle\propto\prod_i\left[\sum_{r=0}^{k-1}w^{-rp_i}{\hat{D}_i}^r\right]\prod_P\left[\sum_{r=0}^{k-1}\hat{P}^r\right]|\sigma_{v_\bullet}=\tau_{v_\circ}=1\rangle\label{Z2groundstate}\end{equation} where $\hat{D}_i=\hat{P}_{5,7}$ are the pentagon or heptagon plaquette operators in eq.\eqref{pentagonheptagonoperator} at the $i^{th}$ defect, and $\hat{P}=\hat{P}_\bullet,\hat{P}_\circ$ runs over all other even sided plaquettes. Here $p_i$ are integer (half-integer) mod $k$ for $k$ odd (even) that specify the ground state eigenvalues $\hat{D}_i|0\rangle=w^{p_i}|0\rangle$ determined by local defect Hamiltonian \eqref{H57}. All ground states can be generated by Wilson operators \begin{equation}|{\bf m}\rangle=\left(\mathcal{Z}_{even}\right)^{\bf m}|0\rangle\end{equation}\begin{align}\left(\mathcal{Z}_{even}\right)^{\bf m}&=\prod_{j=1}^{N/2-1}\left(\mathcal{Z}_{2j,2j+1}^Y\right)^{m_{2j-1}}\left(\mathcal{Z}_{2j,2j+1}^R\right)^{m_{2j}}\\\left(\mathcal{Z}_{odd}\right)^{\bf m}&=\prod_{j=1}^{N/2-1}\left(\mathcal{Z}_{2j-1,2j}^Y\right)^{m_{2j-1}}\left(\mathcal{Z}_{2j-1,2j}^R\right)^{m_{2j}}\end{align} where ${\bf m}=(m_1,\ldots,m_{N-2})$ are the multi-exponents for Wilson loops. Assuming the branch cuts are chosen to avoid cutting across the odd Wilson loops $\mathcal{Z}_{2j-1,2j}$ such as those in figure~\ref{fig:Z2intersection}, the odd Wilson operators fix the particular ground state $\left(\mathcal{Z}_{odd}\right)^{\bf m}|0\rangle=|0\rangle$. The matrix elements of observables are given by \begin{align}\langle{\bf	m}'|\left(\mathcal{Z}_{odd}\right)^{\bf n}|{\bf	m}\rangle&=e^{i\frac{2\pi}{k}{\bf n}^T\mathbb{I}{\bf m}}\delta_{{\bf m}',{\bf m}}\label{Z2Zoddeigenvalues}\\\langle{\bf	m}'|\left(\mathcal{Z}_{even}\right)^{\bf n}|{\bf m}\rangle&=\delta_{{\bf m}',{\bf m}+{\bf n}}\end{align} Since $\mathbb{I}$ is invertible, ground states $|{\bf m}\rangle$ are completely distinguished by its eigenvalues with respect to $\mathcal{Z}_{odd}$ and therefore form an orthonormal basis for the ground state Hilbert space. The ground state degeneracy (G.S.D.) is given by \begin{equation}G.S.D.=k^{N-2}\label{Z2dimensionGSD}\end{equation} and matches the quantum dimension $d_{[1/2]}=k$ predicted in \eqref{Z2quantumdimension} by counting lattice degree of freedom.

\begin{figure}[ht]
	\centering
	\includegraphics[width=2.5in]{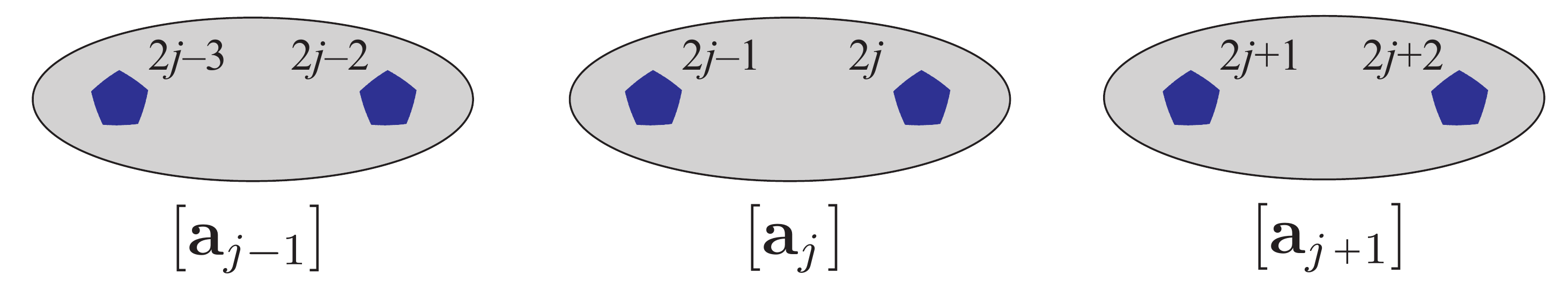}
	\caption{Abelian anyon channel $[{\bf a}_j]=[{\bf a}_\bullet^j,{\bf a}_\circ^j]$ of the $j^{th}$ pair of twofold defects measured by Wilson loops $\mathcal{Z}_{2j-1,2j}^\chi$.}\label{fig:Z2fusion}
\end{figure}
It would be more convenient for later considerations in fusion and braiding if the ground states are labeled by the abelian anyon fusion channels $[{\bf a}_j]=[{\bf a}_\bullet^j,{\bf a}_\circ^j]$ of the $j^{th}$ pair of twofold defects (see figure~\ref{fig:Z2fusion}), where ${\bf a}_\bullet^j=y_1^j{\bf e}_Y+r_1^j{\bf e}_R$ and ${\bf a}_\circ^j=y_2^j{\bf f}^Y+r_2^j{\bf f}^R$ are discrete vectors in the abelian anyon lattice (figure~\ref{fig:abeliananyonlattice}). The $\bullet$-part of the anyon channels ${\bf a}_\bullet=(y_1^1,r_1^1,\ldots,y_1^{N/2-1},r_1^{N/2-1})$ can be read off from \eqref{Z2Zoddeigenvalues} for the ground state $|{\bf m}\rangle$. \begin{equation}{\bf a}_\bullet=-i\sigma_yJ\otimes\mathbb{J}_0{\bf m}\end{equation} Eq.\eqref{Zloopinsideout} ensures a constraint on the $\circ$-part of the abelian fusion channels ${\bf a}_\circ=(y_2^1,r_2^1,\ldots,y_2^{N/2-1},r_2^{N/2-1})$ so that \begin{equation}{\bf a}_\circ=\mathbb{J}_0{\bf m}+{\boldsymbol\theta}_\circ\end{equation} where ${\boldsymbol\theta}_\circ=(\boldsymbol\theta_1,\ldots,\boldsymbol\theta_{N/2-1})$, $\boldsymbol\theta_j=(\theta_j^Y,\theta_j^R)$, depends only on defect species ${\bf l}_i=(l^Y_i,l^R_i)$, for $\Theta^\chi_i=w^{l^\chi_i}$. \begin{equation}\boldsymbol\theta_j={\bf f}-J^{-1}({\bf l}_{2j-1}+{\bf l}_{2j})\end{equation} where ${\bf f}={\bf f}^Y,{\bf f}^R,{\bf f}^B=(1,0),(0,1),(-1,-1)$ are basis vectors in the $\circ$-anyon lattice in figure~\ref{fig:abeliananyonlattice} and $J=J_Y,J_R,J_B$ are matrices in \eqref{Z2JYRB} determined by the color of twofold defects $[1/2]_Y,[1/2]_R,[1/2]_B$.

Under the fusion channel basis, the matrix elements for the Wilson operators are \begin{align}\langle{\bf a}'|\left(\mathcal{Z}_{odd}\right)^{\bf n}|{\bf a}\rangle&=e^{i\frac{2\pi}{k}{\bf n}^T i\sigma_y{\bf a}_\bullet}\delta_{{\bf m}',{\bf m}}\label{Z2groundstatea}\\\langle{\bf a}'|\left(\mathcal{Z}_{even}\right)^{\bf n}|{\bf a}\rangle&=\delta_{{\bf a}'_\bullet,{\bf a}_\bullet-i\sigma_y\mathbb{I}{\bf n}}\end{align} where the abelian fusion channels ${\bf a}=({\bf a}_\bullet,{\bf a}_\circ)$ are constrained by \begin{equation}{\bf a}_\circ=J^{-1}i\sigma_y{\bf a}_\bullet+\boldsymbol\theta_\circ\label{Z2fusionconstraint}\end{equation} Notice that the vacuum fusion channel may not be allowed by \eqref{Z2fusionconstraint}. This is because the $(2j-1)^{th}$ defect is not the anti-partner of the $(2j)^{th}$ in general unless the right hand side of eq.\eqref{Zloopinsideout} or ${\boldsymbol\theta}_\circ$ is trivial, i.e. the species labels satisfy \begin{equation}{\bf l}_{2j-1}+{\bf l}_{2j}=J{\bf f}\end{equation} 

We end this subsection with a cautionary remark on the phase variables of local defect Hamiltonians \eqref{Z33defectham} and \eqref{H57} in a closed system. Similar to $\mathbb{Z}_2$-fermion parity in a closed electronic system, there is a $\mathbb{Z}_3\oplus\mathbb{Z}_3$-{\em anyon parity} in a closed system of threefold defects for $k$ divisible by 3 and a $\mathbb{Z}_k\oplus\mathbb{Z}_k$-{\em anyon parity} in a closed system of twofold defects. They are good quantum numbers that imply global restrictions. The closeness condition \eqref{Z3closeness} requires the species labels $s_i=(s^\bullet_i,s^\circ_i)$ of the threefold twist defects to satisfy \begin{equation}\prod_{i=1}^N\Sigma^{\bullet/\circ}_i=e^{i\frac{2\pi}{3}\sum_{i=1}^Ns_i^{\bullet/\circ}}=1\label{Z3speciescloseness}\end{equation} for $k$ divisible by 3. While for twofold defects, \eqref{Z2closeness} similarly requires the species labels ${\bf l}_i=(l^Y_i,l^R_i)$ to obey \begin{equation}\sum_{i=1}^N{\bf l}_i=\frac{N}{2}J{\bf f}\label{Z2speciescloseness}\end{equation} where ${\bf f}={\bf f}^Y,{\bf f}^R,{\bf f}^B$ depending on the color type of $[1/2]_Y,[1/2]_R,[1/2]_B$. The species ${\bf s}_i$ and ${\bf l}_i$ are however completely determined by phase variables $\phi_i$ of {\em local} defect Hamiltonians \eqref{Z33defectham} and \eqref{H57} in the lattice level, and there are no reasons for the local phases to be restricted by the global closeness conditions \eqref{Z3speciescloseness} or \eqref{Z2speciescloseness}. When the local defect phase variables $\phi_i$ are incompatible with \eqref{Z3speciescloseness} or \eqref{Z2speciescloseness}, there is a topological obstruction in obtaining the absolute lowest energy state. For instance the state in eq.\eqref{Z2groundstate} would be identically zero, $|0\rangle=0$. The system would be forced to have local excitations where energy is not locally minimized. There will be two phases depending on the relative magnitude between $J_\ast$, the energy scale of defect Hamiltonians \eqref{Z33defectham} and \eqref{H57}, and $J$, the underlying energy scale of the original model \eqref{ham1}. For $J_\ast<<J$, excitations would be bounded and localized at defect centers, and eq.\eqref{Z3speciescloseness},\eqref{Z2speciescloseness} would be effectively restored. And therefore the Wilson algebras and ground state degeneracies described before would still persist. If $J_\ast>>J$, there will be delocalized abelian anyon excitations causing an infinite number of ground states. We consider only the former scenario.

\subsection{Word presentation of Wilson algebra}\label{sec:AlphabeticpresentationofWilsonalgebra}
\begin{figure}[ht]
	\centering
	\includegraphics[width=3.5in]{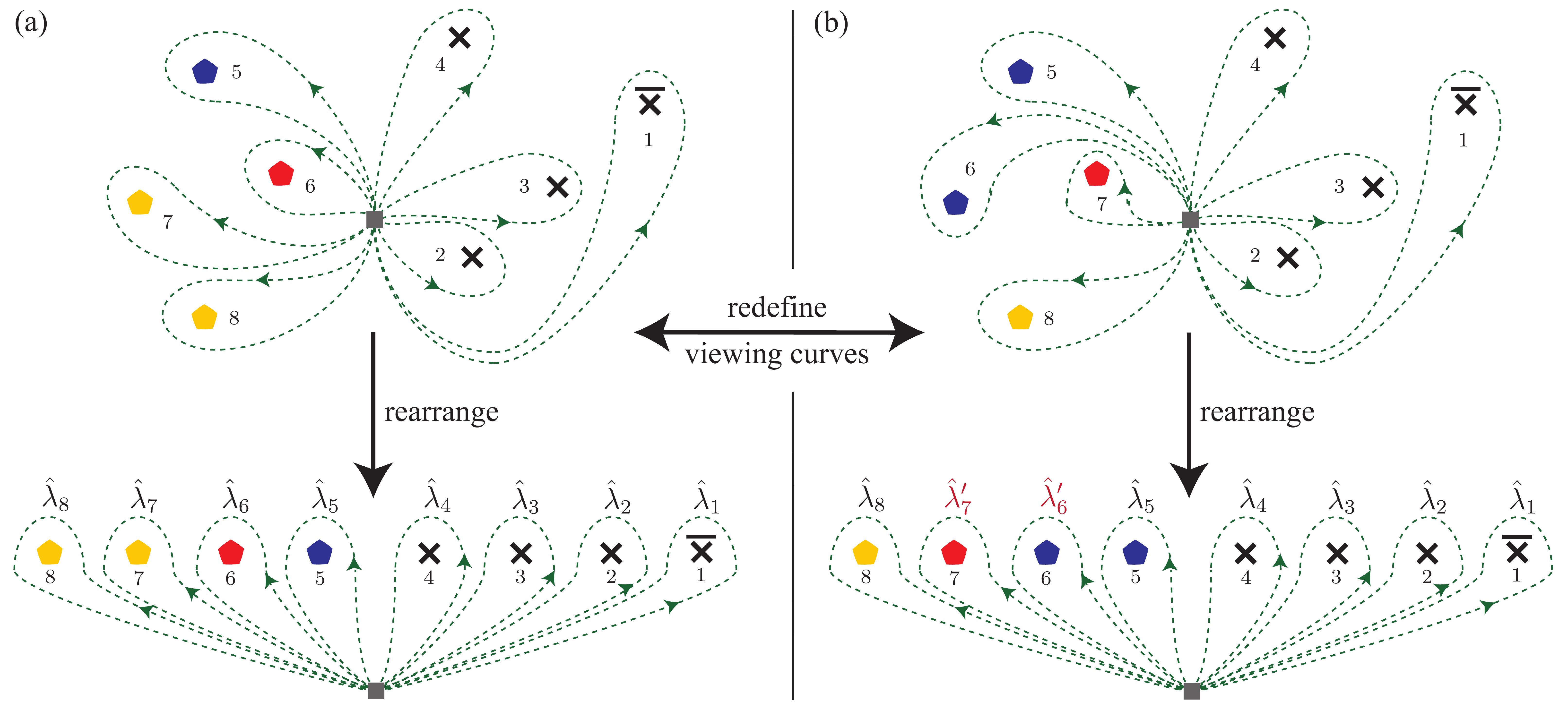}
	\caption{Viewing curve or open Wilson strings (directed dashed lines) $\hat{\lambda}_i$ around the $i^{th}$ twist defect starting and ending at a fixed view point (grey square). (a) Series rearrangement of twist defects without passing defects across viewing curves. (b) Redefining the viewing curves $(\hat{\lambda}_6,\hat{\lambda}_7)\leftrightarrow(\hat{\lambda}'_6,\hat{\lambda}'_7)$ in the same system by passing $\hat{\lambda}_7$ across the $6^{th}$ defect thereby changing the color of the $7^{th}$ one.}\label{fig:multidefects}
\end{figure}
We describe a presentation of the group of Wilson algebra in a system with multi-type twist defects. A defect type is defined by its $S_3$-twisting action on abelian anyon labels (see figure~\ref{fig:anyontwist}) when an abelian anyon is dragged along a path around the defect. Examples of these paths are shown by the viewing curves $\hat{\lambda}_i$ in figure~\ref{fig:multidefects}, each encloses one and only one defect. An abelian anyon $[{\bf a}]=(Y_\bullet)^{y_1}(R_\bullet)^{r_1}(Y_\circ)^{y_2}(R_\circ)^{r_2}$ will be twisted into $\Lambda_i\cdot[{\bf a}]$ according to \eqref{ST} and \eqref{S} if it moves along $\hat{\lambda}_i$, where $\Lambda_i$ is an element in $S_3$ that distinguishes the type of the $i^{th}$ defect by $[1/3]$, $[\overline{1/3}]$ or $[1/2]_\chi$. The defect type or $\Lambda_i$ may depend on the viewing curve $\hat{\lambda}_i$ for a multi-type defect system that contains twofold defects. Defects can be arranged and ordered in series as shown in figure~\ref{fig:multidefects}. A re-ordering of viewing curves will in general alter the defect type. Upon switching the viewing order of the $i^{th}$ and $(i+1)^{th}$ defects, their $S_3$-classifications change according to conjugation \begin{equation}(\Lambda_i,\Lambda_{i+1})\mapsto(\Lambda'_i,\Lambda'_{i+1})=(\Lambda_i^{-1}\Lambda_{i+1}\Lambda_i,\Lambda_i)\label{defectlabelexchange}\end{equation} where this is demonstrated by redefining the viewing curves for the $6^{th}$ and $7^{th}$ defects in figure~\ref{fig:multidefects}. This can be understood in the microscopic lattice level by observing that the $\bullet,\circ$-sublattice type of a threefold defect center or the $YRB$-plaquette color of a twofold defect center depends on the viewing path in the presence of other defects. Twist defects are therefore coarsely classified by the conjugacy classes of $S_3$, which include the class of threefold defects $[1/3],[\overline{1/3}]$ and the class of twofold defects $[1/2]_\chi$. However we do not consider these conjugacy classes as superselection sectors as the $S_3$-symmetry is not gauged, unlike in Ref~\cite{PropitiusBais96, Mochon04, Preskilllecturenotes}.

We denote the {\em alphabet} $[\hat{\lambda}_i]_{\bf a}$ to be an open Wilson string of dragging an abelian anyon ${\bf a}=[{\bf a}_\bullet,{\bf a}_\circ]$ along the viewing curve $\hat{\lambda}_i$ starting from the base point (grey square in figure~\ref{fig:multidefects}). The ordering of ${\bf a}_\bullet$ and ${\bf a}_\circ$ will only affect the Wilson operator by an over phase due to self-intersection, and if necessary, we will adapt the convention that the ${\bf a}_\bullet$-string acts before and sits below the ${\bf a}_\circ$ one. The Wilson algebra of the defect system consists of closed Wilson loops presented in series of the alphabets \begin{align}W(\mathcal{C})=\left[\hat{\lambda}_{i_m}\ldots\hat{\lambda}_{i_2}\hat{\lambda}_{i_1}\right]_{\bf a}=\prod_{r=1}^m\left[\hat{\lambda}_{i_r}\right]_{\prod_{l=1}^{r-1}\Lambda_{i_l}\cdot{\bf a}}\label{alphabetpresentation}\end{align} where $\mathcal{C}$ is the loop of the ordered composition $\hat{\lambda}_{i_m}\ast\ldots\ast\hat{\lambda}_{i_1}$ and the Wilson string begins at the fixed base point as the abelian anyon ${\bf a}$, which moves and transforms along $\mathcal{C}$. In order for $W(\mathcal{C})$ to be closed, the abelian anyon has to close back onto itself and the $S_3$-transformations have to satisfy the closeness relation \begin{equation}\left(\Lambda_{i_m}\ldots\Lambda_{i_2}\Lambda_{i_1}\right)\cdot{\bf a}={\bf a}\label{ATcloseness}\end{equation} The alphabets inside the bracket in the middle of \eqref{alphabetpresentation} has a fixed ordering. Interchanging the order of the alphabets in general will give an entirely different Wilson string that might not even be closed. The product order on the far right of eq.\eqref{alphabetpresentation} however will only contribute an overall $\mathbb{Z}_k$ phase to $W(\mathcal{C})$, and will not affect its intersection with other Wilson operator. Abelian anyon fusion \eqref{abeliananyonfusion} imply the simplification \begin{equation}\left[\hat{\lambda}_i\right]_{\bf a}\left[\hat{\lambda}_i\right]_{\bf b}=\left[\hat{\lambda}_i\right]_{{\bf a}+{\bf b}}\end{equation} For instances, the $\mathbb{Z}_k$-torsion $k{\bf a}=0$ implies $\hat{\lambda}_i^{k\times\mbox{\scriptsize ord}}=1$ for $\mbox{ord}=2,3$ is the order of twofold and threefold defects; the tricolor redundancy requires $\hat{\lambda}_i^3=1$ if the $i^{th}$ defect is a threefold defect, and $\hat{\lambda}_i\hat{\lambda}_j\hat{\lambda}_i\hat{\lambda}_j\hat{\lambda}_i\hat{\lambda}_j=1$ if the $i^{th}$ and $j^{th}$ defects are twofold defects of different colors. If the system of $N$ defects is closed on a compact surface without boundary, we have \begin{equation}\hat{\lambda}_{N}\ldots\hat{\lambda}_2\hat{\lambda}_1=1\end{equation} 

Ignoring the overall $\mathbb{Z}_k$-phase, \eqref{alphabetpresentation} are of the form \begin{equation}W[\{{\bf a}_i\}]=\prod_{i=1}^N\left[\hat{\lambda}_i\right]_{{\bf a}_i},\quad\mbox{for}\quad\sum_{i=1}^N{\bf a}_i=\sum_{i=1}^N\Lambda_i{\bf a}_i\end{equation} The alphabetic presentation of the Wilson loops considered in the previous subsections are summarized in table~\ref{tab:prototypesalphabets}. 
\begin{table}[ht]
\centering
\begin{tabular}{cl}
\hline\hline
Wilson loop prototypes & Word presentation\\\hline\noalign{\smallskip}
$\mathcal{A}^{\chi,\bullet}_{i,i+1}$ & $\left[\hat{\lambda}_{i+1}\right]_{-{\bf e}_\chi}\left[\hat{\lambda}_i\right]_{{\bf e}_\chi}$\\
$\mathcal{A}^{\chi,\circ}_{i,i+1}$ & $\left[\hat{\lambda}_{i+1}\right]_{-{\bf f}^\chi}\left[\hat{\lambda}_i\right]_{{\bf f}^\chi}$\\
$\Sigma^{\bullet/\circ}_i$ & $\left[\hat{\lambda}_i\right]_{\overline{\boldsymbol\kappa}_{\bullet/\circ}}$\\\noalign{\smallskip}
$\mathcal{Z}^{\chi}_{i,i+1}$ & $\left[\hat{\lambda}_{i+1}\hat{\lambda}_i\right]_{{\bf e}_\chi}$\\
$\Theta^{\chi}_i$ & $\left[\hat{\lambda}_i\hat{\lambda}_i\right]_{{\bf e}_\chi}$\\\hline\hline
\end{tabular}
\caption{Alphabetic presentation of the Wilson loop generators in figure~\ref{fig:defectWilsonloop}, \ref{fig:defectlocalloop}, \ref{fig:Z3loopdefinition} and \ref{fig:Z2loopdefinition}. ${\bf e}_\chi$ and ${\bf f}^\chi$ are primitive lattice vectors in the $\bullet$- and $\circ$-anyon triangular lattices in figure~\ref{fig:abeliananyonlattice} respectively. $\overline{\boldsymbol\kappa}_{\bullet/\circ}$ is the 3-fold invariant non-trivial $\bullet/\circ$-anyon in figure~\ref{fig:abeliananyonlattice} when $k$ is divisible by 3.}\label{tab:prototypesalphabets} 
\end{table}

By a slight deformation, one can assume all intersections between two Wilson operators are at the fixed base point. The algebraic relation of Wilson operators are given by \begin{equation}W[\{{\bf a}_i\}]W[\{{\bf b}_i\}]=e^{i\frac{2\pi}{k}\langle\{{\bf a}_i\},\{{\bf b}_i\}\rangle}W[\{{\bf b}_i\}]W[\{{\bf a}_i\}]\label{multidefectalgebra}\end{equation} where the intersection form $\mathbb{I}=\langle\ast,\ast\rangle$ is \begin{equation}\langle\{{\bf a}_i\},\{{\bf b}_i\}\rangle=\sum_{i=1}^N\left({\bf a}_i-\Lambda_i{\bf a}_i\right)\cdot\left[\sum_{j=1}^i{\bf b}_j-\sum_{j=1}^{i-1}\Lambda_j{\bf b}_j\right]\label{multidefectintersection}\end{equation} for ${\bf a}_i=[{\bf a}^i_\bullet,{\bf a}^i_\circ]=[y^i_1{\bf e}_Y+r^i_1{\bf e}_R,y^i_2{\bf f}^Y+r^i_2{\bf f}^R]$ and ${\bf b}_i=[{\bf b}^i_\bullet,{\bf b}^i_\circ]=[y'^i_1{\bf e}_Y+r'^i_1{\bf e}_R,y'^i_2{\bf f}^Y+r'^i_2{\bf f}^R]$, where the anti-symmetric dot product is defined by \begin{align}{\bf a}_i\cdot{\bf b}_j&=({\bf a}_\circ^j)^Ti\sigma_y{\bf b}_\bullet^i-({\bf b}_\circ^j)^Ti\sigma_y{\bf a}_\bullet^i\nonumber\\&=y^i_2r'^j_1-r^i_1y'^j_2-r^i_2y'^j_1+y^i_1r'^j_2\label{anyondotproduct}\end{align} Using the prototype basis identified in table~\ref{tab:prototypesalphabets}, the intersection form \eqref{multidefectintersection} reproduces the intersection matrices \eqref{Z3intersectionmatrix} for threefold defects and \eqref{Z2intersectionmatrix} for twofold defects. As a result of self-intersections such as those in $\Theta_i^\chi$, in general $W[\{{\bf a}_i\}]$ need not be a $k^{th}$ root of unity.

We notice the dot product \eqref{anyondotproduct} and subsequently the intersection form \eqref{multidefectintersection} and Wilson algebra \eqref{multidefectalgebra} are invariant under global symmetry transformation \begin{align}({\bf a}_i,{\bf b}_i)&\to({\bf a}'_i,{\bf b}'_i)=\left(\Lambda\cdot{\bf a}_i,\Lambda\cdot{\bf b}_i\right)\\\Lambda_i&\to\Lambda'_i=\Lambda\Lambda_i\Lambda^{-1}\label{defectdualitytransformation}\end{align} for any $\Lambda$ in $S_3$. The color and sublattice transformations of abelian anyon have already been described in \eqref{ST} and \eqref{S} in the previous section. The symmetry transformation for defects \eqref{defectdualitytransformation} are summarized by the following. \begin{align}\Lambda_3:&\left(\left[\frac{1}{3}\right],\left[\overline{\frac{1}{3}}\right],\left[\frac{1}{2}\right]_\chi\right)\to\left(\left[\frac{1}{3}\right],\left[\overline{\frac{1}{3}}\right],\left[\frac{1}{2}\right]_{\chi+1}\right)\label{defectST}\\\Lambda_{B}:&\left(\left[\frac{1}{3}\right],\left[\overline{\frac{1}{3}}\right],\left[\frac{1}{2}\right]_\chi\right)\to\left(\left[\overline{\frac{1}{3}}\right],\left[\frac{1}{3}\right],\left[\frac{1}{2}\right]_{\Lambda_{B}\cdot\chi}\right)\label{defectS}\end{align} where $\chi=Y,R,B$ are the color types of twofold defects, and $\Lambda_3$ and $\Lambda_B$ are cyclic color permutation and transposition respectively described in \eqref{ST} and \eqref{S}. For $k$ divisible by 3, the threefold defect species labels ${\bf s}=(s_\bullet,s_\circ)\in\mathbb{Z}_3\oplus\mathbb{Z}_3$ transform according to \begin{equation}\Lambda_3:{\bf s}\to{\bf s}'={\bf s},\quad\Lambda_B:{\bf s}\to{\bf s}'=-\sigma_x{\bf s}\label{speciestrans1}\end{equation} The species labels for twofold defects ${\bf l}=(l_Y,l_R)$ transform according to \begin{align}\Lambda_3:{\bf l}\to{\bf l}'&=\left(\Lambda_3^{-1}\right)^T{\bf l}\label{speciestrans2}\\\Lambda_B:{\bf l}\to{\bf l}'&=\left\{\begin{array}{*{20}c}\left(\Lambda_3^{-1}\right)^T{\bf l},&\mbox{for $[1/2]_Y$}\\\Lambda_3^T{\bf l}\hfill,&\mbox{for $[1/2]_R$}\\{\bf l}\hfill,&\mbox{for $[1/2]_B$}\end{array}\right.\label{speciestrans3}\end{align} where $\Lambda_3=\left(\begin{array}{*{20}c}0&-1\\1&-1\end{array}\right)$ represents cyclic color permutation. As a result of \eqref{defectST} and \eqref{defectS}, twist defects do not only transform the labels of an orbiting abelian anyon but also the type of an orbiting twist defect. %This defect braiding property will be discussed in the following section.

\section{Non-Commutative Fusion}\label{sec:defectfusion}

We describe the quantum bases and transformations for twist defects. Our system consists of deconfined abelian anyon excitations of the $\mathbb{Z}_k$-gauge theory as well as semiclassical twist defects that locally violate the $S_3$-symmetry.

Abelian anyons form the basis for all quantum states and measurements. There is a correspondence between anyon operators, non-local Jordan-Wigner strings that leave local excitations at their ends, and excitation states formed when Jordan-Wigner operators act on a ground state. The anyon charge of an excited state can be locally measured by plaquette stabilizers or closed Wilson strings accumulated by dragging a conjugate anyon around it. 

None of the above holds for defects. Twist defects are not excitations of a Hamiltonian describing a topological gauge theory. There are no ``defect operators" that correspond to quantum states by their action on the vacuum. Instead, states are generated by non-contractible closed strings of anyon trajectories around defects. If one switches the type of a defect even within the same conjugacy class, for example $[1/3]\leftrightarrow[\overline{1/3}]$ or $[1/2]_Y\leftrightarrow[1/2]_R$, a closed anyon trajectory may become open by violating eq.\eqref{ATcloseness}. Hence in the semiclassical description, one cannot take superposition between states for different defect configurations --- the probability amplitude for a defect to be a particular $S_3$-element is either 0 or 1. Moreover, defects cannot be used as a tool for measurement. No quantum states are observed by expectation values of unitary operations involving moving a twist defect in a cycle. Therefore, anyons and twist defects are fundamentally different. The distinction stems from the fact that, unlike the underlying $\mathbb{Z}_k$-gauge symmetry, the tricolor and bipartite $S_3$-symmetry is a classical non-dynamical physical symmetry and is {\em weakly} broken~\cite{Kitaev06} by anyon labels. We assume there is only a finite (in particular non-dense) population of twist defects so that the system admits an almost global tricoloration and bipartite structure except along finite length branch cuts (a set with measure zero), and hence a {\em gauging} of $S_3$-symmetry is unnecessary.

Analogous physical examples include any defect heterostructures between superconductors, (anti)-ferromagnets and (fractional) topological insulators~\cite{FuKane08, LindnerBergRefaelStern, ClarkeAliceaKirill, MChen, Vaezi} or strong spin-orbit coupled semiconductors~\cite{SauLutchynTewariDasSarma, OregRefaelvonOppen10, OregSelaStern13}, where the pairing phase, magnetic spin order, band inversion mass gap and fermi energy, are all treated as non-dynamical variables. For vortices in chiral $p+ip$ superconductors~\cite{Volovik99, ReadGreen, Ivanov} and across the topological insulator to superconductor interface~\cite{FuKane08, TeoKane09}, the pairing phase vortex and branch cuts in the fermion sign, are treated as classical objects. In the study of crystalline dislocations and disclinations in topological insulators and superconductors~\cite{RanZhangVishwanath, TeoKane, TeoHughes}, the underlying lattice is also regarded as stationary.

In practice, although an abelian anyon label $[{\bf a}]=[{\bf a}_\bullet,{\bf a}_\circ]$ may be {\em twisted} during a cycle around defects, the change is traceable in the classical level as there are no superpositions of defect configurations. This means there is no need to consider superposition of quantum states of different anyon labels within a $S_3$-multiplet, unlike a discrete gauge theory~\cite{PropitiusBais96, Mochon04, Preskilllecturenotes}. And therefore under the semi-classical defect treatment, the {\em superselection sector} of abelian anyon $[{\bf a}]$ stays unchanged and in particular a {\em $S_3$-orbifold superselection sector}~\cite{BarkeshliWen11} redefinition $[{\bf a}]_{S_3}^{\pm}=\{\Lambda\cdot({\bf a}_\bullet\pm{\bf a}_\circ):\Lambda\in S_3\}$ according to irreducible representation of $S_3\ltimes\mathbb{Z}_k^4$ is unnecessary and inappropriate. Similarly the twist defects should not be regarded as quantum $S_3$ fluxes~\cite{PropitiusBais96, Mochon04, Preskilllecturenotes}, as they are distinguished by their anyon twisting characteristics according to group elements rather than conjugacy classes of $S_3$. Since defects are classical objects, their trajectories are also classical. In particular, although a triple exchange between $[1/2]_Y$ and $[1/2]_R$ or a single exchange between $[1/3]$ and $[\overline{1/3}]$ that go $360^\circ$-around a $[1/2]_\chi$ would leave the defect configuration invariant, there would still be no quantum interference between direct and exchange scattering of different type defects because defect trajectories are classically traceable. There is an underlying lattice or in general gravitation environment that could accidentally measure and distinguish the two scattering paths. And therefore there is no reason to identify defects according to conjugacy classes.
  
In this section, we describe the fusion and braiding characteristics of a defect system. The objects are quantum deconfined abelian anyons and semiclassical $S_3$-twist defects. Two objects can be {\em fused} by projecting out any Wilson strings or defect trajectories that go between them and treating the composite as a single entity. Contrary to conventional topological gauge theory, the composite outcomes or {\em fusion channels} may depend on the order of the two initial objects and the location of the viewpoint. This follows from the non-abelian nature of the group $S_3$. For instance $[1/2]_Y\times[1/2]_R=[1/3]$ but $[1/2]_R\times[1/2]_Y=[\overline{1/3}]$. The objects thus form a {\em fusion category}~\cite{Kitaev06, Turaevbook, BakalovKirillovlecturenotes, Wangbook} with a non-commutative fusion product. We establish a basis convention for the quantum states involved in a fusion or splitting process. This gives us a consistent convention for the $F$-symbols, which are basis transformations between the different fusion orderings $(x\times y)\times z\cong x\times(y\times z)$. 
%We also define unitary braiding operations, called $R$-symbols, that exchange the ordering of objects, $x\times y\cong y\times x$. Since fusion products are not necessarily commutative and objects may not be invariant under exchange as illustrated in \eqref{defectlabelexchange}, the fusion category can only be {\em partially braided} and we only compute the $R$-symbols for commuting objects. Finally, we generalize the notion of {\em topological spin} ($\theta\approx R^2$) to twist defects.

\subsection{Fusion rules}\label{sec:fusionrules}

We are considering the fusion and splitting of the collection: \begin{equation}\mathcal{O}=\left\{[{\bf a}],[1/3]_{\bf s},[\overline{1/3}]_{\bf s},[1/2]_{\chi,{\bf l}}\right\}\label{objectset}\end{equation} Here $[{\bf a}]=[{\bf a}_\bullet,{\bf a}_\circ]$ labels the abelian anyons, for ${\bf a}_\bullet$ and ${\bf a}_\circ$ are 2-dimensional $\mathbb{Z}_k$-coefficient vectors in the triangular anyon lattice in figure~\ref{fig:abeliananyonlattice}, and $1\equiv[0]$ is the vacuum. $[1/3]_{\bf s}$ and $[\overline{1/3}]_{\bf s}$ are threefold defects characterized by its anyon twisting shown in figure~\ref{fig:anyontwist}(a), and ${\bf s}=(s_\bullet,s_\circ)\in\mathbb{Z}_3\oplus\mathbb{Z}_3$ label the species of defects distinguished by local Wilson operators $\Sigma_\bullet,\Sigma_{\circ}$ in figure~\ref{fig:defectlocalloop}(b). Since threefold defects are subdivided into species only when $k$ is divisible by 3, we automatically set \begin{equation}{\bf s}=0,\quad\mbox{for $3\centernot\mid k$}\end{equation} $[1/2]_{\chi,{\bf l}}$ are twofold defects characterized in figure~\ref{fig:anyontwist}(b) for $\chi={Y},{R},{B}$, and ${\bf l}=(l_Y,l_R)$ label the defect species distinguished by local operator $\Theta_Y,\Theta_R$ in figure~\ref{fig:defectlocalloop}(a). Depending on the color $\chi$ of the defect and the eveness or oddness of $k$, $l_Y$ and $l_R$ are either integers or half-integers modulo $k$ so that \begin{equation}e^{2\pi il_{Y/R}}=\Theta_{Y/R}^k=(-1)^{(k-1)(1-\delta^\chi_{Y/R})}\end{equation} Mathematically speaking, the species labels ${\bf s}$ and ${\bf l}$ are irreducible representations of the centralizer subgroup $C_{\mathbb{Z}_k^4}(\Lambda)$ of abelian anyons in $\mathbb{Z}_k^4$ invariant under the twisting action $\Lambda\in S_3$ of the corresponding defect.

Fusion and splitting of objects are described by the equation \begin{equation}x\times y=\sum_{z\in\mathcal{O}}N_{xy}^zz\end{equation} where the fusion matrix $N_x=(N_{xy}^z)$ has non-negative integer entries. $N_{xy}^z$ counts the multiplicity of distinguishable ways the ordered pair $(x,y)$ can be identified together as the object $z$. Identification can be done by projective measurements that send abelian anyons around the objects, a multiple number of times if necessary. The antiparticle $\bar{x}$ of an object $x$ is the unique object so that $N_{x\bar{x}}^1=N_{\bar{x}x}^1=1$ and $N_{xy}^1=0$ whenever $y\neq\bar{x}$.

Fusion between abelian anyons is given by addition on the anyon lattice \begin{equation}[{\bf a}]\times[{\bf b}]=[{\bf a}+{\bf b}].\end{equation} Fusion between abelian anyons and defects changes the species labels. \begin{align} [{\bf a}]\times[1/3]_{\bf s}&=[1/3]_{\bf s}\times[{\bf a}]=[1/3]_{{\bf s}'}\\ [{\bf a}]\times[\overline{1/3}]_{\bf s}&=[\overline{1/3}]_{\bf s}\times[{\bf a}]=[\overline{1/3}]_{{\bf s}'}\\ [{\bf a}]\times[1/2]_{\chi,{\bf l}}&=[1/2]_{\chi,{\bf l}}\times[{\bf a}]=[1/2]_{\chi,{\bf l}'}\end{align} where the species labels of the incoming and outgoing channels are related by including the anyon $[{\bf a}]$ inside the local Wilson opertor $\Sigma_{\bullet/\circ}$ for threefold defects and $\Theta_{Y/R}$ for twofold ones.  \begin{align}s'_\bullet&=s_\bullet-\frac{3}{k}{\boldsymbol\kappa}_\bullet^Ti\sigma_y{\bf a}_\bullet,\quad\mbox{for $3|k$}\\s'_\circ&=s_\circ+\frac{3}{k}{\boldsymbol\kappa}_\circ^Ti\sigma_y{\bf a}_\circ,\quad\mbox{for $3|k$}\\{\bf l}'&={\bf l}+i\sigma_y({\bf a}_\bullet+\Lambda_\chi{\bf a}_\circ)\end{align} for ${\bf a}_\bullet=(y_1,r_1)$, ${\bf a}_\circ=(y_2,r_2)$, ${\boldsymbol\kappa}_\bullet={\boldsymbol\kappa}_\circ=(k/3,-k/3)$ are threefold fixed anyons, $\sigma_y$ the Pauli matrix, $\Lambda_\chi$ is the transposition action of the twofold defect $[1/2]_\chi$ and is represented by $\Lambda_\chi=i\sigma_yJ_\chi$ for $J_\chi$ defined in eq.\eqref{Z2JYRB} or \eqref{lambdachiapp}.

Fusions between threefold defects for $k$ not divisible by 3 are given by \begin{align}[1/3]\times[1/3]&=k^2[\overline{1/3}]\label{33fusion}\\ [\overline{1/3}]\times[\overline{1/3}]&=k^2[1/3]\label{3bar3barfusion}\\ [1/3]\times[\overline{1/3}]&=[\overline{1/3}]\times[1/3]=\sum_{{\bf a}}[{\bf a}]\label{33barfusion}\end{align} The multiplicity $k^2$ in \eqref{33fusion}, \eqref{3bar3barfusion} and the $k^4$ number of abelian anyon channels in \eqref{33barfusion} match the quantum dimension $d_{[1/3]}=d_{[\overline{1/3}]}=k^2$ shown in \eqref{Z3dimension} and \eqref{Z3dimensionGSD}. For $k$ divisible by 3, the fusion rules are decorated with species labels and modified to accommodate the reduced quantum dimensions $d_{[1/3]}=d_{[\overline{1/3}]}=k^2/3$ shown in \eqref{Z3dimension} and \eqref{Z33dimensionGSD}. \begin{align}[1/3]_{{\bf s}_2}\times[1/3]_{{\bf s}_1}&=\frac{k^2}{3}[\overline{1/3}]_{{\bf s}_1+{\bf s}_2}\\ [\overline{1/3}]_{{\bf s}_2}\times[\overline{1/3}]_{{\bf s}_1}&=\frac{k^2}{3}[1/3]_{{\bf s}_1+{\bf s}_2}\\ [1/3]_{{\bf s}_2}\times[\overline{1/3}]_{{\bf s}_1}&=[\overline{1/3}]_{{\bf s}_1}\times[1/3]_{{\bf s}_2}={\sum_{{\bf a}}}'[{\bf a}]\label{33barfusion3}\end{align} where the sum in \eqref{33barfusion3} is taken over the $k^2/3$ anyons that satisfy the constraints \begin{align}\frac{k}{3}(s^1_\bullet+s^2_\bullet)=-{\boldsymbol\kappa}_\bullet^Ti\sigma_y{\bf a}_\bullet,\quad \frac{k}{3}(s^1_\circ+s^2_\circ)={\boldsymbol\kappa}_\circ^Ti\sigma_y{\bf a}_\circ\end{align} so that the eigenvalues for the product of local Wilson operators $\Sigma^1_{\bullet/\circ}\Sigma^2_{\bullet/\circ}$ are preserved. Thus the antiparticle of a threefold defect has the inverse species label. \begin{equation}\overline{[1/3]_{{\bf s}}}=[\overline{1/3}]_{-{\bf s}},\quad\overline{[\overline{1/3}]_{{\bf s}}}=[1/3]_{-{\bf s}}\end{equation}

\begin{figure}[ht]
	\centering
	\includegraphics[width=3in]{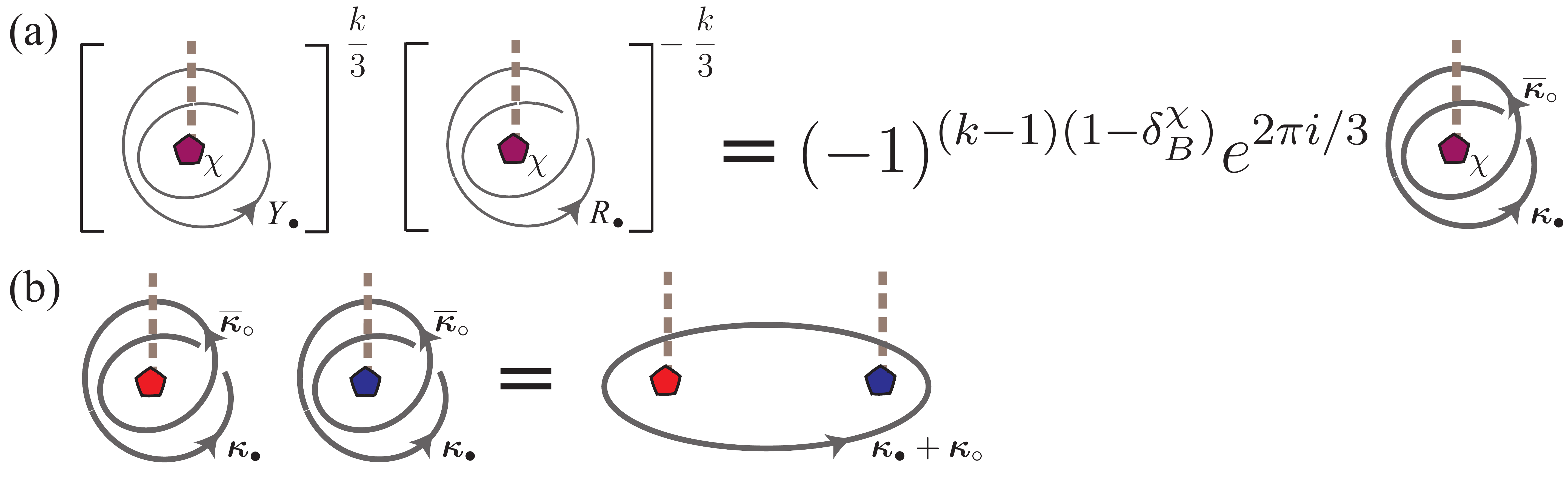}
	\caption{Relation between species labels of the fusion $[1/2]_{\chi,{\bf l}_2}\times[1/2]_{\chi\pm1,{\bf l}_1}=[1/3]_{\bf s}$ or $[\overline{1/3}]_{\bf s}$ for $k$ divisible by 3. (a) Left hand side equals $\Theta_Y^{k/3}\Theta_R^{-k/3}=w^{l_Y-l_R}$, $w=e^{2\pi i/k}$. $\chi={Y},{R},{B}$ is the color of the twofold defect. (b) Right hand side equals $\Sigma_\bullet^{-1}\Sigma_\circ=e^{i\frac{2\pi}{3}(s_\circ-s_\bullet)}$, where ${\boldsymbol\kappa}_\bullet=(Y_\bullet)^{k/3}(R_\bullet)^{-k/3}$ and $\overline{\boldsymbol\kappa}_\circ=(Y_\circ)^{-k/3}(R_\circ)^{k/3}$}\label{fig:thetatosigma}
\end{figure}

Fusion between twofold defects of different colors, is non-commutative. For $k$ not divisible by 3, \begin{align}[1/2]_{\chi,{\bf l}_2}\times[1/2]_{\chi+1,{\bf l}_1}&=[1/3]\label{2c2c+1fusion}\\ [1/2]_{\chi+1,{\bf l}_2}\times[1/2]_{\chi,{\bf l}_1}&=[\overline{1/3}]\label{2c2c-1fusion}\\ [1/2]_{\chi,{\bf l}_2}\times[1/2]_{\chi,{\bf l}_1}&={\sum_{{\bf a}}}'[{\bf a}]\label{2c2cfusion}\end{align} where the sum of \eqref{2c2cfusion} are restricted over the $k^2$ abelian anyons that satisfy eq.\eqref{Z2fusionconstraint}, i.e. \begin{equation}{\bf a}_\circ=J_\chi^{-1}\left(i\sigma_y{\bf a}_\bullet-({\bf l}_1+{\bf l}_2)\right)+{\bf f}^\chi\label{22fusionconstraint}\end{equation} where ${\bf a}_\bullet=(y_1,r_1)$, ${\bf a}_\circ=(y_2,r_2)$, $J_\chi$ are given in \eqref{Z2JYRB} for $\chi={Y},{R},{B}$, $\sigma_y$ is the Pauli matrix, and ${\bf f}^\chi={\bf f}^Y,{\bf f}^R,{\bf f}^B=(1,0),(0,1),(-1,-1)$ respectively are basis vectors in the $\circ$-anyon lattice in figure~\ref{fig:abeliananyonlattice}. The fusion constraint \eqref{22fusionconstraint} is to ensure the eigenvalues of the product of local Wilson operators $\Theta^\chi_1\Theta^\chi_2$ stays unchanged (see figure~\ref{fig:doublelooppair} and eq.\eqref{Zloopinsideout}). This shows that the antiparticle of a twofold defect has a reciprocal but shifted species label. \begin{equation}\overline{[1/2]_{\chi,{\bf l}}}=[1/2]_{\chi,\overline{\bf l}},\quad\overline{\bf l}=-{\bf l}+J_\chi{\bf f}^\chi\label{Z2reciprocalspecies}\end{equation} ($\chi$ is not a summation index.) We define the self-reciprocal species ${\bf l}_0$ so that ${\bf l}_0=\overline{{\bf l}_0}$. This can be chosen regardless of whether $k$ is even or odd so that \begin{align}{\bf l}_0=(l^Y_0,l^R_0)=-i\sigma_y\frac{k+1}{2}{\bf f}^\chi\label{selfreciprocalspecies}\end{align} This is the unique such species for odd $k$ but there are three other self-reciprocal species when $k$ is even, differing from the above ${\bf l}_0$ by $(k/2,0),(0,k/2),(k/2,k/2)$. For $k$ divisible by 3, \eqref{2c2c+1fusion} and \eqref{2c2c-1fusion} are decorated by species labels and become multi-channeled. \begin{align}[1/2]_{\chi,{\bf l}_2}\times[1/2]_{\chi+1,{\bf l}_1}&={\sum_{\bf s}}'[1/3]_{\bf s}\\ [1/2]_{\chi+1,{\bf l}_2}\times[1/2]_{\chi,{\bf l}_1}&={\sum_{\bf s}}'[\overline{1/3}]_{\bf s}\end{align} where the sums are restricted to the 3 species labels ${\bf s}=(s_\bullet,s_\circ)$ that satisfy %\begin{equation}s_\circ-s_\bullet=\left(l_Y^1+l_Y^2\right)-\left(l_R^1+l_R^2\right)+1+\frac{3}{2}(\delta^\chi_R+\delta^\chi_B)\end{equation} 
\begin{align}(-1)^{(k-1)(1-\delta^\chi_Y)}e^{i\frac{2\pi}{3}(s_\circ-s_\bullet)}=e^{i\frac{2\pi}{3}(l_Y^1+l_Y^2-l_R^1-l_R^2)}\end{align} This can be proven by comparing the local Wilson operators in figure~\ref{fig:defectlocalloop}(a) and (b) by a linking process shown in figure~\ref{fig:thetatosigma}.

Fusion between twofold  and threefold defects never commutes. \begin{align}[1/2]_{\chi,{\bf l}}\times[1/3]_{\bf s}&=[\overline{1/3}]_{\bf s}\times[1/2]_{\chi,{\bf l}}\nonumber\\&={\sum_{{\bf l}'}}'[1/2]_{\chi+1,{\bf l}'}\\ [1/3]_{\bf s}\times[1/2]_{\chi+1,{\bf l}}&=[1/2]_{\chi+1,{\bf l}}\times[\overline{1/3}]_{\bf s}\nonumber\\&={\sum_{{\bf l}'}}'[1/2]_{\chi,{\bf l}'}\end{align} where the species labels ${\bf s}$ and ${\bf l}$ will not affect the $k^2$ fusion channels ${\bf l}'$ unless $k$ is divisible by 3, in which the number of fusion channels are reduced to $k^2/3$ and are restricted by \begin{equation}e^{i\frac{2\pi}{3}(l'_Y-l'_R)}=(-1)^{(k-1)(1-\delta^\chi_Y)}e^{i\frac{2\pi}{3}(s_\circ-s_\bullet+l_Y-l_R)}\end{equation} This can be deduced by comparing the local Wilson operator $\Theta_Y^{k/3}\Theta_R^{-k/3}$ around the twofold defects before and after fusing with $[1/3]$. The factors are given by the linking process in figure~\ref{fig:thetatosigma}(a).

The quantum dimension $d_x$ of an object $x$, dictates the scaling behavior of the ground state degeneracies in the thermodynamic limit. These are illustrated in eq.\eqref{Z3dimensionGSD} and \eqref{Z2dimensionGSD} for threefold  and twofold defects. The ground state degeneracy for $\mathcal{N}$ identical object $x$ can also be read off from the fusion rules governed by the fusion matrices $N_x=(N_{xy}^z)$. \begin{align}G.S.D.\left[x^{\mathcal{N}}\right]&=\sum_{\{y_i\}}\sum_{\{\mu_i\}}\left|\vcenter{\hbox{\includegraphics[width=1.5in]{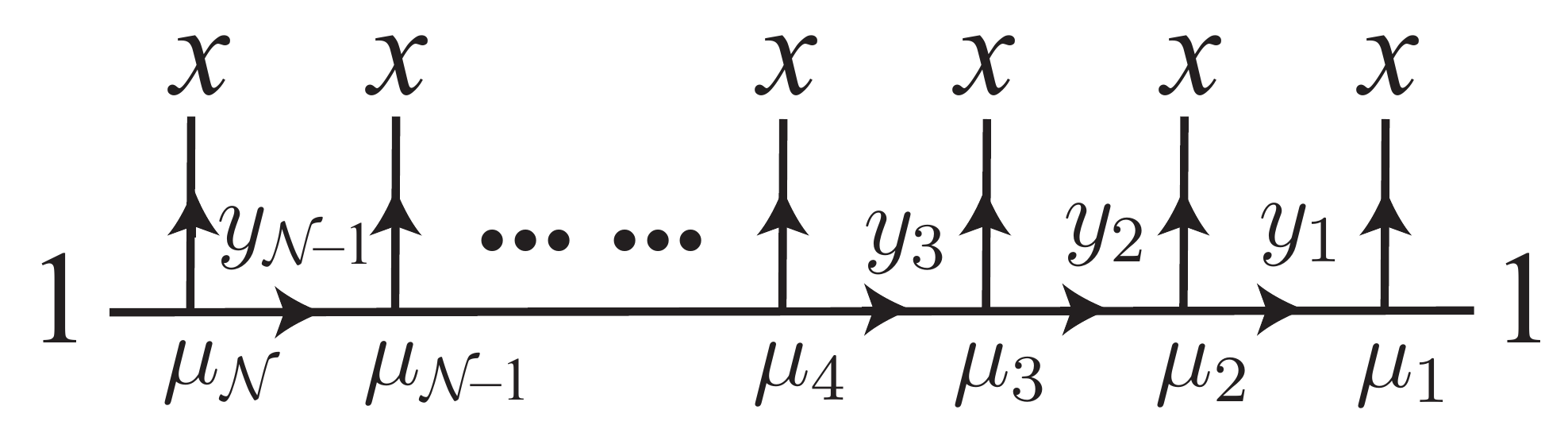}}}\right|\nonumber\\&=\sum_{\{y_i\}}N_{xy_{\mathcal{N}-1}}^{1}N_{xy_{\mathcal{N}-2}}^{y_{\mathcal{N}-1}}\ldots N_{xy_2}^{y_3}N_{xy_1}^{y_2}N_{x1}^{y_1}\nonumber\\&=\left[\left(N_x\right)^{\mathcal{N}}\right]_1^1\propto d_x^{\mathcal{N}}\label{Npower}\end{align} where the ground state degeneracy is counted by the numbers of distinguishable fusion channels and fusion multiplicities labeled by $y_i$ and $\mu_i$, that $\mathcal{N}$ identical objects $x$ can fuse into the vacuum channel. And eq.\eqref{Npower} is non-trivial if $\mathcal{N}$ is a multiple of the {\em period} of the fusion matrix $N_x$, the number of $x$'s required to fuse to the vacuum $x\times x\times\ldots\times x=1$. Using the Perron-Frobenius theorem, the quantum dimension $d_x$ is given by the largest (absolute) eigenvalues of $N_x$. Since the fusion rules preserve the dimensions, \begin{equation}d_xd_y=\sum_zN^z_{xy}d_z\end{equation} and an object has the same dimension as its antiparticle, $d_x=d_{\bar x}$, the quantum dimension can be read off by the square root of number of abelian anyon fusion channels of $x\times\bar{x}$. This matches the lattice Hamiltonian prediction \eqref{Z3dimension} and \eqref{Z2quantumdimension}.

\subsection{Splitting and fusion spaces}\label{sec:splittingspaces}
Splitting spaces have already been defined for abelian anyons $[{\bf a}]\times[{\bf b}]=[{\bf a}+{\bf b}]$ in figure~\ref{fig:abeliananyonsplittingspace}(a), where a particular local basis state is chosen by fixing an orientation and ordering of Jordan-Wigner strings. Although arbitrary, a fixed set of basis states for splitting and fusion is implicit in any braided fusion theory, for a consistent collection of basis transformations ($F$-matrices) and braiding operations ($R$-matrices) to be written down. We explicitly choose a basis set of splitting states using a particular local configuration of superposition of Wilson strings. This provides a direct understanding of fusion channels and multiplicities on the quantum state level, and sets the stage for describing basis transformation and braiding in a multi-defect system.

Consider a non-trivial fusion channel $z$ of $x\times y$ so that $N^z_{xy}\neq0$. The splitting of $z$ into $x$ and $y$ can be defined in the microscopic level by locally inserting lattice points (i.e. rotor spaces) and replacing $z$ by a finer trivalent graph that contains $x$ and $y$. For example any vertex on the honeycomb lattice can be blown up into seven vertices by bulging, i.e. replacing the three nearest hexagons by three octagons surrounding three squares. Repeating as needed, any primitive disclination can be replaced by composite disclinations with the same overall Frank angle (curvature) and translation type (torsion). Suppose $\Omega(z)$ is an $\epsilon$-neighborhood around $z$ and $\Omega^c(z)=\mathbb{R}^2-\Omega(z)$ is the complement environment. A ground state with object $z$ is an entangled sum of tensor products \begin{equation}|GS\rangle_z\propto\sum_{b.c.}|\Omega^c(z);b.c.\rangle\otimes|\Omega(z);b.c.\rangle\end{equation} over possible boundary conditions ($b.c.$). Since the local Hilbert space becomes larger after splitting $z\to x\times y$, the old and new ground states $|GS\rangle_z$ and $|GS\rangle_{x\times y}$ should be related by replacing the local ground state $|\Omega(z)\rangle$ with a splitting state $|\mu\rangle\in V_{xy}^z$ that matches the boundary condition. \begin{align}|GS(\mu)\rangle_{x\times y}&\propto\sum_{b.c.}|\Omega^c(z);b.c.\rangle\otimes\left|\vcenter{\hbox{\includegraphics[width=0.4in]{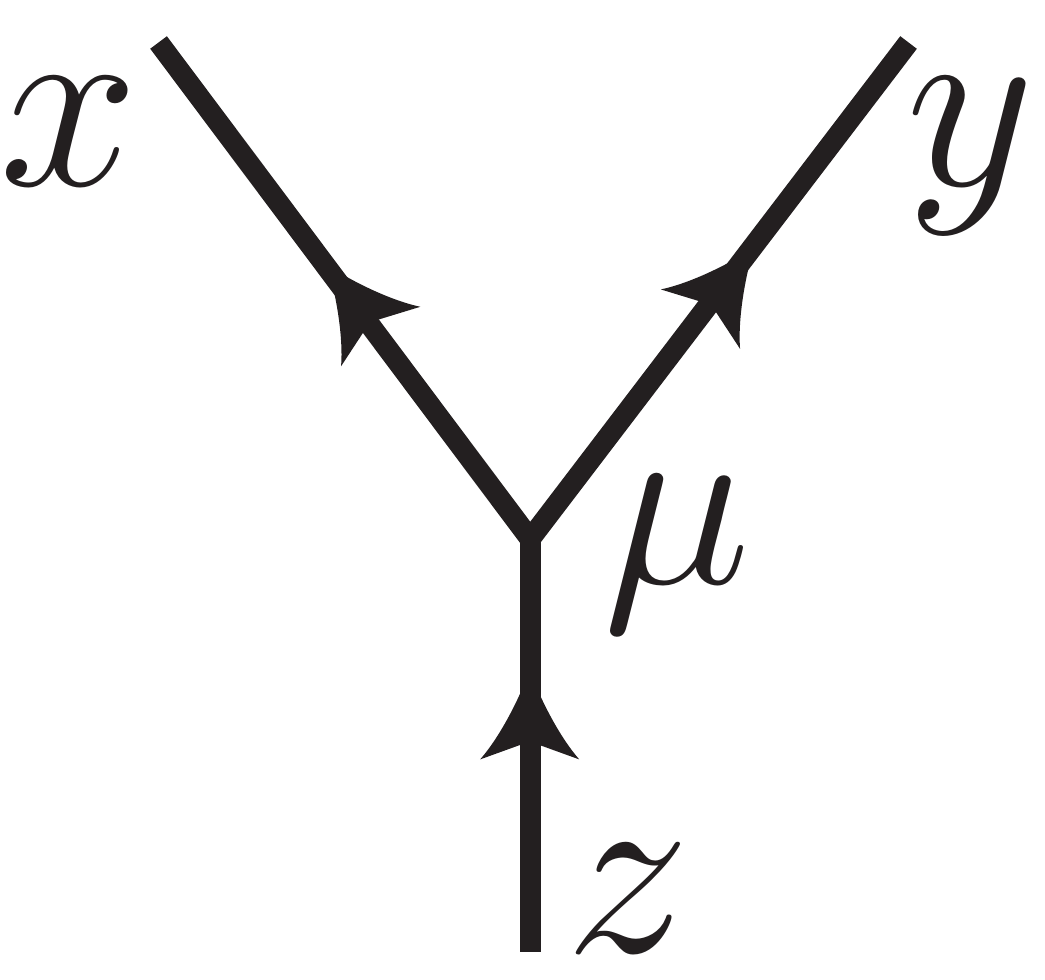}}}\right\rangle_{b.c.}\label{entanglementsum}\\&=\sum_{b.c.}|\Omega^c(z);b.c.\rangle\otimes|\mu;b.c.\rangle\end{align} In general the splitting space $V_{xy}^z$ could be degenerate in which case there are distinct ways for $z$ to split into $x\times y$ and giving rise to multiplicity $N_{xy}^z=\dim(V_{xy}^z)$. These are labeled by $\mu$ and distinguished by Wilson observables. The corresponding splitting state $|\mu\rangle$ is fixed, not only projectively but also with a definite phase, by picking a convention for splitting Jordan-Wigner strings and branch cuts. By definition fusion spaces are adjoints of splitting spaces, diagramatically represented with reversed time ordered arrows and generated by the {\em bra}-state $\langle\mu|$.

\begin{figure}[ht]
	\centering
	\includegraphics[width=3.3in]{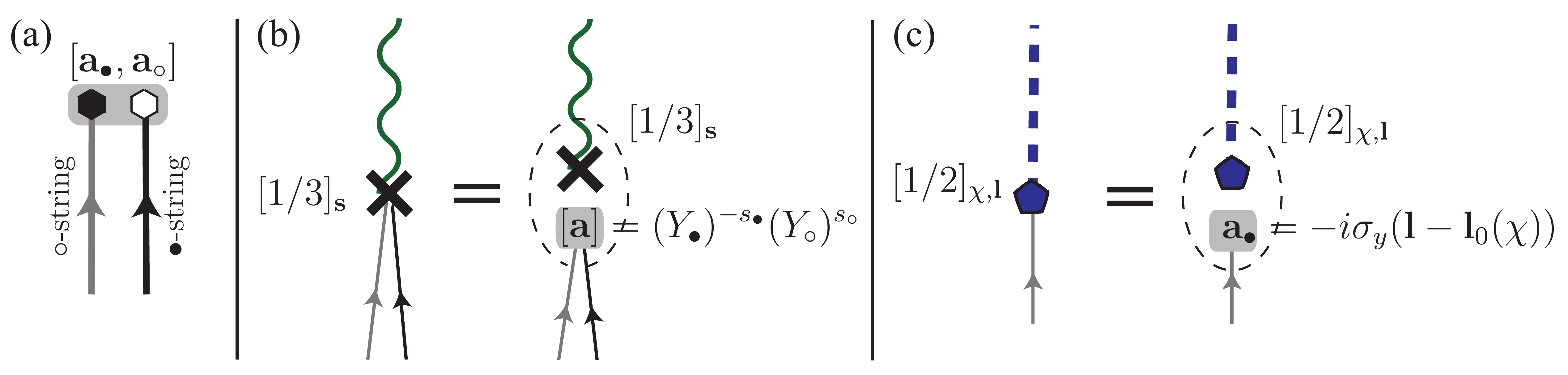}
	\caption{Representation of objects. Solid black and grey lines are $\bullet$- and $\circ$-strings. Curly and dashed lines are threefold and twofold branch cuts. ${\bf s}=(s_\bullet,s_\circ)$ labels $[1/3]$ and $[\overline{1/3}] $for $k$ divisible by 3, for $s_\bullet,s_\circ=-1,0,1$. ${\bf l}=(l_Y,l_R)$ labels $[1/2]_\chi$, for $\chi={Y},{R},{B}$, so that ${\bf l}-{\bf l}_0(\chi)\in\mathbb{Z}_k\oplus\mathbb{Z}_k$, where ${\bf l}_0(\chi)$ is the self-reciprocal species label in \eqref{selfreciprocalspecies}.}\label{fig:objectrep}
\end{figure}
We choose the convention in subsection \ref{sec:AlphabeticpresentationofWilsonalgebra} so that a viewpoint is fixed below an ordered series of objects along a horizontal axes, all branch cuts lives above the horizontal axes, and there is a consistent color and sublattice calibration of abelian anyons on the lower half plane. An object is equipped with \begin{enumerate}[(i)] \item a Jordan-Wigner tail directed from below with its constituent $\circ$-string sitting above (i.e. acting on the vacuum after) and to the left of the $\bullet$-string; \item a $S_3$-branch cut directed from above that matches the defect type of the object.\end{enumerate} Representations of general objects are shown in figure~\ref{fig:objectrep}. The corresponding local state is given by acting the Wilson operator on the particular {\em bare} ground state \begin{align}|GS\rangle_0&\propto\prod_i\left[\sum_{r=0}^{k-1}w^{-rp_i}\hat{D}_i^r\right]\prod_{P}\left[\sum_{r=0}^{k-1}\hat{P}^r\right]|\sigma_{v_\bullet}=\tau_{v_\circ}=1\rangle\label{baregroundstate}\end{align} where $\hat{D}_i$ is the defect operator for the $i^{th}$-defect, the phase $w^{-p_i}$ ensures $\Sigma_{\bullet/\circ}=1$ at a {\em bare} $[1/3]$-defect and $\Theta^{Y/R}=w^{l^{Y/R}_0(\chi)}$ at a {\em bare} $[1/2]_\chi$-defect. The plaquette operator product is taken over both sublattice types $\hat{P}_\bullet$ and $\hat{P}_\circ$, and $\bullet,\circ$-vertex sublattice types are assigned with according to the particular branch cut configuration and calibrated from the fixed viewpoint.

\begin{figure*}[ht]
	\centering
	\includegraphics[width=6in]{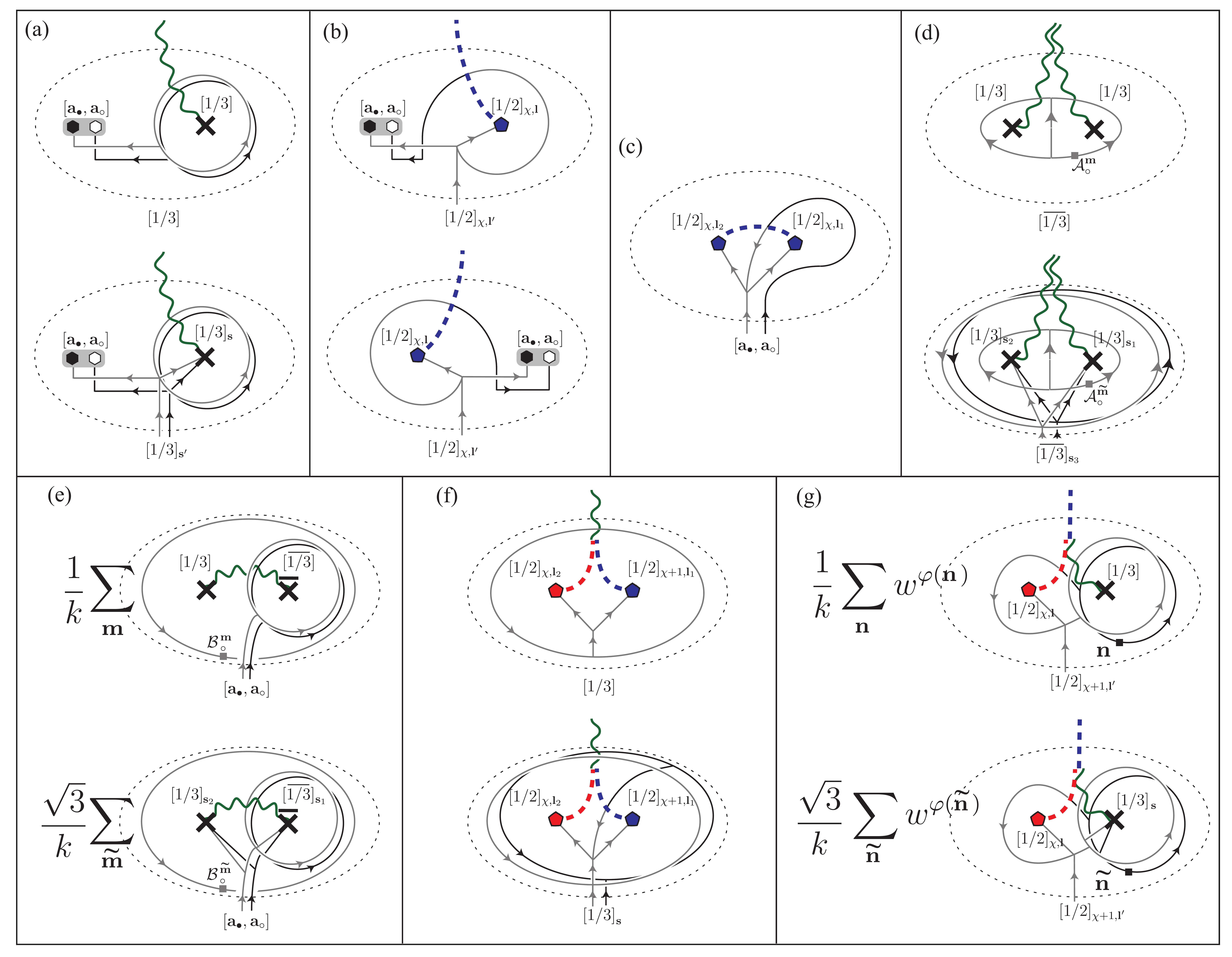}
	\caption{Wilson string configurations for splitting states in eq.\eqref{splittingstatesdefinition}. Crosses and pentagons represent threefold and twofold defects. threefold defects are equipped with a species label ${\bf s}$ only when $k$ is a multiple of 3. Solid black and grey lines are Wilson $\bullet$- and $\circ$-strings. All $\circ$-strings act after and sit above all $\bullet$-ones except $\mathcal{B}^{\bf m}_\circ$ in (e). All closed black $\bullet$-strings act trivially on $|GS\rangle_0$ in eq.\eqref{baregroundstate} while closed grey $\circ$-strings are non-trivial on $|GS\rangle_0$ unless contractible. The phase $\varphi({\bf n})$ in (g) is given by eq.\eqref{splittingphasefigg} and \eqref{splittingphasefigg2}, for $w=e^{2\pi i/k}$.}\label{fig:splittingspaces}
\end{figure*}

Splitting states $|\mu\rangle\in V_{xy}^z$ are obtained by acting a prescribed Wilson operator configuration (figure~\ref{fig:splittingspaces}) onto the particular bare ground state \eqref{baregroundstate} restricted to $\Omega(z)$. \begin{equation}\left|\vcenter{\hbox{\includegraphics[width=0.4in]{Y3string.pdf}}}\right\rangle=\mathcal{O}\left(\vcenter{\hbox{\includegraphics[width=0.4in]{Y3string.pdf}}}\right)|GS\rangle_0\label{splittingstatesdefinition}\end{equation} The prescribed Wilson operator configurations $\mathcal{O}$ are chosen and shown in figure~\ref{fig:splittingspaces}. The anyon label of each Wilson string is set to match that of individual objects $x,y,z$ as defined in figure~\ref{fig:objectrep}. So that different fusion channels $z$ correspond different Wilson string configurations and orthogonal splitting states.  Closed Wilson loops $\mathcal{A}^{\bf m}_\circ,\mathcal{B}^{\bf m}_\circ$ are the two prototypes shown in figure~\ref{fig:defectWilsonloop}(a),(b), and the multi-exponents notation ${\bf m}=(m_1,m_2)$ is adapted so that for example $\mathcal{A}_\circ^{\bf m}=(\mathcal{A}_\circ^Y)^{m_1}(\mathcal{A}_\circ^R)^{m_2}$ is observed at the viewpoint fixed at the grey square. The multi-exponents ${\bf m}=(m_1,m_2)$ live on some 2D triangular $\mathbb{Z}_k$ lattice such as those in figure~\ref{fig:abeliananyonlattice} that labels abelian anyons, and when $k$ is divisible by 3, $\widetilde{\bf m}=(\tilde{m}_1,\tilde{m}_2)$ live on a reduced lattice as shown in figure~\ref{fig:secondBZ}. 

Multiplicity occurs for the splitting $[1/3]\times[1/3]=d_{[1/3]}[\overline{1/3}]$ shown in figure~\ref{fig:splittingspaces}(d) and the degeneracy is generated by the non-trivial Wilson loops $\mathcal{A}^{\bf m}_\circ$ defined in section \ref{sec:Z3twistdefects}. It might be more physically natural to label the degenerate state $|{\bf m}\rangle=\mathcal{A}_\circ^{\bf m}|GS\rangle_0$ according to eigenvalues ${\boldsymbol\alpha}=I_0{\bf m}$ of the observable $\mathcal{A}_\bullet^{\bf n}=e^{i\frac{2\pi}{k}{\bf n}^T{\boldsymbol\alpha}}$, where the intersection matrix $I_0$ was defined in \eqref{Z3intersectionmatrix}. Or when $k$ is divisible by 3, the eigenvalues are given by $\widetilde{\boldsymbol\alpha}=I_0\widetilde{\bf m}+{\bf s}_0$, where ${\bf s}_0$ is related to the species labels for the $[1/3]_{\bf s}$ as discussed in eq.\eqref{shiftvectora0} and \eqref{Z33Wilsonmatrix1}.

The normalized sum in figure~\ref{fig:splittingspaces}(e) for splitting $[1/3]\times[\overline{1/3}]=\sum_{\bf a}[{\bf a}]$ is to ensure the splitting state is an eigenstate of $\mathcal{B}^{\bf m}_{\bullet/\circ}$, closed Wilson loops surround the threefold defect pair that measure the overall abelian anyonic fusion channel $[{\bf a}]$. No such summation is required for figure~\ref{fig:splittingspaces}(c) for splitting $[1/2]_\chi\times[1/2]_\chi=\sum'_{\bf a}[{\bf a}]$ since the splitting state is automatically an eigenstate for $\mathcal{Z}_\bullet^{\bf m}$ ($\bullet$-loops that enclose both twofold defects) thanks to the definition \eqref{baregroundstate} for $|GS\rangle_0$, while any $\mathcal{Z}_\circ^{\bf m}$ can be turn inside out into some $\mathcal{Z}_\bullet^{{\bf m}'}$ as shown in figure~\ref{fig:defectWilsonloop}(c), figure~\ref{fig:doublelooppair} and eq.\eqref{Zloopinsideout}.

The normalized sum in figure~\ref{fig:splittingspaces}(g) for splitting $[1/2]_{\chi,{\bf l}}\times[1/3]=[1/2]_{\chi+1,{\bf l}'}$ is to ensure the overall Wilson loops $\Theta'_{Y/R}$ defined in figure~\ref{fig:defectlocalloop}(a) or figure~\ref{fig:Z2loopdefinition} surrounding both the twofold and threefold defects are in the condensate and can be absorbed by the local ground state. The phase factor $\varphi({\bf n})$ ensures that the splitting state carries eigenvalues $\Theta'_{Y/R}=w^{l'_{Y/R}}$, $w=e^{2\pi i/k}$, for ${\bf l}'=(l'_Y,l'_R)$ is the overall species label. It is given by \begin{align}\varphi_{\left[\frac{1}{2}\right]_\chi\times\left[\frac{1}{3}\right]}({\bf n})&={\bf n}^Ti\sigma_y\left[\frac{k}{2}{\bf f}^{\chi-1}+\frac{1}{2}\left(\Lambda_{\chi+1}-\Lambda_\chi\right){\bf n}\right]\label{splittingphasefigg}\end{align} The splitting state for the reversed order $[1/3]\times[1/2]_{\chi+1,{\bf l}}=[1/2]_{\chi,{\bf l}'}$ is defined using a mirror image of figure~\ref{fig:splittingspaces}(g) with the modified phase factor \begin{align}\varphi_{\left[\frac{1}{3}\right]\times\left[\frac{1}{2}\right]_\chi}({\bf n})&={\bf n}^Ti\sigma_y\left[\frac{k}{2}{\bf f}^{\chi+1}+\frac{1}{2}\left(\Lambda_{\chi}-\Lambda_{\chi-1}\right){\bf n}\right]\label{splittingphasefigg2}\end{align} \eqref{splittingphasefigg} and \eqref{splittingphasefigg2} are phase differences between the overall double Wilson loop and the local one that encloses only the constituent one. These are evaluated in appendix~\ref{sec:doubleloop}. For $k$ is divisible by 3, the summation is taken only over the $k^2/3$ vectors $\tilde{\bf n}$ in the reduced triangular lattice (figure~\ref{fig:secondBZ}).

\subsection{Basis Transformation}\label{sec:Fmoves}
In a defect system, ground states are labeled by eigenvalues of observables. Examples were given by the eigenstate $|{\boldsymbol\alpha}\rangle$ in \eqref{Z3groundstatealpha} of a threefold defect system measured by $\mathcal{A}_\bullet^{\bf n}=e^{i\frac{2\pi}{k}{\bf n}^T{\boldsymbol\alpha}}$, and the eigenstate $|{\bf a}\rangle$ in \eqref{Z2groundstatea} of a twofold defect system with abelian anyon fusion channel $[{\bf a}]=[{\bf a}_\bullet,{\bf a}_\circ]$ observed by $(\mathcal{Z}_{odd})^{\bf n}=e^{i\frac{2\pi}{k}{\bf n}^Ti\sigma_y{\bf a}_\bullet}$. In a general multi-defect system, one can pick an arbitrary ordering of the defect series (see figure~\ref{fig:multidefects}) and label ground states according to a complete set of maximally commuting observables generated by the Wilson loop prototypes in figure~\ref{fig:defectWilsonloop} between defect pairs, such as $\mathcal{B}^{\chi,\bullet/\circ}$ and $\mathcal{Z}^{\chi}$ that observe abelian anyon fusion channel, and $\mathcal{A}^{\chi,\bullet/\circ}$ that measures fusion degeneracy.

\begin{figure}[ht]
	\centering
	\includegraphics[width=3in]{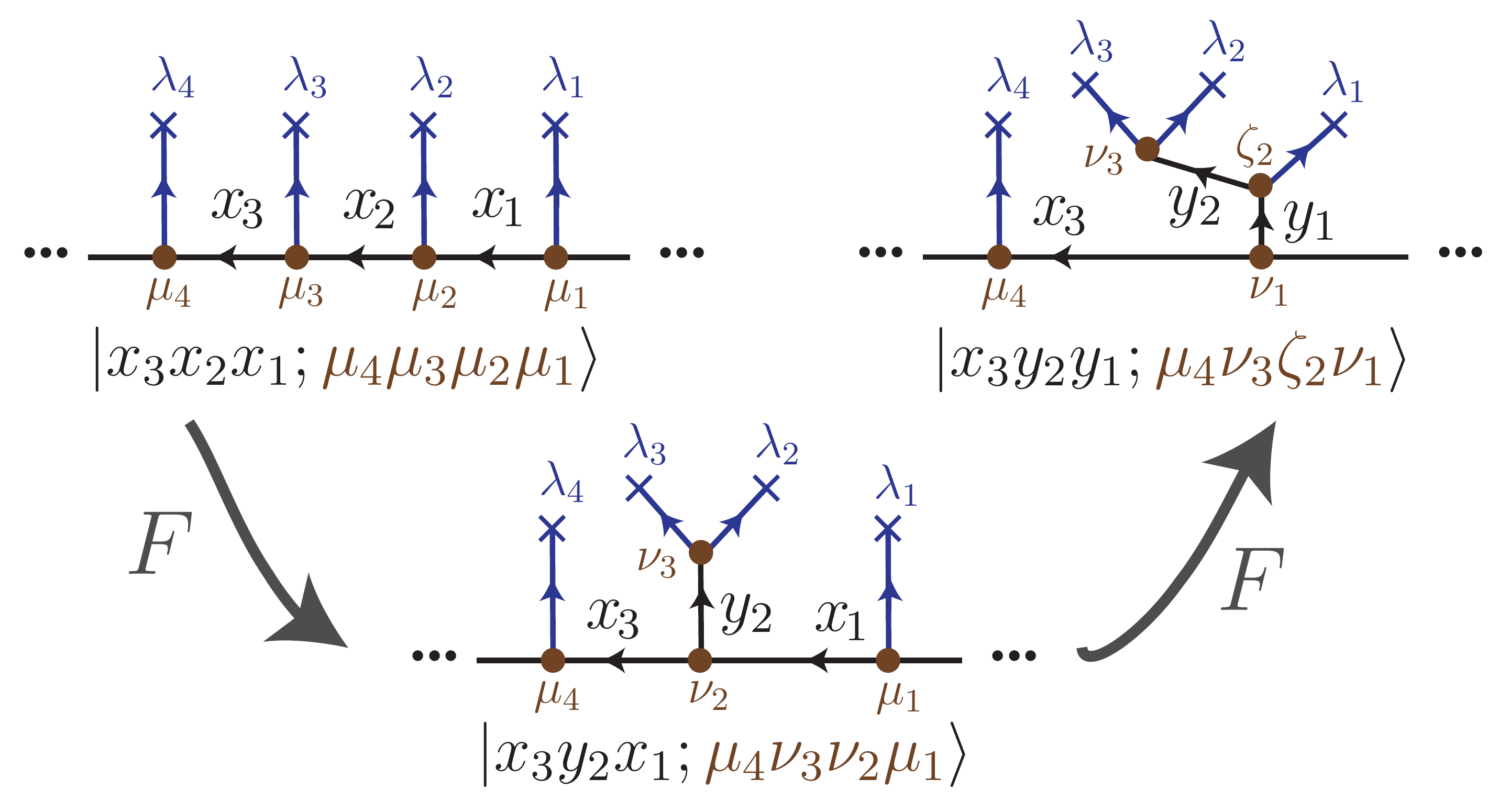}
	\caption{Basis transformation of defect states by composition of fundamental $F$-moves. $\lambda_i$ are objects, $x_i,y_i$ are admissible fusion channels, and $\mu_i,\nu_i,\zeta_i$ label splitting space degeneracies.}\label{fig:Fmoves}
\end{figure}
A particular choice of maximally commuting observables can be represented by a fusion/splitting tree (examples are shown in figure~\ref{fig:Fmoves}) where directed branches and trivalent vertices are specified by fusion/splitting channels $x_i$ and fusion/splitting degeneracy $\mu_j$ respectively. Fusion channels are labeled by objects in \eqref{objectset}, which are either abelian anyon $[{\bf a}]$ or twist defects $[1/2]$ or $[1/3]$. All fusion channels should be admissible so that $N_{xy}^z$ at each trivalent vertex is non-zero, i.e. $x\times y\to z$ is allowed. Fusion degeneracy only occurs for $[1/3]\times[1/3]=d_{[1/3]}[\overline{1/3}]$, in which case the vertex is labeled by splitting state $|{\boldsymbol\alpha}\rangle$ measured by Wilson operator $\mathcal{A}_\bullet^{\bf n}=e^{i\frac{2\pi}{k}{\bf n}^T{\boldsymbol\alpha}}$ and generated by $\mathcal{A}_\circ^{\bf m}$, ${\bf m}=I_0^{-1}\boldsymbol\alpha$, shown in the splitting diagram figure~\ref{fig:splittingspaces}(d). Ground states are therefore of one to one correspondence to the set of admissible internal fusion channels and degeneracy states, and are denoted by $|\{x_i\};\{\mu_j\}\rangle$ and shown in figure~\ref{fig:Fmoves}. Note that the state $|\{x_i\};\{\mu_j\}\rangle$ is {\em not} only projectively defined. Its $U(1)$-phase is well-defined by acting on the particular ground state $|GS\rangle_0$ in eq.\eqref{baregroundstate} with the prescribed Wilson string configuration for individual splitting states chosen in figure~\ref{fig:splittingspaces}. %This feature is unique in an exact solvable lattice model. Projectivization might be required in an effective low energy theory.

A different set of maximally commuting observables represented by another fusion/splitting tree gives rise to a different representation of ground states $|\{y_p\};\{\nu_q\}\rangle$. Since these are energy eigenstates of the same defect system, they are related by some unitary basis transformation \begin{equation}|\{x_i\};\{\mu_i\}\rangle=\sum_{\{y_p\},\{\nu_q\}}F^{\{y_p\},\{\nu_q\}}_{\{x_i\},\{\mu_j\}}|\{y_p\};\{\nu_q\}\rangle\end{equation} This can be broken down into a sequence of fundamental moves, known as $F$-symbols, each involves a rearrangement of three adjacent branches that reorders fusion by associativity $(x\times y)\times z\cong x\times(y\times z)$. An example of such a sequence is shown in figure~\ref{fig:Fmoves}.

Consider a system containing the ordered objects $\lambda_1,\lambda_2,\lambda_3$ which fuse to the overall object $\lambda_4$. Similar to \eqref{entanglementsum}, ground state can be expressed as an entangled sum of tensor products between local ground state around the objects and the environment with matching boundary conditions. With fixed boundary condition, local ground states are tensor products of splitting states \begin{align}\left|\vcenter{\hbox{\includegraphics[width=0.5in]{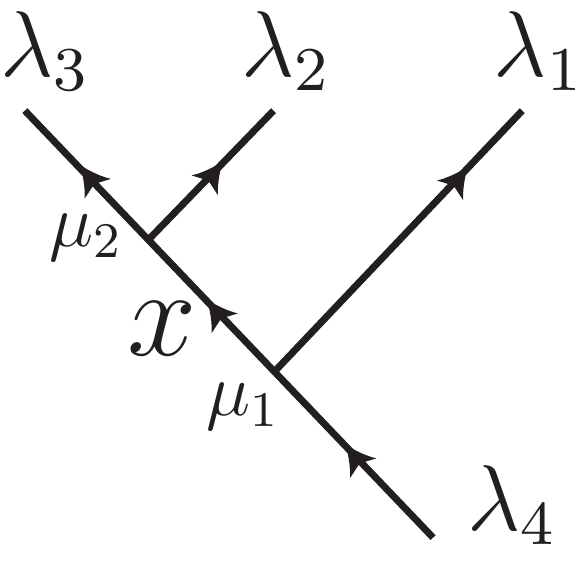}}}\right\rangle&=\left|\vcenter{\hbox{\includegraphics[width=0.3in]{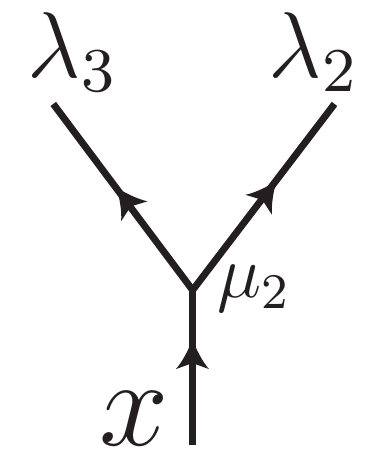}}}\right\rangle\otimes\left|\vcenter{\hbox{\includegraphics[width=0.3in]{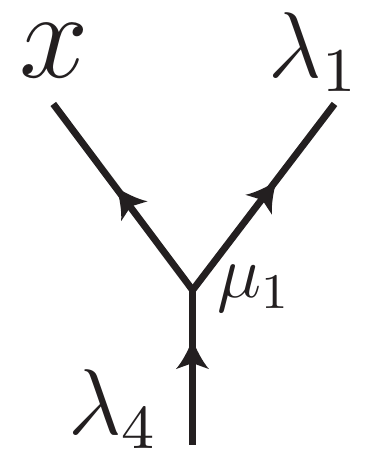}}}\right\rangle\\\left|\vcenter{\hbox{\includegraphics[width=0.5in]{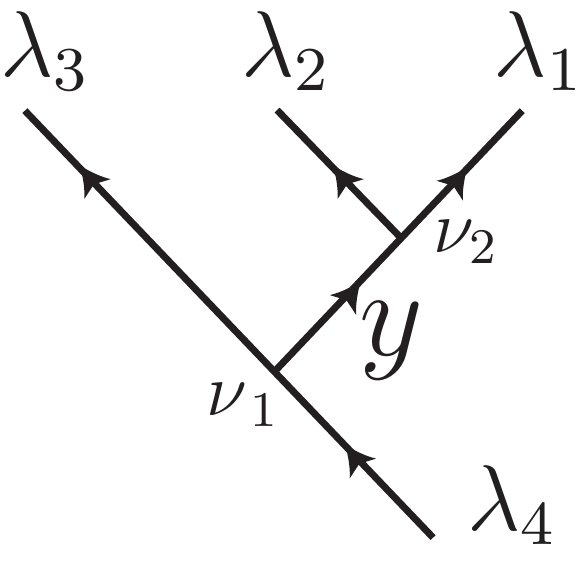}}}\right\rangle&=\left|\vcenter{\hbox{\includegraphics[width=0.3in]{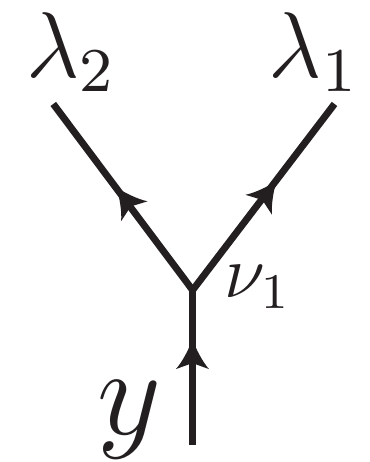}}}\right\rangle\otimes\left|\vcenter{\hbox{\includegraphics[width=0.3in]{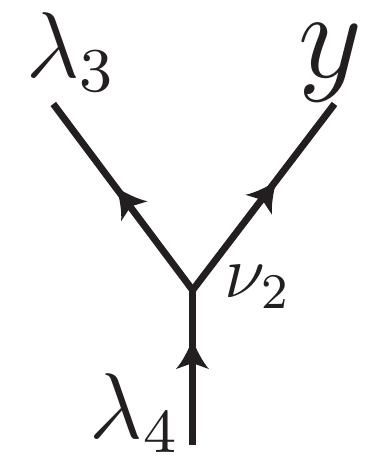}}}\right\rangle\end{align} defined by acting on the bare ground state $|GS\rangle_0$ in \eqref{baregroundstate} and \eqref{splittingstatesdefinition} with Wilson operators prescribed by figure~\ref{fig:splittingspaces}. These two ground states are related by the unitary basis transformaion \begin{align}\left|\vcenter{\hbox{\includegraphics[width=0.5in]{F1.pdf}}}\right\rangle=\sum_{y,\nu_1,\nu_2}\left[F^{\lambda_3\lambda_2\lambda_1}_{\lambda_4}\right]_{x,\mu_2,\mu_1}^{y,\nu_1,\nu_2}\left|\vcenter{\hbox{\includegraphics[width=0.5in]{F2.pdf}}}\right\rangle\end{align} where $F$-matrix entries are given by the inner product \begin{align}\left[F^{\lambda_3\lambda_2\lambda_1}_{\lambda_4}\right]_{x,\mu_1,\mu_2}^{y,\nu_1,\nu_2}=\left\langle\vcenter{\hbox{\includegraphics[width=0.5in]{F2.pdf}}}\right.\left|\vcenter{\hbox{\includegraphics[width=0.5in]{F1.pdf}}}\right\rangle\label{Fdefinition}\end{align} The overall basis transformation between any two fusion trees or maximally commuting sets of observables is independent from the sequence of $F$-moves in between. This cocycle consistency is ensured by the pentagon equation ``$FF=FFF$" and MacLane's coherence theorem (see ref.[\onlinecite{Kitaev06, MacLanebook}]). Instead of solving the algebraic pentagon equation, we compute the $F$-matrices directly from definition \eqref{Fdefinition} with the gauge degree of freedom fixed by the prescribed splitting states in figure~\ref{fig:splittingspaces}. An exhaustive list of $F$-matrices for $k$ not divisible by 3 is given in table~\ref{tab:Fsymbols} in appendix~\ref{sec:Fsymbols}. Here we demonstrate a few simple examples for the purpose of illustration.

Consider the basis transformation of two abelian anyons sandwiching a threefold defect, ${\bf a}\times[1/3]\times{\bf b}$. According to figure~\ref{fig:splittingspaces}(a), the splitting states tensor products are represented by the union of two splitting diagrams. \begin{align}\left|\vcenter{\hbox{\includegraphics[width=0.5in]{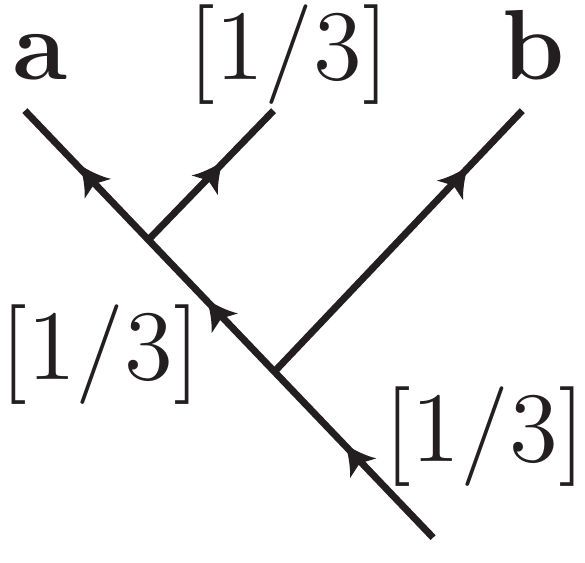}}}\right\rangle&=\left[\vcenter{\hbox{\includegraphics[width=1.5in]{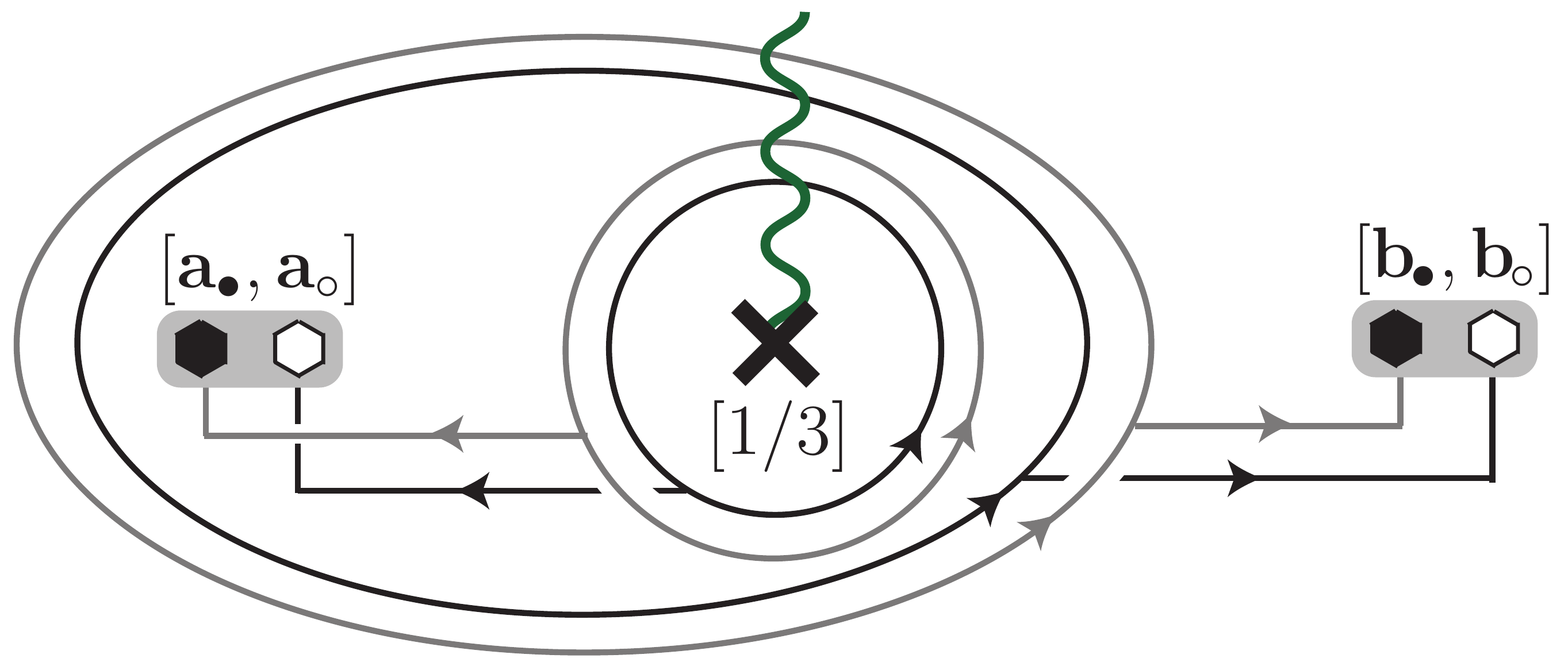}}}\right]|GS\rangle_0\nonumber\\&=\left[\vcenter{\hbox{\includegraphics[width=1.2in]{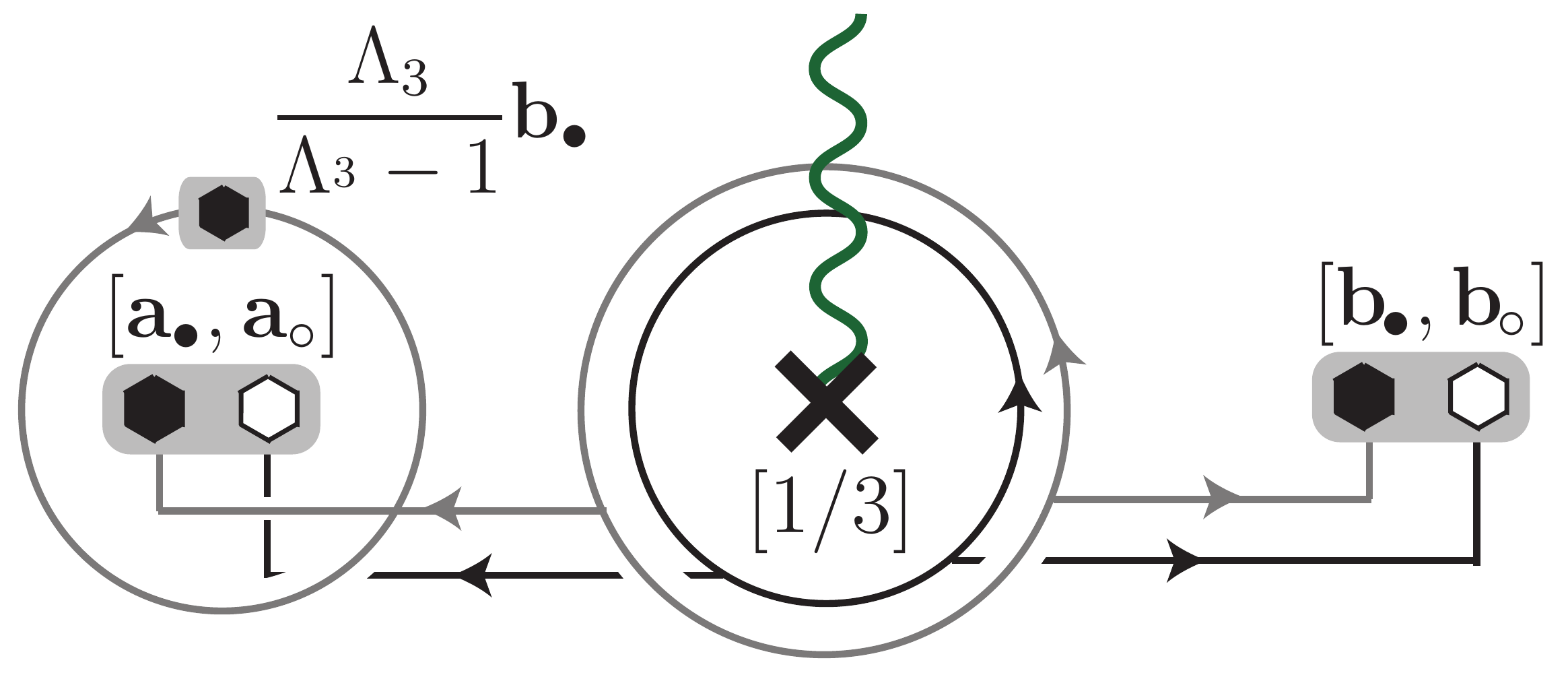}}}\right]|GS\rangle_0\\\left|\vcenter{\hbox{\includegraphics[width=0.5in]{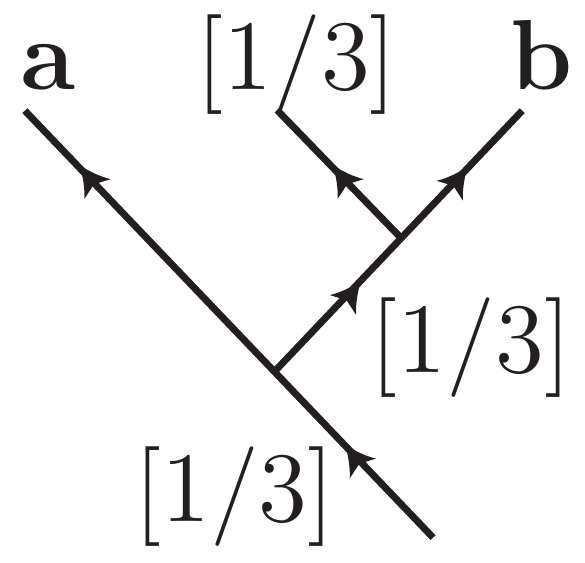}}}\right\rangle&=\left[\vcenter{\hbox{\includegraphics[width=1.5in]{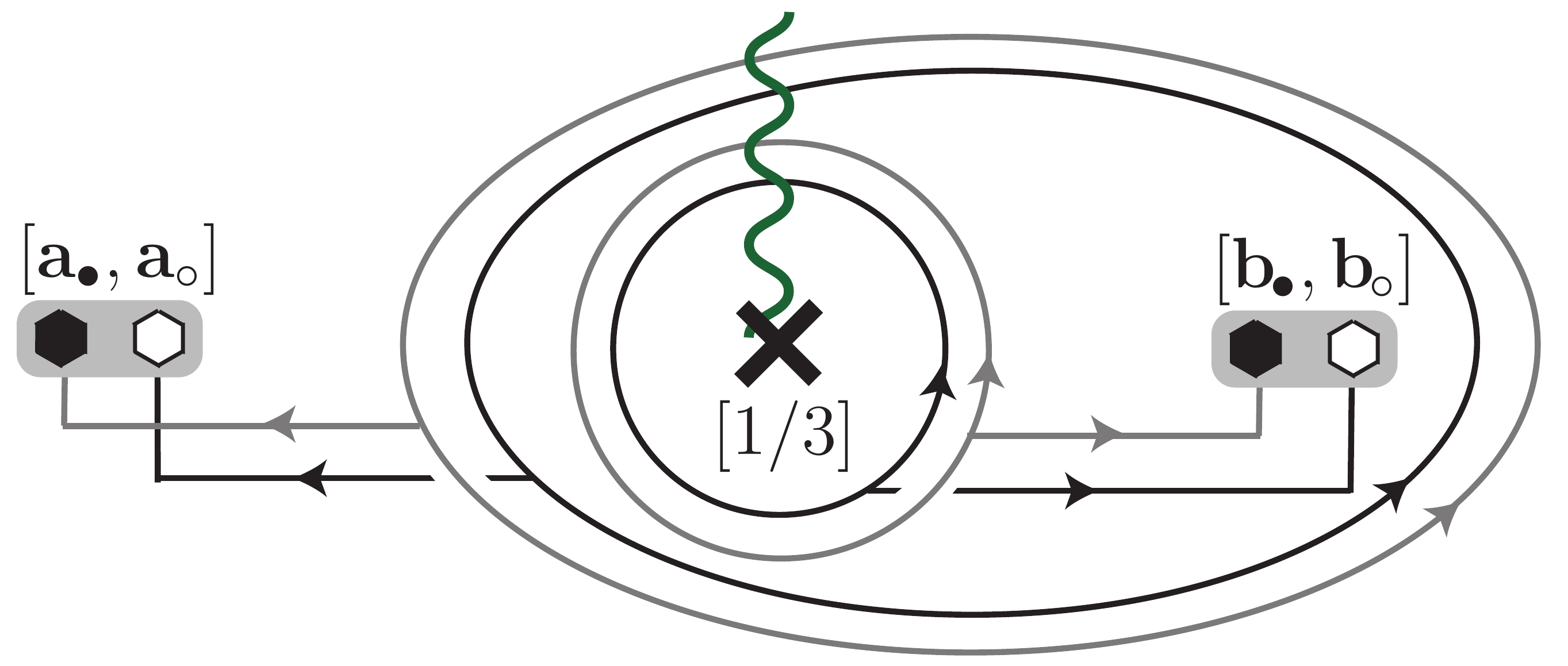}}}\right]|GS\rangle_0\nonumber\\&=\left[\vcenter{\hbox{\includegraphics[width=1.2in]{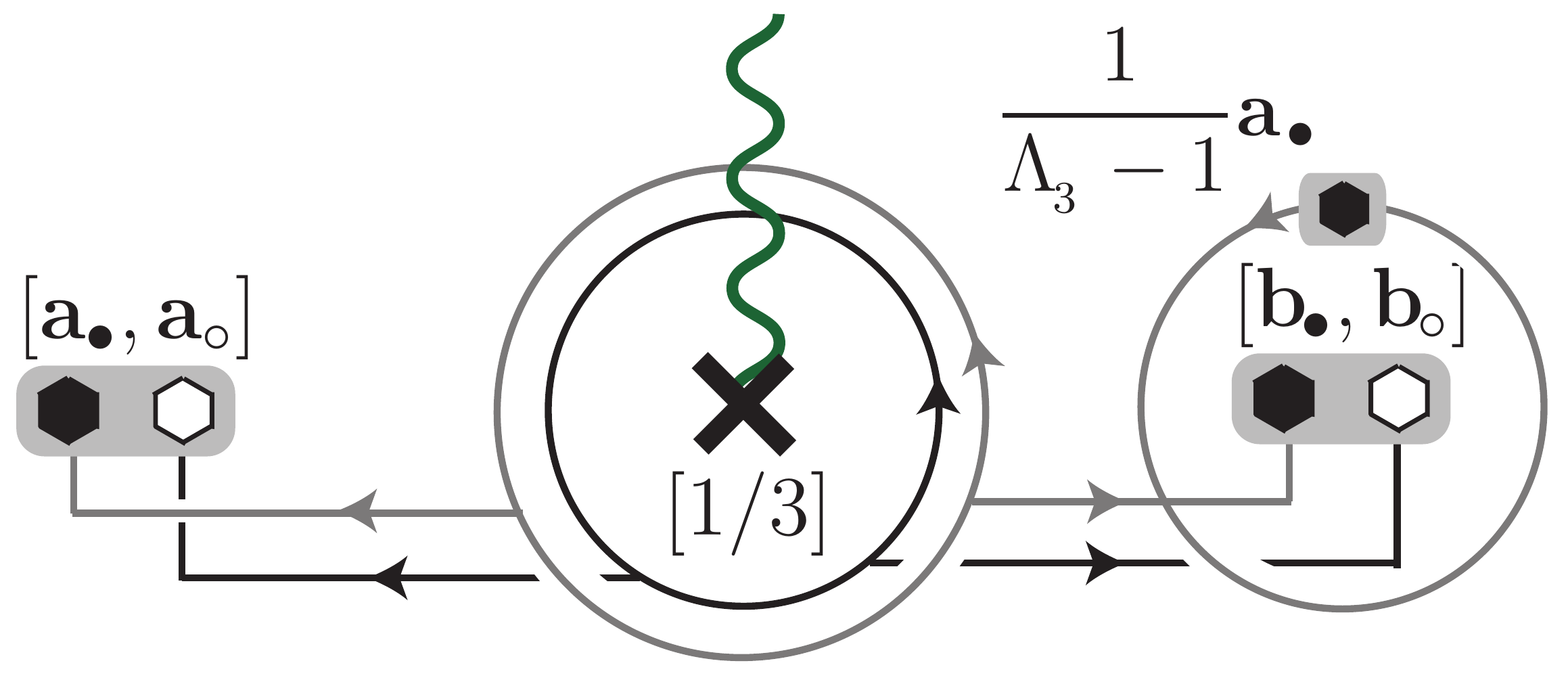}}}\right]|GS\rangle_0\end{align}
where the deformation of Wilson strings are facilitated by the Wilson loop condensate in the ground state. The $F$-symbol is given by their overlap and is an abelian phase. \begin{align}F^{{\bf a}[1/3]{\bf b}}_{[1/3]}&=\left[\vcenter{\hbox{\includegraphics[width=0.5in]{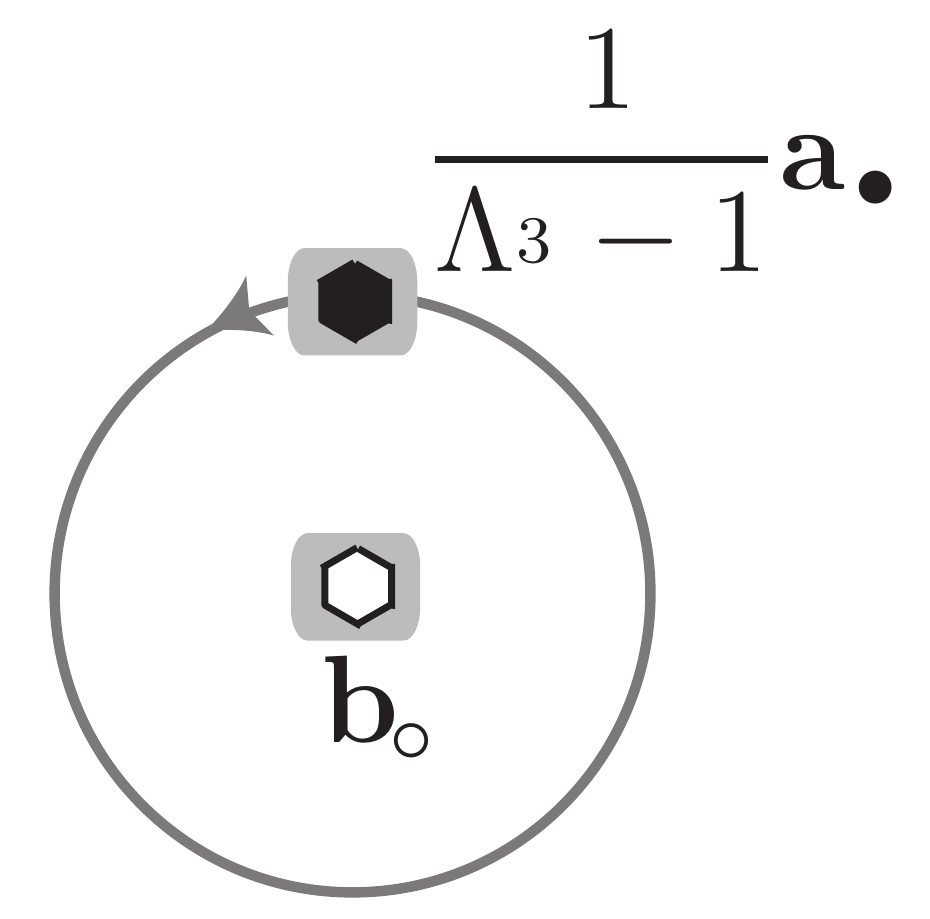}}}\right]^\dagger\left[\vcenter{\hbox{\includegraphics[width=0.5in]{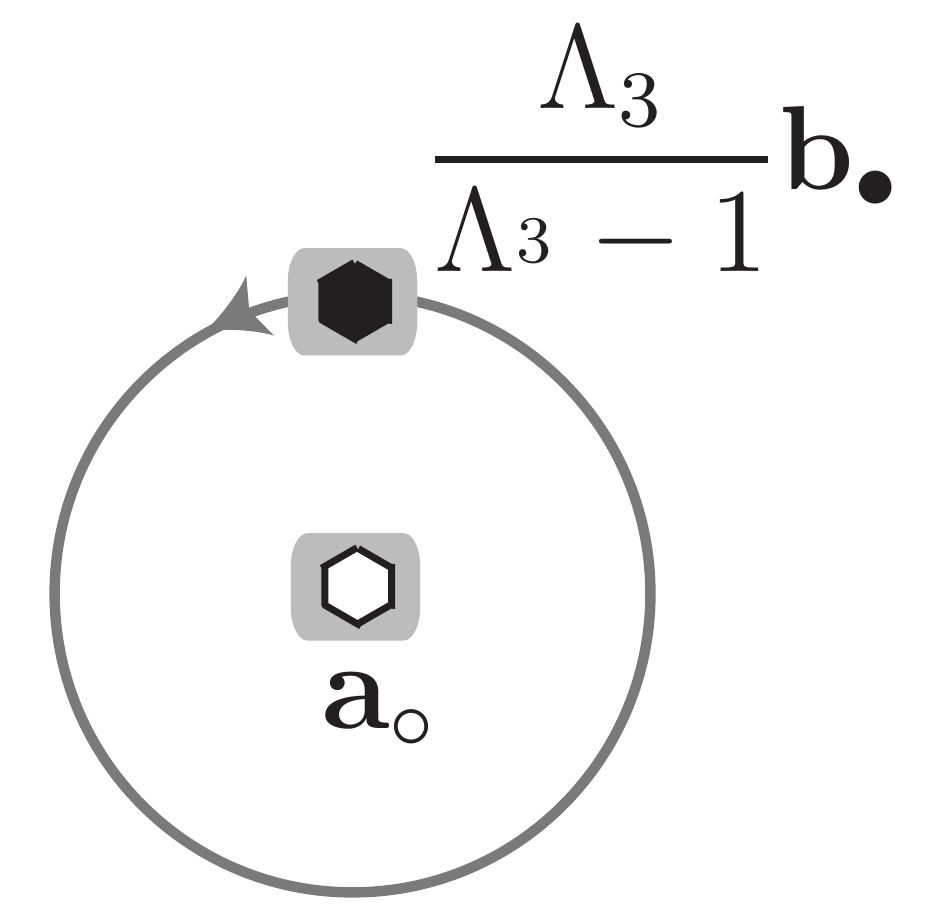}}}\right]\nonumber\\&=w^{{\bf a}_\circ^Ti\sigma_y\frac{\Lambda_3}{\Lambda_3-1}{\bf b}_\bullet-{\bf b}_\circ^Ti\sigma_y\frac{1}{\Lambda_3-1}{\bf a}_\bullet}\end{align} where ${\bf a}_\bullet,{\bf a}_\circ,{\bf b}_\bullet,{\bf b}_\circ$ are $\mathbb{Z}_k$-valued 2-dimensional vectors living on the anyon lattices (figure~\ref{fig:abeliananyonlattice}). $\Lambda_3$ is the $2\times2$ matrix $\left(\begin{array}{*{20}c}0&-1\\1&-1\end{array}\right)$ representing cyclic color permutation, and $(\Lambda_3-1)$ is invertible with $\mathbb{Z}_k$-entries only when $k$ is not a multiple of 3.

Next we consider the transformation for $[\overline{1/3}]\times[1/3]\times[1/3]$. The splitting states tensor product for first fusing  $[\overline{1/3}]\times[1/3]$ is represented by gluing figure~\ref{fig:splittingspaces}(a) and (e). \begin{align}\left|\vcenter{\hbox{\includegraphics[width=0.6in]{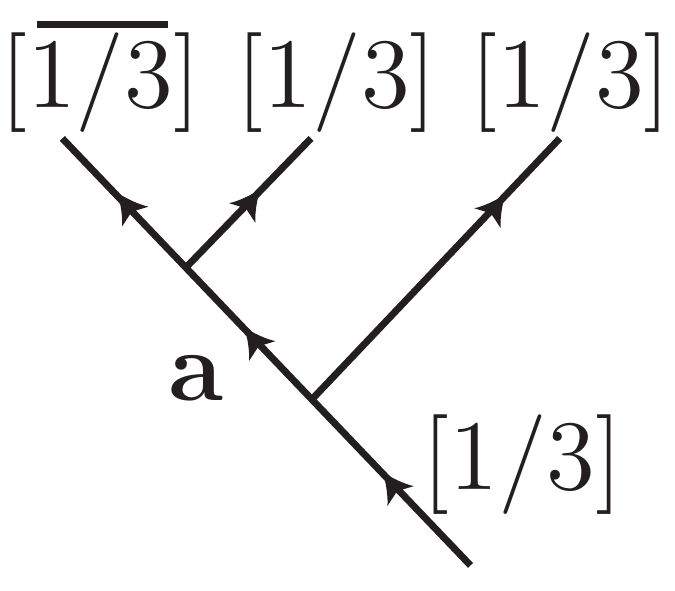}}}\right\rangle&=\frac{1}{k}\sum_{\bf n}\left[\vcenter{\hbox{\includegraphics[width=1.2in]{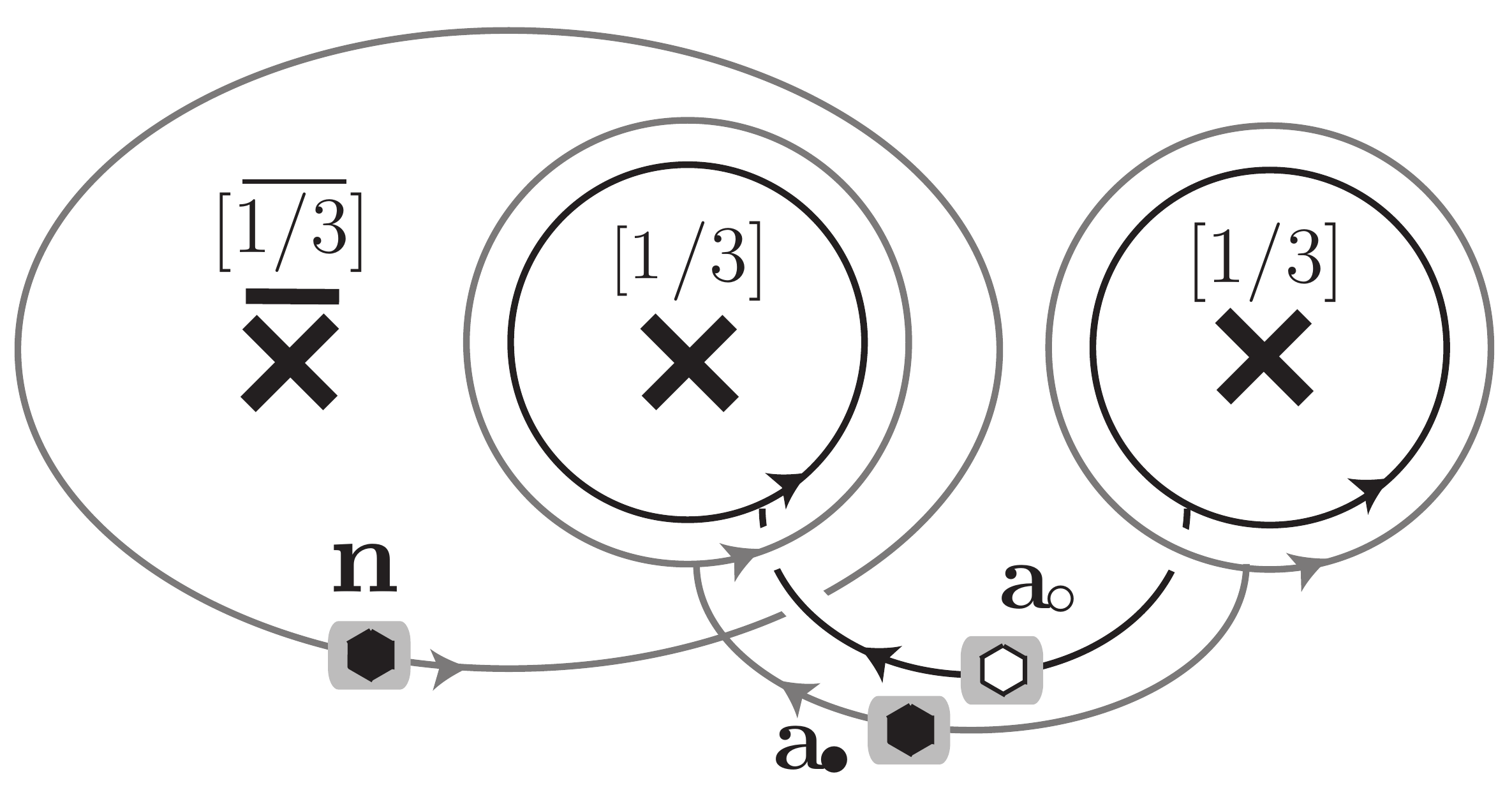}}}\right]|GS\rangle_0\nonumber\\&=\frac{1}{k}\sum_{\bf n}\left[\vcenter{\hbox{\includegraphics[width=1.2in]{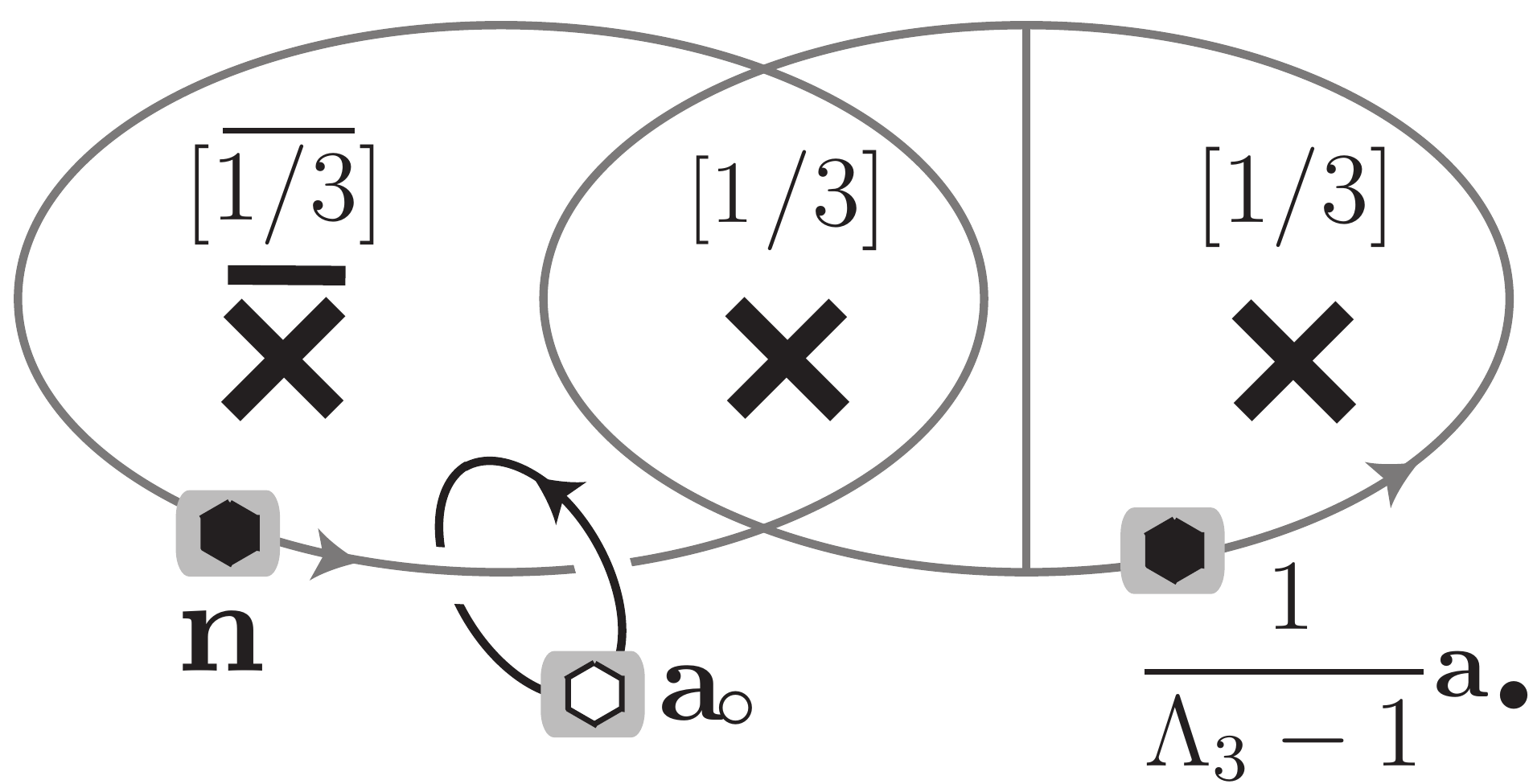}}}\right]|GS\rangle_0\end{align} where the black dumbbell-shaped Wilson loop can be absorbed by the ground state $|GS\rangle_0$ leaving a link between ${\bf a}_\circ$ and ${\bf n}$. The splitting states tensor product for first fusing $[{1/3}]\times[1/3]$ is given by \begin{align}\left|\vcenter{\hbox{\includegraphics[width=0.6in]{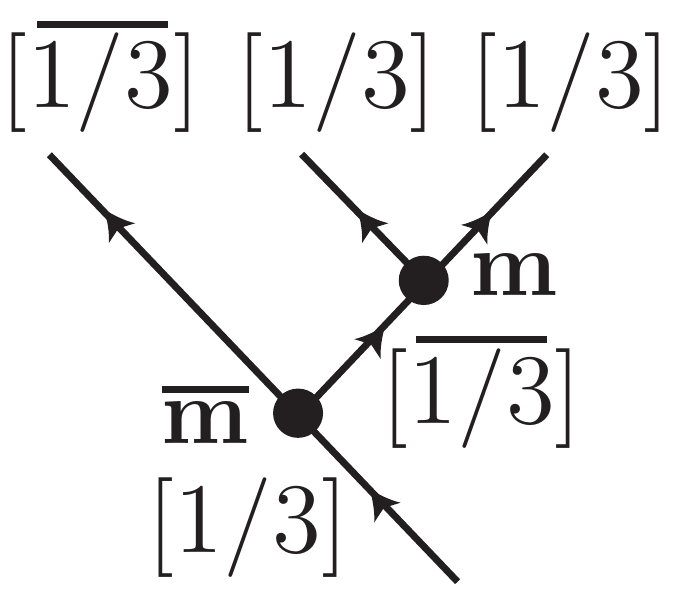}}}\right\rangle&=\left[\vcenter{\hbox{\includegraphics[width=1.2in]{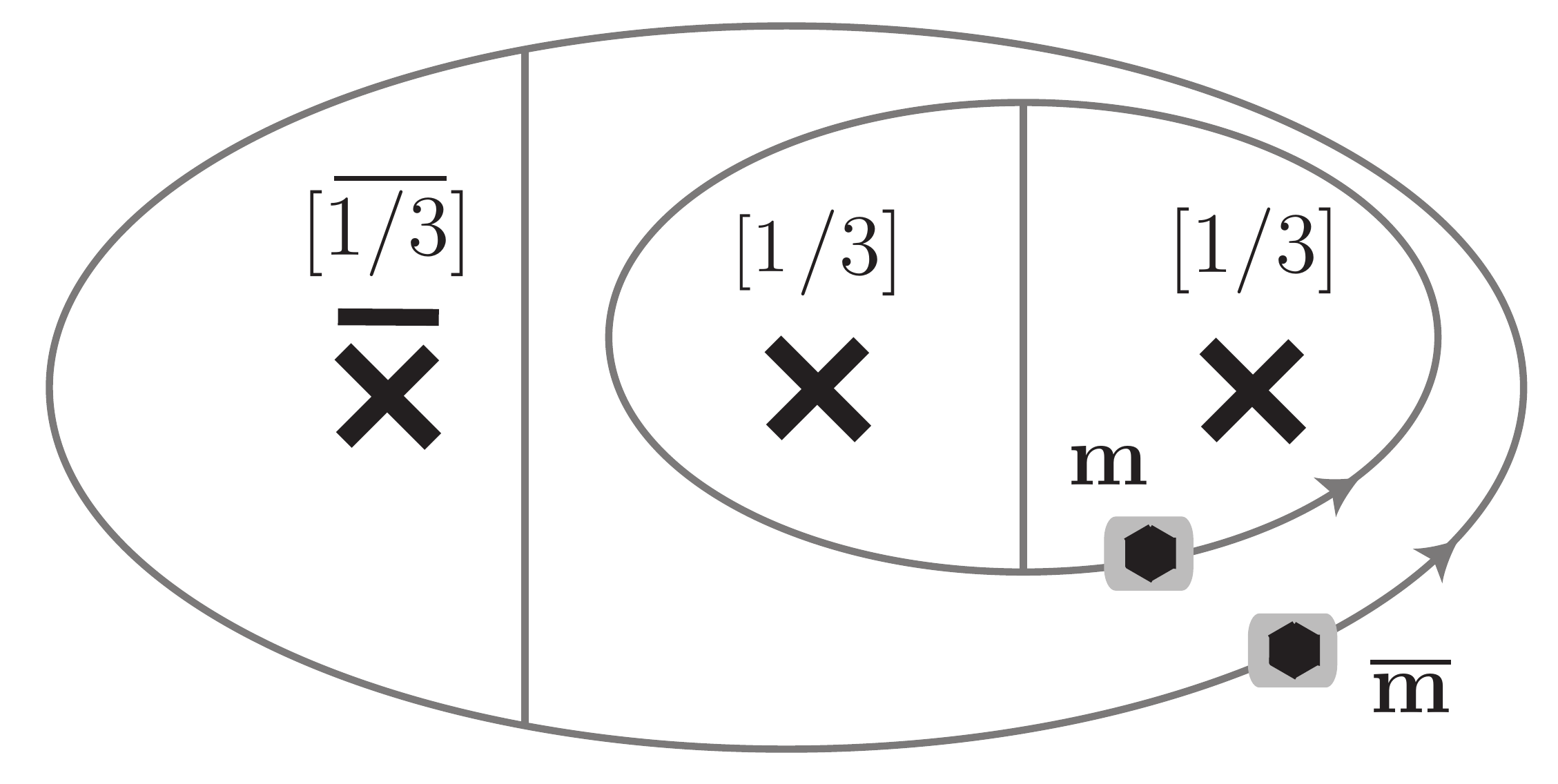}}}\right]|GS\rangle_0\nonumber\\&=\left[\vcenter{\hbox{\includegraphics[width=1.2in]{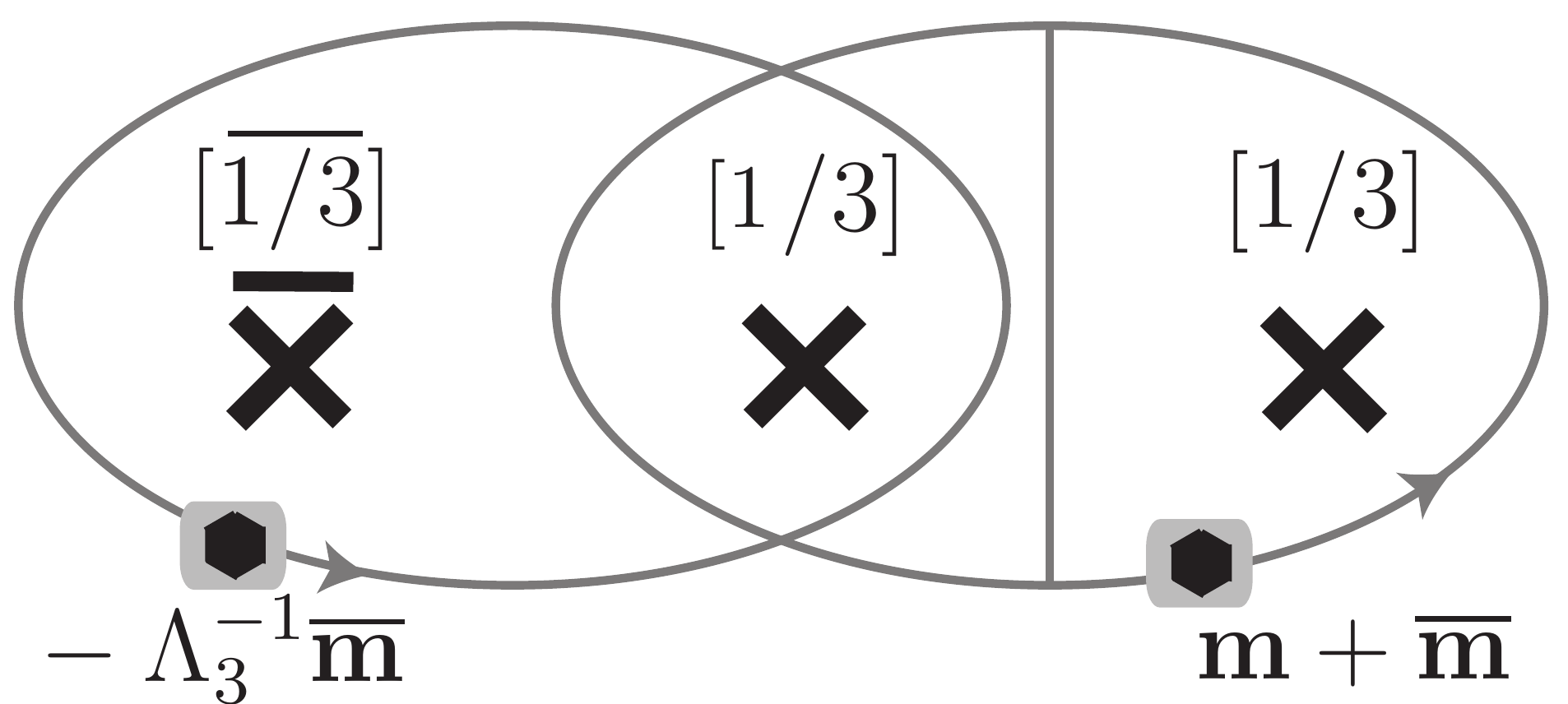}}}\right]|GS\rangle_0\end{align} where the fusion degeneracies are labeled by ${\bf m}=I_0^{-1}\boldsymbol\alpha$ and $\overline{\bf m}=\bar{I}_0^{-1}\overline{\boldsymbol\alpha}$, for $I_0=-\bar{I}_0=i\sigma_y\Lambda_3(\Lambda_3-1)$ is the intersection matrix in \eqref{I0I1}. The $k^4\times k^4$ $F$-matrix is given by the overlap between the two splitting states. \begin{align}\left[F^{[\overline{1/3}][1/3][1/3]}_{[1/3]}\right]_{\bf a}^{\overline{\bf m}{\bf m}}&=\frac{1}{k}\sum_{\bf n}\left[\vcenter{\hbox{\includegraphics[width=0.5in]{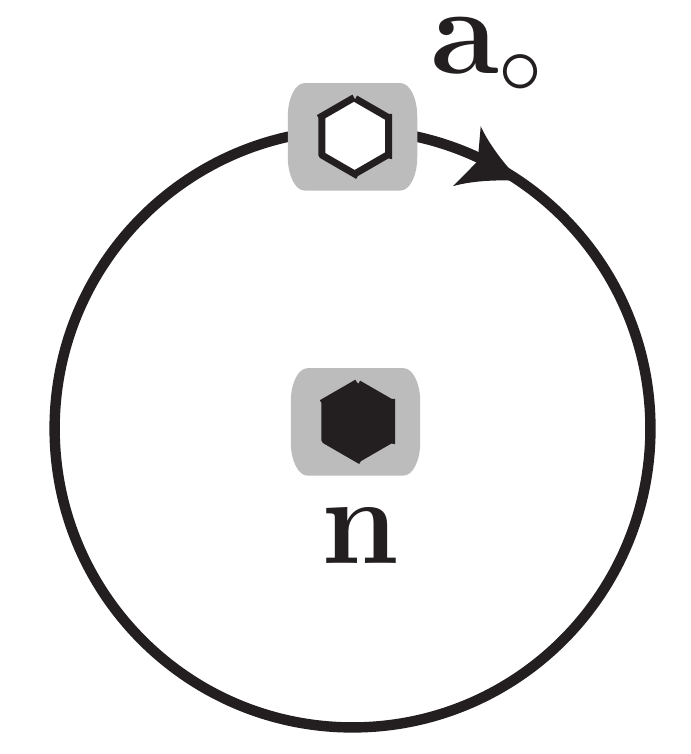}}}\right]\delta^{\frac{1}{\Lambda_3-1}{\bf a}_\bullet}_{{\bf m}+\overline{\bf m}}\delta^{\bf n}_{-\Lambda_3^{-1}\overline{\bf m}}\nonumber\\&=\frac{1}{k}w^{\overline{\boldsymbol\alpha}^T\frac{1}{\Lambda_3-1}{\bf a}_\circ}\delta^{\boldsymbol\alpha-\overline{\boldsymbol\alpha}}_{i\sigma_y\Lambda_3{\bf a}_\bullet}\end{align}

Lastly we consider the transformation for twofold defects of the same color type, $[1/2]_{\chi,{\bf l}_3}\times[1/2]_{\chi,{\bf l}_2}\times[1/2]_{\chi,{\bf l}_1}=[1/2]_{\chi,{\bf l}}$. The splitting state tensor products are given by pasting together the Wilson strings in figure~\ref{fig:splittingspaces}(b) and (c). Fusing the first pair gives  \begin{align}\left|\vcenter{\hbox{\includegraphics[width=0.7in]{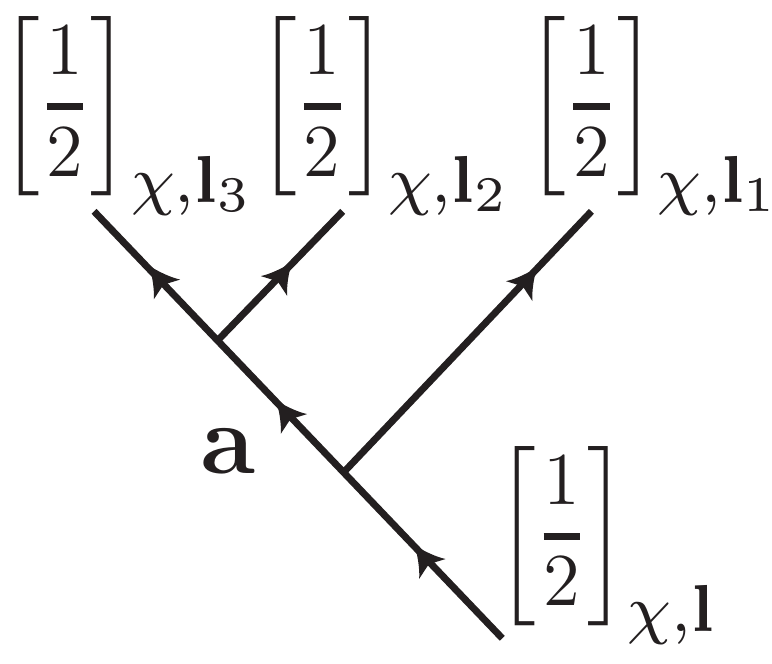}}}\right\rangle&=\left[\vcenter{\hbox{\includegraphics[width=1.4in]{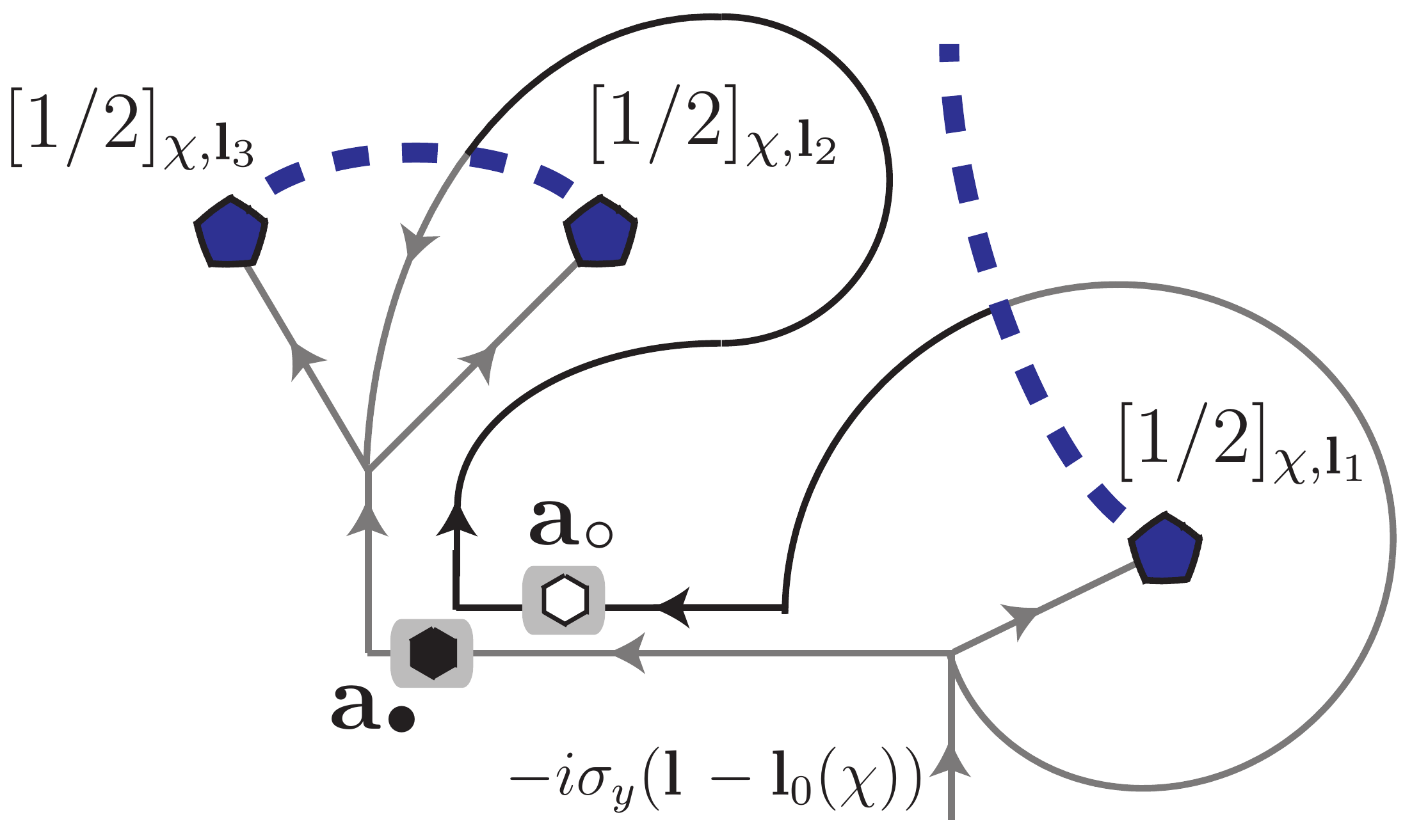}}}\right]|GS\rangle_0^L\nonumber\\&=\left[\vcenter{\hbox{\includegraphics[width=1.2in]{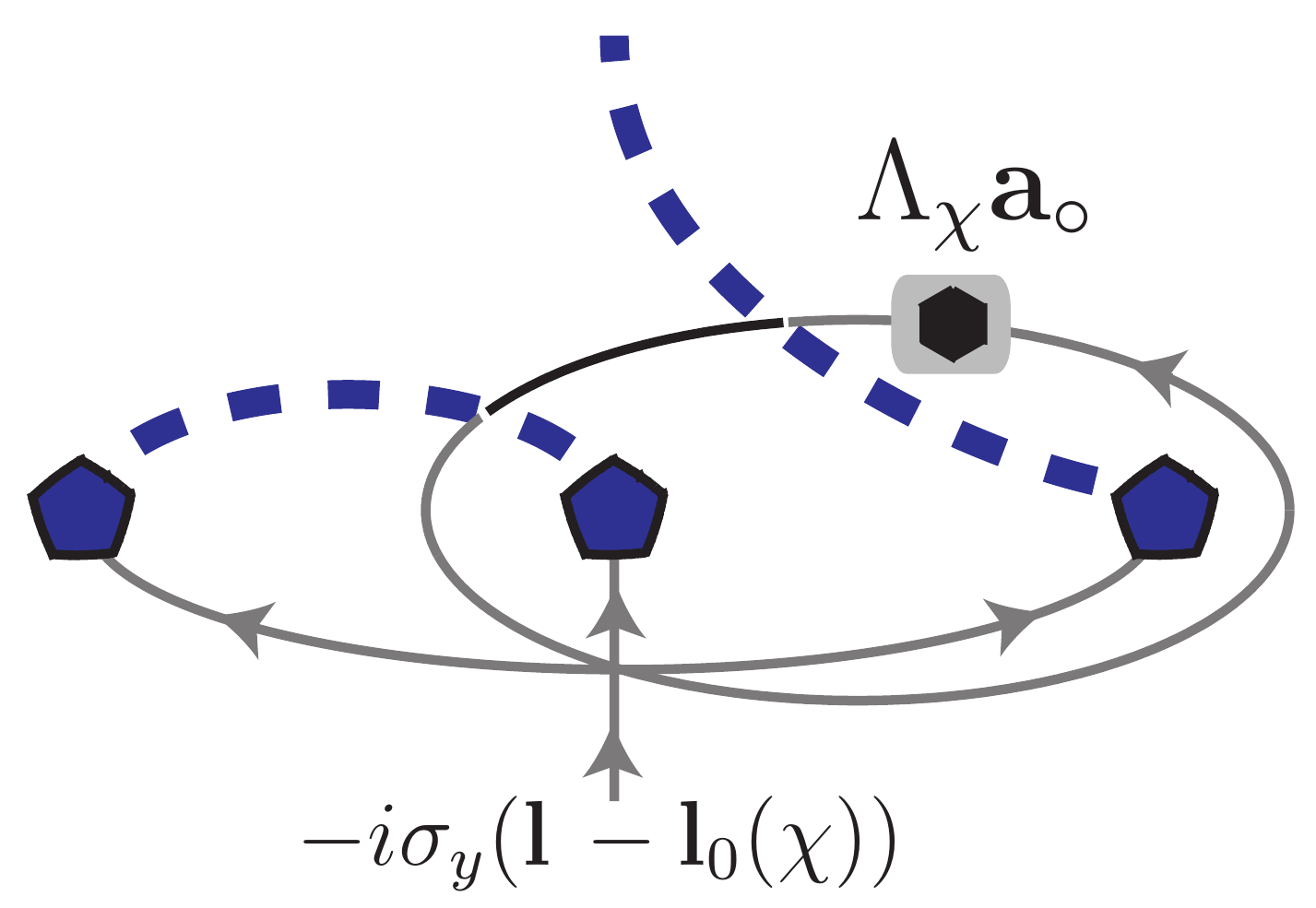}}}\right]|GS\rangle_0^L\label{F222GS1}\end{align} where any grey string attaching to a defect $[1/2]_{\chi,{\bf l}_i}$ brings an abelian anyon $-i\sigma_y({\bf l}_i-{\bf l}_0(\chi))$ into the defect, for $i=1,2,3$ (see figure~\ref{fig:objectrep}(c)), and $\Lambda_\chi=i\sigma_yJ_\chi$ is the $2\times2$ matrix representing the transposition that characterizes $[1/2]_\chi$ (see eq.\eqref{Z2JYRB} or \eqref{lambdachiapp}). Fusing the last pair gives \begin{align}\left|\vcenter{\hbox{\includegraphics[width=0.7in]{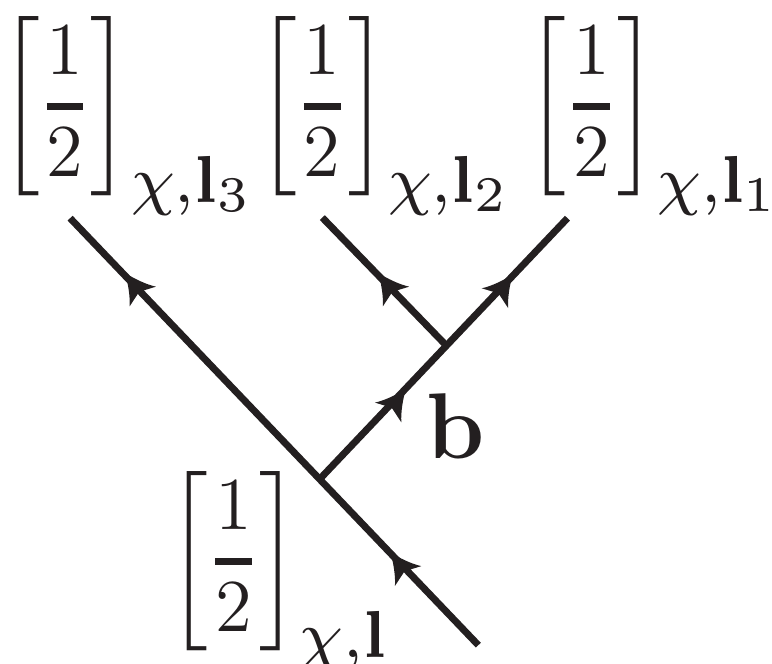}}}\right\rangle&=\left[\vcenter{\hbox{\includegraphics[width=1.4in]{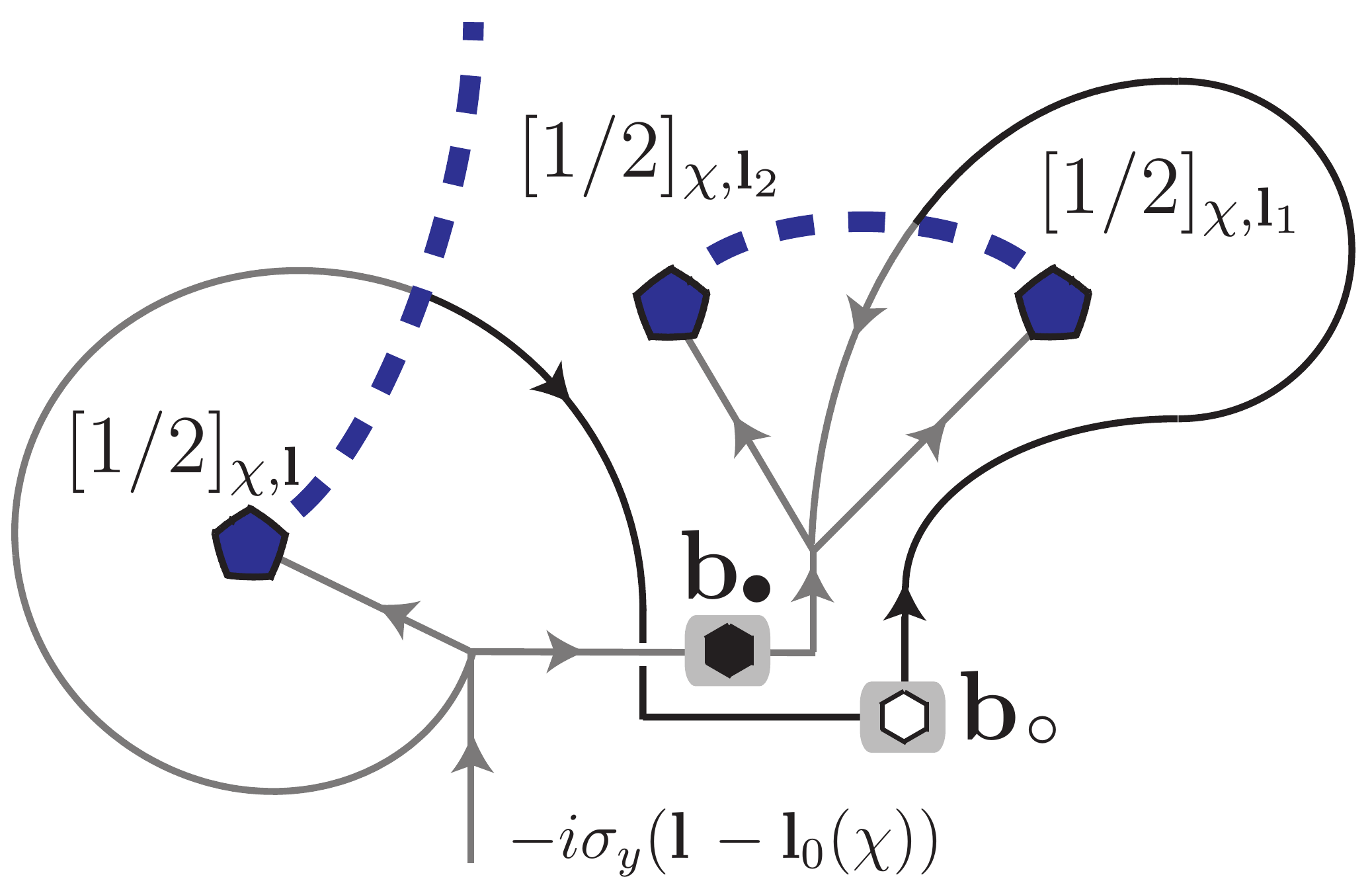}}}\right]|GS\rangle_0^R\nonumber\\&=\left[\vcenter{\hbox{\includegraphics[width=1.2in]{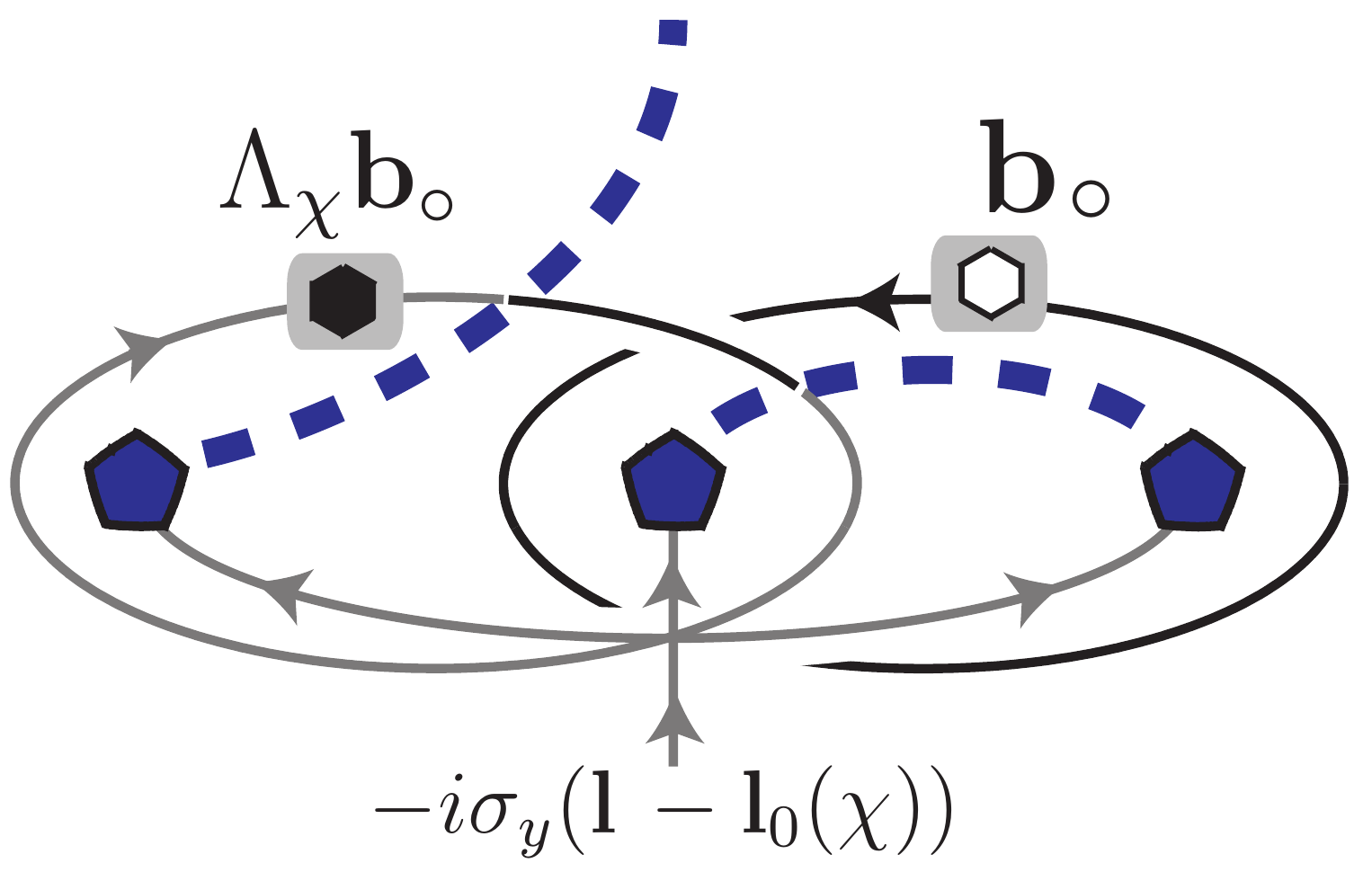}}}\right]|GS\rangle_0^R\label{F222GS2}\end{align} where the black loop ${\bf b}_\circ$ can be absorbed and eliminated by the ground state condensate. Notice the bare ground states $|GS\rangle_0^{L/R}$ for the two cases in eq.\eqref{F222GS1} and \eqref{F222GS2} depend on different branch cut configurations and are therefore distinct from each other. Wilson loops circling the first two defects are in the condensate for $|GS\rangle_0^L$ but send the $|GS\rangle_0^R$ to a new ground state. The two bare ground states are related by \begin{align}|GS\rangle_0^R=\frac{1}{k}\sum_{\bf m}\left[\vcenter{\hbox{\includegraphics[width=1in]{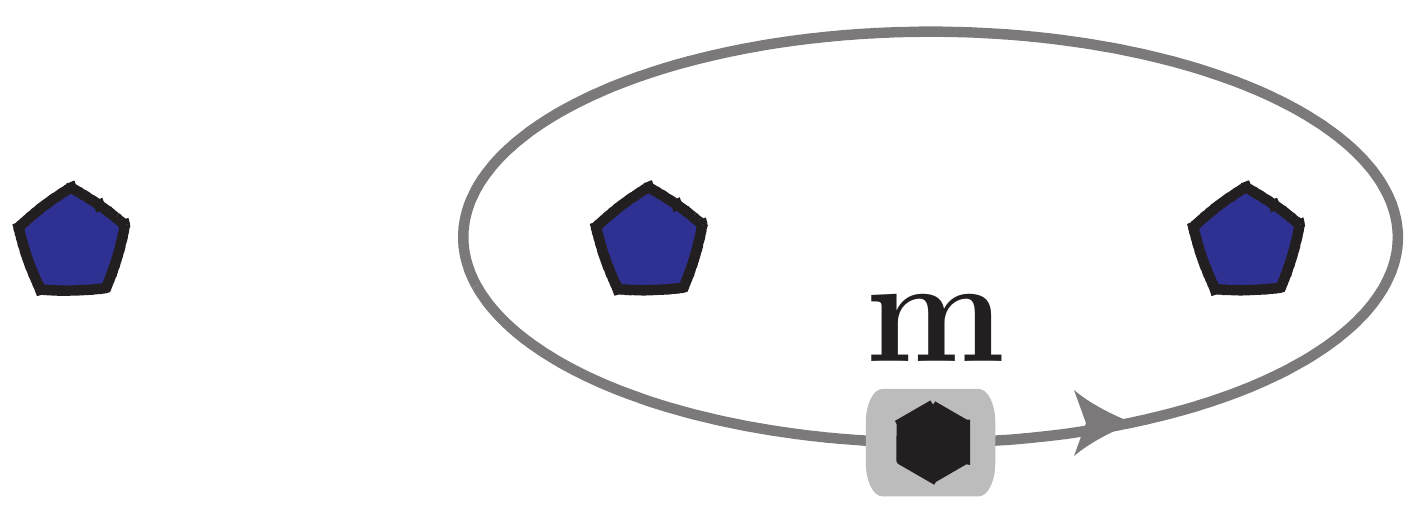}}}\right]|GS\rangle_0^L\end{align} and in particular $_0^R\langle GS|GS\rangle_0^L=1/k$. The $F$-matrix is given by the overlap between \eqref{F222GS1} and \eqref{F222GS2}. \begin{align} \left[F^{\left[\frac{1}{2}\right]_{\chi}\left[\frac{1}{2}\right]_{\chi}\left[\frac{1}{2}\right]_{\chi}}_{\left[\frac{1}{2}\right]_{\chi,{\bf l}}}\right]_{\bf a}^{\bf b}&={}_0^R\langle GS|\left[\vcenter{\hbox{\includegraphics[width=1in]{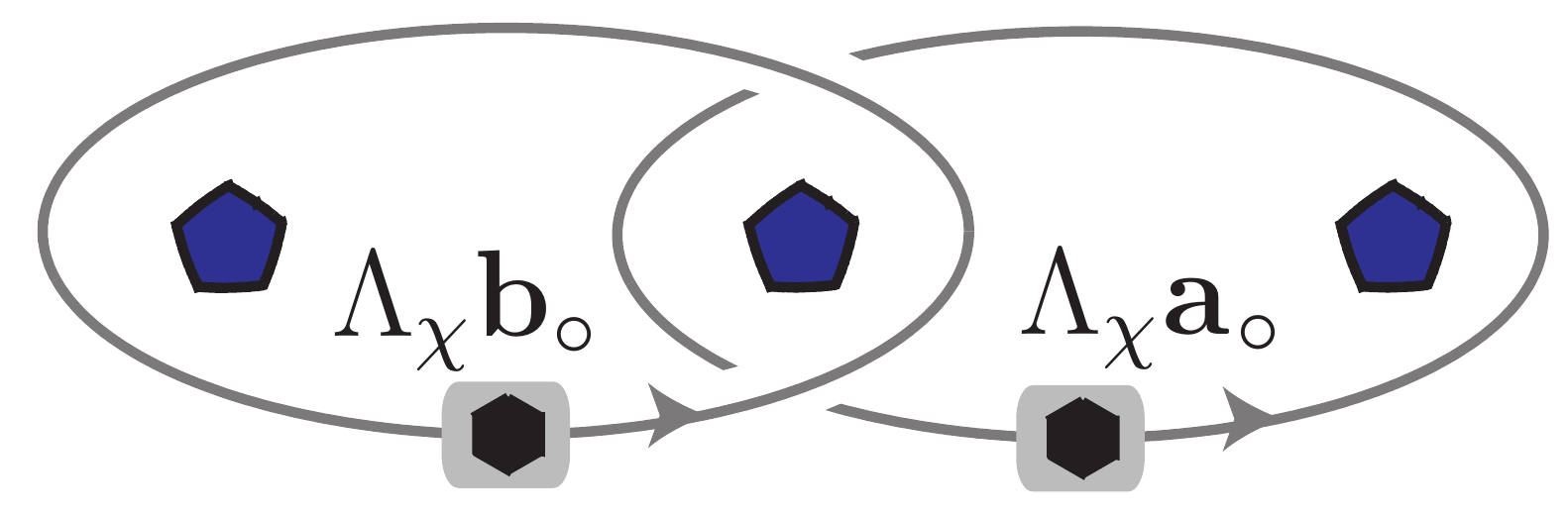}}}\right]|GS\rangle_0^L\nonumber\\&=\frac{1}{k}w^{{\bf a}_\circ^Ti\sigma_y\Lambda_\chi{\bf b}_\circ}\label{F222intext}\end{align} where the two loops can be passed across each other, leaving a phase and absorbed in the bare ground states $|GS\rangle_0^{L/R}$. Note that the intermediate abelian channels are restricted by the species labels \begin{align}{\bf l}&={\bf l}_1+{\bf l}_2+{\bf l}_3-3{\bf l}_0(\chi)\nonumber\\&={\bf l}_1+i\sigma_y({\bf a}_\bullet+\Lambda_\chi{\bf a}_\circ)={\bf l}_3+i\sigma_y({\bf b}_\bullet+\Lambda_\chi{\bf b}_\circ)\end{align} and therefore the $F$-matrix is of dimension $k^2$.

Quantum dimension of an object $x$ can be read off by the first entry of a $F$-matrix, $1/d_x=|[F^{x\bar xx}_x]^0_0|$. \begin{align}\frac{1}{d_{[1/3]}}&=\left[F^{\left[\frac{1}{3}\right]\left[\overline{\frac{1}{3}}\right]\left[\frac{1}{3}\right]}_{\left[\frac{1}{3}\right]}\right]^0_0=\frac{1}{k^2}\label{F3GSD}\\\frac{1}{d_{[1/2]_\chi}}&=\left[F^{\left[\frac{1}{2}\right]_{\chi,{\bf l}}\left[\frac{1}{2}\right]_{\chi,{\overline{\bf l}}}\left[\frac{1}{2}\right]_{\chi_{\bf l}}}_{\left[\frac{1}{2}\right]_{\chi,{\bf l}}}\right]^0_0=\frac{1}{k}\label{F2GSD}\end{align} where $0$ is the vacuum, the reciprocal species for twofold defect is given by $\overline{\bf l}=-{\bf l}+2{\bf l}_0(\chi)$ as shown in \eqref{Z2reciprocalspecies} so that the intermediate vacuum channel is admissible, and the full $F$-matrix for threefold defects can be found in table~\ref{tab:Fsymbols} in appendix~\ref{sec:Fsymbols}. This matches the prediction from the fusion matrices $N_x=(N_{xy}^z)$ in eq.\eqref{Npower}, ground state degeneracies in \eqref{Z3dimensionGSD} and \eqref{Z2dimensionGSD}, and the counting of the degrees of freedom \eqref{Z3dimension} and \eqref{Z2quantumdimension} at the lattice level. We note that first entries of the $F$-matrices in eq.\eqref{F3GSD} and \eqref{F2GSD} is purely real, implying the triviality of any bending phase or Frobenius-Schur indicator for a self-reciprocal defect, $\varkappa_x=+1$.

\section{Defect Exchange and Braiding}\label{sec:defectexchangebraiding}

We are interested in unitary transformations induced by exchanging and braiding twist defects. These operations can be interpreted in the continuum limit by moving objects adiabatically along some braiding trajectories in real space, or treated on the lattice level as a basis transformation induced by changing the viewing order of objects while keeping the lattice fixed. An example of this was shown in figure~\ref{fig:multidefects} where the ordering of the $6^{th}$ and $7^{th}$ defects are switched by a redefinition of viewing curves without physically moving them. After exchanging a pair, the $S_3$-type and species label of one of them may be twisted by the action of the other according to conjugation \eqref{defectlabelexchange} and transformations \eqref{speciestrans1}, \eqref{speciestrans2} or \eqref{speciestrans3} respectively. Due to this non-commutativity, the set of objects \eqref{objectset} does not admit complete braiding in the sense of a conventional braided fusion category~\cite{Kitaev06, Turaevbook, BakalovKirillovlecturenotes, Wangbook}. In this article we only consider operations that leave object labels and frames invariant, i.e. braiding between objects that mutually commute in $S_3$. These include exchanging (i) a pair of abelian anyons, which is well known and already discussed in figure~\ref{fig:abeliananyonsplittingspace} and eq.\eqref{Rabeliananyon}, (ii) a pair of threefold defects, and (iii) a pair of twofold defects of the same color. We find that {\em partial} braiding is sufficient for defining topological spin as a discrete characterization of defect exchange statistics.

Counter-clockwise exchange of commuting objects $x$ and $y$ defines a unitary operation $R^{xy}_z:V^{xy}_z\to V^{yx}_z$ between splitting spaces so that \begin{align}\left|\vcenter{\hbox{\includegraphics[width=0.45in]{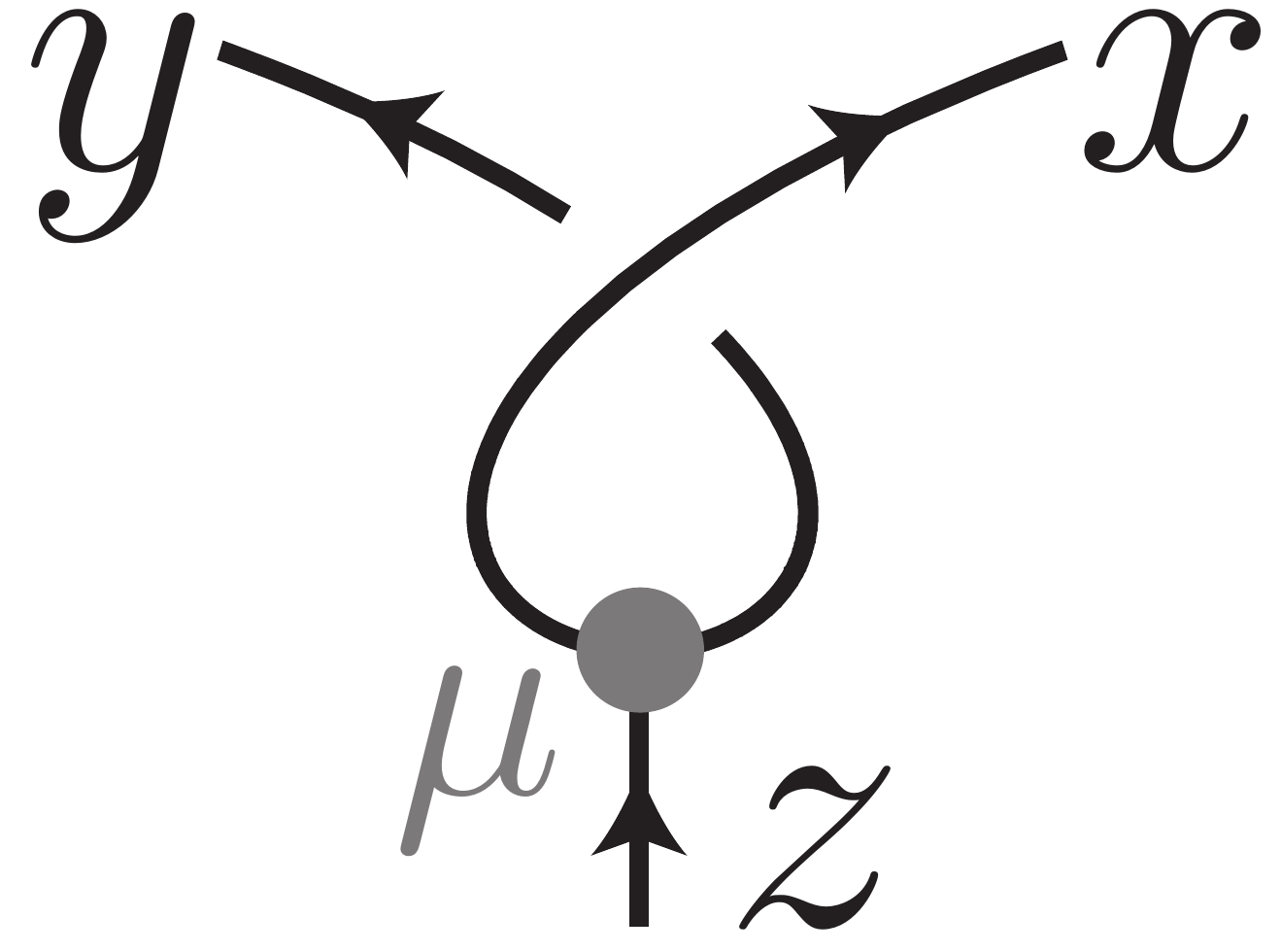}}}\right\rangle=\sum_\nu\left[R^{xy}_z\right]^\nu_\mu\left|\vcenter{\hbox{\includegraphics[width=0.45in]{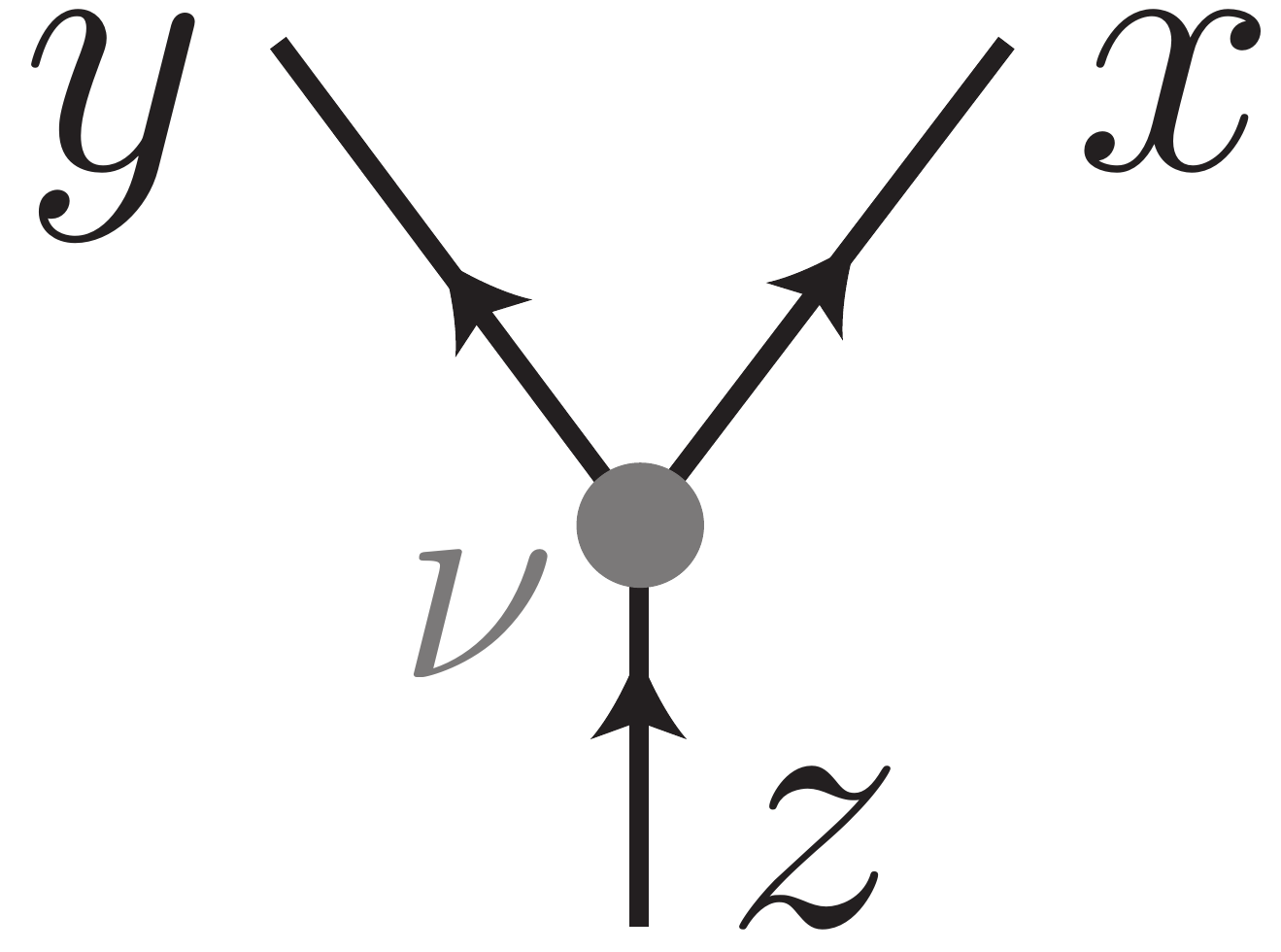}}}\right\rangle\end{align} The $R$-matrix can be computed by counter-clockwise rotating Wilson strings in the splitting states and re-expressing them in the canonical form defined in figure~\ref{fig:splittingspaces} through deformation and unlinking. The simplest example is given by exchange of a pair of abelian anyons $R^{{\bf a}{\bf b}}_{{\bf a}+{\bf b}}=w^{{\bf a}_\circ^Ti\sigma_y{\bf b}_\bullet}$ shown in figure~\ref{fig:abeliananyonsplittingspace} and eq.\eqref{Rabeliananyon}. 

Since they depend on the choice of splitting states, $R$-matrices are in general gauge dependent. In fact if $x\neq y$, $180^\circ$ exchange is not a cyclic evolution as the system does not closed back onto itself. Exchange between identical objects $R^{xx}_z$ and $360^\circ$ braiding between different objects $R^{xy}_zR^{yx}_z$ are however gauge invariant quantities (or gauge covariant if there are fusion degeneracies). 

We define the topological spin of an object $x$ by its exchange statistics \begin{align}\theta_x=\frac{1}{d_x}\vcenter{\hbox{\includegraphics[width=0.5in]{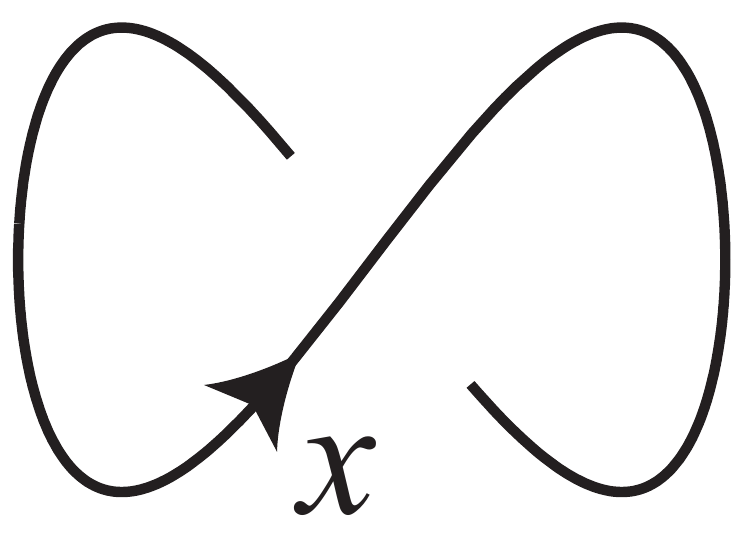}}}=\frac{1}{d_x}\sum_{z}d_z\mbox{Tr}\left(R^{xx}_z\right)\label{topologicalspin}\end{align} This is a rational $U(1)$ phase, $\theta_x^m=1$ for some integer $m$, accumulated by $\mbox{ord}(x)\times360^\circ$ rotation of the defect,\cite{ord360rotation} where $\mbox{ord}(x)$ is the order of the group element $x$ in the symmetry group or equivalently the minimal number of copies of $x$'s that fuse to the overall vacuum channel. Note that a single $360^\circ$ rotation does not give a topologically protected phase because the defect system does not go back to its initial configuration. This can be seen by rotating the branch cut attaching to a twofold defect \begin{align}\vcenter{\hbox{\includegraphics[width=0.1in]{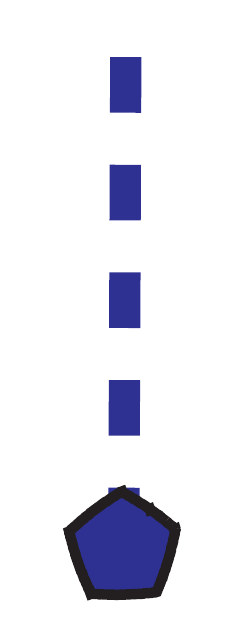}}}\xrightarrow{360^\circ}\vcenter{\hbox{\includegraphics[width=0.4in]{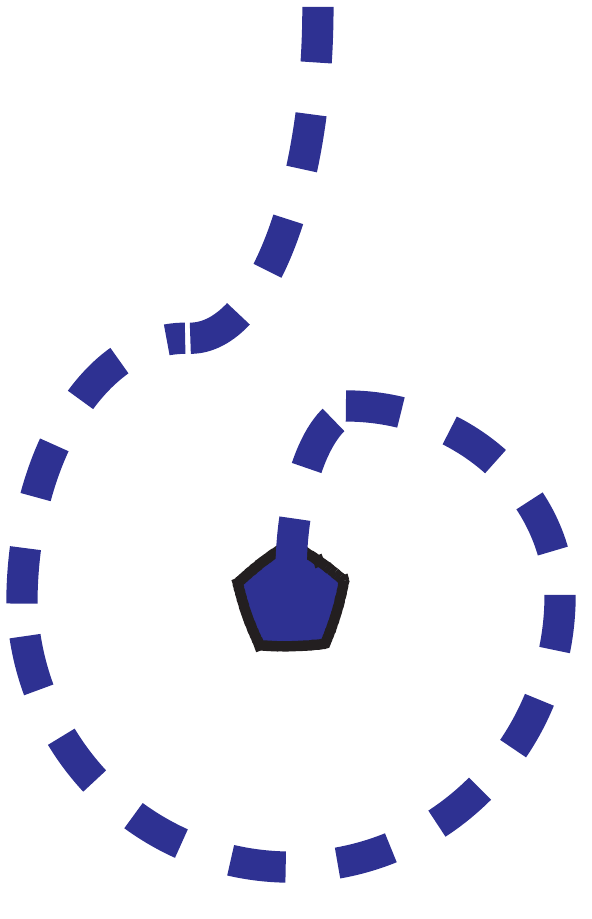}}}\xrightarrow{360^\circ}\vcenter{\hbox{\includegraphics[width=0.4in]{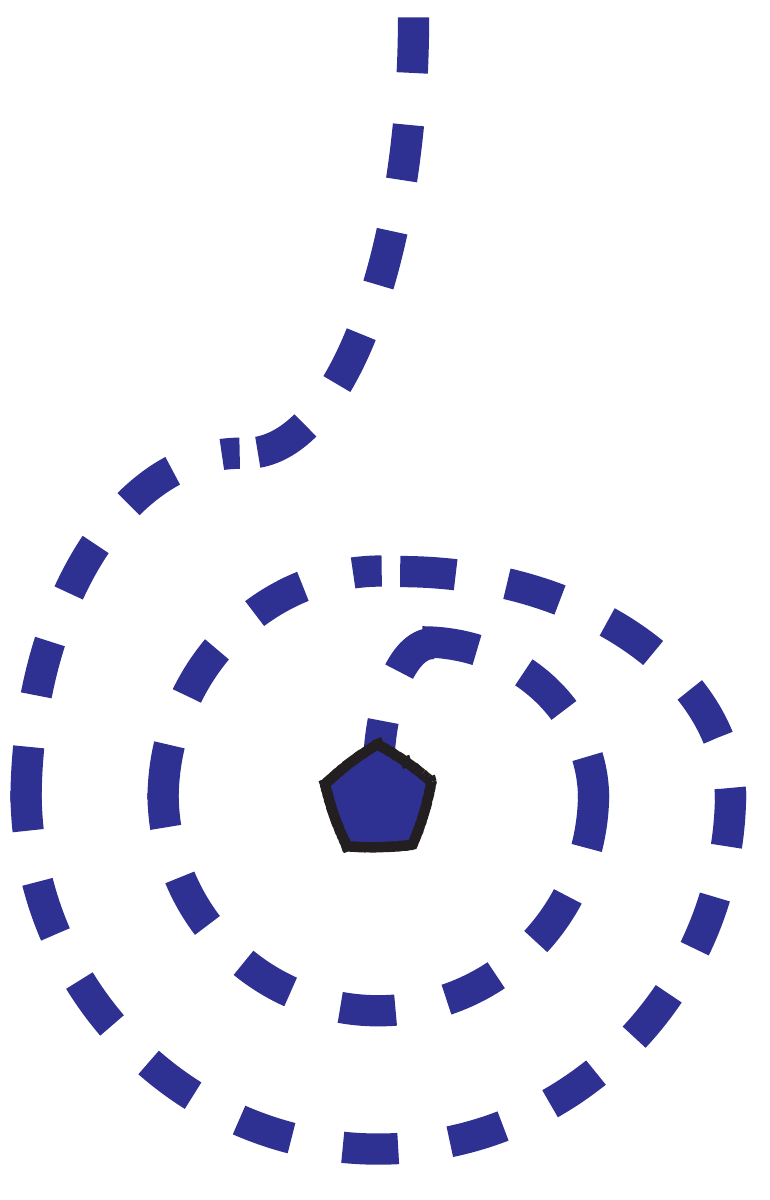}}}=\vcenter{\hbox{\includegraphics[width=0.1in]{720spincut2.pdf}}}\end{align} where a single twist is irremovable but a double twist is cancelable since a pair of twofold branch cuts annihilate.

Given defects $\lambda_{{\bf l}_1}$ and $\lambda_{{\bf l}_2}$ of the same symmetry twisting type $\lambda$ in $S_3$, we define the defect braiding quantity \begin{align}S^\lambda_{{\bf l}_1{\bf l}_2}=\frac{1}{\mathcal{D}_\lambda}\vcenter{\hbox{\includegraphics[width=0.6in]{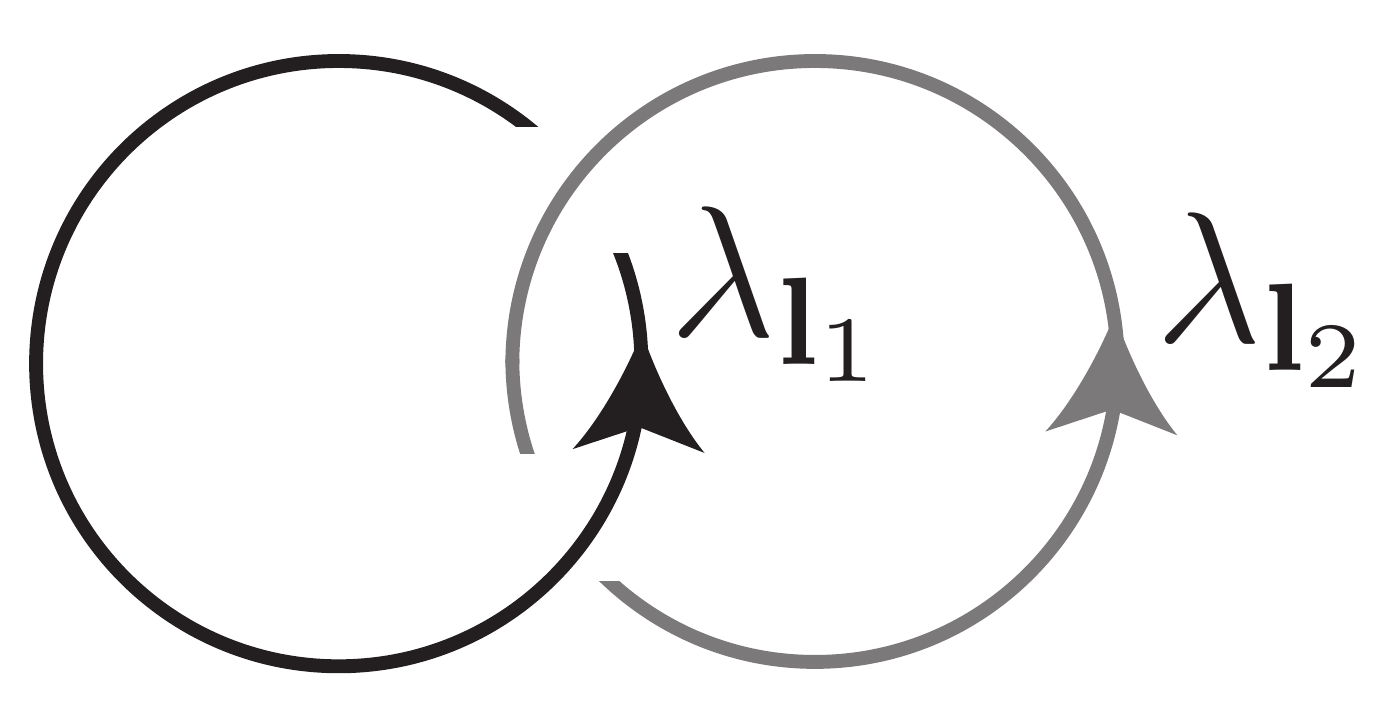}}}=\frac{1}{\mathcal{D}_\lambda}\sum_{\bf a}d_{\bf a}\mbox{Tr}\left(R^{\overline{\lambda_{{\bf l}_2}}\lambda_{{\bf l}_1}}_{\bf a}R^{\lambda_{{\bf l}_2}\overline{\lambda_{{\bf l}_2}}}_{\bf a}\right)\label{defectSmatrix}\end{align} where the normalization $\mathcal{D}_\lambda^2=\sum_{{\bf l}}d_{\lambda_{\bf l}}^2$ is defined by summing over all species labels, and is identical to the total quantum dimension of the underlying abelian anyon system $\mathcal{D}_\lambda=\mathcal{D}=\sqrt{\sum_{\bf a}d^2_{\bf a}}=k^2$. Eq.\eqref{defectSmatrix} is a generalization of the topological $S$-matrix that characterizes mutual braiding of semiclassical defects of the same symmetry type. Together with the topological $T$-matrix \begin{equation}T^\lambda_{{\bf l}_1{\bf l}_2}=\delta_{{\bf l}_1{\bf l}_2}\theta_{\lambda_{{\bf l}_1}}\label{defectTmatrix}\end{equation} that characterizes self-exchange defined in \eqref{topologicalspin}, they form a unitary representation of a novel set of {\em defect modular transformations}. For instance, if $\lambda=1$ is the trivial defect, its species labels are the abelian anyons ${\bf a}$, and $S^1,T^1$ are the braiding \eqref{anyonfullbraiding} and exchange \eqref{anyonexchangespin} matrices of the underlying abelian theory that unitarily represent the modular group \begin{align}SL(2;\mathbb{Z})=\left\langle\begin{array}{*{20}c}S^1=t_xt_yt_x,\\T^1=t_x\end{array}\left|\begin{array}{*{20}c}t_xt_yt_x=t_yt_xt_y,\\(t_xt_yt_x)^4=1\end{array}\right.\right\rangle\end{align} where $t_x$ and $t_y$ identify with Dehn twists along the two cycles of a torus, and the matrices $S^1$ and $S^1T^1$ identify with $90^\circ$ and $120^\circ$ rotations respectively. However for non-trivial defect $\lambda$, modular transformations are in general restricted to a congruent subgroup that fixes one cycle, and the $S^\lambda$ and $T^\lambda$ matrices have different geometric interpretations.%It is evident for instance along a 2D toric interface between topological insulator and superconductor in 3D where the longitudinal and meridian cycles are distinguished.

\subsection{Statistics of threefold defects}\label{sec:statisticsZ3}
For simplicity, we only study defects when $k$ is not a multiple of 3. The $R$-matrix for exchanging a pair of $[1/3]$ defects is evaluated by deforming the Wilson loop $\mathcal{A}_\circ^{\bf m}$ in the splitting state for $[1/3]\times[1/3]$ in figure~\ref{fig:splittingspaces}(d). \begin{align}\left|\vcenter{\hbox{\includegraphics[width=0.6in]{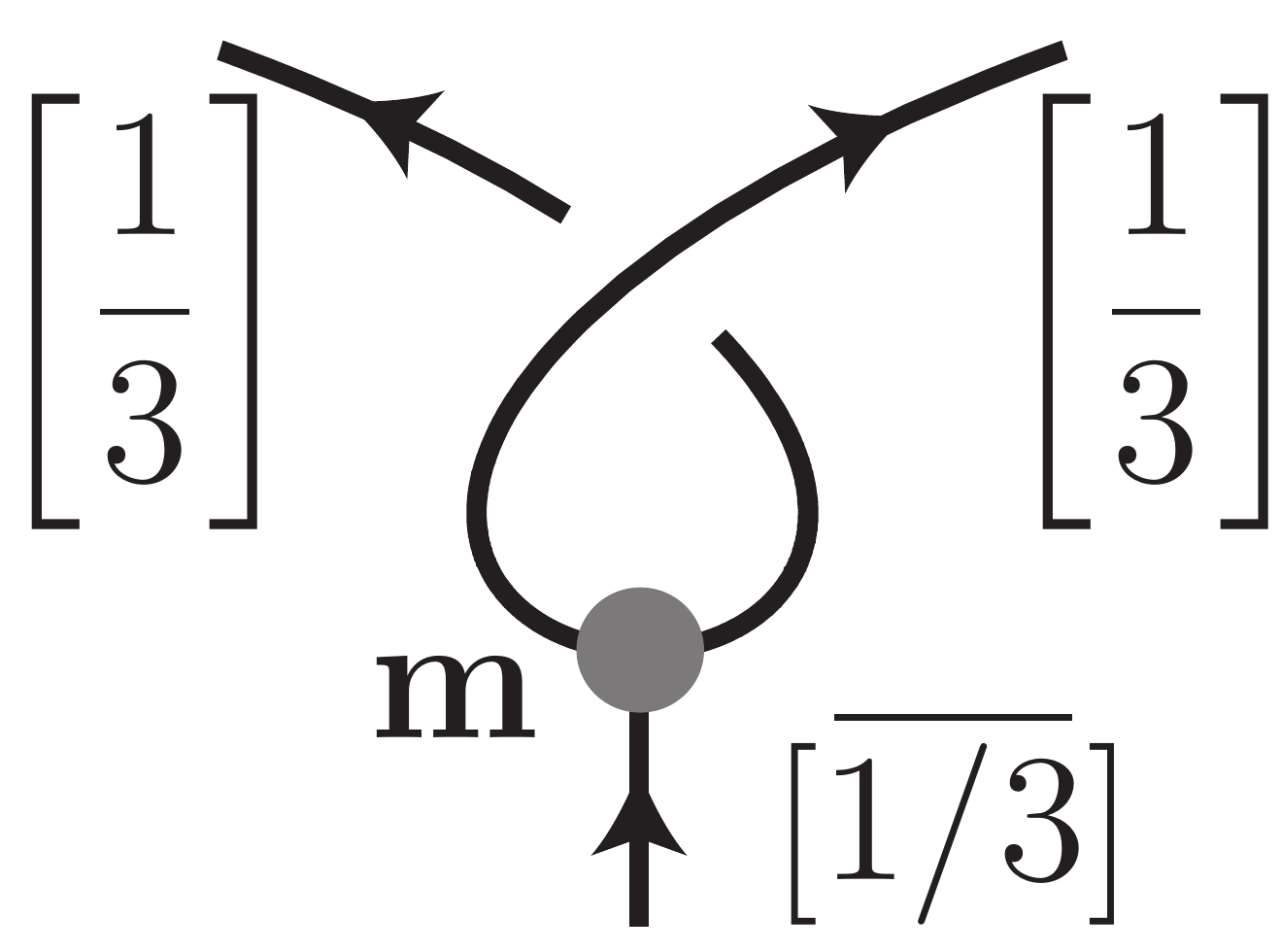}}}\right\rangle&=\left[\vcenter{\hbox{\includegraphics[width=0.7in]{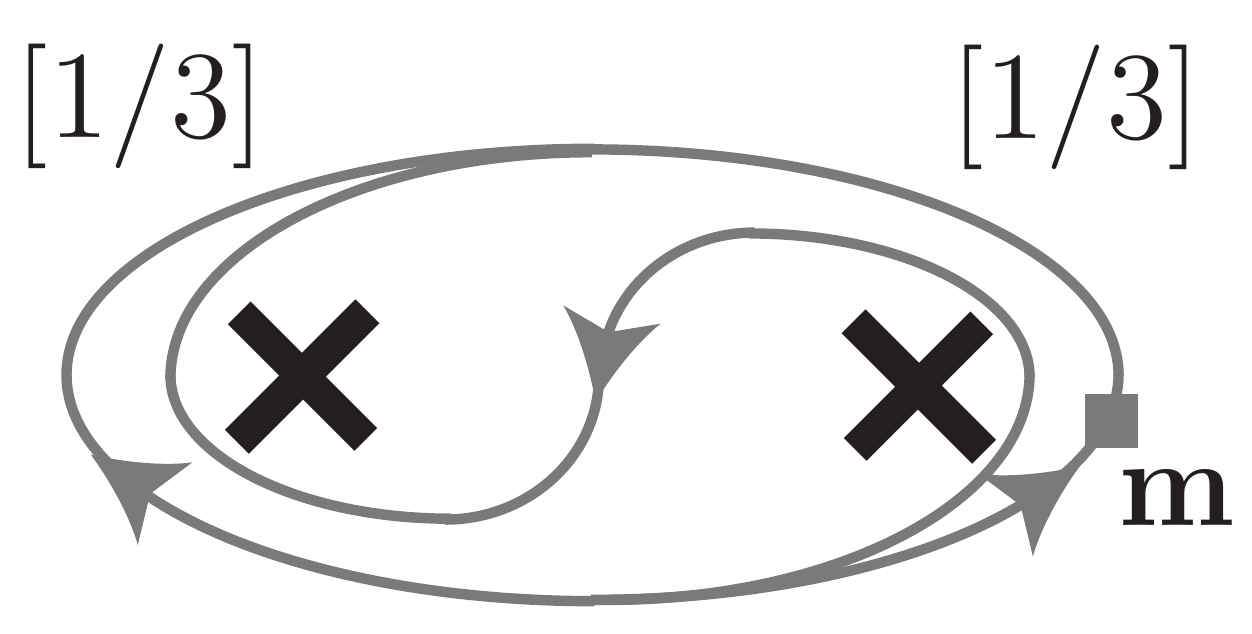}}}\right]|GS\rangle_0\nonumber\\&=\left[\vcenter{\hbox{\includegraphics[width=0.7in]{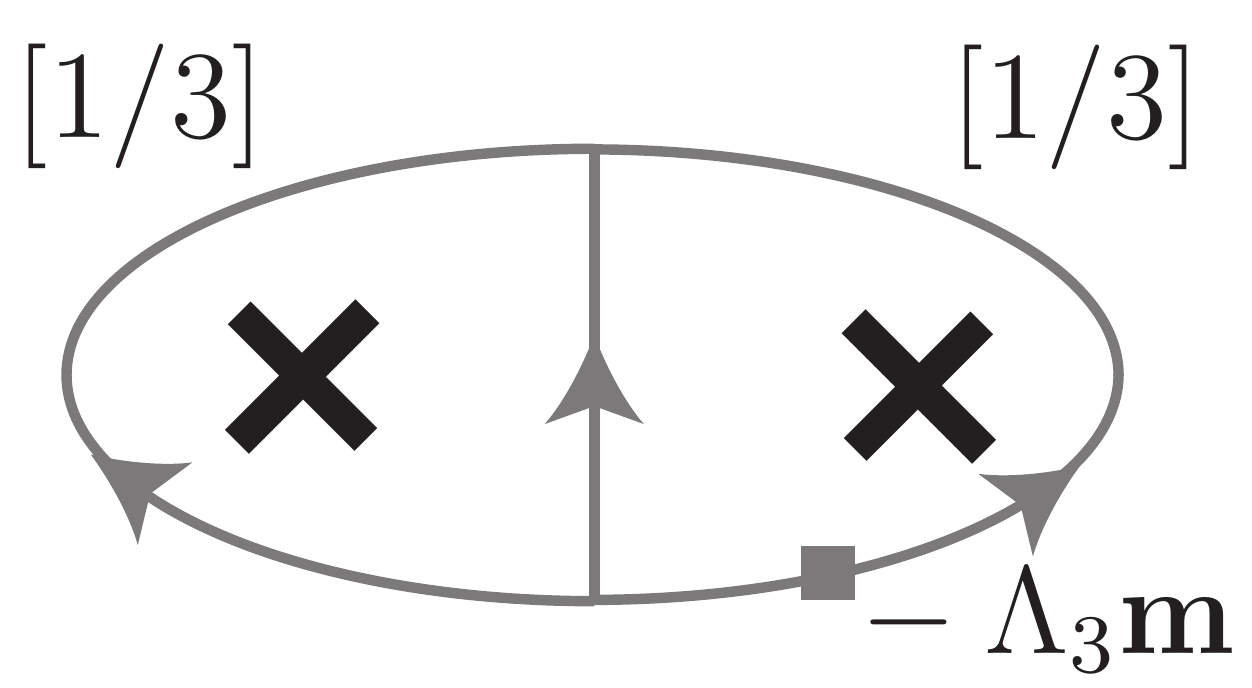}}}\right]|GS\rangle_0=\left|\vcenter{\hbox{\includegraphics[width=0.6in]{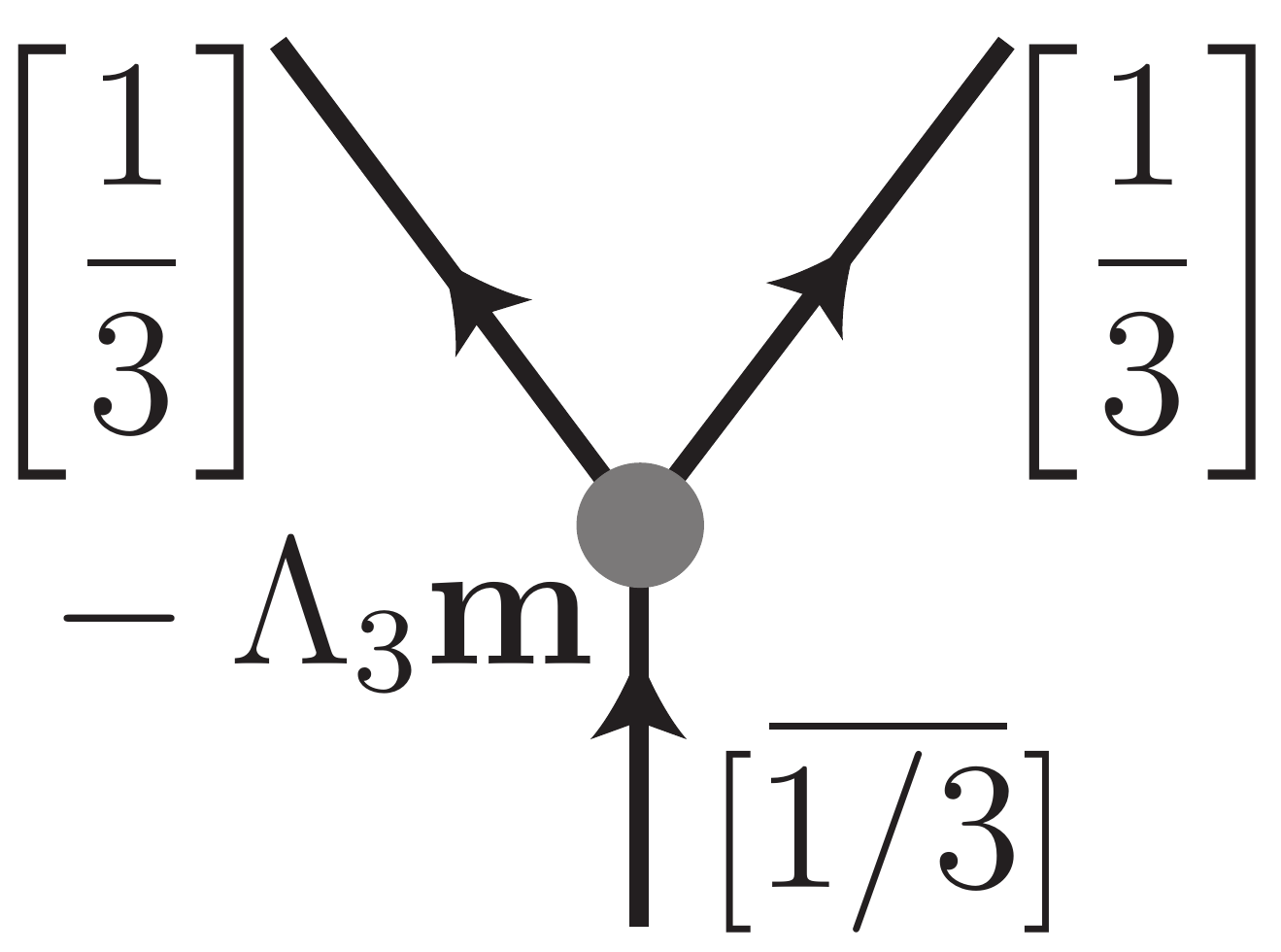}}}\right\rangle\end{align} Recall $\boldsymbol\alpha=I_0{\bf m}$, the $R$-matrix is given by the overlap \begin{align}\left[R^{\left[\frac{1}{3}\right]\left[\frac{1}{3}\right]}_{[\overline{1/3}]}\right]^{\boldsymbol\beta}_{\boldsymbol\alpha}=\left\langle\vcenter{\hbox{\includegraphics[width=0.6in]{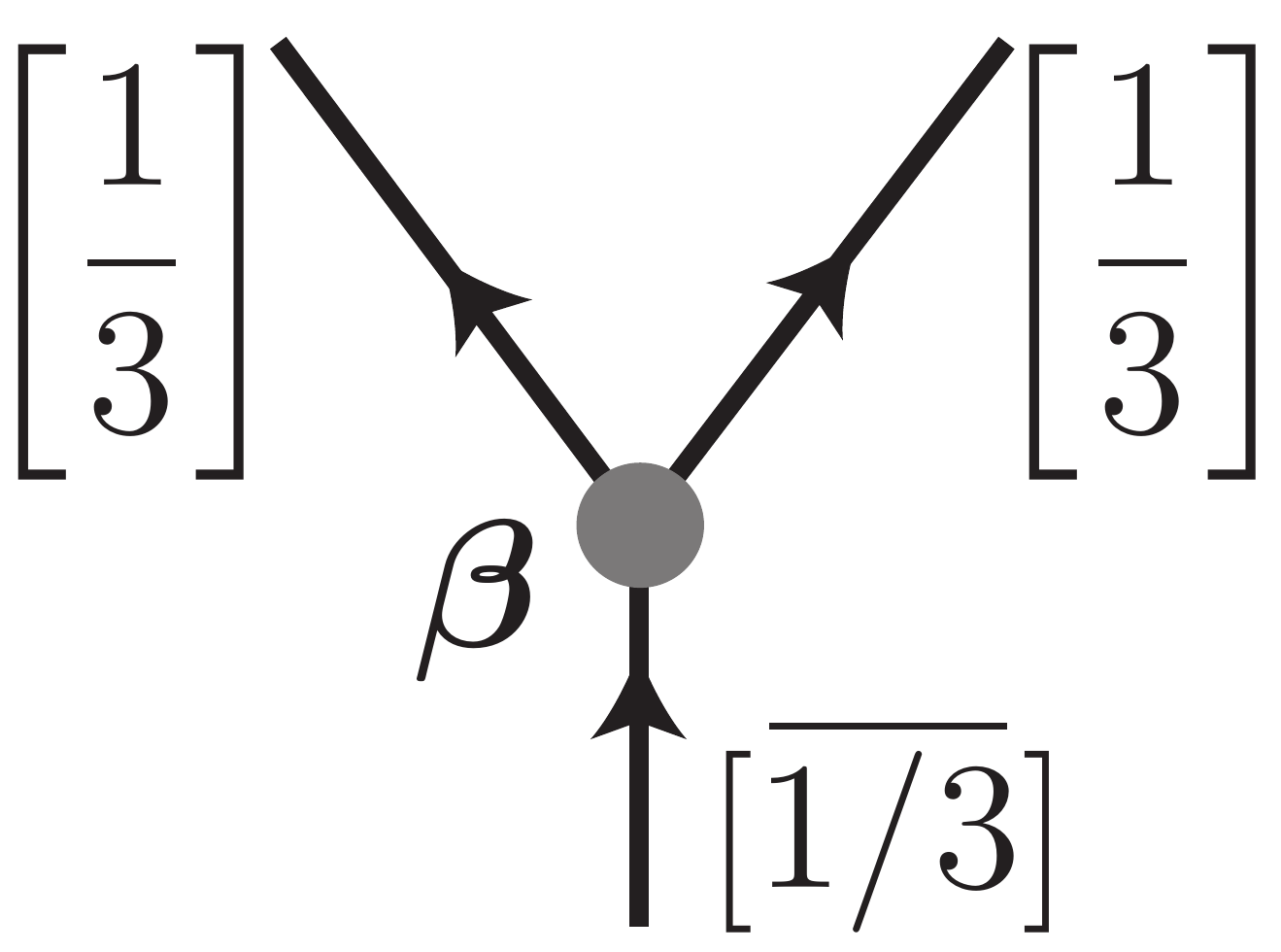}}}\right.\left|\vcenter{\hbox{\includegraphics[width=0.6in]{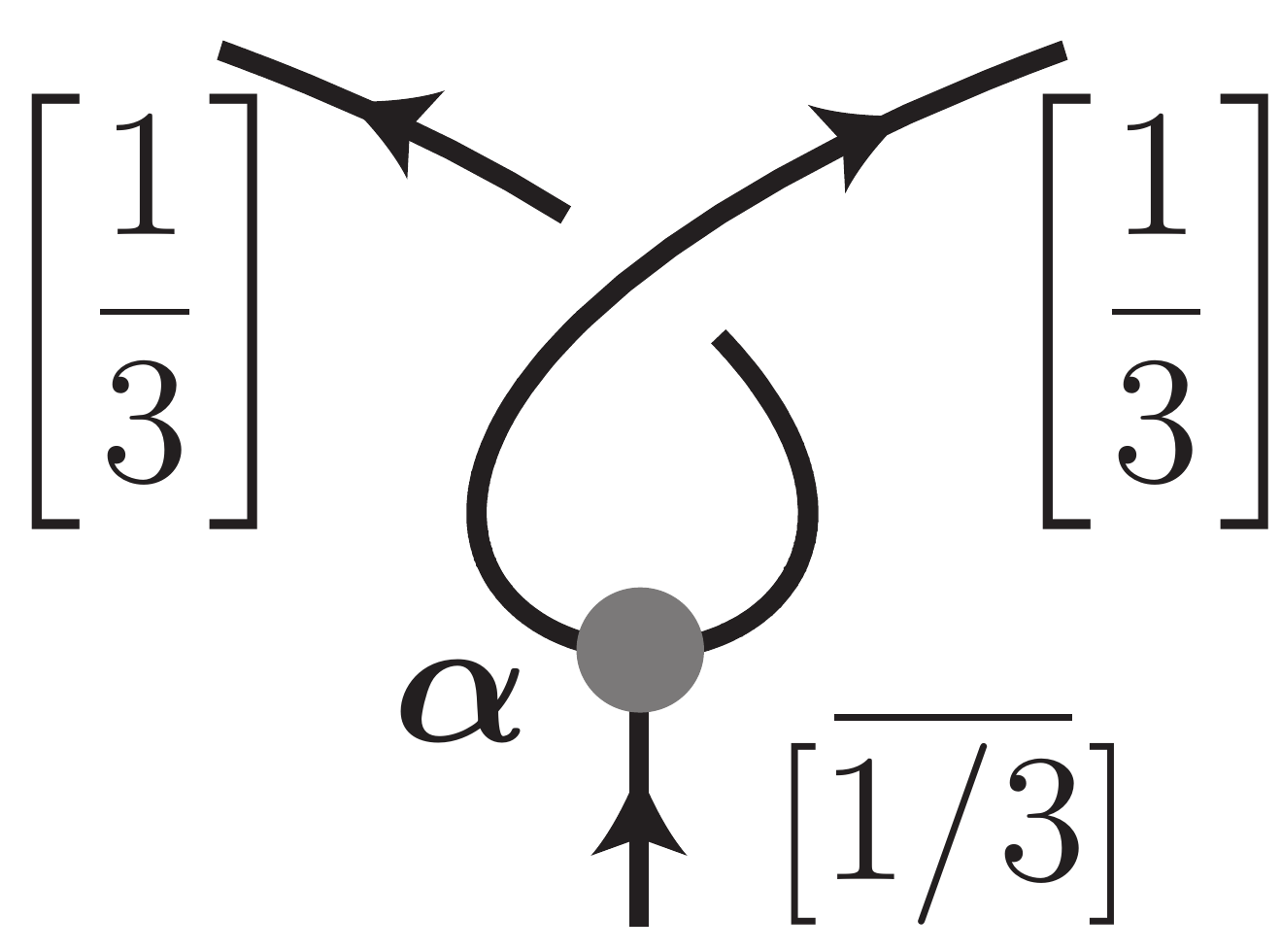}}}\right\rangle=\delta^{\boldsymbol\beta}_{-(\Lambda_3^{-1})^T\boldsymbol\alpha}\label{R33eqn}\end{align} The matrix $R^{[\overline{1/3}][\overline{1/3}]}_{[1/3]}$ for the antiparticle can be read off by replacing $\Lambda_3$ by $\Lambda_3^{-1}$ in \eqref{R33eqn}. According to \eqref{topologicalspin}, threefold defects carry trivial spin. \begin{equation}\theta_{[1/3]}=\mbox{Tr}\left(R^{\left[\frac{1}{3}\right]\left[\frac{1}{3}\right]}_{[\overline{1/3}]}\right)=1\end{equation} This can be understood by seeing there is no orientation frame or any Wilson string attached to a threefold defect.

The phase of exchanging $[1/3]$ and its antiparticle $[\overline{1/3}]$ is obtained by rotating the splitting states for $[1/3]\times[\overline{1/3}]$ in figure~\ref{fig:splittingspaces}(e). \begin{align}\left|\vcenter{\hbox{\includegraphics[width=0.6in]{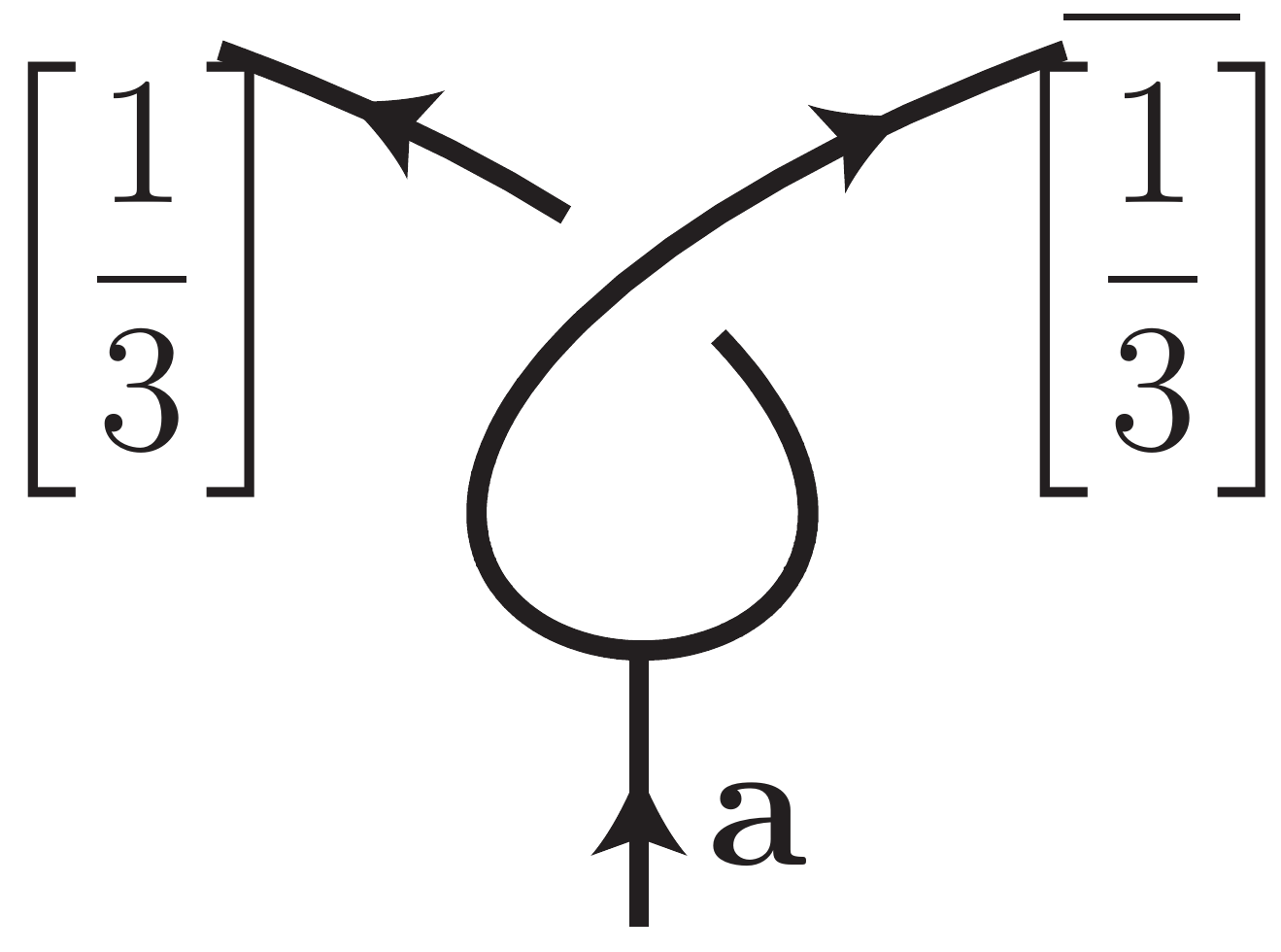}}}\right\rangle&=\frac{1}{k}\sum_{\bf m}\left[\vcenter{\hbox{\includegraphics[width=1in]{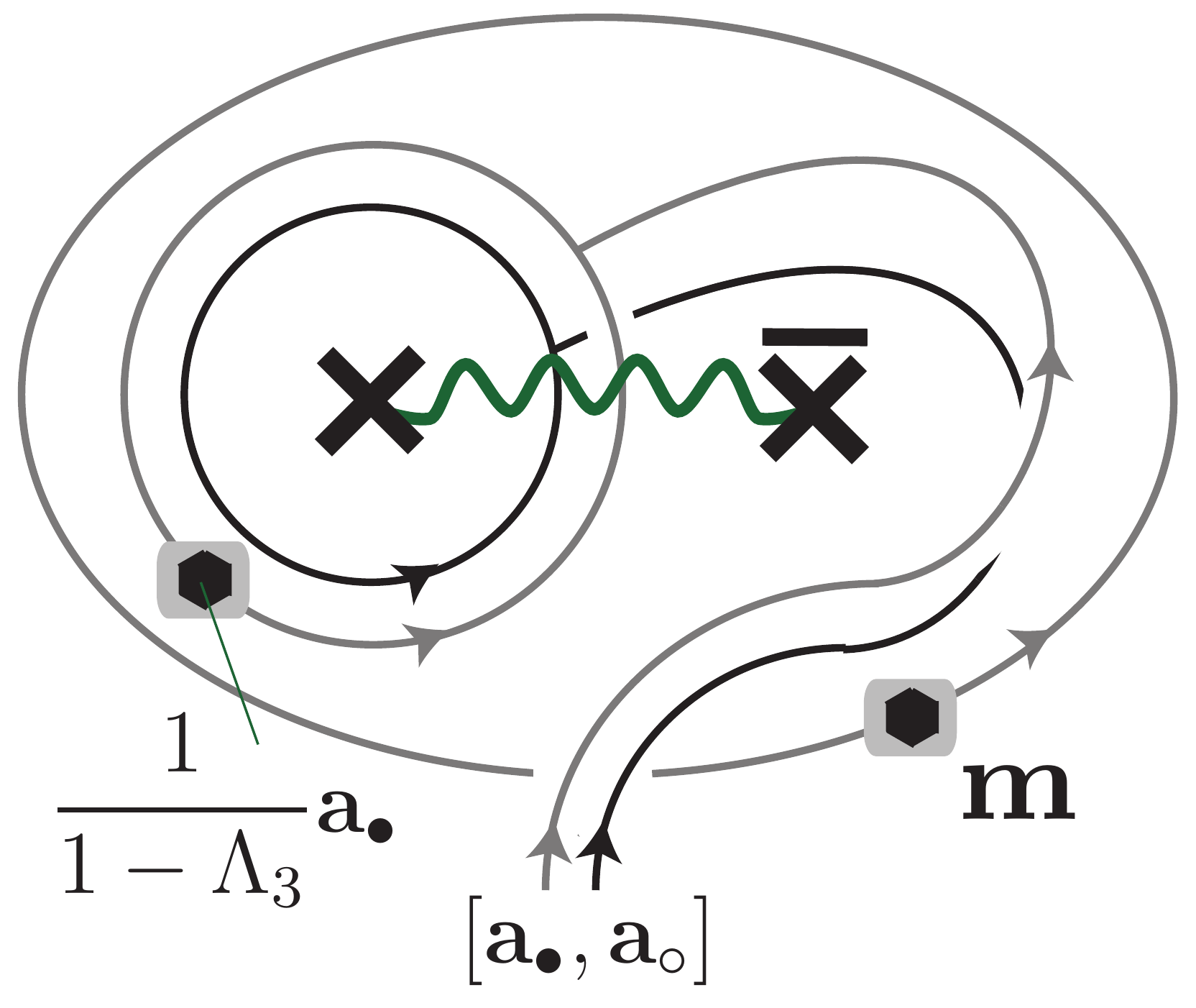}}}\right]|GS\rangle_0\nonumber\\&=\frac{1}{k}\sum_{\bf m}\left[\vcenter{\hbox{\includegraphics[width=1in]{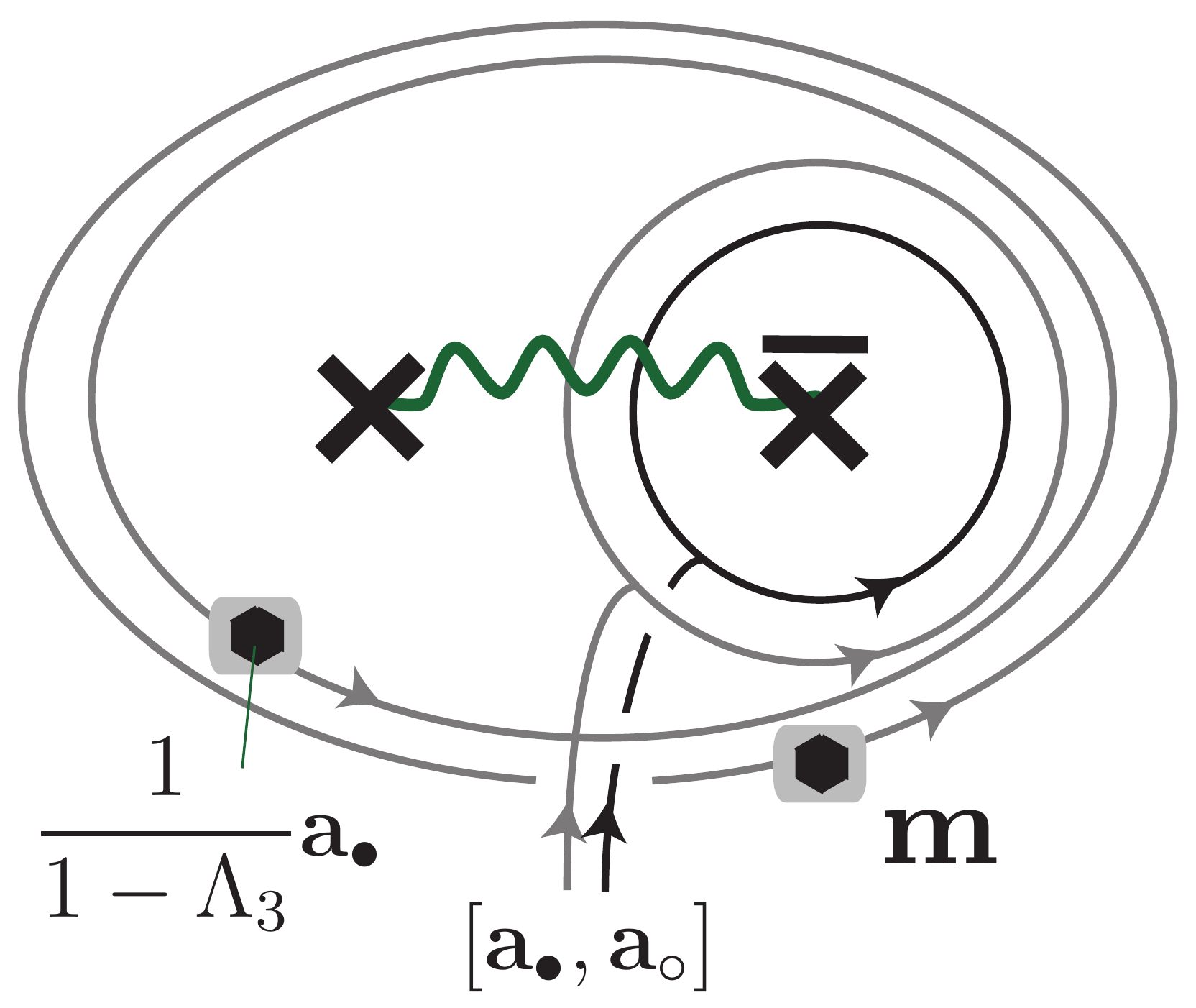}}}\right]|GS\rangle_0\nonumber\\&=\left[\vcenter{\hbox{\includegraphics[width=0.5in]{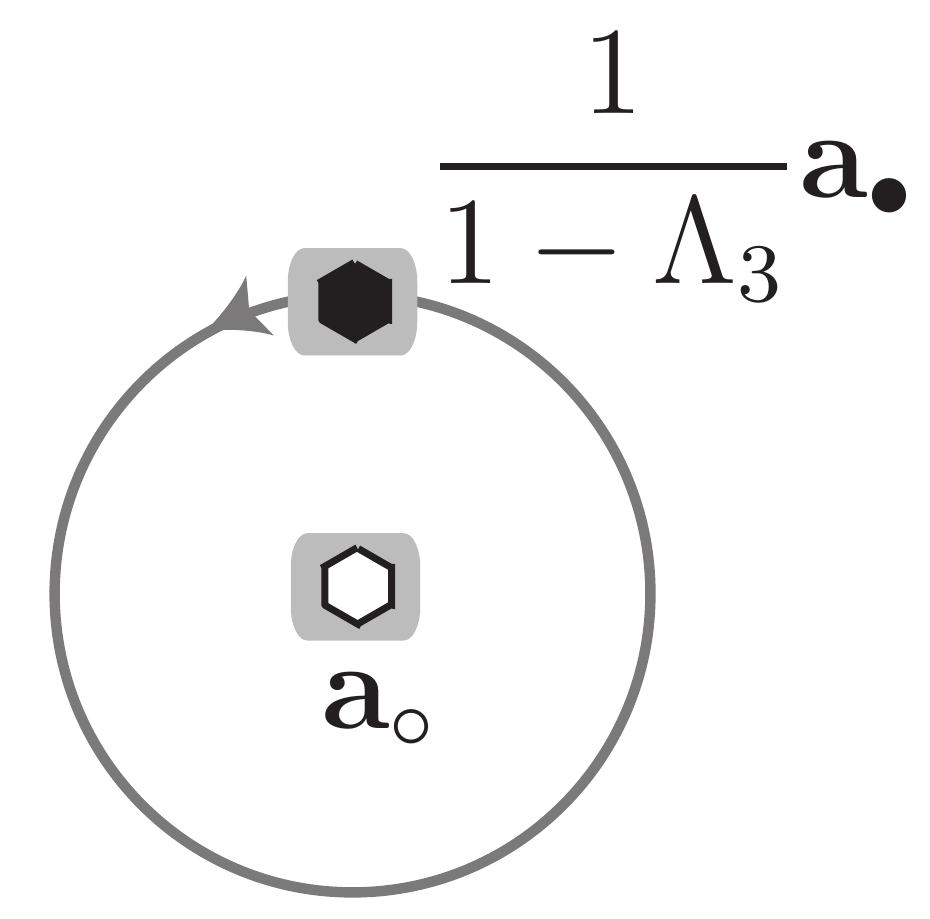}}}\right]\left|\vcenter{\hbox{\includegraphics[width=0.6in]{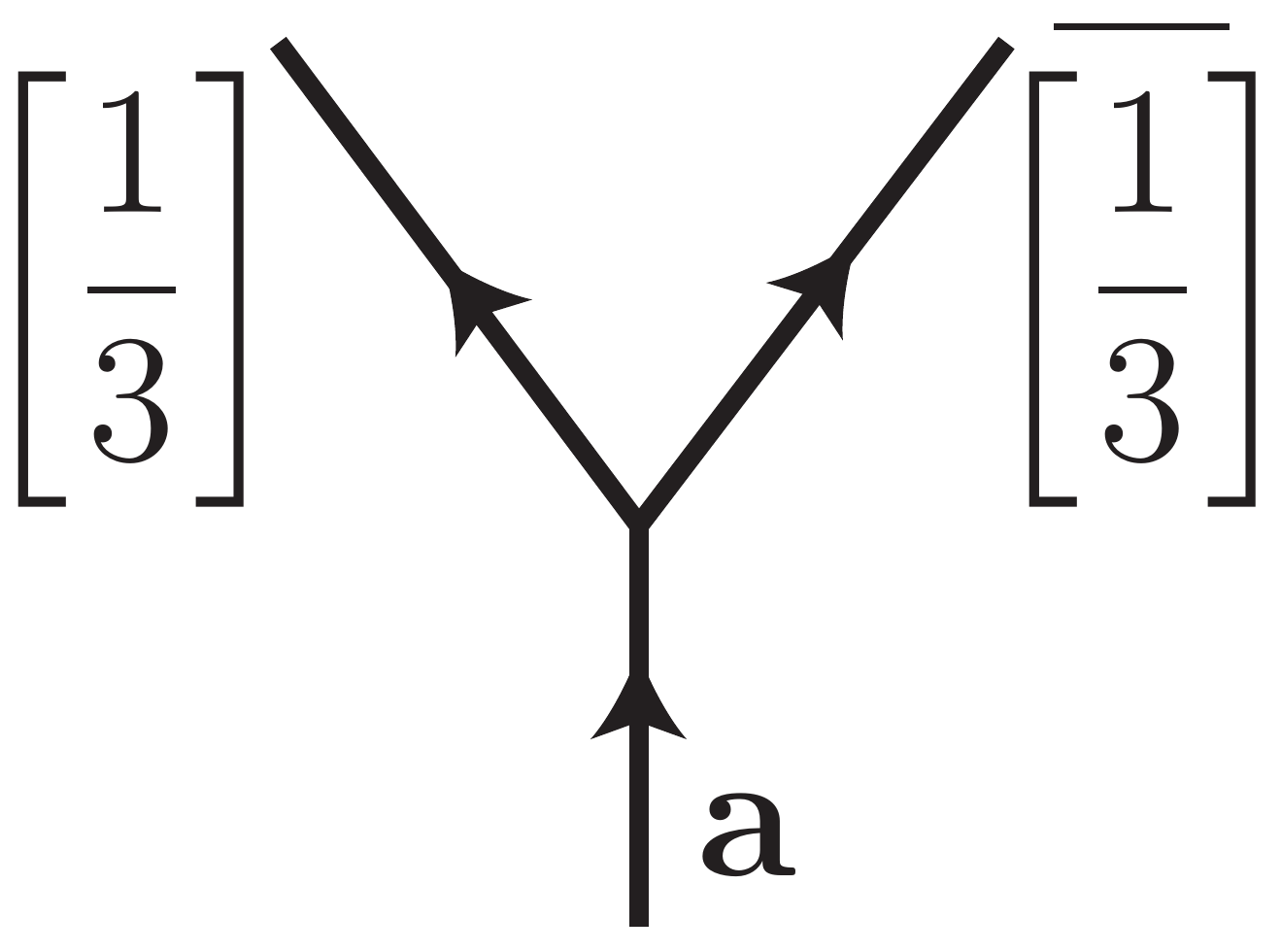}}}\right\rangle\end{align} And hence the $R$-symbol is given by the intersection phase \begin{equation}R^{\left[\overline{\frac{1}{3}}\right]\left[\frac{1}{3}\right]}_{\bf a}=w^{{\bf a}_\circ^Ti\sigma_y\frac{1}{1-\Lambda_3}{\bf a}_\bullet},\quad R^{\left[\frac{1}{3}\right]\left[\overline{\frac{1}{3}}\right]}_{\bf a}=w^{{\bf a}_\circ^Ti\sigma_y\frac{\Lambda_3}{\Lambda_3-1}{\bf a}_\bullet}\end{equation} 

In particular the $360^\circ$ braiding between $[1/3]$ and $[\overline{1/3}]$ equals the topological spin of their overall abelian anyon fusion channel. \begin{align}\vcenter{\hbox{\includegraphics[width=0.6in]{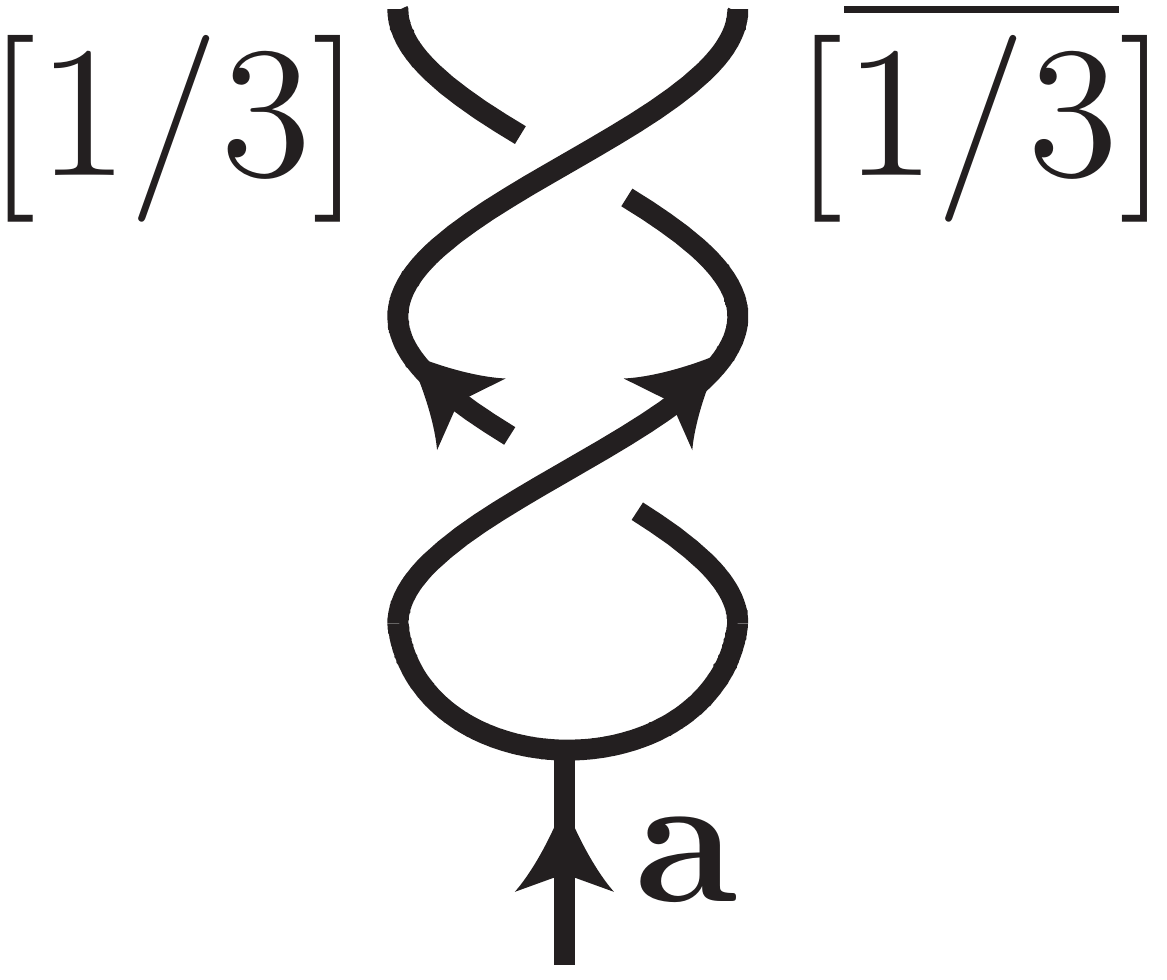}}}&\equiv R^{[\overline{\frac{1}{3}}][\frac{1}{3}]}_{\bf a}R^{[\frac{1}{3}][\overline{\frac{1}{3}}]}_{\bf a}=w^{{\bf a}_\circ^Ti\sigma_y{\bf a}_\bullet}=\theta_{\bf a}\end{align} %This agrees with the ``{\em pair of pants}" argument that relates spins and the monodromy operator. \begin{align}\vcenter{\hbox{\includegraphics[width=0.6in]{braid33b}}}=\vcenter{\hbox{\includegraphics[width=0.6in]{pants33b}}}=\frac{\theta_{\bf a}}{\theta_{[1/3]}\theta_{[\overline{1/3}]}}\end{align}
The $360^\circ$ braiding between a pair of $[1/3]$'s gives a threefold transformation of the degenerate splitting space $[1/3]\times[1/3]=k^2[\overline{1/3}]$. \begin{align}\vcenter{\hbox{\includegraphics[width=0.6in]{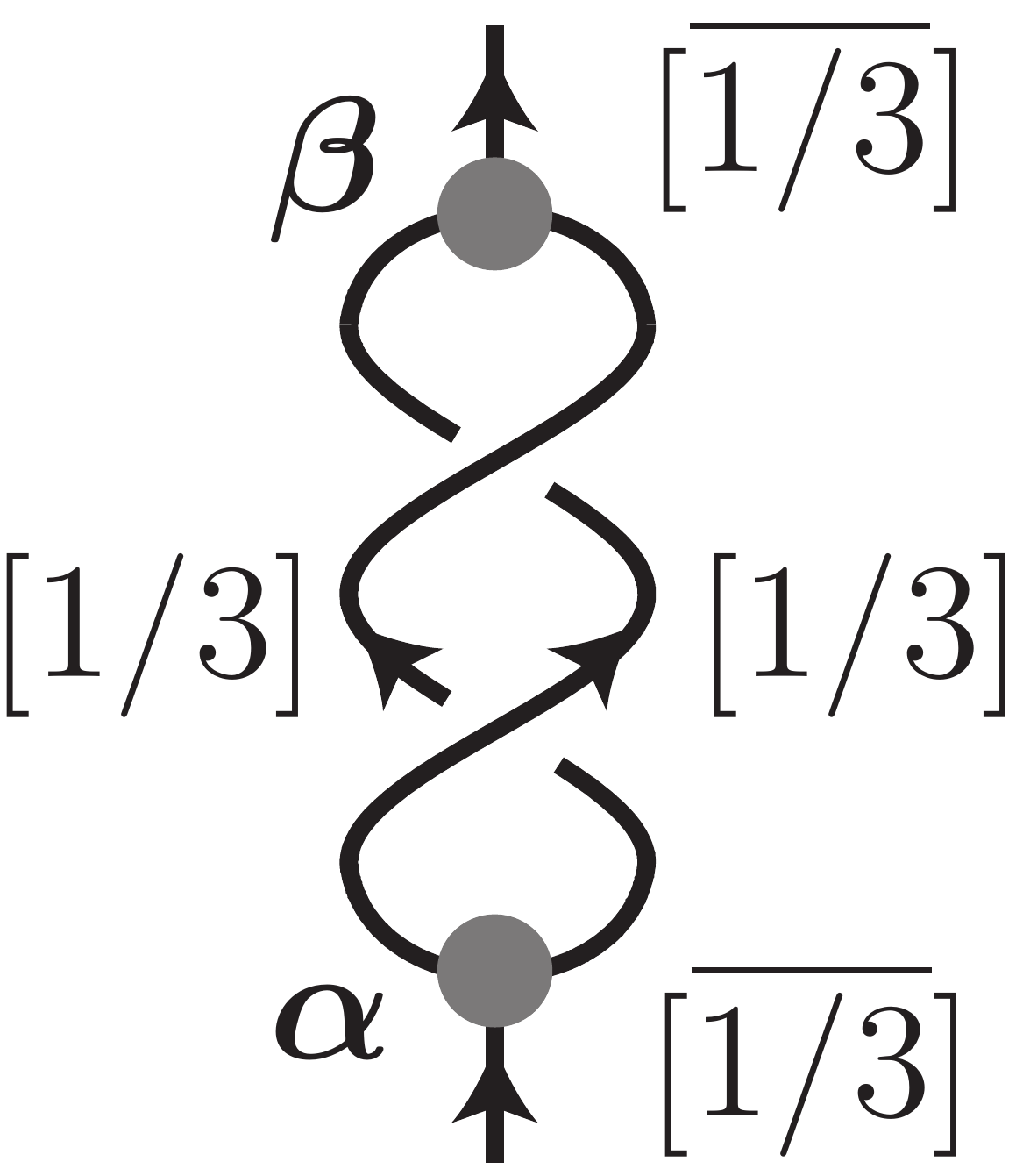}}}&\equiv\left[R^{[\frac{1}{3}][\frac{1}{3}]}_{\overline{1/3}}R^{[\frac{1}{3}][\frac{1}{3}]}_{\overline{1/3}}\right]_{\boldsymbol\alpha}^{\boldsymbol\beta}=\delta^{\boldsymbol\beta}_{\Lambda_3^T\boldsymbol\alpha}\label{Z3fullbraiding}\end{align} This process can be untwisted by dragging the viewpoint once around the overall $[\overline{1/3}]$ defect thus changing the color of the Wilson loop $\mathcal{A}_\circ^{\bf m}\to\mathcal{A}_\circ^{\Lambda_3^{-1}{\bf m}}$ between the pair of $[1/3]$ and sending $\boldsymbol\alpha=\mathbb{I}_0{\bf m}\to\Lambda_3^T\boldsymbol\alpha$. We note this is a unique feature for semiclassical twist defect when the color degree of freedom is non-dynamical.

\subsection{Statistics of twofold defects}\label{sec:statisticsZ2}
The exchange phase between a pair of twofold defects of the same color $\chi$ is given by rotating the splitting state in figure~\ref{fig:splittingspaces}(c). \begin{align}\left|\vcenter{\hbox{\includegraphics[width=0.7in]{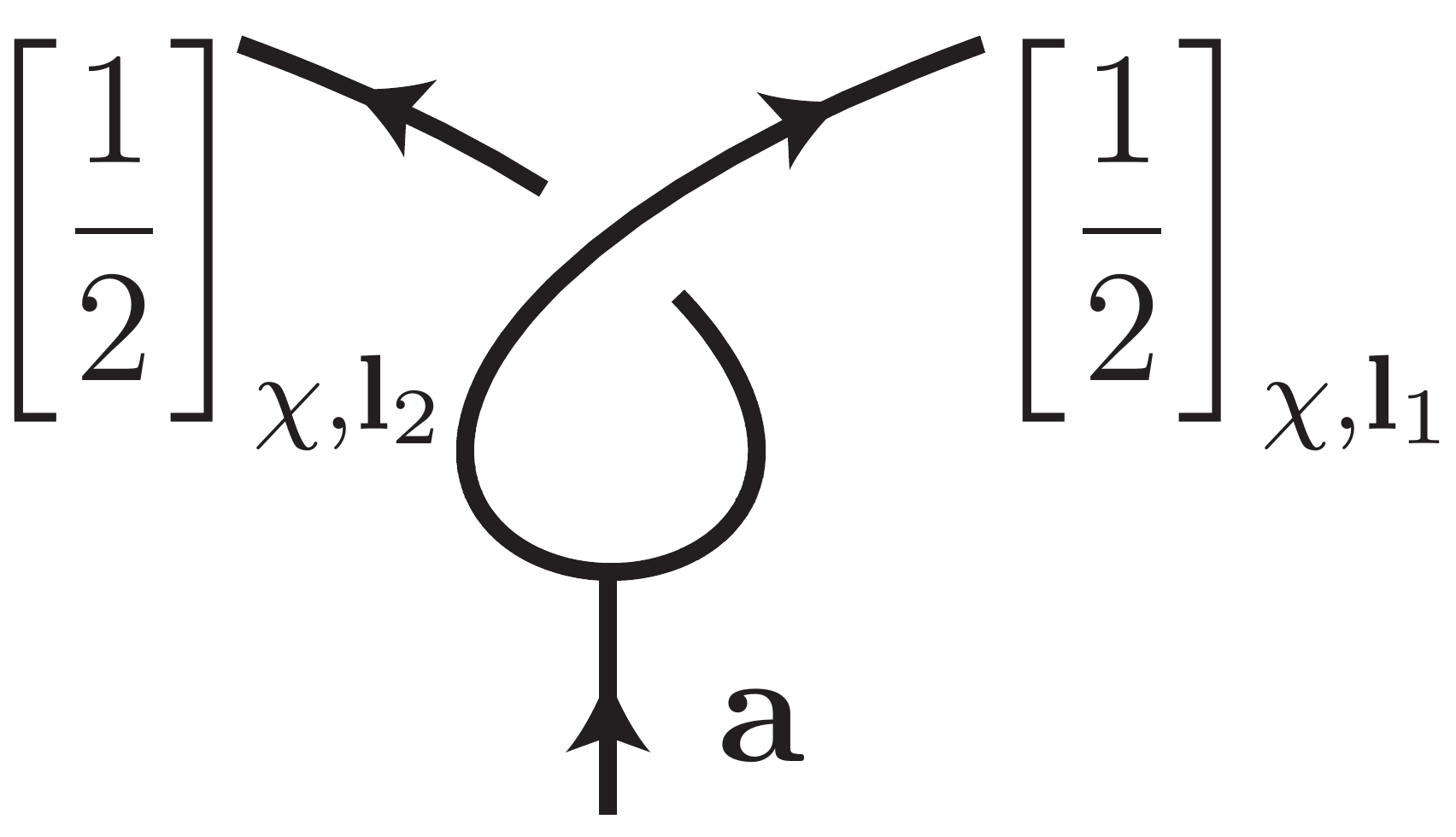}}}\right\rangle&=\left[\vcenter{\hbox{\includegraphics[width=1in]{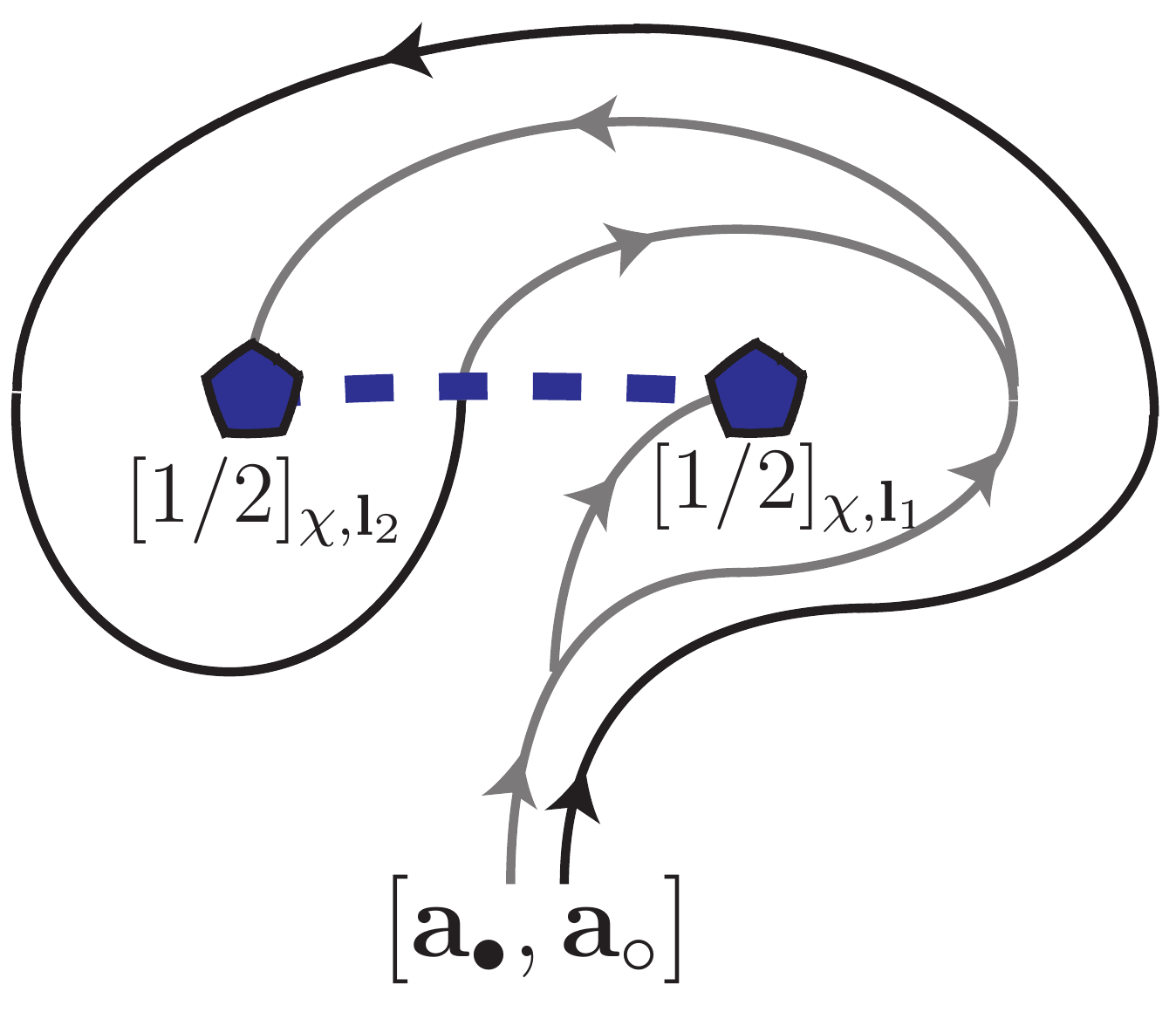}}}\right]|GS\rangle_0\nonumber\\&=\left[\vcenter{\hbox{\includegraphics[width=0.6in]{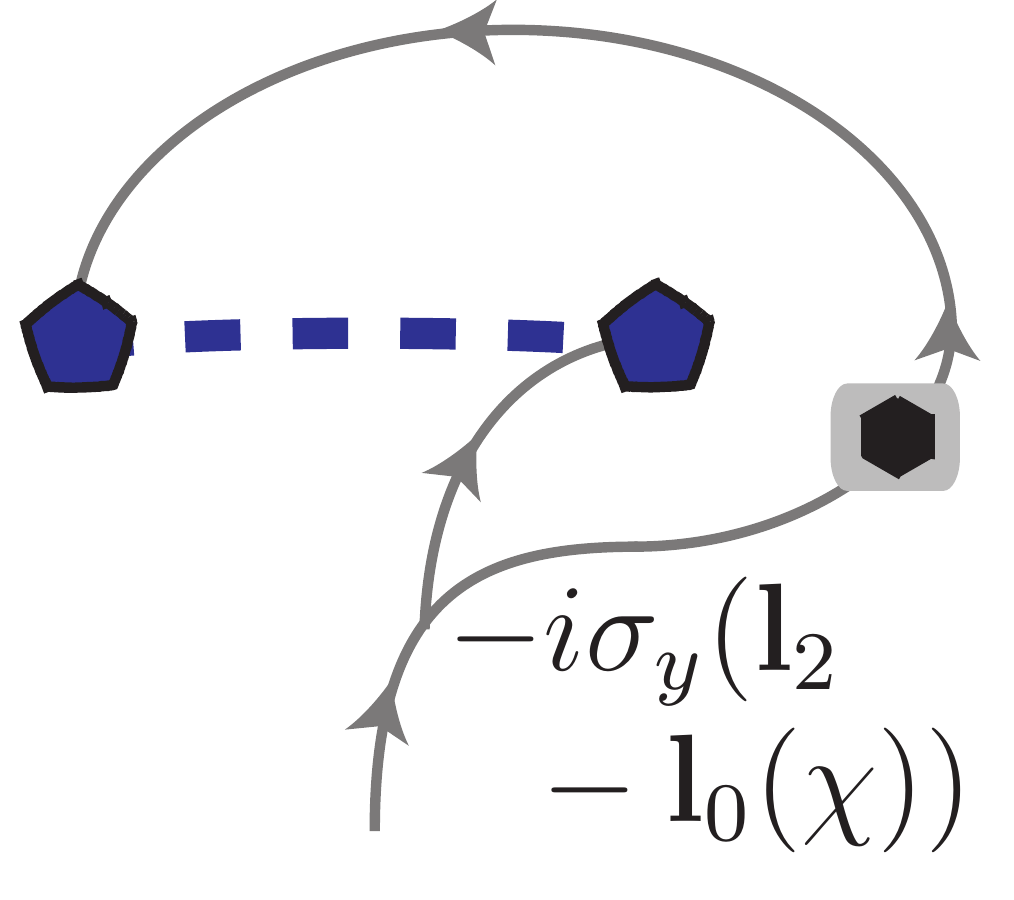}}}\right]\left[\vcenter{\hbox{\includegraphics[width=0.8in]{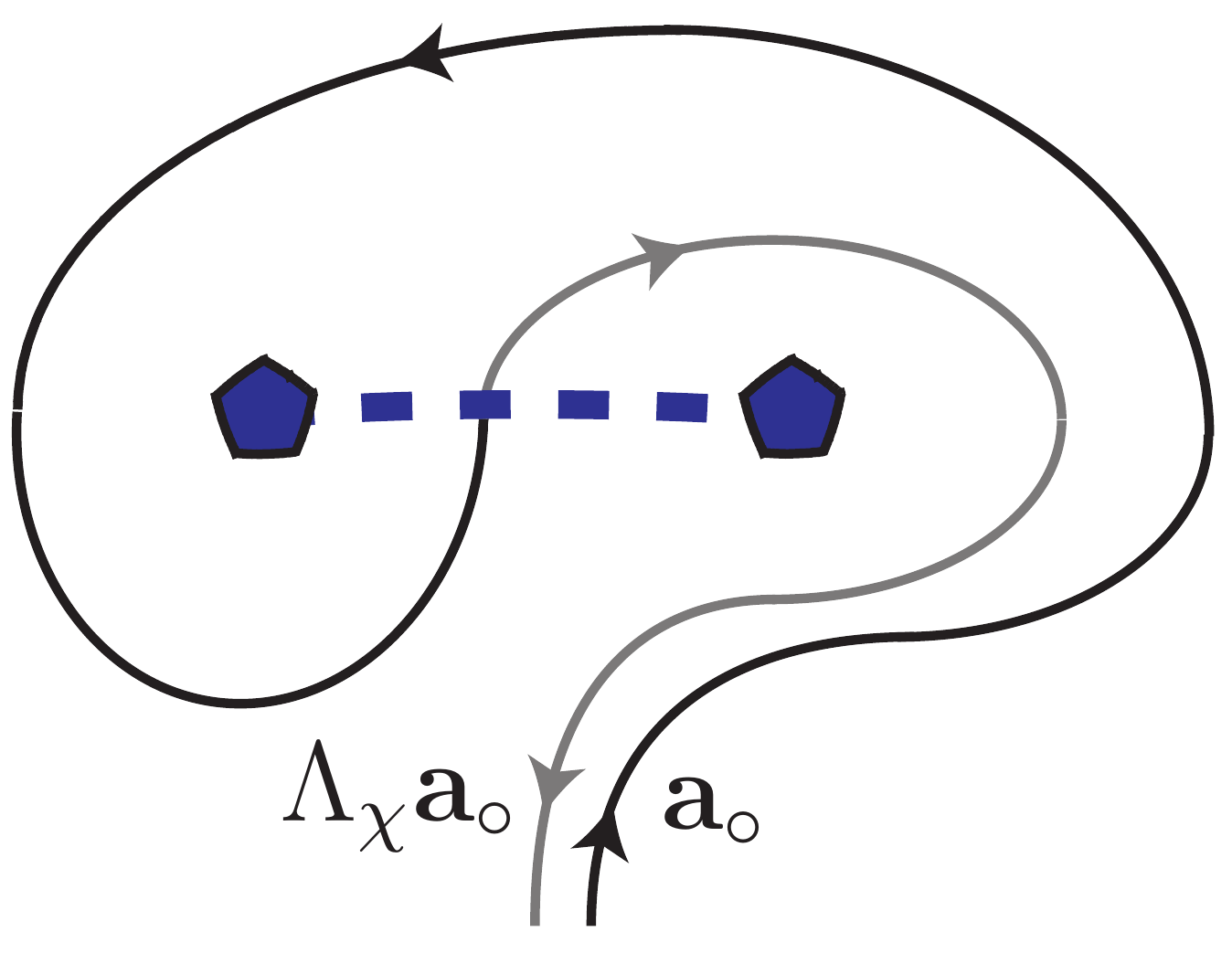}}}\right]|GS\rangle_0\nonumber\\&=\left[\vcenter{\hbox{\includegraphics[width=0.6in]{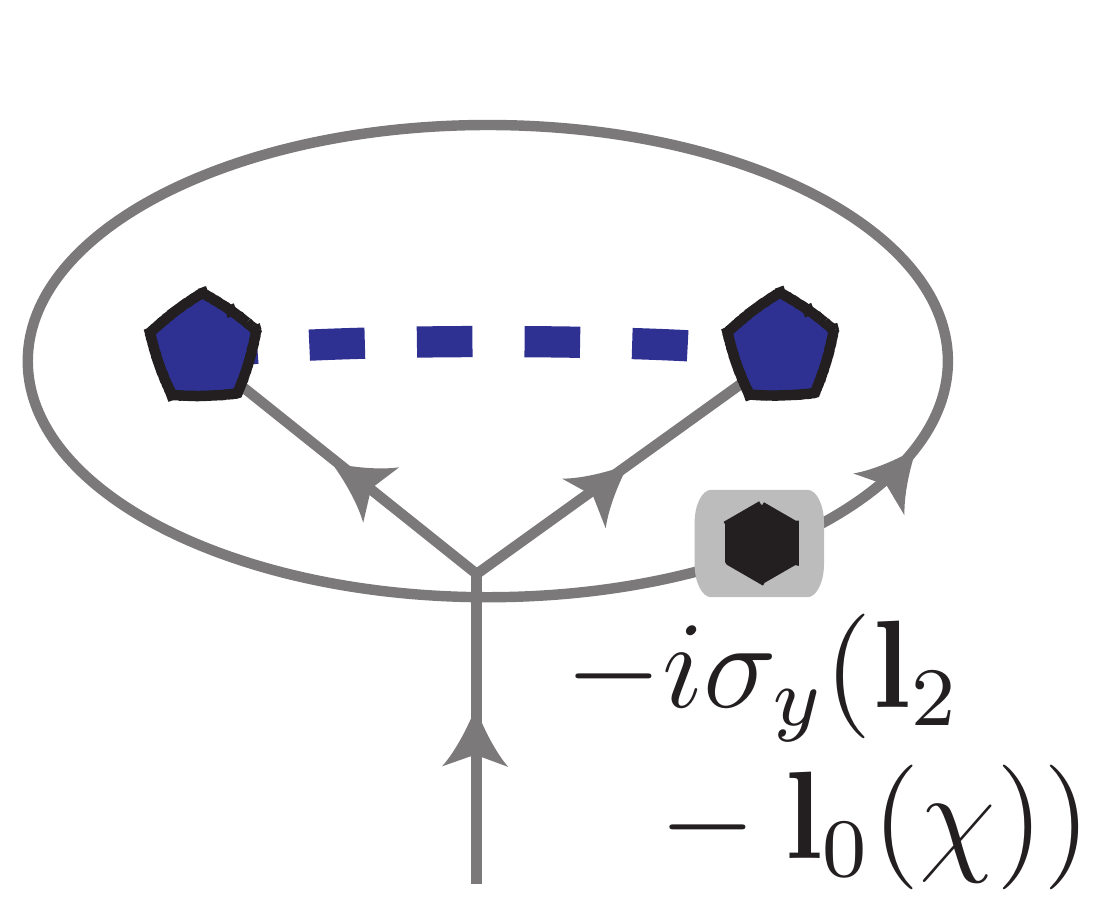}}}\right]\left[\vcenter{\hbox{\includegraphics[width=0.7in]{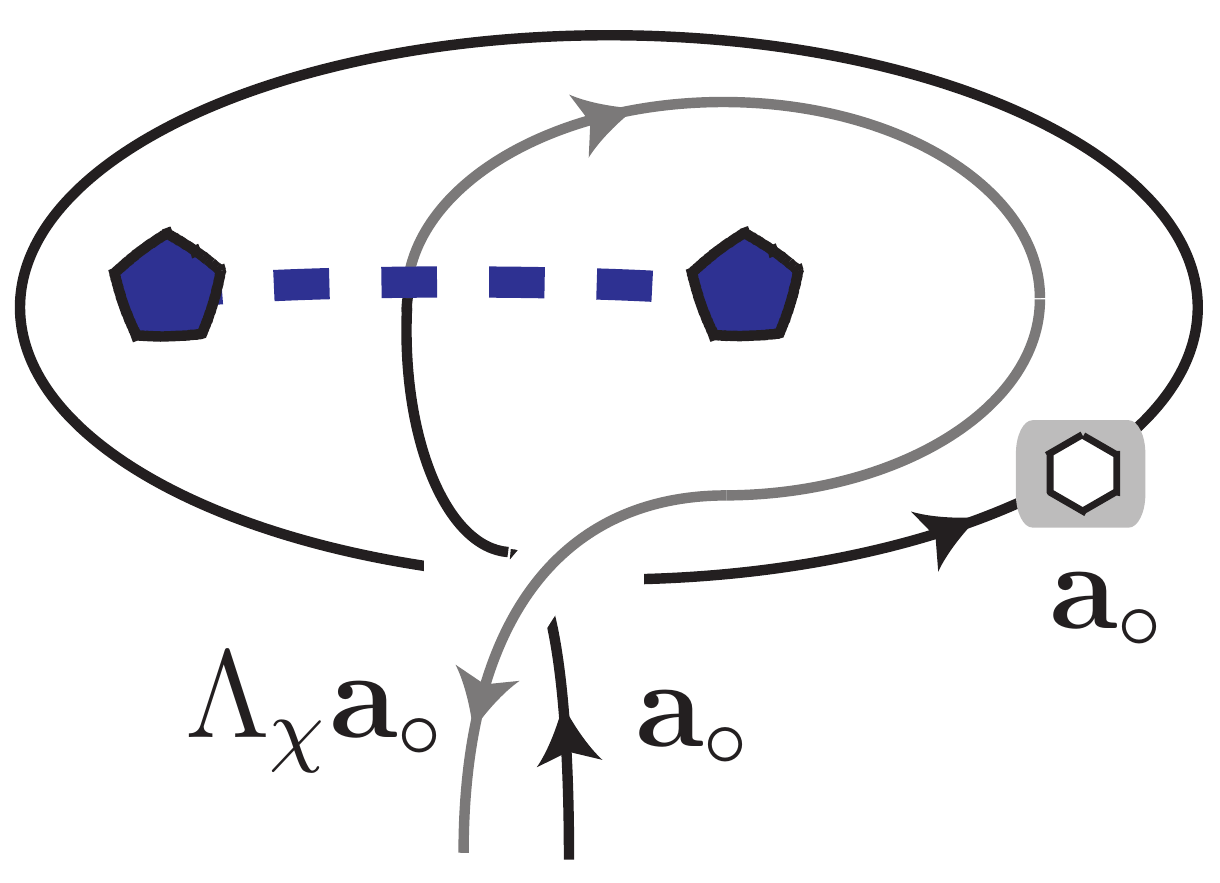}}}\right]|GS\rangle_0\end{align} The ${\bf a}_\circ$ loop on the right can be directly absorbed in the bare ground state condensate while the $-i\sigma_y({\bf l}_2-{\bf l}_0(\chi))$ one on the left are absorbed after crossing a black string. The remaining Wilson strings can be re-expressed in the form of the original splitting state configuration in figure~\ref{fig:splittingspaces}(c) by adding a local double Wilson loop at the right defect. The $R$-symbol is therefore given by the abelian phase \begin{align}R^{\left[\frac{1}{2}\right]_{\chi,{\bf l}_1}\left[\frac{1}{2}\right]_{\chi,{\bf l}_2}}_{\bf a}&=\left[\vcenter{\hbox{\includegraphics[width=0.6in]{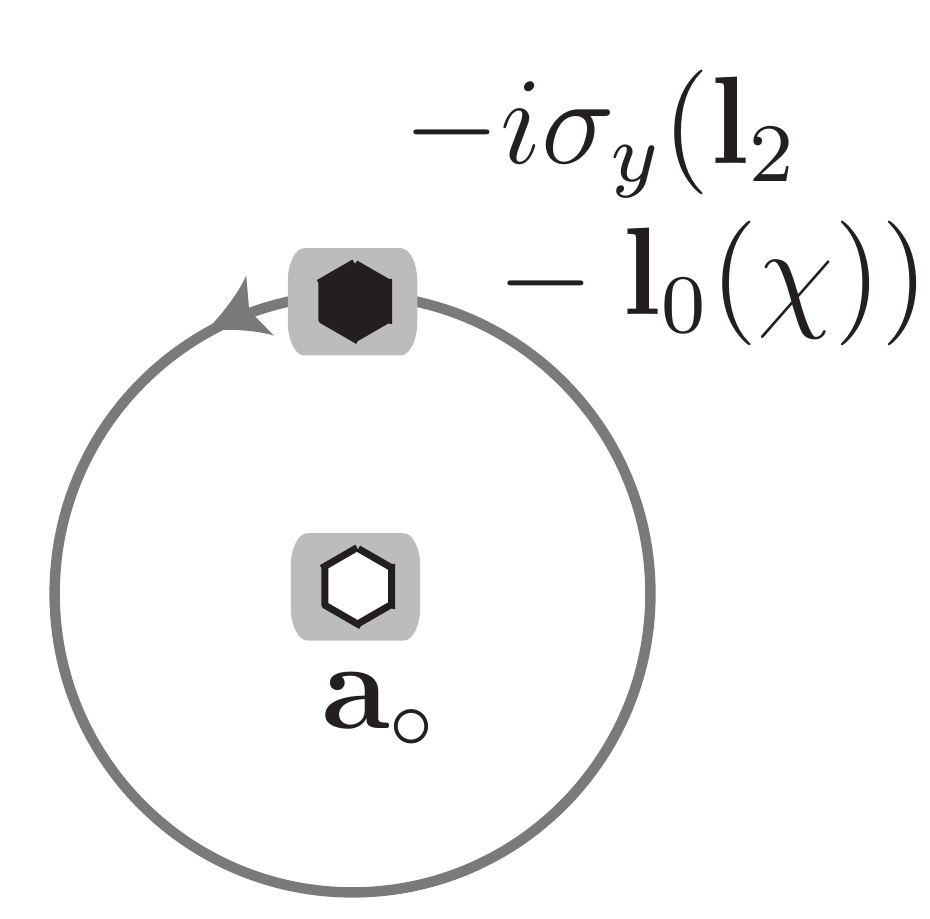}}}\right]\left[\vcenter{\hbox{\includegraphics[width=0.4in]{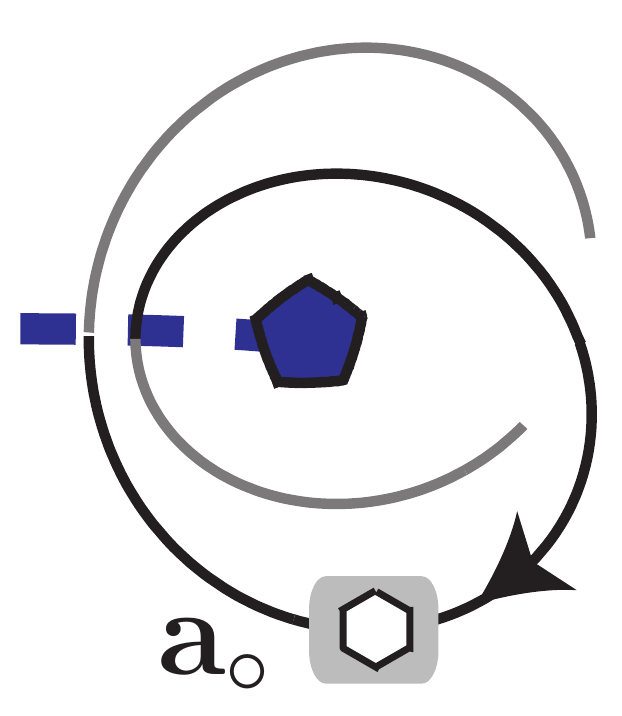}}}\right]\nonumber\\&=w^{{\bf a}_\circ^T\left[{\bf l}_2+\frac{1}{2}i\sigma_y({\bf f}^\chi-\Lambda_\chi{\bf a}_\circ)\right]}\label{Z2exchange}\end{align} where ${\bf f}^\chi=(1,0),(0,1),(-1,-1)$ for $\chi=Y,R,B$ respectively, and $\Lambda_\chi=i\sigma_yJ_\chi$ is the transposition matrix defined in \eqref{Z2JYRB} or \eqref{lambdachiapp}. The phase of the double Wilson loop on the bare twofold defect are evaluated by counting self-intersections and is illustrated in \eqref{doubleloopbarephase} in appendix~\ref{sec:doubleloop}. The topological spin of a twofold defect depends on its species label ${\bf l}$ and is given by the sum \eqref{topologicalspin} over exchange \eqref{Z2exchange}. \begin{align}\theta_{\left[\frac{1}{2}\right]_{\chi,{\bf l}}}&=\frac{1}{k}\sum_{\bf a}R^{\left[\frac{1}{2}\right]_{\chi,{\bf l}}\left[\frac{1}{2}\right]_{\chi,{\bf l}}}_{\bf a}=w^{\frac{1}{2}{\bf l}^T\left[i\sigma_y\Lambda_\chi^T{\bf l}+{\bf f}^\chi\right]}\label{Z2topologicalspin}\end{align} where the summation is taken over the $k^2$ possible overall abelian anyon fusion channel restricted by \eqref{22fusionconstraint} or $i\sigma_y({\bf a}_\bullet+\Lambda_\chi{\bf a}_\circ)=2({\bf l}-{\bf l}_0(\chi))$, and the closed form solution is obtained by using the identity $\sum_{r=0}^{k-1}w^{r(r+m)/2}=\sqrt{k}w^{(k-m^2)/8}$ for $m\equiv k$ mod 2. The spin can also be understood by the orientation frame provided by the Wilson string ${\bf c}=-i\sigma_y({\bf l}-{\bf l}_0(\chi))$ attaching to a twofold defect (see figure~\ref{fig:objectrep}(c)). Upon $4\pi$-rotation, the string is dragged around the defect and can be untwisted by a local double loop. And therefore \begin{align}\vcenter{\hbox{\includegraphics[width=0.4in]{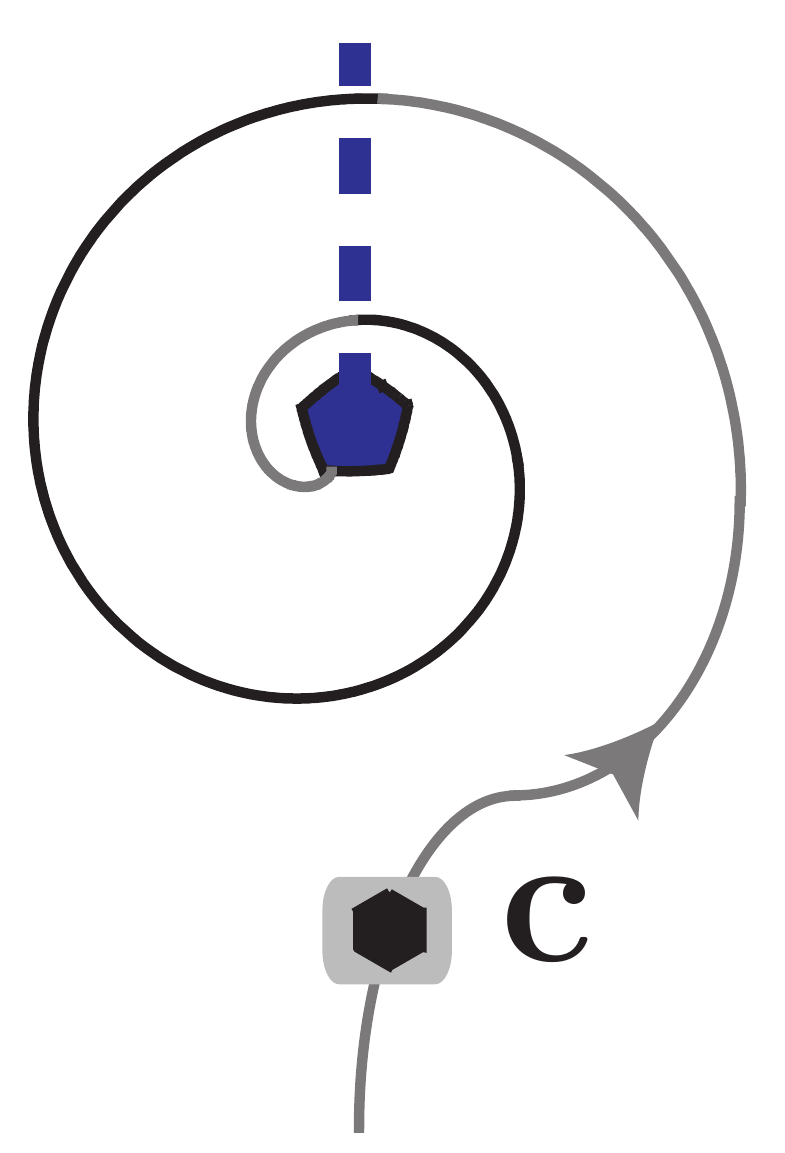}}}&=\vcenter{\hbox{\includegraphics[width=0.4in]{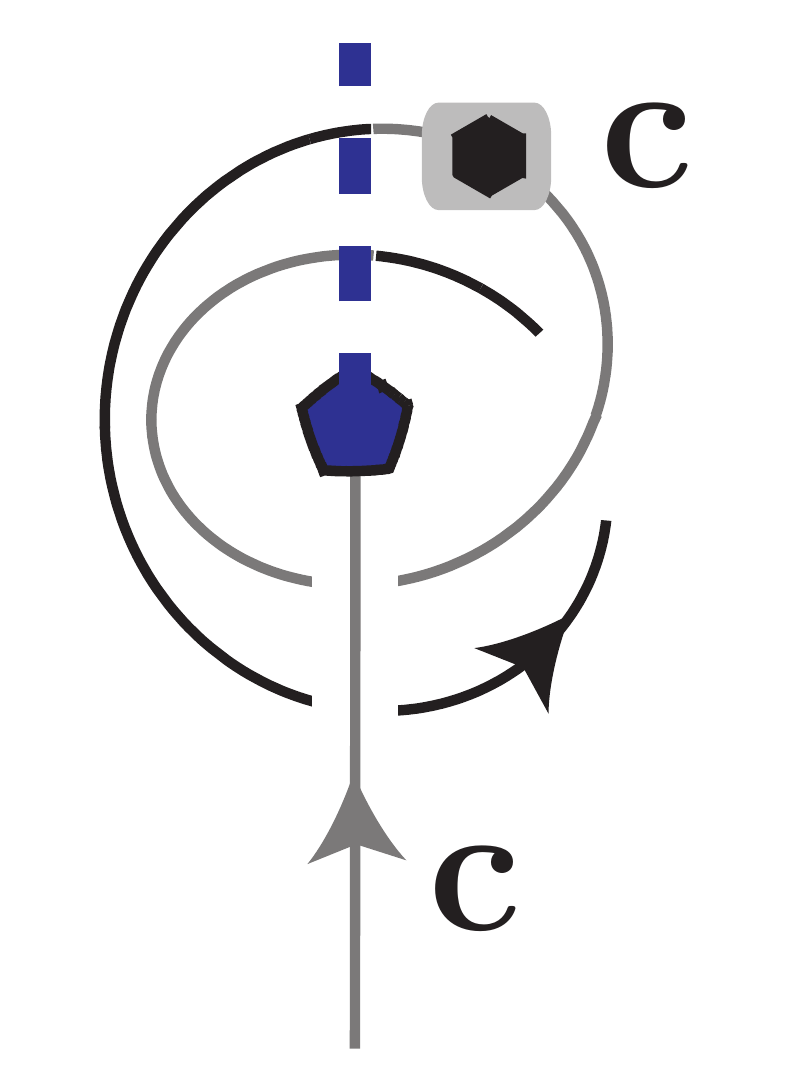}}}=\theta_{\left[\frac{1}{2}\right]_{\chi,{\bf l}}}\vcenter{\hbox{\includegraphics[width=0.15in]{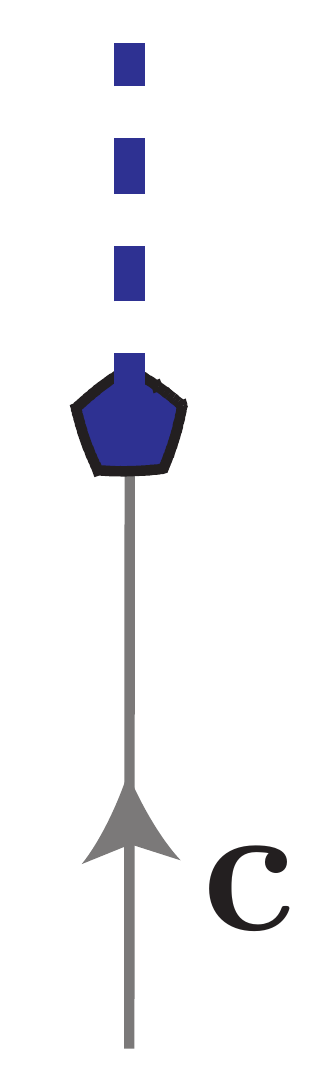}}}\end{align} \begin{align}\theta_{\left[\frac{1}{2}\right]_{\chi,{\bf l}}}&=\vcenter{\hbox{\includegraphics[width=0.4in]{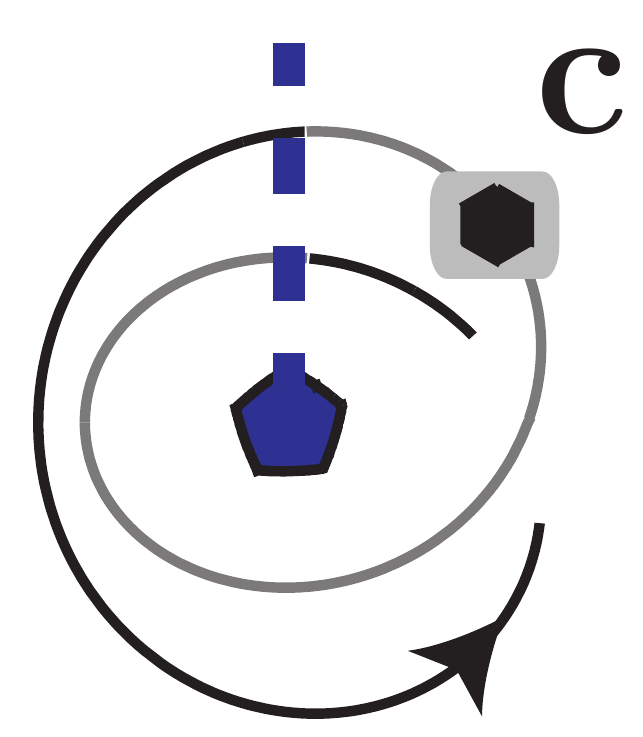}}}=w^{(\Lambda_\chi{\bf c})^Ti\sigma_y\left[\frac{k}{2}{\bf f}^{\chi}+\frac{1}{2}{\bf c}\right]}\label{Z2topologicalspin2}\end{align} which matches \eqref{Z2topologicalspin} exactly (see eq.\eqref{doubleloopbarephase} in appendix~\ref{sec:doubleloop}). 

Eq.\eqref{Z2topologicalspin} or \eqref{Z2topologicalspin2} are invariant under species conjugation $\overline{\bf l}=-{\bf l}-i\sigma_y{\bf f}^{\chi}=-{\bf l}+2{\bf l}_0(\chi)$ or $\overline{\bf c}=-{\bf c}$, and therefore twofold defect carries the same topological spin as its antiparticle. %Moreover, if the species label is self-conjugate so that $2{\bf l}_0=-i\sigma_y{\bf f}^\chi$, i.e. ${\bf c}_0=0$, then the defect is its own antiparticle and has trivial topological spin, which matches the trivial exchange phase at the overall vacuum fusion channel. \begin{equation}\theta_{\left[\frac{1}{2}\right]_{\chi,{\bf l}_0}}=R^{\left[\frac{1}{2}\right]_{\chi,{\bf l}_0}\left[\frac{1}{2}\right]_{\chi,{\bf l}_0}}_1=1\end{equation}
The bare twofold defect with self conjugate label ${\bf l}={\bf l}_0(\chi)=-\frac{k+1}{2}i\sigma_y{\bf f}^\chi$ or ${\bf c}=0$ carries trivial spin. %We notice that the topological spin $\theta_{[1/2]_{\chi,{\bf l}}}$ is {\em fractionalized} so that $\theta^k=-1$ when $k$ is even and $({\bf f}^\chi)^T({\bf l}-{\bf l}_0(\chi))$ is odd.

The $360^\circ$ braiding between a pair of twofold defects gives a phase that identifies the spin of the overall abelian fusion channel. \begin{align}\vcenter{\hbox{\includegraphics[width=0.6in]{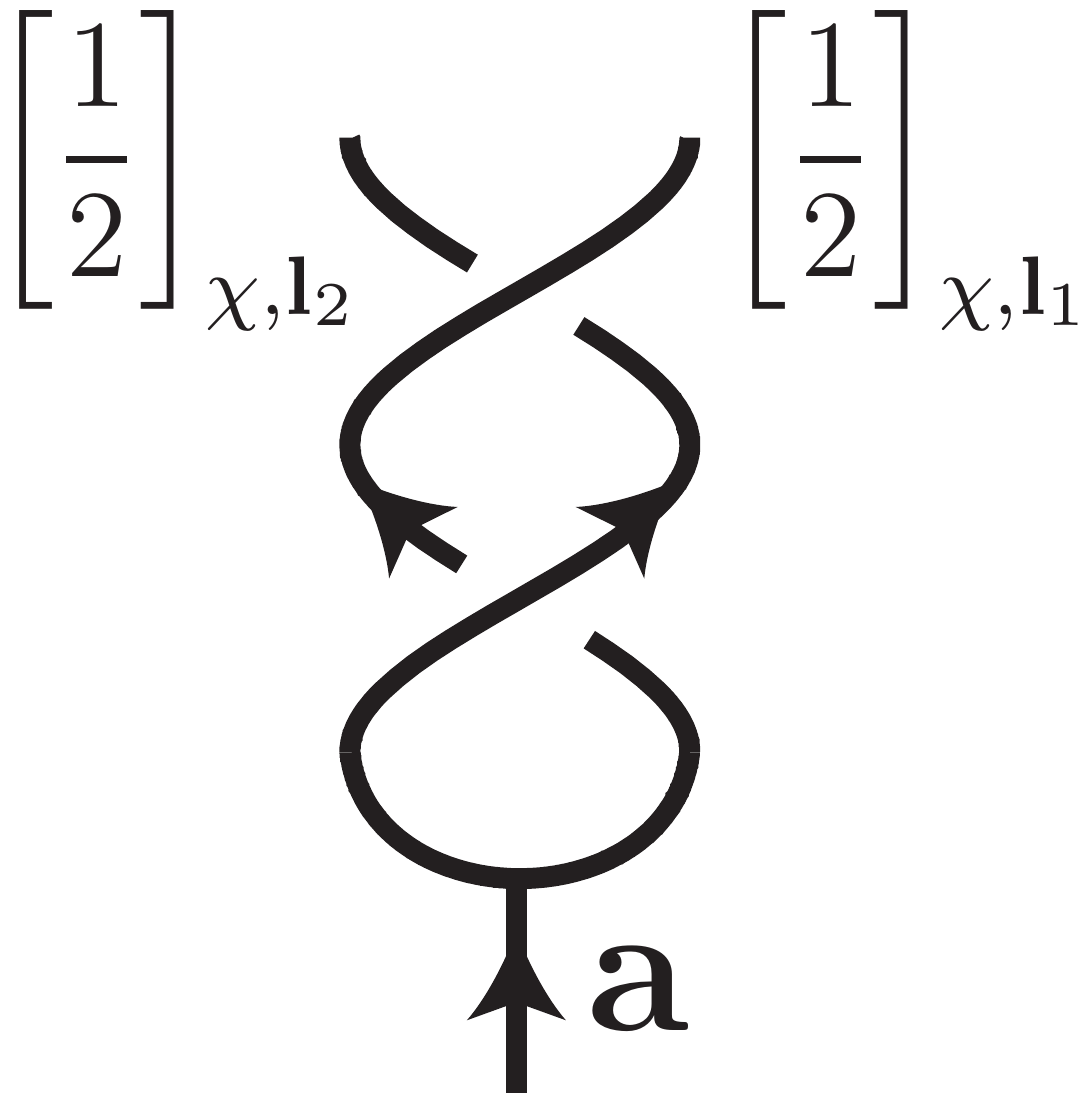}}}&\equiv R^{\left[\frac{1}{2}\right]_{\chi,{\bf l}_1}\left[\frac{1}{2}\right]_{\chi,{\bf l}_2}}_{\bf a}R^{\left[\frac{1}{2}\right]_{\chi,{\bf l}_2}\left[\frac{1}{2}\right]_{\chi,{\bf l}_1}}_{\bf a}=\theta_{\bf a}\nonumber\\&=w^{{\bf a}_\circ^Ti\sigma_y{\bf a}_\bullet}=w^{{\bf a}_\circ^T\left[{\bf l}_1+{\bf l_2}-2{\bf l}_0(\chi)-i\sigma_y\Lambda_\chi{\bf a}_\circ\right]}\label{Z2fullbraiding}\end{align} where the overall abelian channel is restricted by \eqref{22fusionconstraint} or $i\sigma_y({\bf a}_\bullet+\Lambda_\chi{\bf a}_\circ)={\bf l}_1+{\bf l}_2-2{\bf l}_0(\chi)$.

\subsection{Defect modular transformations}\label{sec:defectmodulartransformation}
We compute the topological $S$ and $T$ matrices for twofold defects ($S=T=1$ for threefold defects), interpret them geometrically as restricted Dehn twists on a decorated torus, and study the group structures for small $k$. From definition \eqref{defectSmatrix} and the $360^\circ$ braiding \eqref{Z2fullbraiding}, the $S$-matrix has $k^2\times k^2$ entries \begin{align}S_{{\bf l}_1{\bf l}_2}&=\frac{1}{k^2}\sum_{\bf a}R^{\left[\frac{1}{2}\right]_{\chi,\overline{{\bf l}_2}}\left[\frac{1}{2}\right]_{\chi,{\bf l}_1}}_{\bf a}R^{\left[\frac{1}{2}\right]_{\chi,{\bf l}_1}\left[\frac{1}{2}\right]_{\chi,\overline{{\bf l}_2}}}_{\bf a}\nonumber\\&=\frac{1}{k^2}\sum_{{\bf a}_\circ}w^{{\bf a}_\circ^T\left[{\bf l}_1-{\bf l}_2-i\sigma_y\Lambda_\chi{\bf a}_\circ\right]}\label{Ssum}\end{align} where $w=e^{2\pi i/k}$ and $\overline{{\bf l}_2}=-{\bf l}_2+2{\bf l}_0(\chi)$ is the reciprocal species label. The quadratic Gaussian sum can be expressed in a closed form depending on $k$ modulo 4. Let $\delta{\bf l}={\bf l}_1-{\bf l}_2$. For $k$ odd, \begin{align}S_{{\bf l}_1{\bf l}_2}&=\frac{1}{k}w^{\left(\frac{k+1}{2}\right)^2\delta{\bf l}^Ti\sigma_y\Lambda_\chi^T\delta{\bf l}}\label{Z2Smatrixkodd}\end{align} for $k\equiv0$ mod 4, \begin{align}S_{{\bf l}_1{\bf l}_2}&=\left\{\begin{array}{*{20}c}\frac{2}{k}w^{\frac{1}{4}\delta{\bf l}^Ti\sigma_y\Lambda_\chi^T\delta{\bf l}},&\mbox{for $\Lambda_\chi\delta{\bf l}\equiv\{0,0\}$ mod 2}\\0\hfill,&\mbox{otherwise}\hfill\end{array}\right.\label{Z2Smatrixk0}\end{align} and for $k\equiv2$ mod 4, \begin{align}S_{{\bf l}_1{\bf l}_2}&=\left\{\begin{array}{*{20}c}\frac{2}{k}w^{\frac{1}{4}\delta{\bf l}^Ti\sigma_y\Lambda_\chi^T\delta{\bf l}},&\mbox{for $\Lambda_\chi\delta{\bf l}\equiv\{1,1\}$ mod 2}\\0\hfill,&\mbox{otherwise}\hfill\end{array}\right.\label{Z2Smatrixk2}\end{align} The $T$ matrix is diagonal with entries given by topological spins $\theta_{[1/2]_{\chi,{\bf l}}}$ evaluated in \eqref{Z2topologicalspin2}.

\begin{figure}[ht]
\centering
\includegraphics[width=3in]{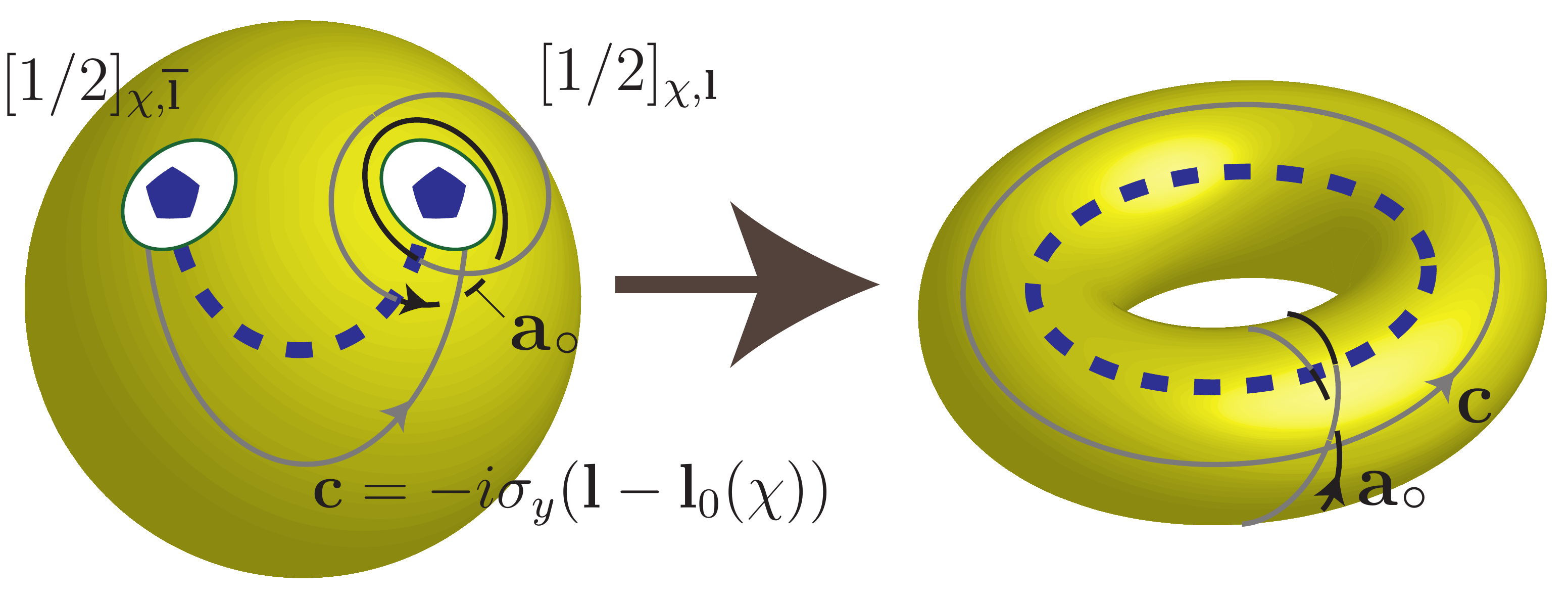}
\caption{Gluing a conjugate pair of twofold defects on a punctured sphere into a torus with a branch cut (blue dashed line) along the longitudinal cycle.}\label{fig:modulartorus}
\end{figure}
\begin{figure}[ht]
\centering
\includegraphics[width=2.7in]{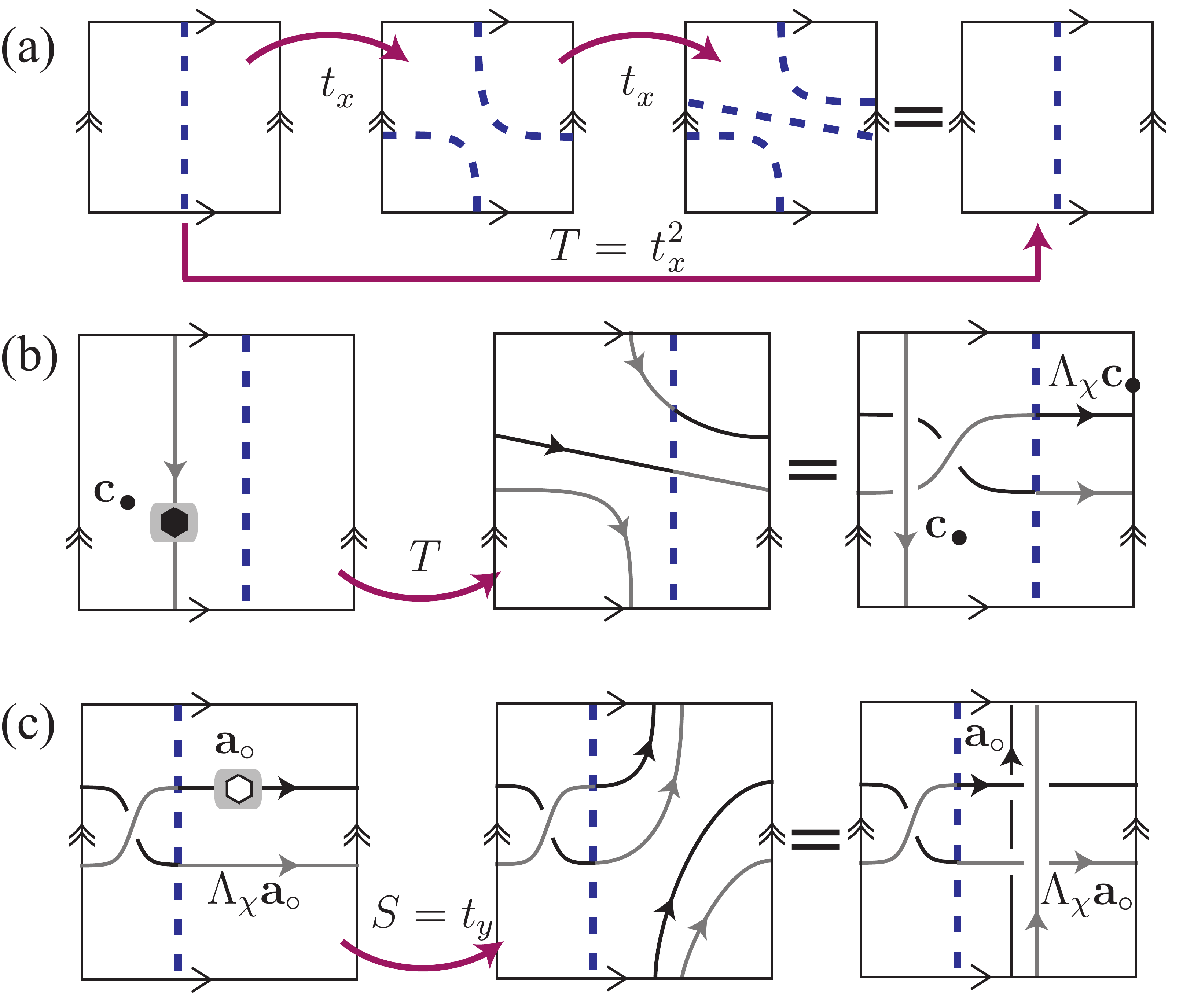}
\caption{Dehn twists. (a) Double Dehn twist $T$ along the horizontal direction that leaves the branch cut invariant. (b) $T$-action on a vertical Wilson loop $W({\bf c}_\bullet)$. (c) Dehn twist $S$ along the vertical direction and its action on a double Wilson loop $\Theta({\bf a}_\circ)$.}\label{fig:dehntwist}
\end{figure}
The topological $S$ and $T$-matrices can be geometrically understood as two Dehn twists on a torus with a twofold branch cut along a non-trivial cycle (see figure~\ref{fig:modulartorus} and \ref{fig:dehntwist}). This can be constructed by cutting out a pair of twofold defects with conjugate species labels on a sphere and then pasting the holes together. Equivalently one can also imagine splitting a conjugate pair of twofold defects on a torus and move one of them along a cycle before fusing back to vacuum. (a similar construction was considered for a threefold color branch cut in subsection~\ref{sec:obstructiontoglobaltricolorability}.) Abelian anyons dragged along a meridian cycle or cutting across the twofold branch cut undergo the twofold twisting $({\bf a}_\bullet,{\bf a}_\circ)\to(\Lambda_\chi{\bf a}_\circ,\Lambda_\chi{\bf a}_\bullet)$. Wilson loops on the branch cut decorated torus are generated by (i) $W({\bf c})$, trajectory of an anyon ${\bf c}_\bullet$ going once along the longitudinal $y$-direction, and (ii) $\Theta({\bf a}_\circ)$, trajectory of an anyon ${\bf a}_\circ$ going twice along the meridian $x$-direction. 

Starting with the normalized {\em bare} ground state on the torus with an arbitrarily fixed branch cut \begin{equation}|\emptyset\rangle=\frac{1}{\mathcal{N}}\prod_P\left(\sum_{r=0}^{k-1}\hat{P}^r\right)|\sigma_{v_\bullet}=\tau_{\circ}=+1\rangle\label{modularbareGS}\end{equation} we choose a complete set of $k^2$ ground states \begin{equation}|{\bf c}_\bullet\rangle=W({\bf c}_\bullet)\left[\frac{1}{k}\sum_{{\bf a}_\circ}w^{-{\bf a}_\circ^Ti\sigma_y\left(\frac{k}{2}{\bf f}^\chi+\frac{1}{2}\Lambda_\chi{\bf a}_\circ\right)}\Theta({\bf a}_\circ)\right]|\emptyset\rangle\label{modularbasis}\end{equation} The sum in the bracket makes sure it is an eigenstate of all meridian double loops $\Theta({\bf a}_\circ)$. The phase in the sum originates from self-intersection of the double Wilson loop (see eq.\eqref{doubleloopbarephase} in appendix~\ref{sec:doubleloop}) so that $\langle0|\Theta({\bf a}_\circ)|0\rangle=w^{{\bf a}_\circ^Ti\sigma_y\left(\frac{k}{2}{\bf f}^\chi+\frac{1}{2}\Lambda_\chi{\bf a}_\circ\right)}$. The state $|{\bf c}_\bullet\rangle$ corresponds to the conjugate pair of twofold defects with species label ${\bf l}$ and longitudinal string \begin{equation}{\bf c}_\bullet=-i\sigma_y({\bf l}-{\bf l}_0(\chi))\label{Z2stringspecies}\end{equation} attaching them (see figure~\ref{fig:modulartorus}).

Since the meridian cycle is semiclassically distinct from the longitudinal one, the system does not possesses full $SL(2;\mathbb{Z})$ modular symmetry of the torus. In particular the branch cut is altered by Dehn twist along the meridian $x$-direction but invariant under a double twist because a pair of parallel twofold branch cuts annihilate (see figure~\ref{fig:dehntwist}(a)). The torus decorated with a branch cut therefore admit a courser set of modular transformations generated by Dehn twists $S=t_y$ in the longitudinal $y$-direction and double twist $T=t_x^2$ in the meridian $x$-direction. 

The action of the two Dehn twists on Wilson loops are shown in figure~\ref{fig:dehntwist}(a) and (b). The $T$-transformation leaves all double Wilson loops $\Theta({\bf a}_\circ)$ unchanged while sending \begin{equation}W({\bf c}_\bullet)\to TW({\bf c}_\bullet)T^\dagger=W({\bf c}_\bullet)\Theta(\Lambda_\chi{\bf c}_\bullet)\end{equation} The $S$-transformation leaves all longitudinal Wilson loops $W({\bf c})$ invariant while changing \begin{align}\Theta({\bf a}_\circ)\to S\Theta({\bf a}_\circ)S^\dagger=W(-\Lambda_\chi{\bf a}_\circ)\Theta({\bf a}_\circ)W(-{\bf a}_\circ)\end{align} The branch cut is left unaltered under both $S$ and $T$. The $T$-matrix is diagonal in the basis \eqref{modularbasis} with entries matching the topological spins \eqref{Z2topologicalspin2} for twofold defects. \begin{align}T_{{\bf c}_1{\bf c}_2}&=\langle{\bf c}_1|T|{\bf c}_2\rangle=\langle\emptyset|\Theta(\Lambda_\chi{\bf c}_2)|\emptyset\rangle T_0\delta_{{\bf c}_1{\bf c}_2}\nonumber\\&=w^{(\Lambda_\chi{\bf c}_2)^Ti\sigma_y\left[\frac{k}{2}{\bf f}^\chi+\frac{1}{2}{\bf c}_2\right]}T_0\delta_{{\bf c}_1{\bf c}_2}\label{Z2modularTmatrix}\end{align} where the arbitrary eigenvalue $T_0=\langle0|T|0\rangle$ is set to 1. The $S$-matrix is given by the overlap \begin{align}S_{{\bf c}_1{\bf c}_2}&=\langle{\bf c}_1|S{\bf c}_2\rangle\nonumber\\&=\frac{S_\phi}{k^2}\sum_{{\bf a}_\circ{\bf b}_\circ}w^{{\bf a}_\circ^Ti\sigma_y\left[\frac{k}{2}{\bf f}^\chi+\frac{1}{2}\Lambda_\chi{\bf a}_\circ\right]-{\bf b}_\circ^Ti\sigma_y\left[\frac{k}{2}{\bf f}^\chi+\frac{1}{2}\Lambda_\chi{\bf b}_\circ\right]}\nonumber\\&\quad\times\langle\emptyset|\Theta({\bf a}_\circ)^\dagger W({\bf c}_2-{\bf c}_1-\Lambda_\chi{\bf b}_\circ)\Theta({\bf b}_\circ)W(-{\bf b}_\circ)|\emptyset\rangle\end{align} where the arbitrary eigenvalue $S_\phi=\langle\emptyset |S|\emptyset\rangle$ can be set to 1. The longitudinal Wilson loop $W$ can be passed across the double meridian loop $\Theta$, leaving an intersection phase $w^{{\bf b}_\circ^Ti\sigma_y\left[({\bf c}_2-{\bf c}_1)-\Lambda_\chi{\bf b}_\circ\right]}$ before being absorbed by the condensate in the bare ground state $|\emptyset\rangle$. And since \begin{equation}\langle\emptyset|\Theta({\bf a}_\circ)^\dagger\Theta({\bf b}_\circ)|\emptyset\rangle=\delta_{{\bf a}_\circ{\bf b}_\circ}\end{equation} the $S$-matrix has entries \begin{equation}S_{{\bf c}_1{\bf c}_2}=\frac{1}{k^2}\sum_{{\bf a}_\circ}w^{{\bf a}_\circ^Ti\sigma_y\left[({\bf c}_2-{\bf c}_1)-\Lambda_\chi{\bf a}_\circ\right]}\label{Z2modularSmatrix}\end{equation} which matches \eqref{Ssum} exactly through the identification \eqref{Z2stringspecies} that relates species label ${\bf l}$ and longitudinal string ${\bf c}$.

A torus decorated with a threefold color branch cut was discussed in subsection~\ref{sec:obstructiontoglobaltricolorability}, and a similar modular subgroup applies except that the $T$-transformation is given by a triple Dehn twist. It is straightforward to check that, even when $k$ is divisible by 3 and there is a 9 fold ground state degeneracy, both $S$ and $T$ acts trivially on Wilson loops and therefore on the ground state. In particular this explains the trivial exchange and braiding statistics of threefold defects.

Next we demonstrate the group structure for defect modular transformations. In the pure anyonic case without a branch cut, the modular $T$, $S$ and $ST$ transformations represent a Dehn twist, $90^\circ$ and $120^\circ$ rotations of the torus respectively. None of these hold for non-trivial branch cut of a finite order $N=\mbox{ord}(\lambda)>0$. The mapping class group of a decorated torus that fixes the branch cut as a homological cycle ${\bf e}_1=(1,0)$ in $H_1(T^2;\mathbb{Z}_N)=\mathbb{Z}_N^2$ is known as a congruence subgroup \begin{equation}\Gamma_0(N)=\left\{A\in SL(2;\mathbb{Z}):A{\bf e}_1\equiv{\bf e}_1\mbox{ mod $N$}\right\}\end{equation} which is a non-normal subgroup of $SL(2;\mathbb{Z})$ with finite index. For example the modular subgroups for twofold and threefold defects are finitely presented (using the Reidemeister-Schreier algorithm) by \begin{align}\Gamma_0(2)&=\langle S=t_x,T=t_y^2|(ST)^2=-1\rangle\\\Gamma_0(3)&=\langle S=t_x,T=t_y^3|(ST)^3=1\rangle\end{align} Congruence subgroups have already studied in physical context of {\em modular duality} in quantum Hall plateau transition~\cite{LutkenRoss92, FradkinKivelson96, BayntunBurgessDolanLee11} and abelian gauge theory~\cite{WittenSduality}.

%However, the defect $S$ and $T$ matrices do not necessarily obey the group relation for general $k$. While $\Gamma_0(3)$ is trivially represented by threefold defects, the topological $T$ and $S$ matrices \eqref{Z2modularTmatrix} and \eqref{Z2modularSmatrix} for twofold defects do not form a representation for $\Gamma_0(2)$, and the group relation $(ST)^2=-1$ is in general not even projectively satisfied. This is due to the internal self-intersecting structure of the double Wilson loop $\Theta({\bf a}_\circ)$ so that $180^\circ$ rotation, sending $\Theta({\bf a}_\circ)\to(ST)^2\Theta({\bf a}_\circ)(ST)^{-2}=w^{2{\bf a}_\circ^Ti\sigma_y\Lambda_\chi{\bf a}_\circ}\Theta(-{\bf a}_\circ)$, no longer commutes with either $S$ or $T$ in general. 

In addition to the congruent relations, we also find that $S$ and $T$ have finite orders that depend on $k$. For the twofold defects considered in this article, $S^k=1$ and $T^k=1$ for odd $k$ or $T^{2k}=1$ for even $k$. We demonstrate the different group structures for small $k$. We assume the twofold defects have color $\chi=B$. The $S$ and $T$ matrices are decomposed into tensor products \begin{equation}S_{{\bf l}_1{\bf l}_2}=S^Y_{l^Y_1l^Y_2}\otimes S^R_{l^R_1l^R_2},\quad T_{{\bf l}_1{\bf l}_2}=T^Y_{l^Y_1l^Y_2}\otimes T^R_{l^R_1l^R_2}\end{equation} where ${\bf l}=(l^Y,l^R)$ are in $\mathbb{Z}_k^2$ for $k$ odd or $(\mathbb{Z}_k+1/2)^2$ for $k$ even. From \eqref{Z2Smatrixkodd}, \eqref{Z2Smatrixk0} and \eqref{Z2Smatrixk2}, the entries of $S^{Y/R}$ depend on $\delta l=l_1-l_2$ and are given by \begin{align}S^Y_{l_1l_2}=\left\{\begin{array}{*{20}c}\frac{1}{\sqrt{k}}w^{\left(\frac{k+1}{2}\right)^2\delta l^2},&\mbox{for $k$ odd}\hfill\\\sqrt{\frac{2}{k}}w^{\frac{1}{4}\delta l^2}\hfill,&\mbox{for $\delta l\equiv k/2$ and $k$ even}\\0\hfill,&\mbox{otherwise}\hfill\end{array}\right.\end{align} and $S^R=(S^Y)^\dagger$. $T^{Y/R}$ are diagonal with topological spin entries \eqref{Z2topologicalspin} \begin{equation}T^Y_{l_1l_2}=\delta_{l_1l_2}w^{\frac{1}{2}l_1(l_1-1)},\quad T^R_{l_1l_2}=\delta_{l_1l_2}w^{-\frac{1}{2}l_1(l_1+1)}\end{equation}

\subsubsection{Ising-type doublet at \texorpdfstring{$k=2$}{k=2}}
For $k=2$, the lattice model consists of spins and has the same topological content of two copies of Kitaev toric code. Twofold defects have similar fusion and statistics properties to uncoupled tensor pairs of Majorana fermions (or Ising anyons). The tensor components of $S=S^Y\otimes S^R$ and $T=T^Y\otimes T^R$ are given by \begin{align}S^Y&=(S^R)^\dagger=e^{i\pi/4}\sigma_x\\T^Y&=e^{i\pi/8}e^{i(\pi/4)\sigma_z},\quad T^R=e^{-i\pi/8}e^{i(\pi/4)\sigma_z}\end{align} The allowed topological spins from $4\pi$ rotation are $1,\pm i$. The four twofold defects can therefore be statistically identified with the following pairs of conventional Ising anyons \begin{equation}\begin{array}{*{20}c}[1/2]_{{\bf l}=(+1/2,+1/2)}=[\sigma_+]\otimes[\sigma_+]\\{[1/2]}_{{\bf l}=(-1/2,-1/2)}=[\sigma_-]\otimes[\sigma_-]\\{[1/2]}_{{\bf l}=(+1/2,-1/2)}=[\sigma_+]\otimes[\sigma_-]\\{[1/2]}_{{\bf l}=(-1/2,+1/2)}=[\sigma_-]\otimes[\sigma_+]\end{array}\end{equation} where the $2\pi$-topological spins of the Ising anyons $\sigma_\pm$ are $\theta_{\sigma_\pm}=e^{\mp i\pi/8}$. The full $4$-dimensional $S$ and $T$ matrices represent the dihedral group \begin{equation}D_4=\langle S,T|S^2=T^4=(ST)^2=1\rangle\end{equation} 
This forms a representation of the congruence subgroup $\Gamma_0(2)$.

\subsection{Non-abelian unitary braiding operations}\label{sec:Bmatrices}
\begin{figure}[ht]
	\includegraphics[width=2in]{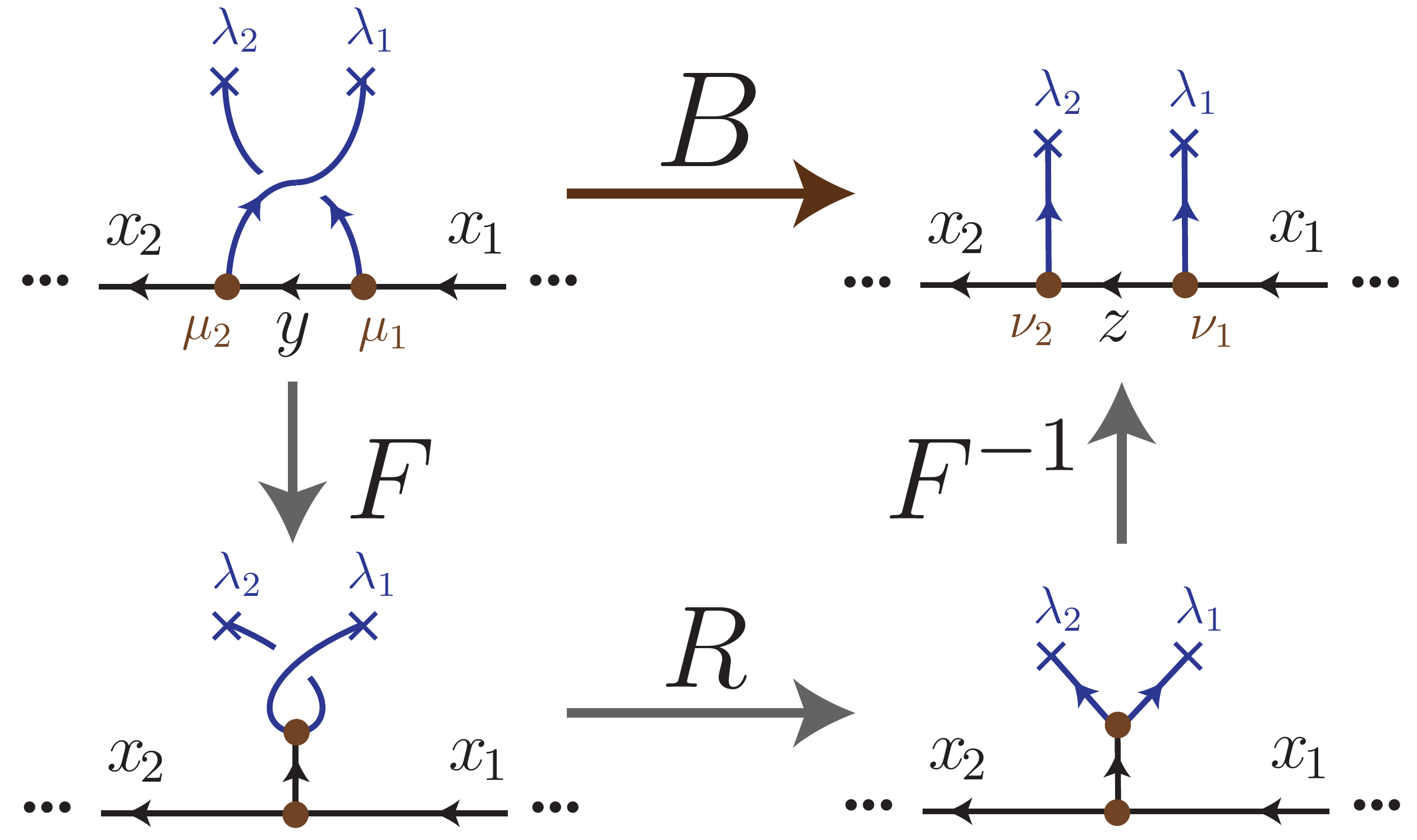}
	\caption{Braiding operation $B$ generated as a sequence of $F$ and $R$-moves. $\lambda_i$ are defects of the same $S_3$-type. Ground states are labeled by the intermediate channels $x_i,y,z$ and possible vertex degeneracies $\mu_i,\nu_i$.}\label{fig:Bmove}
\end{figure}
We evaluate the fundamental unitary exchange operations, called $B$-symbols, in a system where all defects are of the same $S_3$-type. Each $B$-move represents a counter-clockwise permutation of a pair of adjacent defects, and is a transformation between ground states labeled on the same fusion tree, i.e. eigenstates of the same maximal set of commuting observables. It can be generated by a sequence of $F$ and $R$-moves as shown in figure~\ref{fig:Bmove} so that \begin{align}\left|\vcenter{\hbox{\includegraphics[width=0.5in]{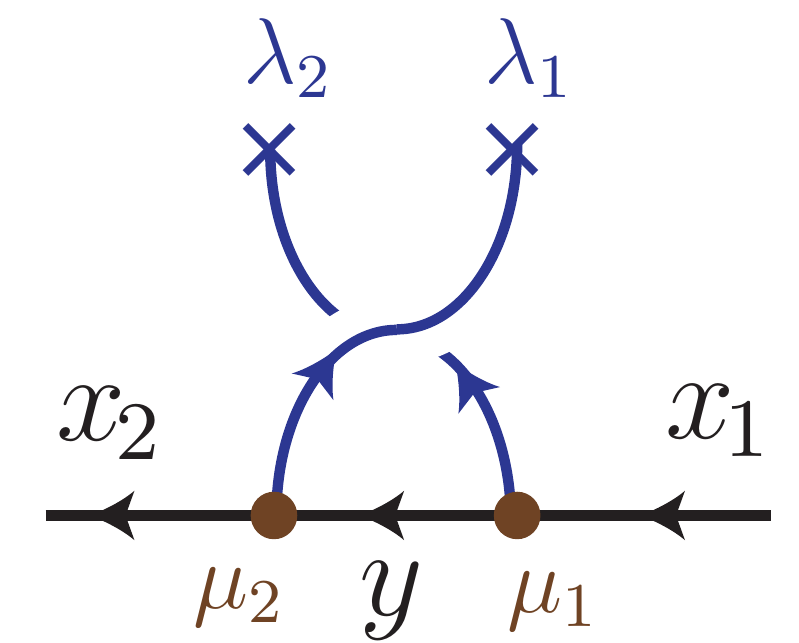}}}\right\rangle=\sum_{z,\nu_1,\nu_2}\left[B^{\lambda_2\lambda_1}_{x_2x_1}\right]_{y,\mu_2,\mu_1}^{z,\nu_1,\nu_2}\left|\vcenter{\hbox{\includegraphics[width=0.5in]{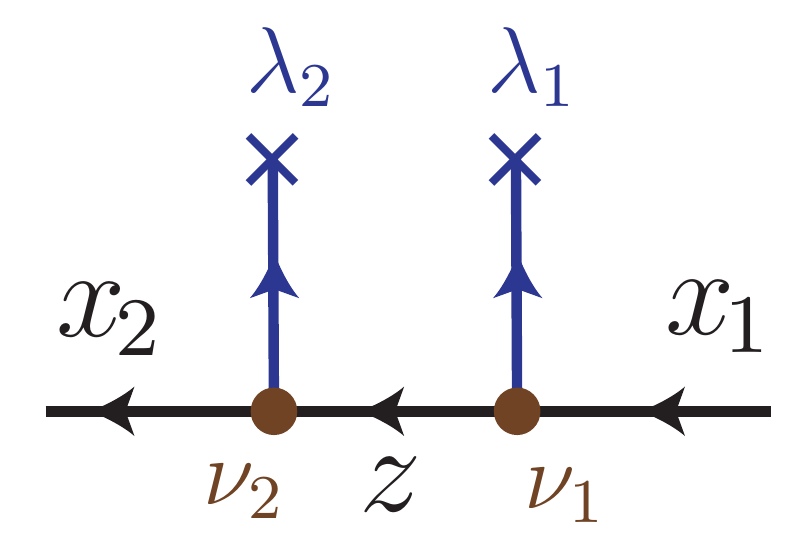}}}\right\rangle\label{Bmatrixdef1}\end{align} where the $B$-matrix is defined by \begin{align}\left[B^{\lambda_2\lambda_1}_{x_2x_1}\right]_{y,\mu_2,\mu_1}^{z,\nu_1,\nu_2}&=\sum_{w\gamma_1\gamma_2\gamma_3}\left(\left[F^{x_2\lambda_2\lambda_1}_{x_1}\right]_{z,\nu_1\nu_2}^{w,\gamma_1,\gamma_3}\right)^\ast\nonumber\\&\quad\quad\left[R^{\lambda_1\lambda_2}_w\right]_{\gamma_2}^{\gamma_3}\left[F^{x_2\lambda_1\lambda_2}_{x_1}\right]_{y,\mu_2,\mu_1}^{w,\gamma_1,\gamma_2}\label{Bmatrixdef2}\end{align}
The exchange operations $B^{\lambda_{i+1}\lambda_i}$ form the building blocks of the braid group of an ordered series of defects $\lambda_N,\ldots,\lambda_1$. 
\begin{figure}[ht]
	\includegraphics[width=1.8in]{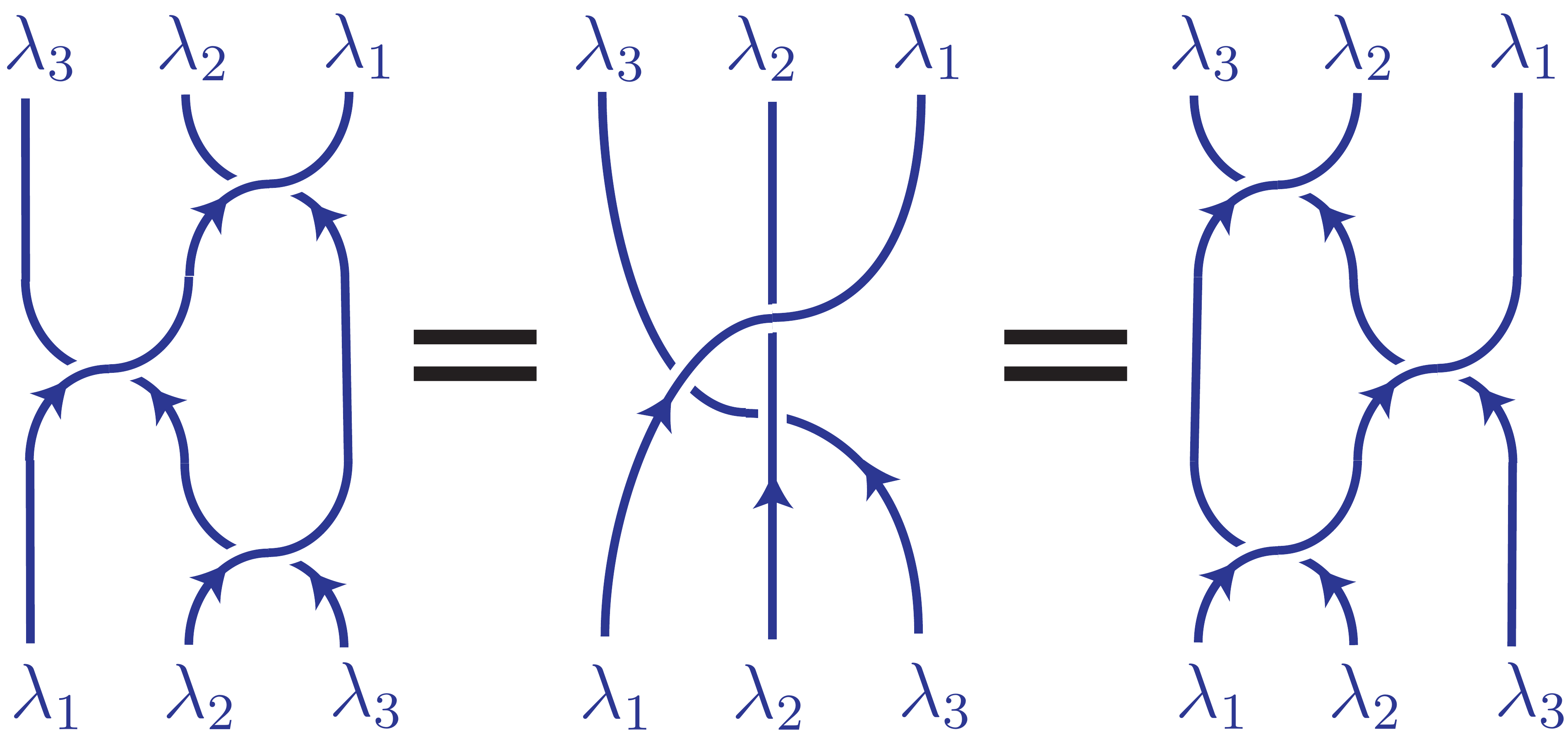}
	\caption{Yang-Baxter identity}\label{fig:YangBaxter}
\end{figure}
They obey the Yang-Baxter equation (see figure~\ref{fig:YangBaxter}) that characterizes braids. \begin{align}B^{\lambda_2\lambda_1}B^{\lambda_3\lambda_1}B^{\lambda_3\lambda_2}=B^{\lambda_3\lambda_2}B^{\lambda_3\lambda_1}B^{\lambda_2\lambda_1}\end{align} where the summation over intermediate channels and vertex degeneracies are suppressed. Defects $\lambda_i,\lambda_j$ are distinguishable when they have distinct species labels. A braiding operation is robustly represented only when the initial and final species labels configuration are identical.

Classical braid groups are further restricted by a compactification relation if the system lives on a closed sphere~\cite{JacakSitkoWieczorekWojsbook}, or equivalently the overall fusion channel of the defect system is the vacuum. For instance one expect the largest braid that moves an object once around all others to be trivial as it should be contractible on the other side of the sphere (see figure~\ref{fig:contractsphereloop}). \begin{align}\mathcal{M}&\equiv b^{2,1}\ldots b^{N,N-1}b^{N,N-1}\ldots b^{2,1}\label{braidoncearound}\\&=\vcenter{\hbox{\includegraphics[width=1in]{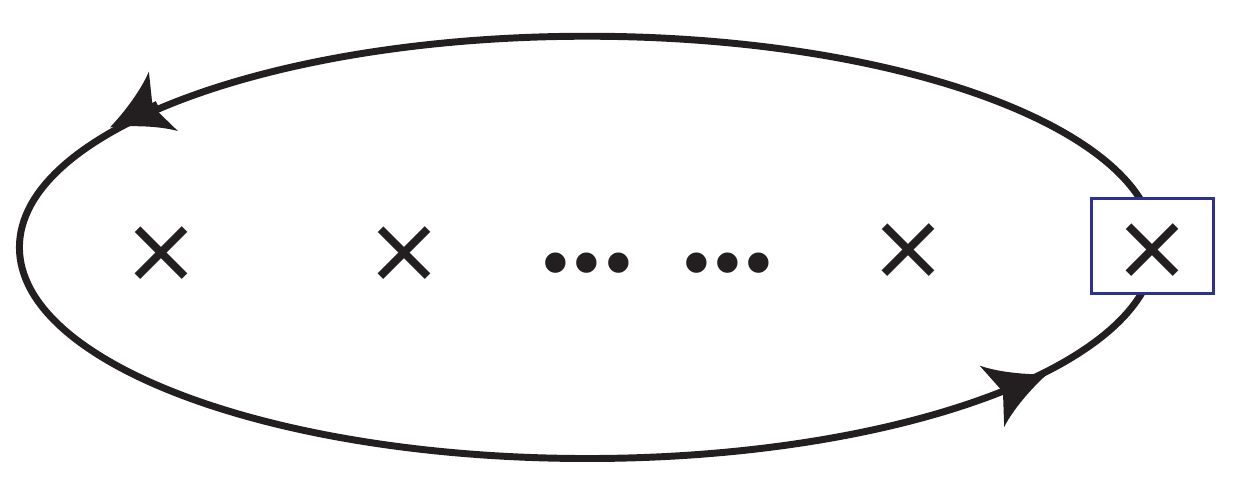}}}=1\label{braidoncearound2}\end{align} where $b^{i+1,i}$ is the exchange braid between the $(i+1)^{th}$ and $i^{th}$ defect to the right. Moreover the full braid involving rotating the whole system by $720^\circ$ should be contractible by the ``Dirac lasso trick" as it corresponds to the trivial element in $\pi_1(SO(3))=\mathbb{Z}_2$. And therefore the braid group is further restricted by (Fadell-Neuwirth) \begin{align}\mathcal{F}\equiv(b^{2,1}\ldots b^{N,N-1})^{2N}=\left[\vcenter{\hbox{\includegraphics[width=1in]{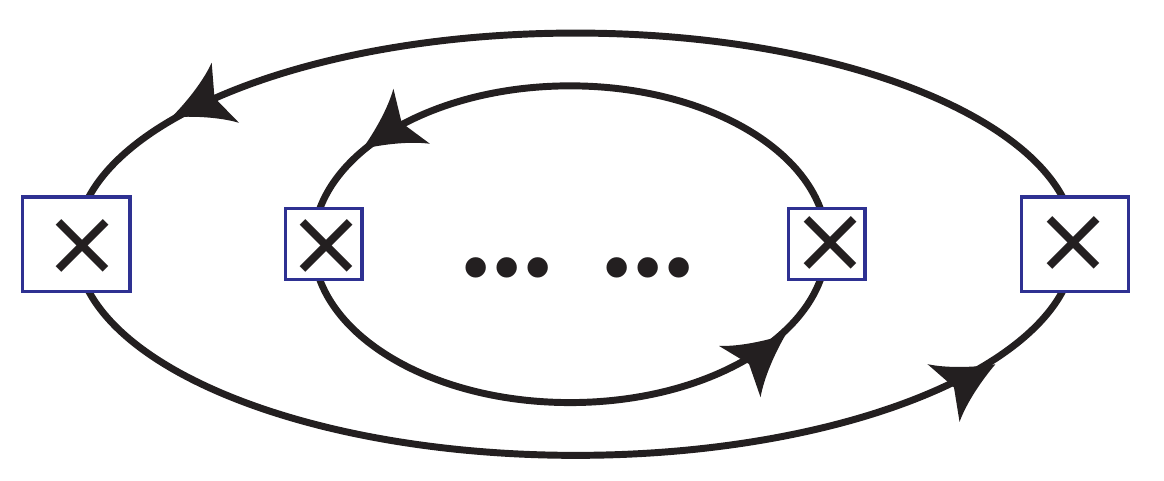}}}\right]^4=1\label{braidfull720}\end{align} We show the compactification relation \eqref{braidoncearound2} is in general only {\em projectively} satisfied for defects, so that $\mathcal{M}$ is non-trivial and generates a central cyclic subgroup that extends the braid group on a sphere.

\subsubsection{Threefold defects}
The three primitive exchange operations \eqref{Bmatrixdef1} for threefold defects can be evaluated by substituting the $F$-symbols $F^{{\bf a}[1/3][1/3]}_{[\overline{1/3}]}$, $F^{[1/3][1/3][1/3]}_{\bf a}$, $F^{[\overline{1/3}][1/3][1/3]}_{[1/3]}$ (see table~\ref{tab:Fsymbols}) and the $R$-symbol $R^{[1/3][1/3]}_{[\overline{1/3}]}$ (see \eqref{R33eqn}) into eq.\eqref{Bmatrixdef2} (again we assume $k$ is not divisible by 3). They are given by \begin{align}\left[B^{[1/3][1/3]}_{{\bf a}[\overline{1/3}]}\right]_{\boldsymbol\alpha}^{\boldsymbol\beta}&=w^{(\boldsymbol\alpha-\boldsymbol\beta)^T\frac{\Lambda_3}{\Lambda_3-1}{\bf a}_\circ}\delta^{i\sigma_y{\bf a}_\bullet-\Lambda_3^T\boldsymbol\beta}_{\boldsymbol\alpha}\label{Z3Bmatrix1}\\\left[B^{[1/3][1/3]}_{[1/3]{\bf a}}\right]_{\boldsymbol\alpha}^{\boldsymbol\beta}&=w^{(\boldsymbol\alpha-\boldsymbol\beta)^T\frac{\Lambda_3}{\Lambda_3-1}{\bf a}_\circ}\delta^{i\sigma_y{\bf a}_\bullet-\Lambda_3^T\boldsymbol\beta}_{\boldsymbol\alpha}\label{Z3Bmatrix2}\\\left[B^{[1/3][1/3]}_{[\overline{1/3}][1/3]}\right]_{\bf a}^{\bf b}&=\frac{1}{k^2}w^{({\bf a}_\circ-{\bf b}_\circ)^Ti\sigma_y\frac{1}{1-\Lambda_3}(\Lambda_3{\bf a}_\bullet+{\bf b}_\bullet)}\label{Z3Bmatrix3}\end{align} where $w=e^{2\pi i/k}$, and ground states are labeled by vertex degeneracies $\boldsymbol\alpha,\boldsymbol\beta$ and intermediate abelian channels ${\bf a},{\bf b}$.

In a closed system of $N=3n$ threefold defects, they fuse to the overall trivial vacuum and can be compactified on a sphere. The $k^{2(N-2)}$ ground state degeneracy is labeled by \begin{align}|{\bf a}_i;\boldsymbol\alpha_j\rangle=\left|\vcenter{\hbox{\includegraphics[width=2in]{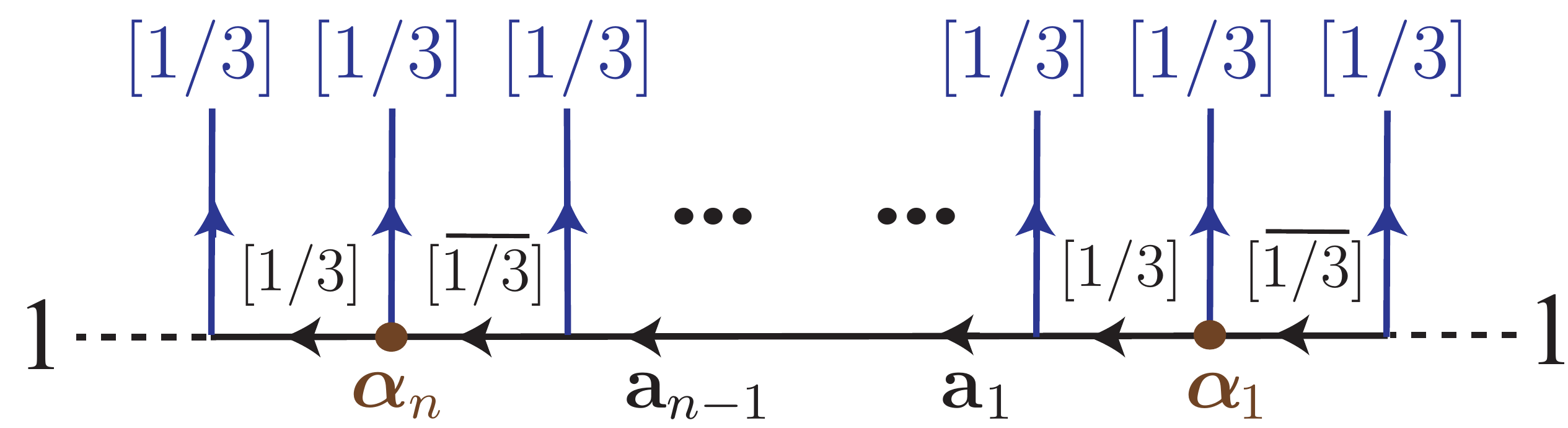}}}\right\rangle\end{align} The defects are indistinguishable as they are not equipped species labels when $k$ is not divisible by 3. The braiding operations \eqref{Z3Bmatrix1}, \eqref{Z3Bmatrix2} and \eqref{Z3Bmatrix3} generate a group $\mathcal{G}_N$ that is non-abelian when $n\geq2$. They {\em projectively} represent the braid group $\mathcal{B}_N(\mathbb{S}^2)$ of $N$ elements on a sphere by a non-trivial $\mathbb{Z}_3$ central extension. \begin{equation}1\to\mathbb{Z}_3\to G_N\to\mathcal{B}_N(\mathbb{S}^2)\to1\label{Z3centralextension}\end{equation} where the center $\mathbb{Z}_3=\langle\mathcal{M}|\mathcal{M}^3=1\rangle$ measures the violation of \eqref{braidoncearound2}.

\begin{figure}[ht]
	\includegraphics[width=2.7in]{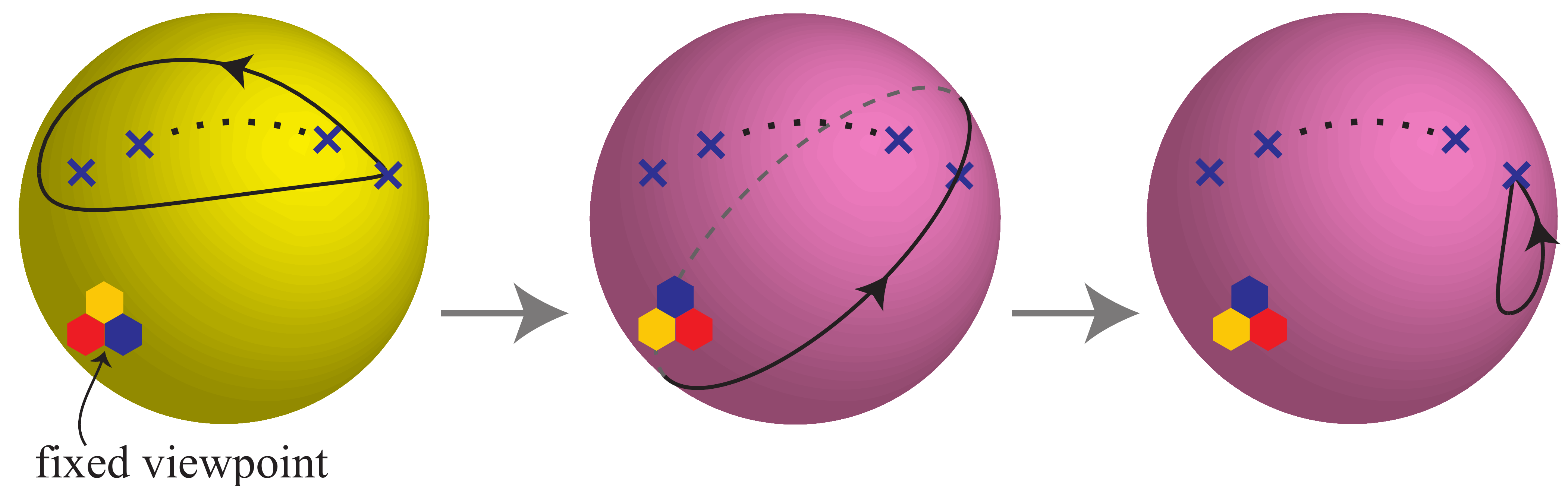}
	\caption{Contraction of the large braid $\mathcal{M}$ in \eqref{braidoncearound2} on a sphere. Color definition is cyclicly permutated after passing the braiding path across the fixed viewpoint.}\label{fig:contractsphereloop}
\end{figure}
The large braid $\mathcal{M}$ in eq.\eqref{braidoncearound} can be understood as a global cyclic color permutation on $\mathbb{S}^2$. The trajectory of the first defect leaves behind a color branch cut. As the braid, and consequently the branch cut, is contracted on the other side of the sphere, it passes through the fixed viewpoint and cyclicly changes the color definition (see figure~\ref{fig:contractsphereloop}). The large braid operation is therefore represented by the matrix \begin{equation}\langle{\bf a}_i;\boldsymbol\alpha_j|\mathcal{M}|{\bf b}_i;\boldsymbol\beta_j\rangle=\prod_{ij}\delta_{{\bf a}_i}^{\Lambda_3{\bf b}_i}\delta_{\boldsymbol\alpha_i}^{\Lambda_3^T\boldsymbol\beta_j}\end{equation} which has order 3 and commute with all braiding operations as \eqref{Z3Bmatrix1}, \eqref{Z3Bmatrix2} and \eqref{Z3Bmatrix3} are symmetric under color permutation.

We demonstrate this in the abelian case when $N=3$. The $k^2$ ground state degeneracy is labeled by the vertex degeneracy $\boldsymbol\alpha$. The group of braiding operations $\mathcal{G}_3\cong\mathbb{Z}_6$ is generated with a single element $(b^{21})_{\boldsymbol\alpha}^{\boldsymbol\beta}=(b^{32})_{\boldsymbol\alpha}^{\boldsymbol\beta}=\delta_{\boldsymbol\alpha}^{-\Lambda_3^T\boldsymbol\beta}$ from \eqref{Z3Bmatrix1} and \eqref{Z3Bmatrix2} with ${\bf a}=0$. The large braid is given by \begin{equation}\mathcal{M}=b^{21}b^{32}b^{32}b^{21}=\delta_{\boldsymbol\alpha}^{\Lambda_3^T\boldsymbol\beta}\end{equation} which has order 3. Moreover as $\mathcal{F}^{1/2}=(b^{21}b^{32})^3=1$, the \eqref{braidfull720} is satisfied.

The central extension \eqref{Z3centralextension} is a consequence of the semiclassical nature of color symmetry. Although the Hamiltonian \eqref{ham1} does not depend on color definition explicitly and there is no order parameter that explicitly breaks the color symmetry, the defect system still remembers the {\em weak} symmetry breaking~\cite{Kitaev06} by anyon labeling and this is revealed by $\mathcal{M}\neq1$.

\subsubsection{Twofold defects}
The two primitive exchange operations \eqref{Bmatrixdef1} for twofold defects can be evaluated by substituting the $F$-symbols $F^{{\bf a}[1/2][1/2]}_{\bf b}=1$, $F^{[1/2][1/2][1/2]}_{[1/2]}$ (see eq.\eqref{F222intext} and table~\ref{tab:Fsymbols}) and the $R$-symbol $R^{[1/2][1/2]}_{\bf a}$ (see \eqref{Z2exchange}) into eq.\eqref{Bmatrixdef2}. They are give by \begin{align}&B^{[1/2]_{\chi,{\bf l}_1}[1/2]_{\chi,{\bf l}_2}}_{{\bf a}{\bf b}}=w^{({\bf b}_{\circ}-{\bf a}_{\circ})^T\left[{\bf l}_1+\frac{1}{2}i\sigma_y({\bf f}^{\chi}-\Lambda_{\chi}({\bf b}_{\circ}-{\bf a}_{\circ}))\right]}\label{Z2Bmatrix1}\\&\left[B^{[1/2]_{\chi,{\bf l}_1}[1/2]_{\chi,{\bf l}_2}}_{[1/2]_{\chi,{\bf l}_3}[1/2]_{\chi,{\bf l}}}\right]_{\bf a}^{\bf b}=\frac{1}{k}w^{\frac{1}{2}{\bf q}^T\left[i\sigma_y\Lambda_{\chi}^T{\bf q}+{\bf f}^{\chi}\right]}\label{Z2Bmatrix2}\end{align} where ${\bf q}={\bf l}_1+i\sigma_y\Lambda_{\chi}({\bf a}_{\circ}-{\bf b}_{\circ})$, the abelian channels ${\bf a},{\bf b}$ label the ground state, and \eqref{Z2Bmatrix2} is evaluated by a Gaussian sum similarly encountered in the topological spin \eqref{Z2topologicalspin}. 

In a closed system with $N=2n$ twofold defect, the $k^{N-2}$ ground state degeneracy is labeled by \begin{align}|{\bf a}_i\rangle=\left|\vcenter{\hbox{\includegraphics[width=1.5in]{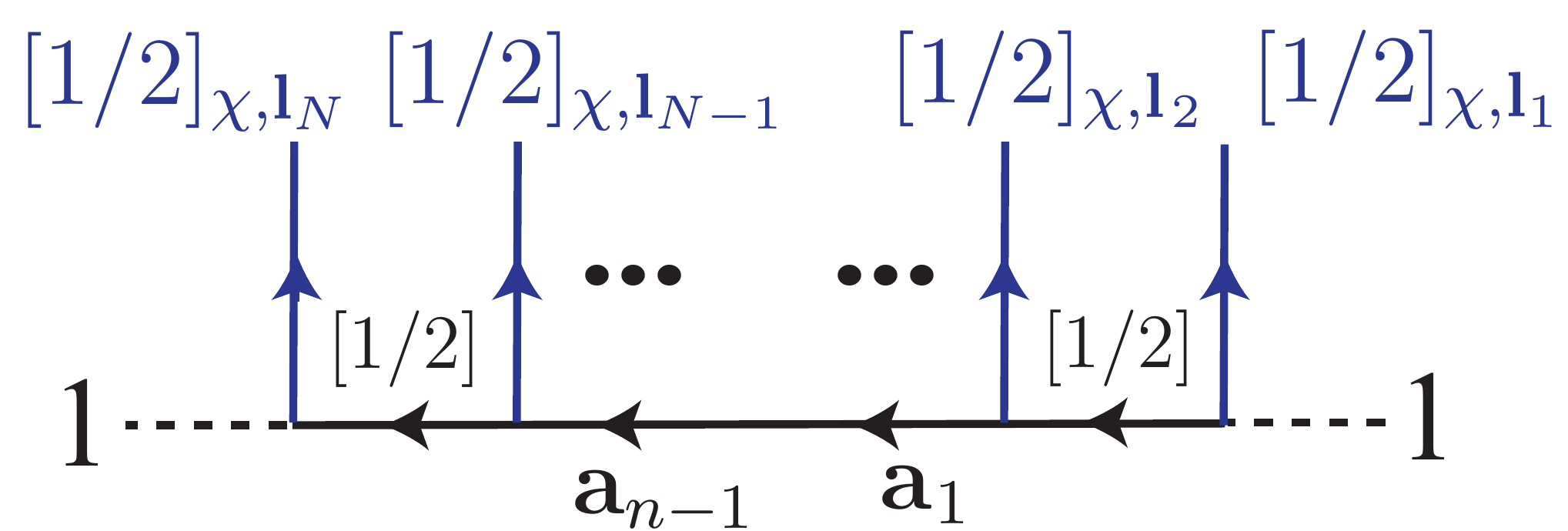}}}\right\rangle\end{align} where the abelian channels are restricted by \begin{equation}{\bf l}_{2i}+{\bf l}_{2i-1}-2{\bf l}_0(\chi)=i\sigma_y\left[({\bf a}^\bullet_{i-1}-{\bf a}^\bullet_{i})+\Lambda_\chi({\bf a}^\circ_{i-1}-{\bf a}^\circ_{i})\right]\end{equation} (or equivalently eq.\eqref{Z2fusionconstraint}) so that they are completely determined by ${\bf a}_i^\circ$, and the species labels are restricted by the compactification relation \begin{equation}\sum_{i=1}^N{\bf l}_i=N{\bf l}_0(\chi)\end{equation} (or equivalently \eqref{Z2speciescloseness}). Since defects are distinguishable by their species labels ${\bf l}_i$, a braid operation is closed only when it leaves the labels configuration $({\bf l}_N,\ldots,{\bf l}_1)$ unchanged. We assume braiding is adiabatic enough $\tau>>1/J_\ast$ so that species mutation can be ignored, where $J_\ast$ is the energy scale in the lattice defect Hamiltonian \eqref{H57} relevant to anyon tunneling between defects.

We demonstrate the simplest case $N=4$, so that braiding operations are non-abelian on the $k^2$ degenerate ground states labeled by restricted abelian channel ${\bf a}$. Let ${\bf c}_i=-i\sigma_y({\bf l}_i-{\bf l}_0(\chi))$ be the anyon string that is attached to the $i^{th}$ defect, for $i=1,2,3,4$. The large braid $\mathcal{M}$ defined in \eqref{braidoncearound} can be evaluated to have the matrix element \begin{equation}\langle{\bf a}|\mathcal{M}|{\bf b}\rangle=\theta_{[1/2]_{\chi,{\bf l}_1}}\theta_{[1/2]_{\chi,{\bf l}_3}}w^{{\bf b}_{\circ}^Ti\sigma_y({\bf c}_1+{\bf c}_4)}\delta^{\Lambda_\chi({\bf c}_3+{\bf c}_4)-{\bf b}_\circ}_{{\bf a}_\circ}\end{equation} where the spins $\theta_{[1/2]_{\chi,{\bf l}}}$ can be found in \eqref{Z2topologicalspin}.

Contrary to the threefold defect case where its order is given by that of the defect, the large braid operation for twofold defects does not square to the identity. Instead, carrying out the operation twice detects the product of the $4\pi$-spins of individual defects. \begin{equation}\mathcal{M}^2=\theta_{[1/2]_{\chi,{\bf l}_1}}\theta_{[1/2]_{\chi,{\bf l}_2}}\theta_{[1/2]_{\chi,{\bf l}_3}}\theta_{[1/2]_{\chi,{\bf l}_4}}\in\mathbb{Z}_k\end{equation} This suggests that the defect system encoding more semiclassical information than just the twofold transposition symmetry. This is related to the quasi-3D nature of the double loop $\Theta_{{\bf a}_\circ}$, whose phase depends on the ordering of the self-intersection (see appendix~\ref{sec:doubleloop}).

\section{Discussion and Speculation}\label{sec:discussion}

We conclude with some remarks on further implications and unaddressed issues. 

In this article, we have demonstrated some consequences of the ungauged symmetry of topological defects in a toy model. 
It would be interesting to see how defects behave in more realistic settings, such as (fractional) Majorana fermions at SC-FM heterostructures, especially with respect to their semiclassical nature inherited from the winding of a non-dynamical order parameter. A theoretical situation has been considered in which the order parameter disorders and the system enters a Coulomb phase, where defect excitations become quantum deconfined~\cite{FreedmanHastingsNayakQi} but leave a gapless fermionic {\em hopfion}~\cite{RanHosurVishwanath11} degree of freedom. In addition, there is recent work on general quasi-topological phases~\cite{BondersonNayak12} and their {\em metaplectic} anyonic excitations~\cite{HastingsNayakWang12} in the presence of gapless modes. Similar arguments could be made in the twist defect context by {\em gauging} the underlying $S_3$-symmetry and proliferating the twistless square and octagon defects so that the non-abelian twist defects become deconfined after a phase transition. This type of construction has been field theoretically applied to quantum Hall states~\cite{BarkeshliWen10, BarkeshliWen11, BarkeshliWen12} and other twist defect systems~\cite{MesarosKimRan13}. Using the bulk-boundary correspondence, this new topological phase by gauging should be related to a $S_3$ orbifold of the edge conformal field theory. Similar treatments were applied to fractional quantum Hall states~\cite{BarkeshliWen10} as well as symmetry protected phases~\cite{LuVishwanath13}. Gauging $S_3$-symmetry in our model would turn twist defects into non-abelian fluxes labeled by conjugacy classes and abelian anyons into non-abelian superselection sectors, thereby making fusion commutative. Presumably full modular $SL(2;\mathbb{Z})$ invariance would be restored, but the exact mechanism is unknown. With a non-simple group like $S_3$, there could be an intermediate stage in the gauging, where a non-abelian symmetric topological phase exists.

As mentioned in section~\ref{sec:lowenergyeffectivefieldtheory}, although the model we used carries the same anyon types as two copies of the $\mathbb{Z}_k$ toric code, it cannot be decomposed into their tensor product without violating the $S_3$-symmetry. The 4-component $U(1)$ effective field theory however contains a 2-component $S_3$-invariant part, which corresponds to a two-generator $S_3$-closed unsplit subgroup in the discrete gauge group $\mathbb{Z}_k^4$ and is modular (braiding non-degenerate) only when $k$ is odd. When $k$ is 2, a generalized string-net construction has been proposed by Bombin~\cite{Bombin11} which would circumvent this and carry colored fraction Majorana fermions in the form of twist defects labeled also by the symmetry group $S_3$. It would be interesting to investigate the similarities and differences in fusion and braiding between these two constructions. On the other hand, if charge-flux $\mathbb{Z}_2$-duality is the only concern, one could simply consider a single copy of the $\mathbb{Z}_k$ Kitaev toric code, where all braiding and statistics phenomena would apply~\cite{TeoRoyXiaoappearsoon} although fusion would be commutative.

A tremendous effort has already been expended on quasi-1D topological systems such as the AKLT chain~\cite{Haldanespinchain, AKLT}, anyonic quantum spin chains~\cite{FeiguinTrebstLudwigTroyerKitaevWangFreedman07, Fendley12, GilsArdonneTrebstHuseLudwigTroyerWang13}, classification of symmetry protected phases~\cite{ChenGuLiuWen11, GuWen12, ChenGuLiuWen12}, and gapped edges of fractional topological insulators~\cite{MotrukTurnerBergPollmann13, BondesanQuella13}. Numerical analysis of a defect chain could prove useful for understanding defect correlation and the two dimensional defect lattice. Twist defect ground state degeneracy could be lifted or gapped by adding non-local Wilson operators in the Hamiltonian that mimic coupling and anyon tunneling between neighboring defects. It would be interesting to investigate the transitions between different gapped phases, as well as non-local anyonic transport by a pumping process driven by the defect phase parameter discussed in section~\ref{sec:twistdefect} and appendix~\ref{sec:Fsymbols}.

Coupling of defect arrays could give rise to effective braiding without actually moving the lattice defects, which are immovable on the microscopic level without crystal distortion. This has been proposed in measurement-only topological quantum computation~\cite{BondersonFreedmanNayak08} and applied in (fractional) Majorana fermions in SC-FM heterostructures~\cite{AliceaOregRefaeOppenFisher11, LindnerBergRefaelStern, ClarkeAliceaKirill} and twist defects~\cite{BarkeshliJianQi}. Note that the crucial ingredient in the measurement-only approach is the data of $F$-symbols, that we present in section \ref{sec:Fsymbols}.

Topological entanglement entropy has proved to be a useful computational tool ~\cite{KitaevPreskill06}. There has been recent work on entanglement in twist defect systems~\cite{BrownBartlettDohertyBarrett13}. It would be very interesting to see if the modified modular invariance in section~\ref{sec:defectmodulartransformation} is revealed by ground state entanglement~\cite{ZhangGroverTurnerOshikawaVishwanath12}.

{\em Projectiveness} in braiding occurs when the absolute phase has a non-universal dynamical or geometric component, and is thus non-measurable. It has been seen in topological defect systems in 3D~\cite{FreedmanHastingsNayakQiWalkerWang}, fractional Majorana excitations~\cite{LindnerBergRefaelStern, ClarkeAliceaKirill} and twist defects~\cite{YouWen, BarkeshliJianQi}. Certain quantities are even unprojectively defined, such as the topological spin which is related to the local species label of the twist defect. An important consequence of the modified spin-statistics is the violation of the ``pair of pants" or ribbon relation that equates a full braid of two objects with a fixed fusion channel, with their respective topological spins \begin{equation}\sum_\nu\left[R^{xy}_z\right]^\nu_\mu\left[R^{yx}_z\right]^\lambda_\nu=\frac{\theta_z}{\theta_x\theta_y}\delta^\lambda_\mu.\label{ribbonrelation}\end{equation} This equation is inapplicable to semiclassical defects as topological spin is no longer identified with a $360^\circ$ twist but rather multiple twists depending on the order of the defect. An example of this violation was seen for full braiding of a pair of twofold defects in eq.\eqref{Z2fullbraiding}. The ribbon relation \eqref{ribbonrelation} may be violated even when all objects involved are spinless as illustrated by a pair of threefold defects in eq.\eqref{Z3fullbraiding}. 
%It follows that the defect braiding $S$-matrix can no longer be expressed as a weighted combination of spins, which in turn leads to the non-modular group structure of the $S$ and $T$-operations. Although there is a geometric understanding of this group as a restricted congruence subgroup of the modular group ($\Gamma_0(2)$ for twofold defects), for general $\mathbb{Z}_k$-rotor models the subgroup may not even be projectively represented and it is unclear how the algebraic structure of the braiding $S$ and $T$ is related to geometrical one. We suspect that this is a remnant of the quasi-three dimensional structure of the double loop (figure \ref{fig:defectlocalloop}) that distinguishes species. 
From the nilpotent property of $S$ and $T$-matrices for twofold defects, one might expect an analogous rigidity~\cite{Ocneanu, Kitaev06} against perturbation in a generalized defect braided fusion category.

Finally, the braiding structure of semiclassical topological defects is incomplete. 
A complete mathematical description for $S_3$ twist defects should include braiding between pairs of non-commuting objects such as $[1/3]$ and $[1/2]_\chi$, $[1/2]_\chi$ and $[1/2]_{\chi'}$ for $\chi\neq\chi'$, as well as $[{\bf a}]$ and $[1/3]$ or $[{\bf a}]$ and $[1/2]$. They do not admit braiding in the conventional sense as the object labels will change by conjugacy after a cycle as shown in section~\ref{sec:AlphabeticpresentationofWilsonalgebra}. In fact our collection of $R$-symbols cannot be treated as a self-consistent basis transformation as the intermediate step of the hexagon relation $RFR=FRF$ would involve an exchange of a pair of non-commuting objects. There is, however, no apparent obstruction to defining multiple braiding cycles such that the final incoming and outgoing labels of the objects are matched.

{\bf Acknowledgements}
We thank Eduardo Fradkin, Taylor Hughes, Shinsei Ryu and Michael Stone for very helpful advice at the ICMT. JT also thank Xiao-Liang Qi for insightful discussions at the IAS workshop at HKUST. This work was supported by the Simons Foundation (JT) and National Science Foundation through grant DMR 09-03291 (AR) and DMR-1064319 (XC).

%

%and twofold defect species label ${\bf l}=(l_Y,l_R)\in\mathbb{Z}_k\oplus\mathbb{Z}_k$ for $k$ odd or $(\mathbb{Z}+1/2)_k\oplus(\mathbb{Z}+1/2)_k$ for $k$ even transform according to \begin{align}ST:&\left({\bf s},{\bf l}\right)\to\left({\bf s},(ST)\cdot{\bf l}\right)\\S:&\left({\bf s},{\bf l}\right)\to\left({\bf s},(ST)\cdot{\bf l}\right)\end{align}

% simple example of intersection form
			% independent from branch cut but depend on base point
		% se
% ground state degeneracy for series of defects of the same type
		% Riemann-Hurwitz (appendix?)
% labeling of loops using open circle notation (alphabets), group relation (ababab=1) and intersection

%%%%%% order parameter field by locally binding twist defects
%Composite defects are treated as a single entity in the continuum by not allowing local detail measurements involving Wilson strings passing in between and distinguishing constituent defects. 

\appendix

%\section{Cohomological Algebra and Central Extension}\label{appendix1}

\section{Branch Cover description of Twist Defects}\label{appendix2}
In the continuum description, a topological defect is a pointlike object attached to a branch cut. In this appendix, we describe an alternative geometric picture for the non-local Wilson loops in terms of \emph{covering spaces}. This gives a counting argument for the defect ground state degeneracy, however fails to provide the correct Wilson algebra structure governed by the intersection matrices \eqref{Z3intersectionmatrix} and \eqref{Z2intersectionmatrix}. The covering construction is based on the {\em genon} description proposed in ref.[\onlinecite{BarkeshliQi, BarkeshliJianQi}], and is similar to the classical branch cover description of a Riemann surface. 

Anyon labeling is a multivalued function that jumps discontinuously across branch cuts. A $S_3$-covering space can be constructed with six sheets, represented by the labels $Y_{\bullet/\circ},R_{\bullet/\circ},B_{\bullet/\circ}$, sewn together at the branch cuts. On the covering space anyon labeling becomes single-valued everywhere except at the defect sites, which are analogues of ramification points. 

\begin{figure}[ht]
	\centering
	\includegraphics[width=3in]{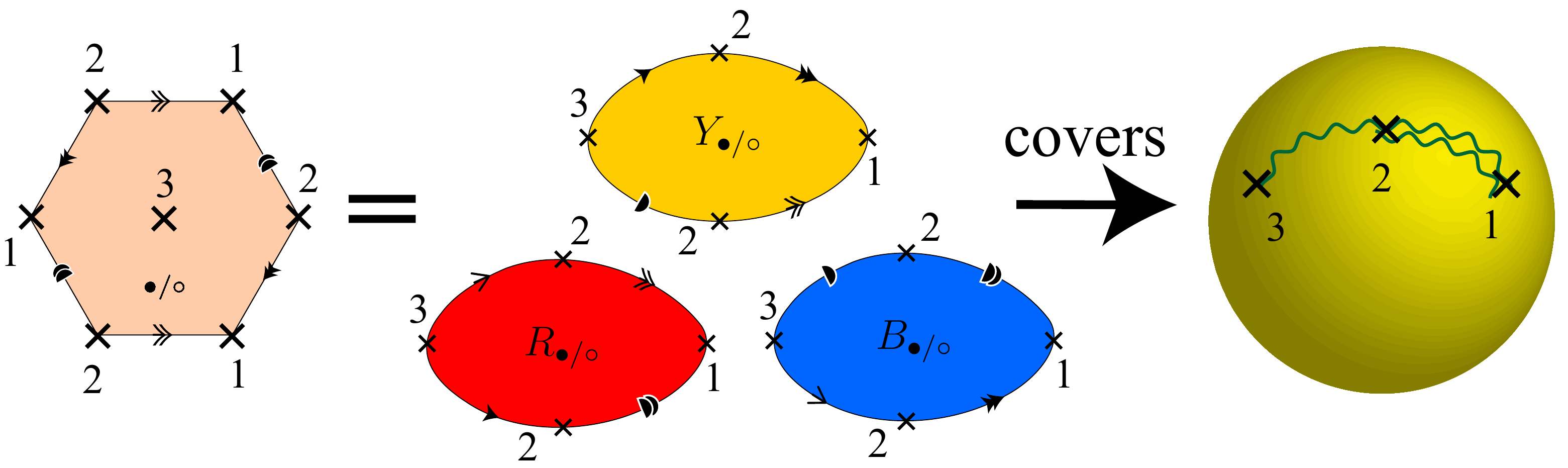}
\caption{Three threefold defects on a sphere. The covering space has six sheets labeled by color and sublattice type. They are glued by identifying similar edges. The result is a pair of disjoint tori distinguished by $\bullet,\circ$.}
\label{fig:glueingthird}
\end{figure}

We illustrate the covering construction with a couple of examples when $k$ is not divisible by 3. Firstly, on a closed sphere with three threefold defects $[1/3]$, there are two branch cuts connecting them, shown as wavy lines in figure \ref{fig:glueingthird}. The glueing convention is forced by distinguishing the cyclic permutation $\Lambda_3$ branch cut (single wavy line or single arrow) from the $\Lambda_3^2$ one (double wavy line or double arrow). The result is a disconnected pair of tori labeled by the remaining anyons $\bullet,\circ$. The ground state degeneracy (GSD) is counted by the set of maximally commuting Wilson loops, i.e. $k^2$. Although the GSD matches that of the $\mathbb{Z}_k$ Kitaev toric code, the intersection matrix \eqref{Z3intersectionmatrix} determining the defect Wilson algebraic structure is different.

A useful formula for counting the genus of the covering space $\Sigma$ over the sphere $S^2$, is the Riemann-Hurwitz theorem, \begin{align}E(\Sigma)=E(S^2)d-\sum_p(e_p-1)\label{eq:reim}\end{align} where $E$ is the Euler characteristic which in turn determines the genus $g$ for a connected component by $E=2-2g$ and $d$ is the number of sheets. $e_{p}$ is the \emph{ramification
index} of the branch point $p$ and is defined so that a neighborhood of $p$ lifts to $d-(e_{p}-1)$ copies in the cover. Ordinary points lift to $d$ copies, one in each sheet, so $e_p = 1$. A point defect ``pinches'' some of the sheets together and hence has $e_p>1$. For a set of threefold defects, one can ignore the doubling from the unmixed $\bullet$ and $\circ$ so that $d=3$ and $e_p=3$ at each defect point $p$. %For twofold defects (see below), we must take a full six sheeted cover so $d=6,e_p=4$. 
Applying \eqref{eq:reim} for $N=3n$ threefold defects, one get a $\bullet,\circ$-doublet of tori, each have Euler characteristics $E=6-2N$ and genus $g=N-2$. The ground state degeneracy is therefore $k^{2g}=k^{2(N-2)}$ which matches eq.\eqref{Z3dimensionGSD}. 

\begin{figure}[ht]
\centering
	\includegraphics[width=2.5in]{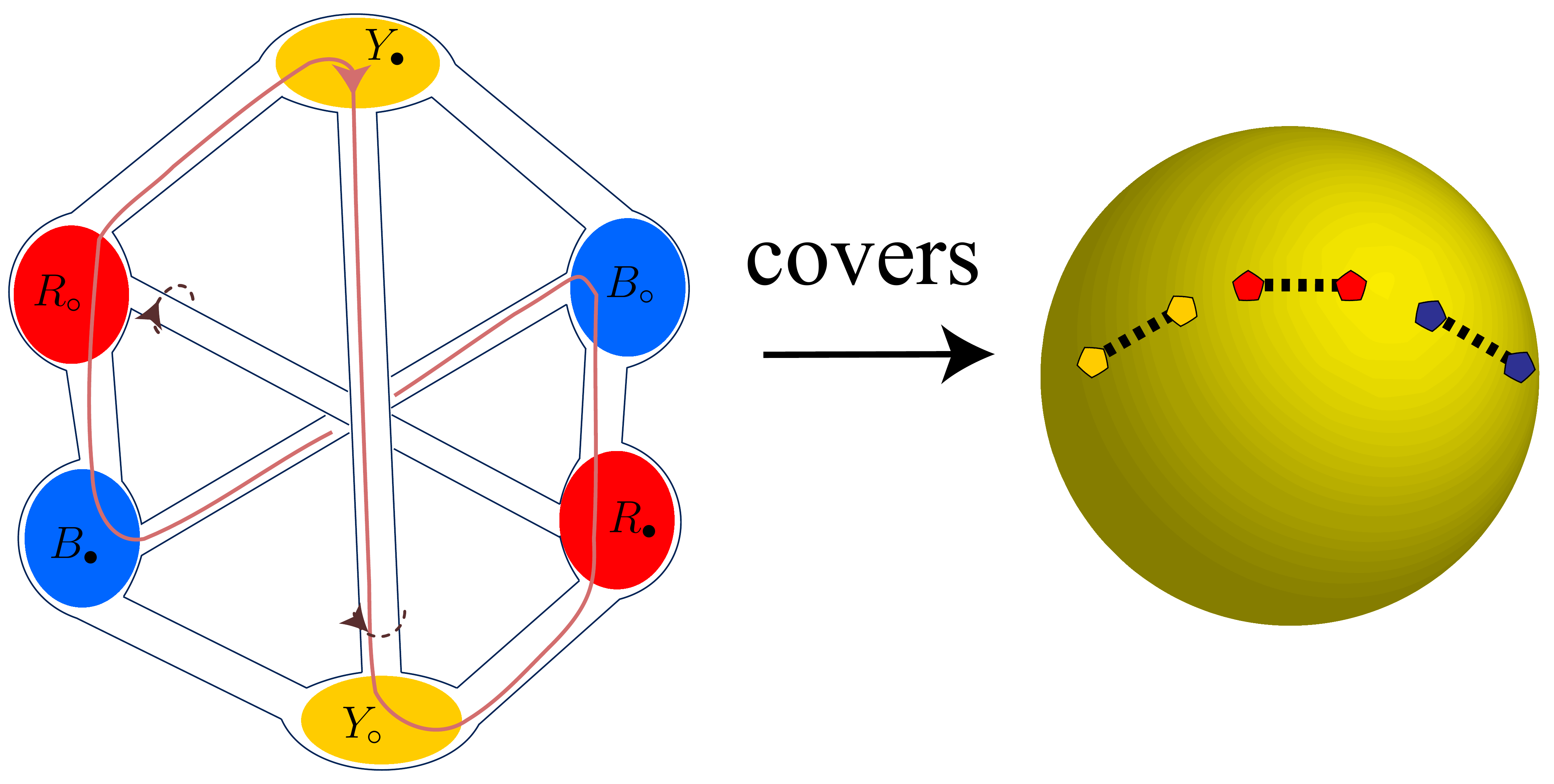}
	\caption{Six twofold defects, two of each color, on a sphere. The covering space has genus $4$. However, there are $4$ loops that are non-contractible in the cover but trivial in the base space --- the sum of the small dashed loops, the longer solid line, and 60$^\circ$ rotations of these two.}\label{fig:glueinghalf}
\end{figure}

Next we consider a set of non-commuting twofold defects, a pair each of $[1/2]_Y$,$[1/2]_R$ and $[1/2]_B$. Neither colors nor sublattice labels are fixed, and the system is lifted to a connected six-sheeted cover. Each of the six covering spheres has three connecting tubes (figure~\ref{fig:glueinghalf}), corresponding to the three brach cuts. Eq.\eqref{eq:reim} gives a genus $g=4$ cover surface which would seem to imply that there are eight independent Wilson loops. However, the key difference from the previous example is that some non-contractible 
loops in the cover are actually \emph{trivial} in the system due to the color redundancy $Y_{\bullet/\circ}\times R_{\bullet/\circ}\times B_{\bullet/\circ}=1$ and non-commutativity of the defects. By an appropriate quotient (see figure~\ref{fig:glueinghalf}), the ground state degeneracy is $k^2$.

\section{Wilson Structure of threefold defects when 3 divides \texorpdfstring{$k$}{k}}\label{sec:Z33Wilsonalgebraappendix}

This is an appendix to section~\ref{sec:Z3twistdefects}.

\begin{figure}[ht]
	\centering
	\includegraphics[width=3in]{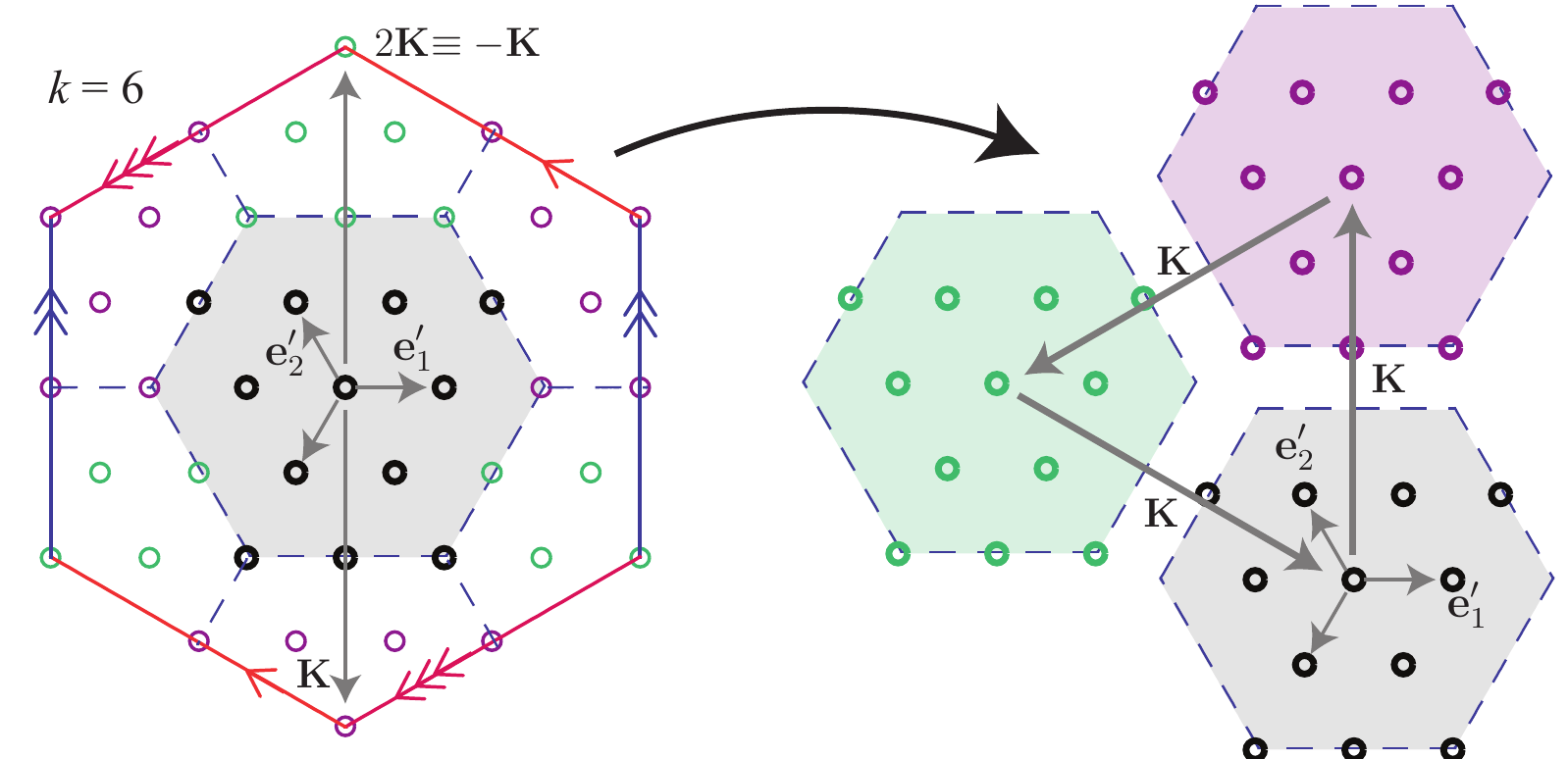}
\caption{Triangular lattice representing the integer mod $k$ powers of $\circ$-Wilson operators $\left(\mathcal{A}_{i,i+1}^{Y,\circ}\right)^{n_{2i-1}}\left(\mathcal{A}_{i,i+1}^{R,\circ}\right)^{n_{2i}}$ by lattice points ${\bf n}_i=n_{2i-1}{\bf e}'_1+n_{2i}{\bf e}'_2$. When $k$ is divisible by 3, the long lattice vector ${\bf K}$ labels the good quantum number $\Sigma^{\circ}_{i,i+1}$ in \eqref{Sigmaii+1}, and divides the lattice into three {\em bands} (colored by black, green and purple circles). $\circ$-Wilson operators on different bands related by the ${\bf K}$ vector are identical up to a $\mathbb{Z}_3$-phase determined by the eigenvalue of $\Sigma_{i,i+1}^\circ$.}\label{fig:secondBZ}
\end{figure}
Being a fixed $\mathbb{Z}_3$ phase, eq.\eqref{Sigmaii+1} constrains the eigenvalues of Wilson operators $\mathcal{A}_{i,i+1}^{Y,\bullet},\mathcal{A}_{i,i+1}^{R,\bullet}$. The allowed eigenvalues are represented by the highlighted sublattice in figure~\ref{fig:3foldsublattice}(b) with longer primitive lattice vectors ${\bf\tilde e}_1,{\bf\tilde e}_2$ and a possibly shifted origin ${\bf s}_0$ when $\Sigma_{i,i+1}^\bullet$ in \eqref{Sigmaii+1} is not 1. The conjugate operators $\mathcal{A}^\circ$ on the other hand live on a reduced triangular lattice with three {\em bands} wrapped onto itself (see figure~\ref{fig:secondBZ} and caption therein). This is analogous to the reduced Brillouin zone for a lattice system with enlarged unit cell with multiple bands wrap onto themselves, since $\mathcal{A}^\circ$ {\em translates} eigenvalues of $\mathcal{A}^\bullet$. Ground states can be enumerated by {\em momentum} ${\bf\tilde m}$ represented by black circles in the reduced triangular lattice in figure~\ref{fig:secondBZ} with Wilson operator matrix elements \begin{align}\langle{\bf\tilde m}'|\left(\mathcal{A}^\bullet\right)^{{\bf n}}|{\bf\tilde m}\rangle&=e^{i\frac{2\pi}{k}{\bf n}^T\left(\mathbb{I}{\bf\tilde m}+{\bf s_0}\right)}\delta_{{\bf\tilde m}',{\bf\tilde m}}\label{Z33Wilsonmatrix-1}\\\langle{\bf\tilde m}'|\left(\mathcal{A}^\circ\right)^{{\bf\tilde n}+{\bf K}}|{\bf\tilde m}\rangle&=\left(\Sigma^\circ\right)^{\frac{3}{k}{\bf K}}\sum_{{\bf K}'}\delta_{{\bf\tilde m}'+{\bf K}',{\bf\tilde m}+{\bf\tilde n}}\label{Z33Wilsonmatrix-2}\end{align} where ${\bf\tilde n}$ is in the reduced triangular lattice represented by black circles in figure~\ref{fig:secondBZ}, and ${\bf K}=\frac{k}{3}(-s_1,s_1,-s_2,s_2\ldots,-s_{N-2},s_{N-2})$, for $s_i=0,1,2$, are vectors annihilated by the intersection matrix $\mathbb{I}$ in \eqref{Z3intersectionmatrix} and are the long lattice vectors ${\bf K}$ in the figure~\ref{fig:secondBZ} that represent the abelian $\mathbb{Z}_3$-phase \begin{equation}\left(\Sigma^\circ\right)^{\frac{3}{k}{\bf K}}=\prod_{i=1}^{N-2}\left(\Sigma_{i,i+1}^\circ\right)^{s_i}\in\mathbb{Z}_3\end{equation} and ${\bf s_0}$ is the shifted origin of figure~\ref{fig:3foldsublattice}(b) fixed by the eigenvalues of $\Sigma^\bullet$ so that \begin{equation}\left(\Sigma^\bullet\right)^{\frac{3}{k}{\bf K}}=\prod_{i=1}^{N-2}\left(\Sigma_{i,i+1}^\bullet\right)^{s_i}=e^{i\frac{2\pi}{k}{\bf K}^T{\bf s}_0}\in\mathbb{Z}_3\label{shiftvectora0}\end{equation} Or one can choose a basis according to the allowed eigenvalues $\widetilde{\boldsymbol\alpha}$ for $\bullet$-Wilson operators on the highlighted sublattice in figure~\ref{fig:3foldsublattice} and the matrix elements for the Wilson operators are \begin{align}\langle\widetilde{\boldsymbol\alpha}'|\left(\mathcal{A}^\bullet\right)^{{\bf n}}|\widetilde{\boldsymbol\alpha}\rangle&=e^{i\frac{2\pi}{k}{\bf n}^T\widetilde{\boldsymbol\alpha}}\delta_{\widetilde{\boldsymbol\alpha}',\widetilde{\boldsymbol\alpha}}\label{Z33Wilsonmatrix1}\\\langle\widetilde{\boldsymbol\alpha}'|\left(\mathcal{A}^\circ\right)^{{\bf\tilde n}+{\bf K}}|\widetilde{\boldsymbol\alpha}\rangle&=\left(\Sigma^\circ\right)^{\frac{3}{k}{\bf K}}\delta_{\widetilde{\boldsymbol\alpha}',\widetilde{\boldsymbol\alpha}+\mathbb{I}{\bf\tilde n}}\label{Z33Wilsonmatrix2}\end{align} The eigenstates $|\widetilde{\boldsymbol\alpha}\rangle$ in \eqref{Z33Wilsonmatrix1}, \eqref{Z33Wilsonmatrix2} can be equated with ground states $|{\bf\tilde m}\rangle$ in \eqref{Z33Wilsonmatrix-1}, \eqref{Z33Wilsonmatrix-2} by the identification $\widetilde{\boldsymbol\alpha}=\mathbb{I}{\bf\tilde m}+{\bf s}_0$.

\section{List of \texorpdfstring{$F$}{F}-symbols}\label{sec:Fsymbols}
  
Here we present the complete list (up to $S_3$-symmetry) of $F$-symbols that consistently generate basis transformations of defect ground states. They are computed with the gauge degree of freedom fixed under the basis choice of splitting spaces in figure~\ref{fig:splittingspaces}. For simplicity, we assume the discrete gauge group \begin{equation}\mathbb{Z}_k=\{w^n:w=e^{i\frac{2\pi}{k}},n\in\mathbb{Z}\}\end{equation} never has 3 torsions, i.e. $k$ is not divisible by 3. 

Objects at the four external and two intermediate channels include 
\begin{enumerate}
	\item abelian anyons labeled by ${\bf a},{\bf b},{\bf c},{\bf d}$, where it can be decomposed into ${\bf a}=({\bf a}_\bullet,{\bf a}_\circ)$ so that ${\bf a}_{\bullet}=(y_1,r_1)$ and ${\bf a}_{\circ}=(y_2,r_2)$ are 2-dimensional integer mod $k$ vectors living on the anyon lattice (figure~\ref{fig:abeliananyonlattice});
	\item threefold defects $[1/3]$ and its anti-particle $[\overline{1/3}]$;
	\item twofold defects $[1/2]_{\chi,{\bf l}}$ with color $\chi=Y,R,B=0,1,2$ (mod $3$) and species label ${\bf l}=(l_Y,l_R)$, where $l_{Y/R}$ are integers (half-integers) mod $k$ if $Y/R=\chi$ (resp. $Y/R\neq\chi$).
\end{enumerate}
Splitting degeneracies at vertices $[1/3]\times[1/3]=k^2[\overline{1/3}]$ are labeled by ${\boldsymbol\alpha}$, ${\boldsymbol\beta}$ (or at $[\overline{1/3}]\times[\overline{1/3}]=k^2[1/3]$ by $\overline{\boldsymbol\alpha}$, $\overline{\boldsymbol\beta}$). These are eigenvalues $\boldsymbol\alpha=I_0{\bf m}$ (or $\overline{\boldsymbol\alpha}=\bar{I}_0\overline{\bf m}$ for anti-defects) of the Wilson operator $\mathcal{A}_{\bullet}^{\bf n}=e^{i{\bf n}^T\boldsymbol\alpha}$ for the $k^2$ splitting states $|{\bf m}\rangle=\mathcal{A}_{\circ}^{{\bf m}}|GS\rangle_0$ of figure~\ref{fig:splittingspaces}(d), where $\boldsymbol\alpha=(\alpha_1,\alpha_2)$ is a 2-dimensional integer mod $k$ vector living on the triangular lattice in figure~\ref{fig:3foldsublattice}, $I_0$ is the intersection matrix defined in eq.\eqref{Z3intersectionmatrix} for $[1/3]$-defects (or $\bar{I}_0=-I_0$ for $[\overline{1/3}]$).

We here summarize notations appear in the $F$-symbols. These include the cyclic color permutation for the threefold defect $[1/3]$
\begin{equation}\Lambda_3=\begin{pmatrix}0&-1\\1&-1\end{pmatrix}\end{equation} The matrix $1-\Lambda_3$ (1 being the $2\times2$ identity matrix) is invertible only when $k$ is not divisible by 3. All $F$-matrices in table~\ref{tab:Fsymbols} that involve threefold defects $[1/3],[\overline{1/3}]$ need to be modified in the case when $3$ divides $k$ and are not computed in this article. Transpositions for the respective twofold defects $[1/2]_Y,[1/2]_R,[1/2]_B$ are given by \begin{equation}\Lambda_Y=\begin{pmatrix}1&-1\\0&-1\end{pmatrix},\quad\Lambda_R=\begin{pmatrix}-1&0\\-1&1\end{pmatrix},\quad\Lambda_B=\begin{pmatrix}0&1\\1&0\end{pmatrix}\label{lambdachiapp}\end{equation} ${\bf f}^{Y}=(1,0)$, ${\bf f}^{R}=(0,1)$, ${\bf f}^{B}=(-1,-1)$ are primitive vectors in the triangular anyon lattice figure~\ref{fig:abeliananyonlattice}, $\sigma_y$ is the $2\times2$ Pauli matrix, and \begin{equation}{\bf l}_0(\chi)=-i\sigma_y\frac{k+1}{2}{\bf f}^{\chi}\end{equation} is the self-reciprocal species label for a bare twofold defect.

The color and species label $\chi,{\bf l}$ for twofold defects are suppressed whenever they do not affect the $F$-symbol. An intermediate or overall abelian channel ${\bf a}=({\bf a}_\bullet,{\bf a}_\circ)$ is restricted by \begin{equation}{\bf l}_1+{\bf l}_2-2{\bf l}_0(\chi)=i\sigma_y({\bf a}_\bullet+\Lambda_\chi{\bf a}_\circ)\end{equation} if it splits into a pair of twofold defects $[1/2]_{\chi,{\bf l}_1},[1/2]_{\chi,{\bf l}_1}$. And similarly by particle-antiparticle duality, an intermediate or overall $[1/2]_{\chi,{\bf l}'}$ channel has its species label determined by \begin{equation}{\bf l}'={\bf l}+i\sigma_y({\bf a}_\bullet+\Lambda_\chi{\bf a}_\circ)\end{equation} if it splits into an abelian anyon ${\bf a}$ and a twofold defect $[1/2]_{\chi,{\bf l}}$. The corresponding delta functions because of these implicit restrictions are suppressed in the table.
 
$F$-symbols that are related by $S_3$-symmetry (for the transformation rules see section~\ref{sec:AlphabeticpresentationofWilsonalgebra}) are only shown once in table~\ref{tab:Fsymbols}. For example, $F_{[\overline{1/3}]}^{{\bf a}[\overline{1/3}]{\bf b}}$ can be obtained by replacing $\Lambda_3$ by $\Lambda_3^{-1}$ in $F_{[1/3]}^{{\bf a}[1/3]{\bf b}}$.

\begin{figure}[ht]
	\includegraphics[width=3in]{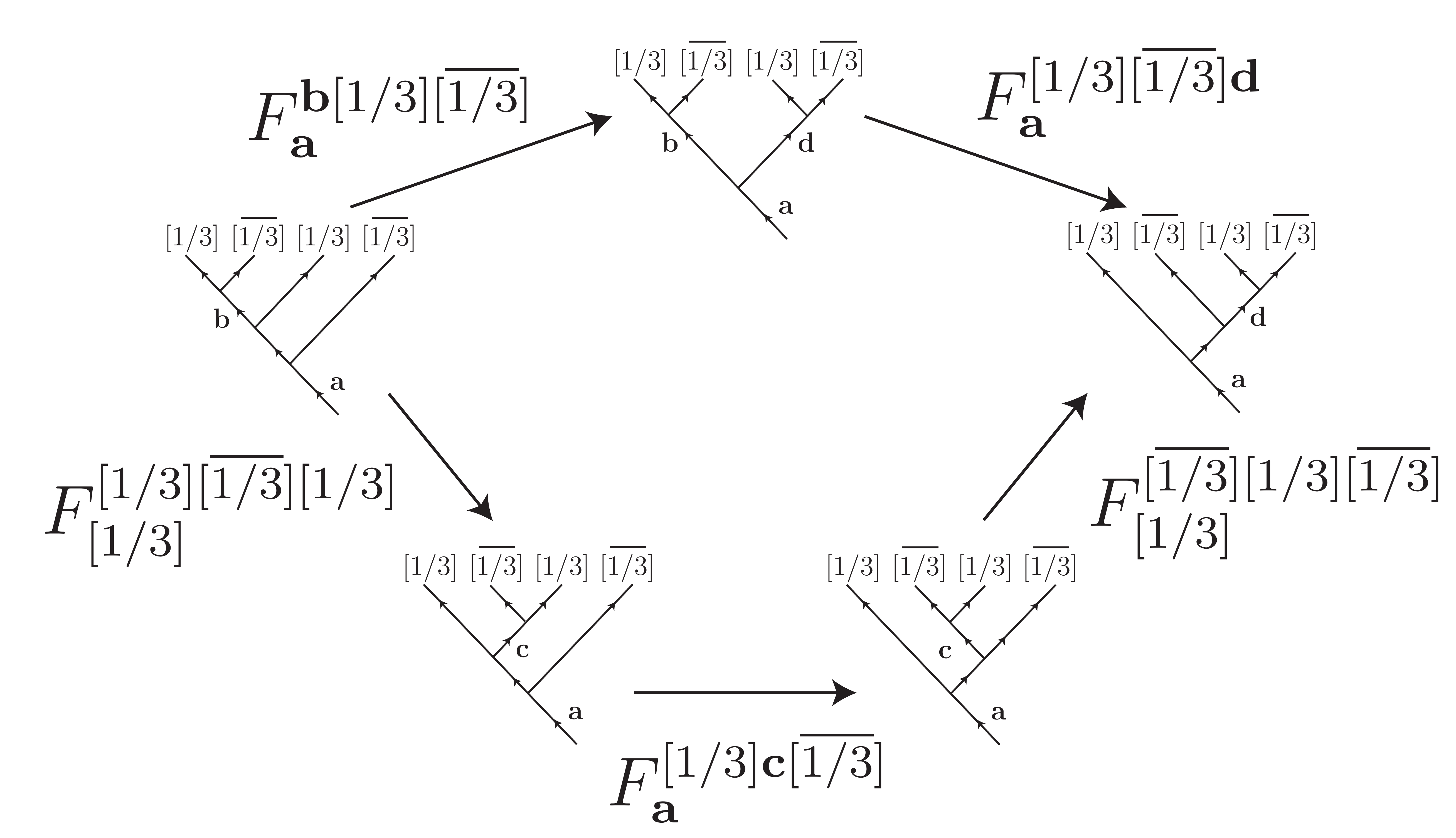}
	\caption{A pentagon identity.}\label{fig:Fpentagon}
\end{figure}
The $F$-symbols satisfy the pentagon identity so that they consistently generate basis transformations that are independent from intermediate steps. For illustration purpose, we demonstrate an example for the fusion $[1/3]\times[\overline{1/3}]\times[1/3]\times[\overline{1/3}]$. Diagram~\ref{fig:Fpentagon} commutes so that the basis transformation is path independent. By substituting the appropriate $F$-symbols in table~\ref{tab:Fsymbols}, we have \begin{align}&F^{{\bf b}[1/3][\overline{1/3}]}_{{\bf a}}F^{[1/3][\overline{1/3}]{\bf d}}_{{\bf a}}\nonumber\\&=\sum_{\bf c}\left[F^{[1/3][\overline{1/3}][1/3]}_{[1/3]}\right]_{\bf b}^{\bf c}F^{[1/3]{\bf c}[\overline{1/3}]}_{{\bf a}}\left[F^{[\overline{1/3}][1/3][\overline{1/3}]}_{[1/3]}\right]_{\bf c}^{\bf d}\nonumber\\&=w^{{\bf d}_\circ^Ti\sigma_y\frac{\Lambda_3}{1-\Lambda_3}{\bf d}_\bullet}\delta_{\bf a}^{{\bf b}+{\bf d}}\end{align}

\begin{longtable*}{ll}
\multicolumn{2}{c}{$F$-matrices}\\\noalign{\smallskip}
\noalign{\smallskip}\hline\hline\\\noalign{\smallskip}
$F^{{\bf a}{\bf b}{\bf c}}_{\bf d}$	
		& $1$ \\\noalign{\smallskip}
$F^{[1/3]{\bf a}{\bf b}}_{[1/3]}$       
		& $w^{{\bf a}_\circ^T i\sigma_y\frac{1}{\Lambda_3-1}{\bf b}_\bullet}$ \\\noalign{\smallskip}
$F^{{\bf a}{\bf b}[1/3]}_{[1/3]}$       
		& $w^{{\bf b}_\circ^T i\sigma_y\frac{\Lambda_3}{1-\Lambda_3}{\bf a}_\bullet}$ \\\noalign{\smallskip}
$F^{{\bf a}[1/3]{\bf b}}_{[1/3]}$       
		& $w^{{\bf a}_\circ^T i\sigma_y\frac{\Lambda_3}{\Lambda_3-1}{\bf b}_\bullet+{\bf b}_\circ^T i\sigma_y\frac{1}{1-\Lambda_3}{\bf a}_\bullet}$ \\\noalign{\smallskip}
$F^{[1/2]{\bf a}{\bf b}}_{[1/2]}$,  $F^{{\bf a}{\bf b}[1/2]}_{[1/2]}$      
		& $1$ \\\noalign{\smallskip}
$F^{{\bf a}[1/2]_\chi{\bf b}}_{[1/2]_\chi}$       
		& $w^{-{\bf a}_\circ^Ti\sigma_y\Lambda_\chi{\bf b}_\circ}$ \\\noalign{\smallskip}
$\left[F^{[1/3][1/3]{\bf a}}_{[\overline{1/3}]}\right]_{\boldsymbol\alpha}^{\boldsymbol\beta}$      
		& $w^{{\boldsymbol\beta}^T\frac{(\Lambda_3)^2}{\Lambda_3-1}{\bf a}_\circ}\delta_{\boldsymbol\alpha+i\sigma_y{\bf a}_\bullet}^{\boldsymbol\beta}$  \\\noalign{\smallskip}
$\left[F^{{\bf a}[1/3][1/3]}_{[\overline{1/3}]}\right]_{\boldsymbol\alpha}^{\boldsymbol\beta}$      
		& $w^{\boldsymbol\alpha^T\frac{\Lambda_3}{\Lambda_3-1}{\bf a}_\circ}\delta_{\boldsymbol\alpha+i\sigma_y(\Lambda_3)^2{\bf a}_\bullet}^{\boldsymbol\beta}$ \\\noalign{\smallskip}
$\left[F^{[1/3]{\bf a}[1/3]}_{[\overline{1/3}]}\right]_{\boldsymbol\alpha}^{\boldsymbol\beta}$      
		& $w^{\boldsymbol\alpha^T\frac{1}{1-\Lambda_3}{\bf a}_\circ+{\bf a}_\circ^Ti\sigma_y\frac{\Lambda_3}{\Lambda_3-1}{\bf a}_\bullet}\delta^{\boldsymbol\beta+i\sigma_y\Lambda_3{\bf a}_\bullet}_{\boldsymbol\alpha}$ \\\noalign{\smallskip}
$F^{[1/3][\overline{1/3}]{\bf a}}_{\bf b}$         
		& $w^{{\bf a}_\circ^Ti\sigma_y\frac{\Lambda_3}{1-\Lambda_3}{\bf b}_\bullet}$ \\\noalign{\smallskip}
$F^{{\bf a}[1/3][\overline{1/3}]}_{\bf b}$         
		& $w^{({\bf a}_\circ-{\bf b}_\circ)^Ti\sigma_y\frac{\Lambda_3}{1-\Lambda_3}{\bf a}_\bullet}$ \\\noalign{\smallskip}
$F^{[1/3]{\bf a}[\overline{1/3}]}_{\bf b}$         
		& $w^{{\bf b}_\circ^Ti\sigma_y\frac{\Lambda_3}{\Lambda_3-1}{\bf a}_\bullet+{\bf a}_\circ^Ti\sigma_y\frac{1}{1-\Lambda_3}{\bf b}_\bullet+{\bf a}_\circ^Ti\sigma_y\frac{1}{1-\Lambda_3}{\bf a}_\bullet}$ \\\noalign{\smallskip}
$F^{[1/2]_{\chi}[1/2]_{\chi}{\bf a}}_{{\bf b}}$, $F^{{\bf a}[1/2]_{\chi}[1/2]_{\chi}}_{{\bf b}}$                
		& $1$ \\\noalign{\smallskip}
$F^{[1/2]_{\chi}{\bf a}[1/2]_{\chi}}_{{\bf b}}$               
		& $w^{-{\bf a}_\circ^Ti\sigma_y\Lambda_\chi{\bf b}_\circ}$ \\\noalign{\smallskip}
$F^{[1/2]_{\chi-1,{\bf l}_2}[1/2]_{\chi,{\bf l}_1}{\bf a}}_{[1/3]}$, $F^{{\bf a}[1/2]_{\chi,{\bf l}_2}[1/2]_{\chi-1,{\bf l}_1}}_{[\overline{1/3}]}$ 
		& $w^{{\bf a}_\circ^Ti\sigma_y\frac{1}{1-\Lambda_3}\left[{\bf a}_\bullet-i\sigma_y({\bf l}_1+{\bf l}_2)+\frac{1}{2}(\Lambda_{\chi}{\bf a}_\circ-{\bf f}^{\chi+1})\right]}$ \\\noalign{\smallskip}
$F^{[1/2]_{\chi+1,{\bf l}_2}[1/2]_{\chi,{\bf l}_1}{\bf a}}_{[\overline{1/3}]}$, $F^{{\bf a}[1/2]_{\chi,{\bf l}_2}[1/2]_{\chi+1,{\bf l}_1}}_{[1/3]}$
		& $w^{{\bf a}_\circ^Ti\sigma_y\frac{\Lambda_3}{\Lambda_3-1}\left[{\bf a}_\bullet-i\sigma_y({\bf l}_1+{\bf l}_2)+\frac{1}{2}(\Lambda_{\chi}{\bf a}_\circ-{\bf f}^{\chi-1})\right]}$ \\\noalign{\smallskip}   
$F^{[1/2]_{\chi}{\bf a}[1/2]_{\chi+1}}_{[1/3]}$, $F^{[1/2]_{\chi}{\bf a}[1/2]_{\chi-1}}_{[\overline{1/3}]}$
		& $1$ \\\noalign{\smallskip}
$F^{{\bf a}[1/2]_\chi[1/3]}_{[1/2]_{\chi+1}}$, $F^{[1/3][1/2]_\chi{\bf a}}_{[1/2]_{\chi-1}}$
		& $1$ \\\noalign{\smallskip}
$F^{[1/2]_{\chi,{\bf l}}[1/3]{\bf a}}_{[1/2]_{\chi+1,{\bf l}'}}$, $F^{{\bf a}[\overline{1/3}][1/2]_{\chi,{\bf l}}}_{[1/2]_{\chi+1,{\bf l}'}}$
		& $w^{{\bf a}_\circ^Ti\sigma_y\frac{1}{\Lambda_3-1}\left[i\sigma_y({\bf l}-{\bf l}')+\frac{1}{2}({\bf f}^{\chi+1}-{\bf f}^{\chi}-\Lambda_{\chi+1}{\bf a}_\circ)\right]}$ \\\noalign{\smallskip}
$F^{[1/2]_{\chi,{\bf l}}[\overline{1/3}]{\bf a}}_{[1/2]_{\chi-1,{\bf l}'}}$, $F^{{\bf a}[1/3][1/2]_{\chi,{\bf l}}}_{[1/2]_{\chi-1,{\bf l}'}}$
		& $w^{{\bf a}_\circ^Ti\sigma_y\frac{\Lambda_3}{1-\Lambda_3}\left[i\sigma_y({\bf l}-{\bf l}')+\frac{1}{2}({\bf f}^{\chi-1}-{\bf f}^{\chi}-\Lambda_{\chi-1}{\bf a}_\circ)\right]}$ \\\noalign{\smallskip}
$F^{[1/2]_{\chi,{\bf l}}{\bf a}[1/3]}_{[1/2]_{\chi+1,{\bf l}'}}$, $F^{[\overline{1/3}]{\bf a}[1/2]_{\chi,{\bf l}}}_{[1/2]_{\chi+1,{\bf l}'}}$
		& $w^{{\bf a}_\circ^Ti\sigma_y\frac{\Lambda_3}{\Lambda_3-1}\left[i\sigma_y({\bf l}-{\bf l}')+\frac{1}{2}({\bf f}^{\chi+1}-{\bf f}^{\chi}-\Lambda_{\chi}{\bf a}_\circ)\right]}$ \\\noalign{\smallskip}
$F^{[1/2]_{\chi,{\bf l}}{\bf a}[\overline{1/3}]}_{[1/2]_{\chi-1,{\bf l}'}}$, $F^{[1/3]{\bf a}[1/2]_{\chi,{\bf l}}}_{[1/2]_{\chi-1,{\bf l}'}}$
		& $w^{{\bf a}_\circ^Ti\sigma_y\frac{1}{1-\Lambda_3}\left[i\sigma_y({\bf l}-{\bf l}')+\frac{1}{2}({\bf f}^{\chi-1}-{\bf f}^{\chi}-\Lambda_{\chi}{\bf a}_\circ)\right]}$ \\\noalign{\smallskip}
$\left[F^{[1/3][1/3][1/3]}_{{\bf a}}\right]_{\boldsymbol\alpha}^{\boldsymbol\beta}$
		& $w^{\boldsymbol\alpha^T\frac{\Lambda_3}{\Lambda_3-1}{\bf a}_\circ}\delta^{\boldsymbol\beta}_{(\Lambda_3^2)^T\boldsymbol\alpha+i\sigma_y{\bf a}_\bullet}$ \\\noalign{\smallskip}
$\left[F^{[\overline{1/3}][1/3][1/3]}_{[1/3]}\right]^{\overline{\boldsymbol\alpha}\boldsymbol\alpha}_{\bf a}$
		& $\frac{1}{k}w^{\overline{\boldsymbol\alpha}^T\frac{1}{\Lambda_3-1}{\bf a}_\circ}\delta^{\boldsymbol\alpha-\overline{\boldsymbol\alpha}}_{i\sigma_y\Lambda_3{\bf a}_\bullet}$\\\noalign{\smallskip}
$\left[F^{[1/3][1/3][\overline{1/3}]}_{[1/3]}\right]_{\boldsymbol\alpha\overline{\boldsymbol\alpha}}^{\bf a}$
		& $\frac{1}{k}w^{(\boldsymbol\alpha+\overline{\boldsymbol\alpha})^T\frac{\Lambda_3}{1-\Lambda_3}{\bf a}_\circ}\delta_{\boldsymbol\alpha-(\Lambda_3^2)^T\overline{\boldsymbol\alpha}}^{-i\sigma_y\Lambda_3{\bf a}_\bullet}$ \\\noalign{\smallskip}
$\left[F^{[1/3][\overline{1/3}][1/3]}_{[1/3]}\right]^{\bf b}_{\bf a}$
		& $\frac{1}{k^2}w^{{\bf b}_\circ^Ti\sigma_y\frac{1}{\Lambda_3-1}({\bf a}_\bullet+{\bf b}_\bullet)+{\bf a}_\circ^Ti\sigma_y\frac{\Lambda_3}{1-\Lambda_3}{\bf b}_\bullet}$ \\\noalign{\smallskip}
$\left[F^{[1/2]_{\chi}[1/2]_{\chi}[1/2]_{\chi}}_{[1/2]_{\chi}}\right]_{\bf a}^{\bf b}$
		& $\frac{1}{k}w^{{\bf a}_\circ^Ti\sigma_y\Lambda_\chi{\bf b_\circ}}$ \\\noalign{\smallskip}
$F^{[1/2]_{\chi\pm1}[1/2]_{\chi}[1/2]_{\chi}}_{[1/2]_{\chi\pm1}}$, $F^{[1/2]_{\chi}[1/2]_{\chi}[1/2]_{\chi\pm1}}_{[1/2]_{\chi\pm1}}$
		& $\frac{1}{k}$ \\\noalign{\smallskip}
\multirow{2}{*}{$F^{[1/2]_{\chi\pm1,{\bf l}_3}[1/2]_{\chi,{\bf l}_2}[1/2]_{\chi\pm1,{\bf l}_1}}_{[1/2]_{\chi\mp1,{\bf l}}}$}
		& $w^{-{\bf p}^Ti\sigma_y\left[\frac{k}{2}{\bf f}^{\chi}+\frac{1}{2}\Lambda_{\chi}{\bf p}\right]}$ \\\noalign{\smallskip}
		& $\quad\quad{\scriptstyle {\bf p}=\frac{1}{\Lambda_{\chi\pm1}-\Lambda_{\chi}}i\sigma_y({\bf l}_1+{\bf l}_2+{\bf l}_3-{\bf l}-3{\bf l}_0(\chi\pm1)-2{\bf l}_0(\chi))}$ \\\noalign{\smallskip}
\multirow{2}{*}{$F^{[1/2]_{\chi\mp1,{\bf l}_3}[1/2]_{\chi\pm1,{\bf l}_2}[1/2]_{\chi,{\bf l}_1}}_{[1/2]_{\chi\pm1,{\bf l}}}$}
		& $w^{{\bf p}^Ti\sigma_y\left[\frac{k}{2}{\bf f}^{\chi}+\frac{1}{2}\Lambda_{\chi}{\bf p}\right]}$ \\\noalign{\smallskip}
		& $\quad\quad{\scriptstyle {\bf p}=\frac{1}{\Lambda_{\chi\pm1}-\Lambda_{\chi}}i\sigma_y({\bf l}_1+{\bf l}_2+{\bf l}_3-{\bf l}+{\bf l}_0(\chi\pm1))}$ \\\noalign{\smallskip}
$\left[F^{[1/2]_{\chi,{\bf l}_2}[1/2]_{\chi,{\bf l}_1}[1/3]}_{[1/3]}\right]_{\bf a}^{[1/2]_{\chi+1,{\bf l}}}$
		& $\frac{1}{k}w^{{\bf a}_\circ^Ti\sigma_y\frac{\Lambda_3}{\Lambda_3-1}\left[i\sigma_y({\bf l}_1+{\bf l}_2)+\frac{1}{2}({\bf f}^{\chi-1}+\Lambda_\chi{\bf a}_\circ)\right]}$ \\\noalign{\smallskip}
$\left[F^{[1/3][1/2]_{\chi,{\bf l}_2}[1/2]_{\chi,{\bf l}_1}}_{[1/3]}\right]^{\bf a}_{[1/2]_{\chi-1,{\bf l}}}$
		& $\frac{1}{k}w^{{\bf a}_\circ^Ti\sigma_y\frac{1}{1-\Lambda_3}\left[i\sigma_y({\bf l}_1+{\bf l}_2)+\frac{1}{2}({\bf f}^{\chi+1}+\Lambda_\chi{\bf a}_\circ)\right]}$ \\\noalign{\smallskip}
\multirow{2}{*}{$\left[F^{[1/2]_{\chi,{\bf l}_2}[1/3][1/2]_{\chi,{\bf l}_1}}_{[\overline{1/3}]}\right]^{[1/2]_{\chi-1,{\bf l}'}}_{[1/2]_{\chi+1,{\bf l}}}$}
		& $\frac{1}{k}w^{-{\bf p}^Ti\sigma_y\left[\frac{k}{2}{\bf f}^{\chi-1}+\frac{1}{2}\Lambda_{\chi-1}{\bf p}\right]}$ \\\noalign{\smallskip}
		& $\quad\quad{\scriptstyle {\bf p}=\frac{\Lambda_3}{\Lambda_3-1}i\sigma_y({\bf l}-{\bf l}'+{\bf l}_1-{\bf l}_2+{\bf l}_0(\chi-1)-{\bf l}_0(\chi+1))}$ \\\noalign{\smallskip}
$\left[F^{[1/2]_{\chi+1,{\bf l}_2}[1/2]_{\chi,{\bf l}_1}[1/3]}_{\bf a}\right]^{[1/2]_{\chi+1,{\bf l}}}_{[\overline{1/3}]}$
		& $\frac{1}{k^2}w^{{\bf a}_\circ^Ti\sigma_y\frac{1}{\Lambda_3-1}\left[i\sigma_y({\bf l}_1+{\bf l}_2)+\frac{1}{2}(\Lambda_{\chi+1}{\bf a}_\circ+{\bf f}^{\chi-1})\right]}$ \\\noalign{\smallskip}
$\left[F^{[1/3][1/2]_{\chi,{\bf l}_2}[1/2]_{\chi-1,{\bf l}_1}}_{\bf a}\right]_{[1/2]_{\chi-1,{\bf l}}}^{[\overline{1/3}]}$
		& $\frac{1}{k^2}w^{{\bf a}_\circ^Ti\sigma_y\frac{\Lambda_3}{1-\Lambda_3}\left[{\bf a}_\bullet+i\sigma_y({\bf l}_1+{\bf l}_2)+\frac{1}{2}(\Lambda_{\chi-1}{\bf a}_\circ+{\bf f}^{\chi+1})\right]}$ \\\noalign{\smallskip}
$F^{[1/2]_\chi[1/3][1/2]_{\chi+1}}_{\bf a}$
		& $\frac{1}{k^2}$ \\\noalign{\smallskip}
$\left[F^{[1/2]_{\chi-1,{\bf l}_2}[1/2]_{\chi,{\bf l}_1}[1/3]}_{[\overline{1/3}]}\right]^{[1/2]_{\chi+1,{\bf l}}}_{[1/3],\boldsymbol\alpha}$
		& $\frac{1}{k}\delta^{(\Lambda_3^2)^T{\bf l}+\Lambda_3^T{\bf l}_1-{\bf l}_2-{\bf l}_0(\chi-1)}_{\boldsymbol\alpha}$ \\\noalign{\smallskip}
$\left[F^{[1/3][1/2]_{\chi,{\bf l}_2}[1/2]_{\chi+1,{\bf l}_1}}_{[\overline{1/3}]}\right]_{[1/2]_{\chi-1,{\bf l}}}^{[1/3],\boldsymbol\alpha}$
		& $\frac{1}{k}\delta_{-\Lambda_3^T{\bf l}+{\bf l}_1-(\Lambda_3^2)^T{\bf l}_2+{\bf l}_0(\chi+1)}^{\boldsymbol\alpha}$ \\\noalign{\smallskip}
\multirow{2}{*}{$\left[F^{[1/2]_{\chi+1,{\bf l}_2}[1/3][1/2]_{\chi,{\bf l}_1}}_{[1/3]}\right]^{[1/2]_{\chi-1,{\bf l}'}}_{[1/2]_{\chi-1,{\bf l}}}$}
		& $\frac{1}{k}w^{{\bf p}^Ti\sigma_y\Lambda_{\chi}{\bf p}}$ \\\noalign{\smallskip}
		& $\quad\quad{\scriptstyle {\bf p}=\frac{1}{\Lambda_3-1}i\sigma_y({\bf l}-{\bf l}'+{\bf l}_1-{\bf l}_2+{\bf l}_0(\chi+1)-{\bf l}_0(\chi))}$ \\\noalign{\smallskip}
\multirow{2}{*}{$\left[F^{[1/3][1/3][1/2]_{\chi,{\bf l}_1}}_{[1/2]_{\chi+1,{\bf l}_2}}\right]^{[1/2]_{\chi-1,{\bf l}}}_{[\overline{1/3}],\boldsymbol\alpha}$}
		& $\frac{1}{k}w^{{\bf p}^Ti\sigma_y\left[\frac{k}{2}{\bf f}^\chi+\frac{1}{2}(\Lambda_{\chi-1}-\Lambda_{\chi+1}){\bf p}\right]}$ \\\noalign{\smallskip}
		& $\quad\quad{\scriptstyle {\bf p}=\frac{1}{1-\Lambda_3}i\sigma_y(\Lambda_3^T\boldsymbol\alpha+{\bf l}+\Lambda_3^T{\bf l}_1+(\Lambda_3^{-1})^T{\bf l}_2-3{\bf l}_0(\chi-1))}$ \\\noalign{\smallskip}
\multirow{2}{*}{$\left[F^{[1/2]_{\chi,{\bf l}_1}[1/3][1/3]}_{[1/2]_{\chi-1,{\bf l}_2}}\right]_{[1/2]_{\chi+1,{\bf l}}}^{[\overline{1/3}],\boldsymbol\alpha}$}
		& $\frac{1}{k}w^{{\bf p}^Ti\sigma_y\left[\frac{k}{2}{\bf f}^{\chi-1}+\frac{1}{2}(\Lambda_{\chi}-\Lambda_{\chi+1}){\bf p}\right]}$ \\\noalign{\smallskip}
		& $\quad\quad{\scriptstyle {\bf p}=\frac{1}{1-\Lambda_3}i\sigma_y(\Lambda_3^T\boldsymbol\alpha-{\bf l}-(\Lambda_3^{-1})^T{\bf l}_1-\Lambda_3^T{\bf l}_2+3{\bf l}_0(\chi+1))}$ \\\noalign{\smallskip}		
$\left[F^{[\overline{1/3}][1/3][1/2]_{\chi,{\bf l}_1}}_{[1/2]_{\chi,{\bf l}_2}}\right]^{[1/2]_{\chi-1,{\bf l}}}_{\bf a}$
		& $\frac{1}{k}w^{{\bf a}_\circ^Ti\sigma_y\frac{1}{\Lambda_3-1}\left[i\sigma_y({\bf l}-{\bf l}_2)+\frac{1}{2}({\bf f}^\chi-{\bf f}^{\chi-1}+\Lambda_\chi{\bf a}_\circ)\right]}$ \\\noalign{\smallskip}
$\left[F^{[1/2]_{\chi,{\bf l}_1}[1/3][\overline{1/3}]}_{[1/2]_{\chi,{\bf l}_2}}\right]_{[1/2]_{\chi+1,{\bf l}}}^{\bf a}$
		& $\frac{1}{k}w^{{\bf a}_\circ^Ti\sigma_y\frac{\Lambda_3}{\Lambda_3-1}\left[i\sigma_y({\bf l}-{\bf l}_1)+\frac{1}{2}({\bf f}^\chi-{\bf f}^{\chi+1}+\Lambda_\chi{\bf a}_\circ)\right]}$ \\\noalign{\smallskip}
\multirow{2}{*}{$\left[F^{[1/3][1/2]_{\chi,{\bf l}_1}[1/3]}_{[1/2]_{\chi,{\bf l}_2}}\right]^{[1/2]_{\chi+1,{\bf l}'}}_{[1/2]_{\chi-1,{\bf l}}}$}
		& $\frac{1}{k^2}w^{{\bf p}^Ti\sigma_y\left[\frac{k}{2}{\bf f}^\chi+\frac{3}{2}\Lambda_{\chi}{\bf p}\right]}$ \\\noalign{\smallskip}
		& $\quad\quad{\scriptstyle {\bf p}=\frac{1}{3}\Lambda_\chi i\sigma_y({\bf l}_1+{\bf l}_2-{\bf l}-{\bf l}'-3{\bf l}_0(\chi))}$ \\\noalign{\smallskip}
\multirow{2}{*}{$\left[F^{[1/3][1/2]_{\chi,{\bf l}_1}[\overline{1/3}]}_{[1/2]_{\chi+1,{\bf l}_2}}\right]^{[1/2]_{\chi-1,{\bf l}'}}_{[1/2]_{\chi-1,{\bf l}}}$}
		& $\frac{1}{k^2}w^{-{\bf p}^Ti\sigma_y\left[\frac{k}{2}{\bf f}^{\chi-1}+\frac{3}{2}\Lambda_{\chi-1}{\bf p}\right]}$ \\\noalign{\smallskip}
		& $\quad\quad{\scriptstyle {\bf p}=\frac{1}{3}\Lambda_{\chi-1}i\sigma_y({\bf l}+{\bf l}'-{\bf l}_1-{\bf l}_2-3{\bf l}_0(\chi-1))}$ \\\noalign{\smallskip}
\noalign{\smallskip}\hline\hline\\\noalign{\smallskip}
\caption{List of admissible $F$-symbols for $k$ not divisible by 3.}\label{tab:Fsymbols}
\end{longtable*}

\section{Local double Wilson loop around twofold defects}\label{sec:doubleloop}

We derive the phase for the local Wilson loop $\Theta({\bf a}_\circ)$ around a twofold twist defect that generalizes the primitive one in figure~\ref{fig:defectlocalloop}. The Wilson loop is defined by dragging a spinless abelian anyon ${\bf a}_\circ$ twice around a twist defect $[1/2]_{\chi}$, for $\chi=Y,R,B$ the color of the defect. It has the word presentation $[\hat\lambda\hat\lambda]_{{\bf a}_\circ}$, for $\hat\lambda$ the alphabet of the twofold defects (see section~\ref{sec:AlphabeticpresentationofWilsonalgebra}). \begin{align}\Theta({\bf a}_\circ)=\vcenter{\hbox{\includegraphics[width=0.5in]{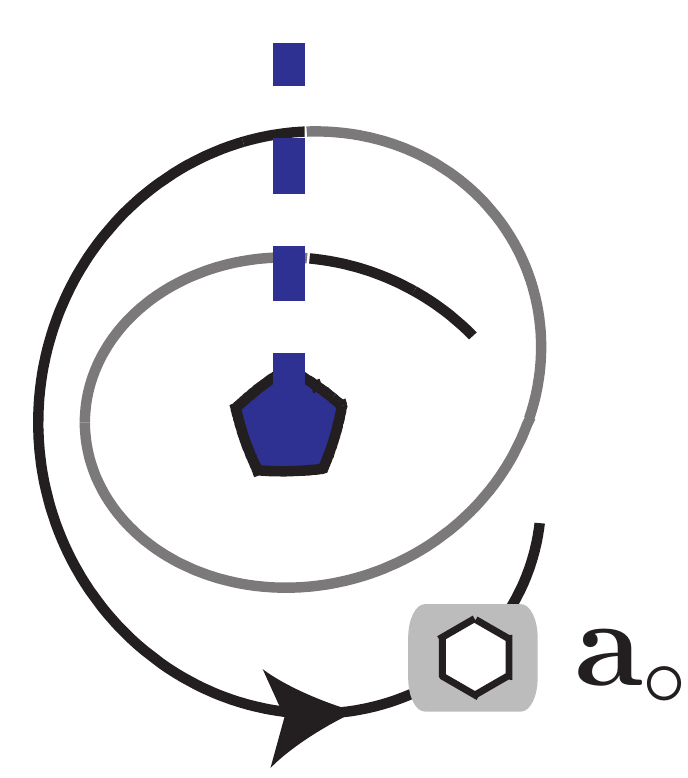}}}=\left[\hat\lambda\hat\lambda\right]_{{\bf a}_\circ}\label{doubleloopdefinition}\end{align} It corresponds to part of the phase of the $R$-symbol $R^{[1/2][1/2]}_{\bf a}$ in eq.\eqref{Z2exchange} and is responsible for the topological spin \eqref{Z2topologicalspin2} for a twofold defect.

\begin{figure}[ht]
	\includegraphics[width=2.8in]{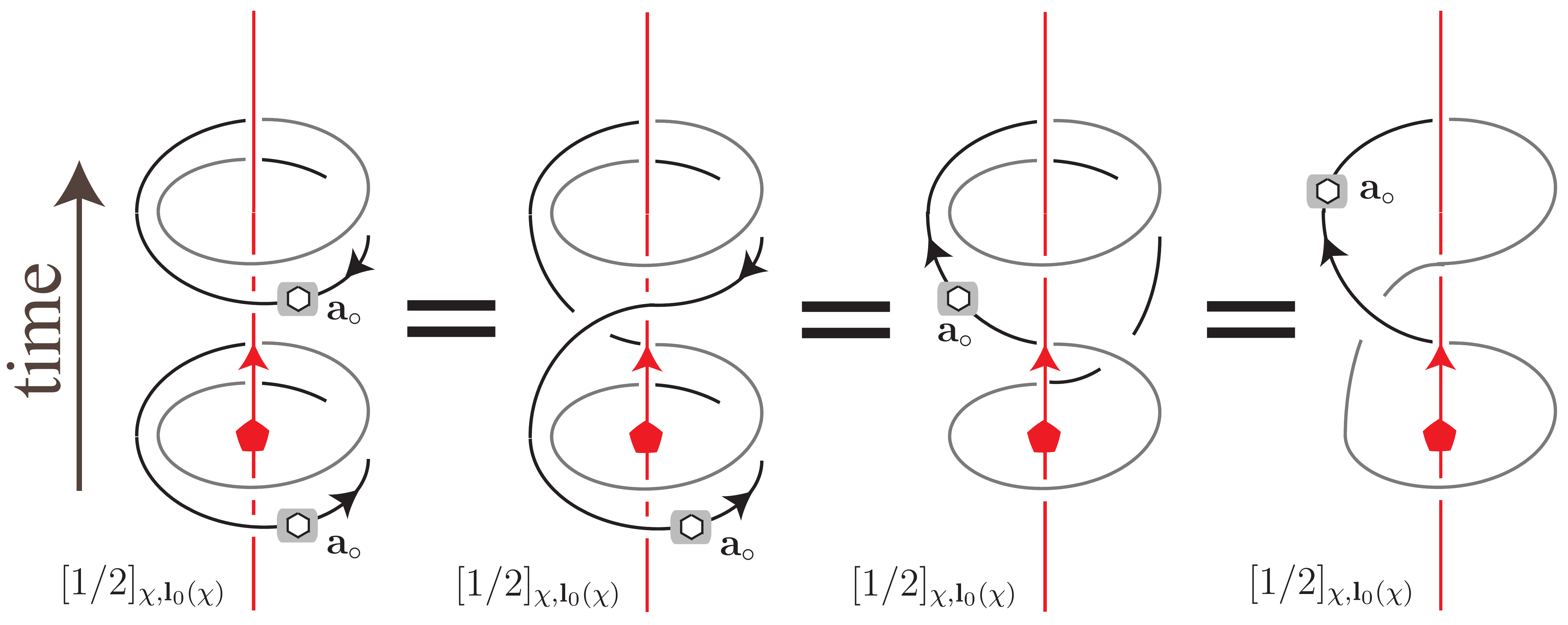}
	\caption{The Whitehead link between defect wordline (red) and anyon trajectory (black/grey) represented by the quadratic phase $\Theta(-{\bf a}_\circ)\Theta({\bf a}_\circ)=w^{{\bf a}_\circ^Ti\sigma_y\Lambda_\chi{\bf a}_\circ}$ in \eqref{whiteheadphase}.}\label{fig:whitehead}
\end{figure}
By breaking down the abelian anyon $[{\bf a}_\circ]=(Y_\circ)^y(R_\circ)^r$ into its primitive components $Y_\circ,R_\circ$, it can be broken down into simpler versions $\Theta_Y,\Theta_R$ seen in figure~\ref{fig:defectlocalloop}(a), which take the eigenvalue of $w^{l_Y},w^{l_R}$ respectively, for $w=e^{2\pi i/k}$, at a twofold defect with species label ${\bf l}=(l_Y,l_R)$. There are extra phases due to self-interersection. \begin{align}\Theta({\bf a}_\circ)\Theta({\bf b}_\circ)&=\left[\vcenter{\hbox{\includegraphics[width=0.5in]{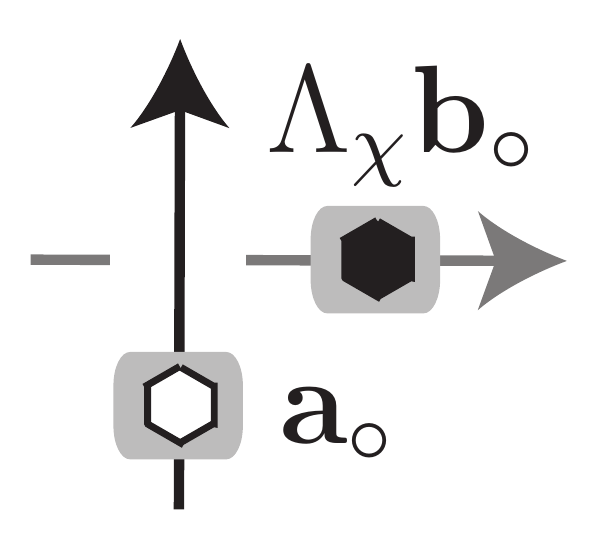}}}\right]\Theta({\bf a}_\circ+{\bf b}_\circ)\nonumber\\&=w^{-{\bf a}_\circ^Ti\sigma_y\Lambda_\chi{\bf b}_\circ}\Theta({\bf a}_\circ+{\bf b}_\circ)\label{doubleloopnonlinear}\end{align} where $\Lambda_\chi$ is the $2\times2$ matrix representing the transposition action of the twist defect $[1/2]_\chi$ (see eq.\eqref{lambdachiapp}). This means the double Wilson loop $\Theta({\bf a}_\circ)$ does not depend linearly on anyon label ${\bf a}_\circ$. This non-planar nature of $\Theta({\bf a}_\circ)$ is most apparent in the quadratic phase upon reordering the self-intersection. \begin{align}\vcenter{\hbox{\includegraphics[width=0.5in]{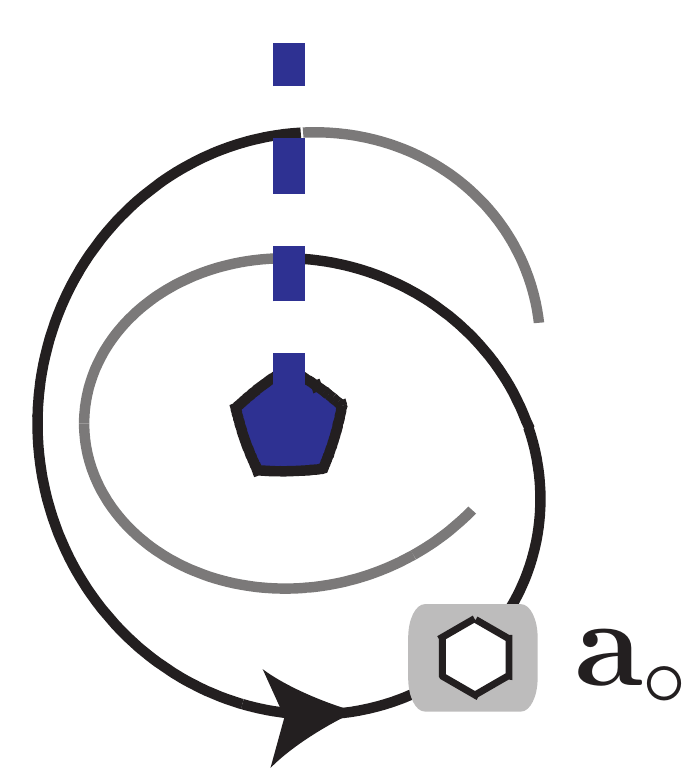}}}&=\Theta(-{\bf a}_\circ)^\dagger=w^{-{\bf a}_\circ^Ti\sigma_y\Lambda_\chi{\bf a}_\circ}\Theta({\bf a}_\circ)\nonumber\\&=w^{-{\bf a}_\circ^Ti\sigma_y\Lambda_\chi{\bf a}_\circ}\vcenter{\hbox{\includegraphics[width=0.5in]{doubleloopdef.pdf}}}\label{whiteheadphase}\end{align} It measures a linking between the world lines of defect and anyon in the $(2+1)$ spacetime  (see figure~\ref{fig:whitehead}), called the Whitehead link. Unlike a conventional link, the Whitehead link does not carry a net linking number but is still topologically non-trivial. It however can be unlinked by crossing the anyon trajectory, results in the quadratic phase in \eqref{whiteheadphase}, but without passing the defect wordline.

The double Wilson loop \eqref{doubleloopdefinition} can be successively be broken down by \eqref{doubleloopnonlinear} and re-expressed in terms of the primitive double Wilson loops. \begin{align}\Theta({\bf a}_\circ)=w^{\frac{1}{2}{\bf a}_\circ^Ti\sigma_y\left[\Lambda_\chi{\bf a}_\circ+{\bf f}^\chi\right]}(\Theta_Y)^y(\Theta_R)^r\end{align} where ${\bf f}^Y=(1,0)$, ${\bf f}^R=(0,1)$, ${\bf f}^B=(-1,-1)$ are the primitive anyon lattice vectors and ${\bf a}_\circ=y{\bf f}^Y+r{\bf f}^R$. At a self-reciprocal {\em bare} twofold defect $[1/2]_{\chi,{\bf l}_0(\chi)}$ where it is not attached with a Wilson string (see figure~\ref{fig:objectrep}), the eigenvalues for the primitive double loops combination $(\Theta_Y)^y(\Theta_R)^r$ is given by $w^{{\bf a}_\circ^T{\bf l}_0(\chi)}$ for ${\bf l}_0=-\frac{k+1}{2}i\sigma_y{\bf f}^\chi$ is the bare species label (see eq.\eqref{selfreciprocalspecies}).

The double Wilson loop $\Theta({\bf a}_\circ)$ thus takes the following eigenvalue on the ground state of a bare twofold defect $[1/2]_{\chi,{\bf l}_0(\chi)}$. \begin{align}\left|\vcenter{\hbox{\includegraphics[width=0.4in]{doubleloopdef.pdf}}}\right\rangle&=w^{\frac{1}{2}{\bf a}_\circ^Ti\sigma_y\left[\Lambda_\chi{\bf a}_\circ+{\bf f}^\chi\right]}(\Theta_Y)^y(\Theta_R)^r\left|\vcenter{\hbox{\includegraphics[width=0.08in]{720spincut2.pdf}}}\right\rangle\nonumber\\&=w^{{\bf a}_\circ^Ti\sigma_y\left[\frac{1}{2}\Lambda_\chi{\bf a}_\circ+\frac{k}{2}{\bf f}^\chi\right]}\left|\vcenter{\hbox{\includegraphics[width=0.08in]{720spincut2.pdf}}}\right\rangle\label{doubleloopbarephase}\end{align} In particular this gives the topological spin for the defect with species ${\bf l}$ by substituting ${\bf a}_\circ=-\Lambda_\chi i\sigma_y({\bf l}-{\bf l}_0(\chi))$, and is shown in \eqref{Z2topologicalspin2}. For a general twofold defect with a Wilson string ${\bf c}_\bullet=-i\sigma_y({\bf l}-{\bf l}_0(\chi))$ attaching to it, the eigenvalues of the double Wilson loop $\Theta({\bf a}_\circ)$ is modified by adding to \eqref{doubleloopbarephase} the intersection phase $w^{{\bf a}_\circ^Ti\sigma_y{\bf c}_\bullet}$.

The eigenvalue \eqref{doubleloopbarephase} also explains the phase factors $\varphi({\bf n})$ in \eqref{splittingphasefigg} and \eqref{splittingphasefigg2} occured in the definition of splitting states for $[1/2]_\chi\times[1/3]=[1/2]_{\chi+1}$ and $[1/3]\times[1/2]_\chi=[1/2]_{\chi-1}$. The Wilson loop labeled by ${\bf n}$ in the weighted sum in the splitting state (see figure~\ref{fig:splittingspaces}(g)) is a combination of the overall double loop that encloses the entire defect pair and the local double loop that only encloses the constituent twofold defect. It takes the eigenvalue $w^{-\varphi({\bf n})}$ for bare defects and is therefore compensated in the weighted sum.


\begin{thebibliography}{145}
\expandafter\ifx\csname natexlab\endcsname\relax\def\natexlab#1{#1}\fi
\expandafter\ifx\csname bibnamefont\endcsname\relax
  \def\bibnamefont#1{#1}\fi
\expandafter\ifx\csname bibfnamefont\endcsname\relax
  \def\bibfnamefont#1{#1}\fi
\expandafter\ifx\csname citenamefont\endcsname\relax
  \def\citenamefont#1{#1}\fi
\expandafter\ifx\csname url\endcsname\relax
  \def\url#1{\texttt{#1}}\fi
\expandafter\ifx\csname urlprefix\endcsname\relax\def\urlprefix{URL }\fi
\providecommand{\bibinfo}[2]{#2}
\providecommand{\eprint}[2][]{\url{#2}}

\bibitem[{\citenamefont{Majorana}(1937)}]{Majorana37}
\bibinfo{author}{\bibfnamefont{E.}~\bibnamefont{Majorana}},
  \bibinfo{journal}{Nuovo Cimento} \textbf{\bibinfo{volume}{5}},
  \bibinfo{pages}{171} (\bibinfo{year}{1937}).

\bibitem[{\citenamefont{Wilczek}(2009)}]{Wilczek09}
\bibinfo{author}{\bibfnamefont{F.}~\bibnamefont{Wilczek}},
  \bibinfo{journal}{Nat. Phys} \textbf{\bibinfo{volume}{5}},
  \bibinfo{pages}{614} (\bibinfo{year}{2009}).

\bibitem[{\citenamefont{Hasan and Kane}(2010)}]{HasanKane10}
\bibinfo{author}{\bibfnamefont{M.~Z.} \bibnamefont{Hasan}} \bibnamefont{and}
  \bibinfo{author}{\bibfnamefont{C.~L.} \bibnamefont{Kane}},
  \bibinfo{journal}{Rev. Mod. Phys.} \textbf{\bibinfo{volume}{82}},
  \bibinfo{pages}{\href{http://link.aps.org/doi/10.1103/RevModPhys.82.3045}{3045}}
  (\bibinfo{year}{2010}).

\bibitem[{\citenamefont{Qi and Zhang}(2011)}]{QiZhangreview11}
\bibinfo{author}{\bibfnamefont{X.-L.} \bibnamefont{Qi}} \bibnamefont{and}
  \bibinfo{author}{\bibfnamefont{S.-C.} \bibnamefont{Zhang}},
  \bibinfo{journal}{Rev. Mod. Phys.} \textbf{\bibinfo{volume}{83}},
  \bibinfo{pages}{\href{http://link.aps.org/doi/10.1103/RevModPhys.83.1057}{1057}}
  (\bibinfo{year}{2011}).

\bibitem[{\citenamefont{Beenakker}(2013)}]{Beenakker11}
\bibinfo{author}{\bibfnamefont{C.~W.~J.} \bibnamefont{Beenakker}},
  \bibinfo{journal}{Annu. Rev. Con. Mat. Phys.} \textbf{\bibinfo{volume}{4}},
  \bibinfo{pages}{113} (\bibinfo{year}{2013}).

\bibitem[{\citenamefont{Alicea}(2012)}]{Alicea12}
\bibinfo{author}{\bibfnamefont{J.}~\bibnamefont{Alicea}},
  \bibinfo{journal}{Rep. Prog. Phys} \textbf{\bibinfo{volume}{75}},
  \bibinfo{pages}{076501} (\bibinfo{year}{2012}).

\bibitem[{\citenamefont{Kitaev}(2003)}]{Kitaev97}
\bibinfo{author}{\bibfnamefont{A.}~\bibnamefont{Kitaev}},
  \bibinfo{journal}{Ann. Phys.} \textbf{\bibinfo{volume}{303}},
  \bibinfo{pages}{2} (\bibinfo{year}{2003}).

\bibitem[{\citenamefont{Ogburn and Preskill}(1999)}]{OgburnPreskill99}
\bibinfo{author}{\bibfnamefont{R.~W.} \bibnamefont{Ogburn}} \bibnamefont{and}
  \bibinfo{author}{\bibfnamefont{J.}~\bibnamefont{Preskill}},
  \bibinfo{journal}{Comp. Sci.} \textbf{\bibinfo{volume}{1509}},
  \bibinfo{pages}{341} (\bibinfo{year}{1999}).

\bibitem[{\citenamefont{Preskill}(2004)}]{Preskilllecturenotes}
\bibinfo{author}{\bibfnamefont{J.}~\bibnamefont{Preskill}},
  \emph{\bibinfo{title}{Topological quantum computation}}
  (\bibinfo{year}{2004}),
  \urlprefix\url{http://www.theory.caltech.edu/people/preskill/ph229/#lecture}.

\bibitem[{\citenamefont{Freedman
  et~al.}(2002{\natexlab{a}})\citenamefont{Freedman, Kitaev, Larsen, and
  Wang}}]{FreedmanKitaevLarsenWang01}
\bibinfo{author}{\bibfnamefont{M.}~\bibnamefont{Freedman}},
  \bibinfo{author}{\bibfnamefont{A.}~\bibnamefont{Kitaev}},
  \bibinfo{author}{\bibfnamefont{M.}~\bibnamefont{Larsen}}, \bibnamefont{and}
  \bibinfo{author}{\bibfnamefont{Z.}~\bibnamefont{Wang}},
  \bibinfo{journal}{Bull. Amer. Math. Soc.} \textbf{\bibinfo{volume}{40}},
  \bibinfo{pages}{31} (\bibinfo{year}{2002}{\natexlab{a}}).

\bibitem[{\citenamefont{Nayak et~al.}(2008)\citenamefont{Nayak, Simon, Stern,
  Freedman, and Das~Sarma}}]{ChetanSimonSternFreedmanDasSarma}
\bibinfo{author}{\bibfnamefont{C.}~\bibnamefont{Nayak}},
  \bibinfo{author}{\bibfnamefont{S.~H.} \bibnamefont{Simon}},
  \bibinfo{author}{\bibfnamefont{A.}~\bibnamefont{Stern}},
  \bibinfo{author}{\bibfnamefont{M.}~\bibnamefont{Freedman}}, \bibnamefont{and}
  \bibinfo{author}{\bibfnamefont{S.}~\bibnamefont{Das~Sarma}},
  \bibinfo{journal}{Rev. Mod. Phys.} \textbf{\bibinfo{volume}{80}},
  \bibinfo{pages}{\href{http://link.aps.org/doi/10.1103/RevModPhys.80.1083}{1083}}
  (\bibinfo{year}{2008}).

\bibitem[{\citenamefont{Wang}(2010)}]{Wangbook}
\bibinfo{author}{\bibfnamefont{Z.}~\bibnamefont{Wang}},
  \emph{\bibinfo{title}{Topological Quantum Computation}}
  (\bibinfo{publisher}{American Mathematics Society}, \bibinfo{year}{2010}).

\bibitem[{\citenamefont{Kane and Mele}(2005{\natexlab{a}})}]{KaneMele2D1}
\bibinfo{author}{\bibfnamefont{C.~L.} \bibnamefont{Kane}} \bibnamefont{and}
  \bibinfo{author}{\bibfnamefont{E.~J.} \bibnamefont{Mele}},
  \bibinfo{journal}{Phys. Rev. Lett.} \textbf{\bibinfo{volume}{95}},
  \bibinfo{pages}{\href{http://link.aps.org/doi/10.1103/PhysRevLett.95.226801}{226801}}
  (\bibinfo{year}{2005}{\natexlab{a}}).

\bibitem[{\citenamefont{Kane and Mele}(2005{\natexlab{b}})}]{KaneMele2D2}
\bibinfo{author}{\bibfnamefont{C.~L.} \bibnamefont{Kane}} \bibnamefont{and}
  \bibinfo{author}{\bibfnamefont{E.~J.} \bibnamefont{Mele}},
  \bibinfo{journal}{Phys. Rev. Lett.} \textbf{\bibinfo{volume}{95}},
  \bibinfo{pages}{\href{http://link.aps.org/doi/10.1103/PhysRevLett.95.146802}{146802}}
  (\bibinfo{year}{2005}{\natexlab{b}}).

\bibitem[{\citenamefont{K\"onig et~al.}(2007)\citenamefont{K\"onig, Wiedmann,
  Br\"une, Roth, Buhmann, Molenkamp, Qi, and Zhang}}]{Molenkamp07}
\bibinfo{author}{\bibfnamefont{M.}~\bibnamefont{K\"onig}},
  \bibinfo{author}{\bibfnamefont{S.}~\bibnamefont{Wiedmann}},
  \bibinfo{author}{\bibfnamefont{C.}~\bibnamefont{Br\"une}},
  \bibinfo{author}{\bibfnamefont{A.}~\bibnamefont{Roth}},
  \bibinfo{author}{\bibfnamefont{H.}~\bibnamefont{Buhmann}},
  \bibinfo{author}{\bibfnamefont{L.}~\bibnamefont{Molenkamp}},
  \bibinfo{author}{\bibfnamefont{X.-L.} \bibnamefont{Qi}}, \bibnamefont{and}
  \bibinfo{author}{\bibfnamefont{S.}~\bibnamefont{Zhang}},
  \bibinfo{journal}{Science} \textbf{\bibinfo{volume}{318}},
  \bibinfo{pages}{766} (\bibinfo{year}{2007}).

\bibitem[{\citenamefont{Moore and Balents}(2007)}]{MooreBalents07}
\bibinfo{author}{\bibfnamefont{J.~E.} \bibnamefont{Moore}} \bibnamefont{and}
  \bibinfo{author}{\bibfnamefont{L.}~\bibnamefont{Balents}},
  \bibinfo{journal}{Phys. Rev. B} \textbf{\bibinfo{volume}{75}},
  \bibinfo{pages}{\href{http://link.aps.org/doi/10.1103/PhysRevB.75.121306}{121306(R)}}
  (\bibinfo{year}{2007}).

\bibitem[{\citenamefont{Roy}(2009)}]{Roy07}
\bibinfo{author}{\bibfnamefont{R.}~\bibnamefont{Roy}}, \bibinfo{journal}{Phys.
  Rev. B} \textbf{\bibinfo{volume}{79}},
  \bibinfo{pages}{\href{http://link.aps.org/doi/10.1103/PhysRevB.79.195322}{195322}}
  (\bibinfo{year}{2009}).

\bibitem[{\citenamefont{Fu et~al.}(2007)\citenamefont{Fu, Kane, and
  Mele}}]{FuKaneMele3D}
\bibinfo{author}{\bibfnamefont{L.}~\bibnamefont{Fu}},
  \bibinfo{author}{\bibfnamefont{C.~L.} \bibnamefont{Kane}}, \bibnamefont{and}
  \bibinfo{author}{\bibfnamefont{E.~J.} \bibnamefont{Mele}},
  \bibinfo{journal}{Phys. Rev. Lett.} \textbf{\bibinfo{volume}{98}},
  \bibinfo{pages}{\href{http://link.aps.org/doi/10.1103/PhysRevLett.98.106803}{106803}}
  (\bibinfo{year}{2007}).

\bibitem[{\citenamefont{Qi et~al.}(2008)\citenamefont{Qi, Hughes, and
  Zhang}}]{QiHughesZhang08}
\bibinfo{author}{\bibfnamefont{X.-L.} \bibnamefont{Qi}},
  \bibinfo{author}{\bibfnamefont{T.~L.} \bibnamefont{Hughes}},
  \bibnamefont{and} \bibinfo{author}{\bibfnamefont{S.-C.} \bibnamefont{Zhang}},
  \bibinfo{journal}{Phys. Rev. B} \textbf{\bibinfo{volume}{78}},
  \bibinfo{pages}{\href{http://link.aps.org/doi/10.1103/PhysRevB.78.195424}{195424}}
  (\bibinfo{year}{2008}).

\bibitem[{\citenamefont{Hsieh et~al.}(2008)\citenamefont{Hsieh, Qian, Wray,
  Xia, Hor, Cava, and Hasan}}]{Hasan08}
\bibinfo{author}{\bibfnamefont{D.}~\bibnamefont{Hsieh}},
  \bibinfo{author}{\bibfnamefont{D.}~\bibnamefont{Qian}},
  \bibinfo{author}{\bibfnamefont{L.}~\bibnamefont{Wray}},
  \bibinfo{author}{\bibfnamefont{Y.}~\bibnamefont{Xia}},
  \bibinfo{author}{\bibfnamefont{Y.~S.} \bibnamefont{Hor}},
  \bibinfo{author}{\bibfnamefont{R.~J.} \bibnamefont{Cava}}, \bibnamefont{and}
  \bibinfo{author}{\bibfnamefont{M.~Z.} \bibnamefont{Hasan}},
  \bibinfo{journal}{Nature} \textbf{\bibinfo{volume}{452}},
  \bibinfo{pages}{970} (\bibinfo{year}{2008}).

\bibitem[{\citenamefont{Kitaev}(2001)}]{Kitaevchain}
\bibinfo{author}{\bibfnamefont{A.~Y.} \bibnamefont{Kitaev}},
  \bibinfo{journal}{Phys.-Usp.} \textbf{\bibinfo{volume}{44}},
  \bibinfo{pages}{131} (\bibinfo{year}{2001}).

\bibitem[{\citenamefont{Volovik}(1999)}]{Volovik99}
\bibinfo{author}{\bibfnamefont{G.~E.} \bibnamefont{Volovik}},
  \bibinfo{journal}{Pisma Zh. Eksp. Teor. Fiz.} \textbf{\bibinfo{volume}{70}},
  \bibinfo{pages}{601} (\bibinfo{year}{1999}).

\bibitem[{\citenamefont{Read and Green}(2000)}]{ReadGreen}
\bibinfo{author}{\bibfnamefont{N.}~\bibnamefont{Read}} \bibnamefont{and}
  \bibinfo{author}{\bibfnamefont{D.}~\bibnamefont{Green}},
  \bibinfo{journal}{Phys. Rev. B} \textbf{\bibinfo{volume}{61}},
  \bibinfo{pages}{\href{http://link.aps.org/doi/10.1103/PhysRevB.61.10267}{10267}}
  (\bibinfo{year}{2000}).

\bibitem[{\citenamefont{Ivanov}(2001)}]{Ivanov}
\bibinfo{author}{\bibfnamefont{D.~A.} \bibnamefont{Ivanov}},
  \bibinfo{journal}{Phys. Rev. Lett.} \textbf{\bibinfo{volume}{86}},
  \bibinfo{pages}{\href{http://link.aps.org/doi/10.1103/PhysRevLett.86.268}{268}}
  (\bibinfo{year}{2001}).

\bibitem[{\citenamefont{Read and Moore}(1991)}]{MooreRead}
\bibinfo{author}{\bibfnamefont{N.}~\bibnamefont{Read}} \bibnamefont{and}
  \bibinfo{author}{\bibfnamefont{G.}~\bibnamefont{Moore}},
  \bibinfo{journal}{Nucl. Phys. B} \textbf{\bibinfo{volume}{360}},
  \bibinfo{pages}{362} (\bibinfo{year}{1991}).

\bibitem[{\citenamefont{Greiter et~al.}(1992)\citenamefont{Greiter, Wen, and
  Wilczek}}]{GreiterWenWilczek91}
\bibinfo{author}{\bibfnamefont{M.}~\bibnamefont{Greiter}},
  \bibinfo{author}{\bibfnamefont{X.-G.} \bibnamefont{Wen}}, \bibnamefont{and}
  \bibinfo{author}{\bibfnamefont{F.}~\bibnamefont{Wilczek}},
  \bibinfo{journal}{Nucl. Phys. B} \textbf{\bibinfo{volume}{374}},
  \bibinfo{pages}{567} (\bibinfo{year}{1992}).

\bibitem[{\citenamefont{Nayak and Wilczek}(1996)}]{NayakWilczek96}
\bibinfo{author}{\bibfnamefont{C.}~\bibnamefont{Nayak}} \bibnamefont{and}
  \bibinfo{author}{\bibfnamefont{F.}~\bibnamefont{Wilczek}},
  \bibinfo{journal}{Nucl. Phys. B} \textbf{\bibinfo{volume}{479}},
  \bibinfo{pages}{529} (\bibinfo{year}{1996}).

\bibitem[{\citenamefont{Fu and Kane}(2008)}]{FuKane08}
\bibinfo{author}{\bibfnamefont{L.}~\bibnamefont{Fu}} \bibnamefont{and}
  \bibinfo{author}{\bibfnamefont{C.~L.} \bibnamefont{Kane}},
  \bibinfo{journal}{Phys. Rev. Lett.} \textbf{\bibinfo{volume}{100}},
  \bibinfo{pages}{\href{http://link.aps.org/doi/10.1103/PhysRevLett.100.096407}{096407}}
  (\bibinfo{year}{2008}).

\bibitem[{\citenamefont{Akhmerov et~al.}(2009)\citenamefont{Akhmerov, Nilsson,
  and Beenakker}}]{AkhmerovNilssonBeenakker09}
\bibinfo{author}{\bibfnamefont{A.~R.} \bibnamefont{Akhmerov}},
  \bibinfo{author}{\bibfnamefont{J.}~\bibnamefont{Nilsson}}, \bibnamefont{and}
  \bibinfo{author}{\bibfnamefont{C.~W.~J.} \bibnamefont{Beenakker}},
  \bibinfo{journal}{Phys. Rev. Lett.} \textbf{\bibinfo{volume}{102}},
  \bibinfo{pages}{\href{http://link.aps.org/doi/10.1103/PhysRevLett.102.216404}{216404}}
  (\bibinfo{year}{2009}).

\bibitem[{\citenamefont{Fu and Kane}(2009)}]{FuKanechargetransport09}
\bibinfo{author}{\bibfnamefont{L.}~\bibnamefont{Fu}} \bibnamefont{and}
  \bibinfo{author}{\bibfnamefont{C.~L.} \bibnamefont{Kane}},
  \bibinfo{journal}{Phys. Rev. Lett.} \textbf{\bibinfo{volume}{102}},
  \bibinfo{pages}{\href{http://link.aps.org/doi/10.1103/PhysRevLett.102.216403}{216403}}
  (\bibinfo{year}{2009}).

\bibitem[{\citenamefont{Law et~al.}(2009)\citenamefont{Law, Lee, and
  Ng}}]{LawLeeNg09}
\bibinfo{author}{\bibfnamefont{K.~T.} \bibnamefont{Law}},
  \bibinfo{author}{\bibfnamefont{P.~A.} \bibnamefont{Lee}}, \bibnamefont{and}
  \bibinfo{author}{\bibfnamefont{T.~K.} \bibnamefont{Ng}},
  \bibinfo{journal}{Phys. Rev. Lett.} \textbf{\bibinfo{volume}{103}},
  \bibinfo{pages}{\href{http://link.aps.org/doi/10.1103/PhysRevLett.103.237001}{237001}}
  (\bibinfo{year}{2009}).

\bibitem[{\citenamefont{Williams et~al.}(2012)\citenamefont{Williams, Bestwick,
  Gallagher, Hong, Cui, Bleich, Analytis, Fisher, and
  Goldhaber-Gordon}}]{GoldhaberGordon12}
\bibinfo{author}{\bibfnamefont{J.~R.} \bibnamefont{Williams}},
  \bibinfo{author}{\bibfnamefont{A.~J.} \bibnamefont{Bestwick}},
  \bibinfo{author}{\bibfnamefont{P.}~\bibnamefont{Gallagher}},
  \bibinfo{author}{\bibfnamefont{S.~S.} \bibnamefont{Hong}},
  \bibinfo{author}{\bibfnamefont{Y.}~\bibnamefont{Cui}},
  \bibinfo{author}{\bibfnamefont{A.~S.} \bibnamefont{Bleich}},
  \bibinfo{author}{\bibfnamefont{J.~G.} \bibnamefont{Analytis}},
  \bibinfo{author}{\bibfnamefont{I.~R.} \bibnamefont{Fisher}},
  \bibnamefont{and}
  \bibinfo{author}{\bibfnamefont{D.}~\bibnamefont{Goldhaber-Gordon}},
  \bibinfo{journal}{Phys. Rev. Lett.} \textbf{\bibinfo{volume}{109}},
  \bibinfo{pages}{\href{http://link.aps.org/doi/10.1103/PhysRevLett.109.056803}{056803}}
  (\bibinfo{year}{2012}).

\bibitem[{\citenamefont{Sau et~al.}(2010)\citenamefont{Sau, Lutchyn, Tewari,
  and Das~Sarma}}]{SauLutchynTewariDasSarma}
\bibinfo{author}{\bibfnamefont{J.~D.} \bibnamefont{Sau}},
  \bibinfo{author}{\bibfnamefont{R.~M.} \bibnamefont{Lutchyn}},
  \bibinfo{author}{\bibfnamefont{S.}~\bibnamefont{Tewari}}, \bibnamefont{and}
  \bibinfo{author}{\bibfnamefont{S.}~\bibnamefont{Das~Sarma}},
  \bibinfo{journal}{Phys. Rev. Lett.} \textbf{\bibinfo{volume}{104}},
  \bibinfo{pages}{\href{http://link.aps.org/doi/10.1103/PhysRevLett.104.040502}{040502}}
  (\bibinfo{year}{2010}).

\bibitem[{\citenamefont{Oreg et~al.}(2010)\citenamefont{Oreg, Refael, and von
  Oppen}}]{OregRefaelvonOppen10}
\bibinfo{author}{\bibfnamefont{Y.}~\bibnamefont{Oreg}},
  \bibinfo{author}{\bibfnamefont{G.}~\bibnamefont{Refael}}, \bibnamefont{and}
  \bibinfo{author}{\bibfnamefont{F.}~\bibnamefont{von Oppen}},
  \bibinfo{journal}{Phys. Rev. Lett.} \textbf{\bibinfo{volume}{105}},
  \bibinfo{pages}{\href{http://link.aps.org/doi/10.1103/PhysRevLett.105.177002}{177002}}
  (\bibinfo{year}{2010}).

\bibitem[{\citenamefont{Mourik et~al.}(2012)\citenamefont{Mourik, Zuo, Frolov,
  Plissard, Bakkers, and Kouwenhoven}}]{Kouwenhoven12}
\bibinfo{author}{\bibfnamefont{V.}~\bibnamefont{Mourik}},
  \bibinfo{author}{\bibfnamefont{K.}~\bibnamefont{Zuo}},
  \bibinfo{author}{\bibfnamefont{S.}~\bibnamefont{Frolov}},
  \bibinfo{author}{\bibfnamefont{S.}~\bibnamefont{Plissard}},
  \bibinfo{author}{\bibfnamefont{E.}~\bibnamefont{Bakkers}}, \bibnamefont{and}
  \bibinfo{author}{\bibfnamefont{L.}~\bibnamefont{Kouwenhoven}},
  \bibinfo{journal}{Science} \textbf{\bibinfo{volume}{336}},
  \bibinfo{pages}{1003} (\bibinfo{year}{2012}).

\bibitem[{\citenamefont{Lindner et~al.}(2012)\citenamefont{Lindner, Berg,
  Refael, and Stern}}]{LindnerBergRefaelStern}
\bibinfo{author}{\bibfnamefont{N.~H.} \bibnamefont{Lindner}},
  \bibinfo{author}{\bibfnamefont{E.}~\bibnamefont{Berg}},
  \bibinfo{author}{\bibfnamefont{G.}~\bibnamefont{Refael}}, \bibnamefont{and}
  \bibinfo{author}{\bibfnamefont{A.}~\bibnamefont{Stern}},
  \bibinfo{journal}{Phys. Rev. X} \textbf{\bibinfo{volume}{2}},
  \bibinfo{pages}{\href{http://link.aps.org/doi/10.1103/PhysRevX.2.041002}{041002}}
  (\bibinfo{year}{2012}).

\bibitem[{\citenamefont{Clarke et~al.}(2012)\citenamefont{Clarke, Alicea, and
  Shtengel}}]{ClarkeAliceaKirill}
\bibinfo{author}{\bibfnamefont{D.~J.} \bibnamefont{Clarke}},
  \bibinfo{author}{\bibfnamefont{J.}~\bibnamefont{Alicea}}, \bibnamefont{and}
  \bibinfo{author}{\bibfnamefont{K.}~\bibnamefont{Shtengel}},
  \bibinfo{journal}{Nature Commun.} \textbf{\bibinfo{volume}{4}},
  \bibinfo{pages}{1348} (\bibinfo{year}{2012}).

\bibitem[{\citenamefont{Cheng}(2012)}]{MChen}
\bibinfo{author}{\bibfnamefont{M.}~\bibnamefont{Cheng}},
  \bibinfo{journal}{Phys. Rev. B} \textbf{\bibinfo{volume}{86}},
  \bibinfo{pages}{\href{http://link.aps.org/doi/10.1103/PhysRevB.86.195126}{195126}}
  (\bibinfo{year}{2012}).

\bibitem[{\citenamefont{Vaezi}(2013)}]{Vaezi}
\bibinfo{author}{\bibfnamefont{A.}~\bibnamefont{Vaezi}},
  \bibinfo{journal}{Phys. Rev. B} \textbf{\bibinfo{volume}{87}},
  \bibinfo{pages}{\href{http://link.aps.org/doi/10.1103/PhysRevB.87.035132}{035132}}
  (\bibinfo{year}{2013}).

\bibitem[{\citenamefont{Levin and Stern}(2009)}]{LevinStern09}
\bibinfo{author}{\bibfnamefont{M.}~\bibnamefont{Levin}} \bibnamefont{and}
  \bibinfo{author}{\bibfnamefont{A.}~\bibnamefont{Stern}},
  \bibinfo{journal}{Phys. Rev. Lett.} \textbf{\bibinfo{volume}{103}},
  \bibinfo{pages}{\href{http://link.aps.org/doi/10.1103/PhysRevLett.103.196803}{196803}}
  (\bibinfo{year}{2009}).

\bibitem[{\citenamefont{Levin and Stern}(2012)}]{LevinStern12}
\bibinfo{author}{\bibfnamefont{M.}~\bibnamefont{Levin}} \bibnamefont{and}
  \bibinfo{author}{\bibfnamefont{A.}~\bibnamefont{Stern}},
  \bibinfo{journal}{Phys. Rev. B} \textbf{\bibinfo{volume}{86}},
  \bibinfo{pages}{\href{http://link.aps.org/doi/10.1103/PhysRevB.86.115131}{115131}}
  (\bibinfo{year}{2012}).

\bibitem[{\citenamefont{Maciejko et~al.}(2010)\citenamefont{Maciejko, Qi,
  Karch, and Zhang}}]{MaciejkoQiKarchZhang10}
\bibinfo{author}{\bibfnamefont{J.}~\bibnamefont{Maciejko}},
  \bibinfo{author}{\bibfnamefont{X.-L.} \bibnamefont{Qi}},
  \bibinfo{author}{\bibfnamefont{A.}~\bibnamefont{Karch}}, \bibnamefont{and}
  \bibinfo{author}{\bibfnamefont{S.-C.} \bibnamefont{Zhang}},
  \bibinfo{journal}{Phys. Rev. Lett.} \textbf{\bibinfo{volume}{105}},
  \bibinfo{pages}{\href{http://link.aps.org/doi/10.1103/PhysRevLett.105.246809}{246809}}
  (\bibinfo{year}{2010}).

\bibitem[{\citenamefont{Swingle et~al.}(2011)\citenamefont{Swingle, Barkeshli,
  McGreevy, and Senthil}}]{SwingleBarkeshliMcGreevySenthil11}
\bibinfo{author}{\bibfnamefont{B.}~\bibnamefont{Swingle}},
  \bibinfo{author}{\bibfnamefont{M.}~\bibnamefont{Barkeshli}},
  \bibinfo{author}{\bibfnamefont{J.}~\bibnamefont{McGreevy}}, \bibnamefont{and}
  \bibinfo{author}{\bibfnamefont{T.}~\bibnamefont{Senthil}},
  \bibinfo{journal}{Phys. Rev. B} \textbf{\bibinfo{volume}{83}},
  \bibinfo{pages}{\href{http://link.aps.org/doi/10.1103/PhysRevB.83.195139}{195139}}
  (\bibinfo{year}{2011}).

\bibitem[{\citenamefont{Levin et~al.}(2011)\citenamefont{Levin, Burnell,
  Koch-Janusz, and Stern}}]{LevinBurnellKochStern11}
\bibinfo{author}{\bibfnamefont{M.}~\bibnamefont{Levin}},
  \bibinfo{author}{\bibfnamefont{F.~J.} \bibnamefont{Burnell}},
  \bibinfo{author}{\bibfnamefont{M.}~\bibnamefont{Koch-Janusz}},
  \bibnamefont{and} \bibinfo{author}{\bibfnamefont{A.}~\bibnamefont{Stern}},
  \bibinfo{journal}{Phys. Rev. B} \textbf{\bibinfo{volume}{84}},
  \bibinfo{pages}{\href{http://link.aps.org/doi/10.1103/PhysRevB.84.235145}{235145}}
  (\bibinfo{year}{2011}).

\bibitem[{\citenamefont{Oreg et~al.}(2013)\citenamefont{Oreg, Sela, and
  Stern}}]{OregSelaStern13}
\bibinfo{author}{\bibfnamefont{Y.}~\bibnamefont{Oreg}},
  \bibinfo{author}{\bibfnamefont{E.}~\bibnamefont{Sela}}, \bibnamefont{and}
  \bibinfo{author}{\bibfnamefont{A.}~\bibnamefont{Stern}},
  \bibinfo{journal}{\href{http://arxiv.org/abs/1301.7335}{arXiv:1301.7335}}
  (\bibinfo{year}{2013}).

\bibitem[{\citenamefont{Kitaev}(2006)}]{Kitaev06}
\bibinfo{author}{\bibfnamefont{A.}~\bibnamefont{Kitaev}},
  \bibinfo{journal}{Ann. Phys.} \textbf{\bibinfo{volume}{321}},
  \bibinfo{pages}{2} (\bibinfo{year}{2006}).

\bibitem[{\citenamefont{Bombin}(2010)}]{Bombin}
\bibinfo{author}{\bibfnamefont{H.}~\bibnamefont{Bombin}},
  \bibinfo{journal}{Phys. Rev. Lett.} \textbf{\bibinfo{volume}{105}},
  \bibinfo{pages}{\href{http://link.aps.org/doi/10.1103/PhysRevLett.105.030403}{030403}}
  (\bibinfo{year}{2010}).

\bibitem[{\citenamefont{Kitaev and Kong}(2012)}]{KitaevKong12}
\bibinfo{author}{\bibfnamefont{A.}~\bibnamefont{Kitaev}} \bibnamefont{and}
  \bibinfo{author}{\bibfnamefont{L.}~\bibnamefont{Kong}},
  \bibinfo{journal}{Commun. Math. Phys.} \textbf{\bibinfo{volume}{313}},
  \bibinfo{pages}{351} (\bibinfo{year}{2012}).

\bibitem[{\citenamefont{You and Wen}(2012)}]{YouWen}
\bibinfo{author}{\bibfnamefont{Y.-Z.} \bibnamefont{You}} \bibnamefont{and}
  \bibinfo{author}{\bibfnamefont{X.-G.} \bibnamefont{Wen}},
  \bibinfo{journal}{Phys. Rev. B} \textbf{\bibinfo{volume}{86}},
  \bibinfo{pages}{\href{http://link.aps.org/doi/10.1103/PhysRevB.86.161107}{161107(R)}}
  (\bibinfo{year}{2012}).

\bibitem[{\citenamefont{You et~al.}(2013)\citenamefont{You, Jian, and
  Wen}}]{YouJianWen}
\bibinfo{author}{\bibfnamefont{Y.-Z.} \bibnamefont{You}},
  \bibinfo{author}{\bibfnamefont{C.-M.} \bibnamefont{Jian}}, \bibnamefont{and}
  \bibinfo{author}{\bibfnamefont{X.-G.} \bibnamefont{Wen}},
  \bibinfo{journal}{Phys. Rev. B} \textbf{\bibinfo{volume}{87}},
  \bibinfo{pages}{\href{http://link.aps.org/doi/10.1103/PhysRevB.87.045106}{045106}}
  (\bibinfo{year}{2013}).

\bibitem[{\citenamefont{Bombin}(2011)}]{Bombin11}
\bibinfo{author}{\bibfnamefont{H.}~\bibnamefont{Bombin}}, \bibinfo{journal}{New
  J. Phys.} \textbf{\bibinfo{volume}{13}}, \bibinfo{pages}{043005}
  (\bibinfo{year}{2011}).

\bibitem[{\citenamefont{Bi et~al.}(2013)\citenamefont{Bi, Rasmussen, and
  Xu}}]{BiRasmussenXu13}
\bibinfo{author}{\bibfnamefont{Z.}~\bibnamefont{Bi}},
  \bibinfo{author}{\bibfnamefont{A.}~\bibnamefont{Rasmussen}},
  \bibnamefont{and} \bibinfo{author}{\bibfnamefont{C.}~\bibnamefont{Xu}},
  \bibinfo{journal}{\href{http://arxiv.org/abs/1304.7272}{arXiv:1304.7272}}
  (\bibinfo{year}{2013}).

\bibitem[{\citenamefont{Mesaros et~al.}(2013)\citenamefont{Mesaros, Kim, and
  Ran}}]{MesarosKimRan13}
\bibinfo{author}{\bibfnamefont{A.}~\bibnamefont{Mesaros}},
  \bibinfo{author}{\bibfnamefont{Y.~B.} \bibnamefont{Kim}}, \bibnamefont{and}
  \bibinfo{author}{\bibfnamefont{Y.}~\bibnamefont{Ran}},
  \bibinfo{journal}{Phys. Rev. B} \textbf{\bibinfo{volume}{88}},
  \bibinfo{pages}{\href{http://link.aps.org/doi/10.1103/PhysRevB.88.035141}{035141}}
  (\bibinfo{year}{2013}).

\bibitem[{\citenamefont{Barkeshli and Qi}(2012)}]{BarkeshliQi}
\bibinfo{author}{\bibfnamefont{M.}~\bibnamefont{Barkeshli}} \bibnamefont{and}
  \bibinfo{author}{\bibfnamefont{X.-L.} \bibnamefont{Qi}},
  \bibinfo{journal}{Phys. Rev. X} \textbf{\bibinfo{volume}{2}},
  \bibinfo{pages}{\href{http://link.aps.org/doi/10.1103/PhysRevX.2.031013}{031013}}
  (\bibinfo{year}{2012}).

\bibitem[{\citenamefont{Barkeshli and Qi}(2013)}]{BarkeshliQi13}
\bibinfo{author}{\bibfnamefont{M.}~\bibnamefont{Barkeshli}} \bibnamefont{and}
  \bibinfo{author}{\bibfnamefont{X.-L.} \bibnamefont{Qi}},
  \bibinfo{journal}{\href{http://arxiv.org/abs/1302.2673}{arXiv:1302.2673}}
  (\bibinfo{year}{2013}).

\bibitem[{\citenamefont{Wang and Ran}(2011)}]{WangRan11}
\bibinfo{author}{\bibfnamefont{F.}~\bibnamefont{Wang}} \bibnamefont{and}
  \bibinfo{author}{\bibfnamefont{Y.}~\bibnamefont{Ran}},
  \bibinfo{journal}{Phys. Rev. B} \textbf{\bibinfo{volume}{84}},
  \bibinfo{pages}{\href{http://link.aps.org/doi/10.1103/PhysRevB.84.241103}{241103(R)}}
  (\bibinfo{year}{2011}).

\bibitem[{\citenamefont{Lu and Ran}(2012)}]{LuRan12}
\bibinfo{author}{\bibfnamefont{Y.-M.} \bibnamefont{Lu}} \bibnamefont{and}
  \bibinfo{author}{\bibfnamefont{Y.}~\bibnamefont{Ran}},
  \bibinfo{journal}{Phys. Rev. B} \textbf{\bibinfo{volume}{85}},
  \bibinfo{pages}{\href{http://link.aps.org/doi/10.1103/PhysRevB.85.165134}{165134}}
  (\bibinfo{year}{2012}).

\bibitem[{\citenamefont{Barkeshli
  et~al.}(2013{\natexlab{a}})\citenamefont{Barkeshli, Jian, and
  Qi}}]{BarkeshliJianQi}
\bibinfo{author}{\bibfnamefont{M.}~\bibnamefont{Barkeshli}},
  \bibinfo{author}{\bibfnamefont{C.-M.} \bibnamefont{Jian}}, \bibnamefont{and}
  \bibinfo{author}{\bibfnamefont{X.-L.} \bibnamefont{Qi}},
  \bibinfo{journal}{Phys. Rev. B} \textbf{\bibinfo{volume}{87}},
  \bibinfo{pages}{\href{http://link.aps.org/doi/10.1103/PhysRevB.87.045130}{045130}}
  (\bibinfo{year}{2013}{\natexlab{a}}).

\bibitem[{\citenamefont{Levin and Gu}(2012)}]{LevinGu12}
\bibinfo{author}{\bibfnamefont{M.}~\bibnamefont{Levin}} \bibnamefont{and}
  \bibinfo{author}{\bibfnamefont{Z.-C.} \bibnamefont{Gu}},
  \bibinfo{journal}{Phys. Rev. B} \textbf{\bibinfo{volume}{86}},
  \bibinfo{pages}{\href{http://link.aps.org/doi/10.1103/PhysRevB.86.115109}{115109}}
  (\bibinfo{year}{2012}).

\bibitem[{\citenamefont{Levin}(2013)}]{Levin13}
\bibinfo{author}{\bibfnamefont{M.}~\bibnamefont{Levin}},
  \bibinfo{journal}{Phys. Rev. X} \textbf{\bibinfo{volume}{3}},
  \bibinfo{pages}{\href{http://link.aps.org/doi/10.1103/PhysRevX.3.021009}{021009}}
  (\bibinfo{year}{2013}).

\bibitem[{\citenamefont{Barkeshli
  et~al.}(2013{\natexlab{b}})\citenamefont{Barkeshli, Jian, and
  Qi}}]{BarkeshliJianQi13}
\bibinfo{author}{\bibfnamefont{M.}~\bibnamefont{Barkeshli}},
  \bibinfo{author}{\bibfnamefont{C.-M.} \bibnamefont{Jian}}, \bibnamefont{and}
  \bibinfo{author}{\bibfnamefont{X.-L.} \bibnamefont{Qi}},
  \bibinfo{journal}{\href{http://arxiv.org/abs/1304.7579}{arXiv:1304.7579}}
  (\bibinfo{year}{2013}{\natexlab{b}}).

\bibitem[{\citenamefont{Barkeshli
  et~al.}(2013{\natexlab{c}})\citenamefont{Barkeshli, Jian, and
  Qi}}]{BarkeshliJianQi13long}
\bibinfo{author}{\bibfnamefont{M.}~\bibnamefont{Barkeshli}},
  \bibinfo{author}{\bibfnamefont{C.-M.} \bibnamefont{Jian}}, \bibnamefont{and}
  \bibinfo{author}{\bibfnamefont{X.-L.} \bibnamefont{Qi}},
  \bibinfo{journal}{\href{http://arxiv.org/abs/1305.7203}{arXiv:1305.7203}}
  (\bibinfo{year}{2013}{\natexlab{c}}).

\bibitem[{\citenamefont{Wen}(2003)}]{Wenplaquettemodel}
\bibinfo{author}{\bibfnamefont{X.-G.} \bibnamefont{Wen}},
  \bibinfo{journal}{Phys. Rev. Lett.} \textbf{\bibinfo{volume}{90}},
  \bibinfo{pages}{\href{http://link.aps.org/doi/10.1103/PhysRevLett.90.016803}{016803}}
  (\bibinfo{year}{2003}).

\bibitem[{\citenamefont{Bombin and Martin-Delgado}(2006)}]{BombinMartin06}
\bibinfo{author}{\bibfnamefont{H.}~\bibnamefont{Bombin}} \bibnamefont{and}
  \bibinfo{author}{\bibfnamefont{M.~A.} \bibnamefont{Martin-Delgado}},
  \bibinfo{journal}{Phys. Rev. Lett.} \textbf{\bibinfo{volume}{97}},
  \bibinfo{pages}{\href{http://link.aps.org/doi/10.1103/PhysRevLett.97.180501}{180501}}
  (\bibinfo{year}{2006}).

\bibitem[{\citenamefont{Kivelson et~al.}(1998)\citenamefont{Kivelson, Fradkin,
  and Emery}}]{KivelsonFradkinEmery98}
\bibinfo{author}{\bibfnamefont{S.~A.} \bibnamefont{Kivelson}},
  \bibinfo{author}{\bibfnamefont{E.}~\bibnamefont{Fradkin}}, \bibnamefont{and}
  \bibinfo{author}{\bibfnamefont{V.~J.} \bibnamefont{Emery}},
  \bibinfo{journal}{Nature} \textbf{\bibinfo{volume}{393}},
  \bibinfo{pages}{550} (\bibinfo{year}{1998}).

\bibitem[{\citenamefont{Wen}(2002)}]{Wenspinliquid02}
\bibinfo{author}{\bibfnamefont{X.-G.} \bibnamefont{Wen}},
  \bibinfo{journal}{Phys. Rev. B} \textbf{\bibinfo{volume}{65}},
  \bibinfo{pages}{\href{http://link.aps.org/doi/10.1103/PhysRevB.65.165113}{165113}}
  (\bibinfo{year}{2002}).

\bibitem[{\citenamefont{Chen et~al.}(2013)\citenamefont{Chen, Gu, Liu, and
  Wen}}]{ChenGuLiuWen11}
\bibinfo{author}{\bibfnamefont{X.}~\bibnamefont{Chen}},
  \bibinfo{author}{\bibfnamefont{Z.-C.} \bibnamefont{Gu}},
  \bibinfo{author}{\bibfnamefont{Z.-X.} \bibnamefont{Liu}}, \bibnamefont{and}
  \bibinfo{author}{\bibfnamefont{X.-G.} \bibnamefont{Wen}},
  \bibinfo{journal}{Phys. Rev. B} \textbf{\bibinfo{volume}{87}},
  \bibinfo{pages}{\href{http://link.aps.org/doi/10.1103/PhysRevB.87.155114}{155114}}
  (\bibinfo{year}{2013}).

\bibitem[{\citenamefont{Gu and Wen}(2012)}]{GuWen12}
\bibinfo{author}{\bibfnamefont{Z.-C.} \bibnamefont{Gu}} \bibnamefont{and}
  \bibinfo{author}{\bibfnamefont{X.-G.} \bibnamefont{Wen}},
  \bibinfo{journal}{\href{http://arxiv.org/abs/1201.2648}{arXiv:1201.2648}}
  (\bibinfo{year}{2012}).

\bibitem[{\citenamefont{Mesaros and Ran}(2013)}]{MesarosRan12}
\bibinfo{author}{\bibfnamefont{A.}~\bibnamefont{Mesaros}} \bibnamefont{and}
  \bibinfo{author}{\bibfnamefont{Y.}~\bibnamefont{Ran}},
  \bibinfo{journal}{Phys. Rev. B} \textbf{\bibinfo{volume}{87}},
  \bibinfo{pages}{\href{http://link.aps.org/doi/10.1103/PhysRevB.87.155115}{155115}}
  (\bibinfo{year}{2013}).

\bibitem[{\citenamefont{Chen et~al.}(2012)\citenamefont{Chen, Gu, Liu, and
  Wen}}]{ChenGuLiuWen12}
\bibinfo{author}{\bibfnamefont{X.}~\bibnamefont{Chen}},
  \bibinfo{author}{\bibfnamefont{Z.-C.} \bibnamefont{Gu}},
  \bibinfo{author}{\bibfnamefont{Z.-X.} \bibnamefont{Liu}}, \bibnamefont{and}
  \bibinfo{author}{\bibfnamefont{X.-G.} \bibnamefont{Wen}},
  \bibinfo{journal}{Science} \textbf{\bibinfo{volume}{338}},
  \bibinfo{pages}{1604} (\bibinfo{year}{2012}).

\bibitem[{\citenamefont{Vishwanath and Senthil}(2013)}]{VishwanathSenthil12}
\bibinfo{author}{\bibfnamefont{A.}~\bibnamefont{Vishwanath}} \bibnamefont{and}
  \bibinfo{author}{\bibfnamefont{T.}~\bibnamefont{Senthil}},
  \bibinfo{journal}{Phys. Rev. X} \textbf{\bibinfo{volume}{3}},
  \bibinfo{pages}{\href{http://link.aps.org/doi/10.1103/PhysRevX.3.011016}{011016}}
  (\bibinfo{year}{2013}).

\bibitem[{\citenamefont{Burnell et~al.}(2013)\citenamefont{Burnell, Chen,
  Fidkowski, and Vishwanath}}]{BurnellChenFidkowskiVishwanath13}
\bibinfo{author}{\bibfnamefont{F.~J.} \bibnamefont{Burnell}},
  \bibinfo{author}{\bibfnamefont{X.}~\bibnamefont{Chen}},
  \bibinfo{author}{\bibfnamefont{L.}~\bibnamefont{Fidkowski}},
  \bibnamefont{and}
  \bibinfo{author}{\bibfnamefont{A.}~\bibnamefont{Vishwanath}},
  \bibinfo{journal}{\href{http://arxiv.org/abs/1302.7072}{arXiv:1302.7072}}
  (\bibinfo{year}{2013}).

\bibitem[{\citenamefont{Turaev}(2010)}]{Turaevbook}
\bibinfo{author}{\bibfnamefont{V.~G.} \bibnamefont{Turaev}},
  \emph{\bibinfo{title}{Quantum Invariants of Knots and 3-Manifolds}}
  (\bibinfo{publisher}{De Gruyter}, \bibinfo{year}{2010}),
  \bibinfo{edition}{2nd} ed.

\bibitem[{\citenamefont{Bakalov and
  Kirillov}(2001)}]{BakalovKirillovlecturenotes}
\bibinfo{author}{\bibfnamefont{B.}~\bibnamefont{Bakalov}} \bibnamefont{and}
  \bibinfo{author}{\bibfnamefont{A.}~\bibnamefont{Kirillov}},
  \emph{\bibinfo{title}{Lectures on tensor categories and modular functor}}
  (\bibinfo{publisher}{American Mathematical Society}, \bibinfo{year}{2001}).

\bibitem[{\citenamefont{Arovas et~al.}(1984)\citenamefont{Arovas, Schrieffer,
  and Wilczek}}]{ArovasSchriefferWilczek84}
\bibinfo{author}{\bibfnamefont{D.}~\bibnamefont{Arovas}},
  \bibinfo{author}{\bibfnamefont{J.~R.} \bibnamefont{Schrieffer}},
  \bibnamefont{and} \bibinfo{author}{\bibfnamefont{F.}~\bibnamefont{Wilczek}},
  \bibinfo{journal}{Phys. Rev. Lett.} \textbf{\bibinfo{volume}{53}},
  \bibinfo{pages}{\href{http://link.aps.org/doi/10.1103/PhysRevLett.53.722}{722}}
  (\bibinfo{year}{1984}).

\bibitem[{\citenamefont{Wilczek}(1990)}]{Wilczekbook}
\bibinfo{author}{\bibfnamefont{F.}~\bibnamefont{Wilczek}},
  \emph{\bibinfo{title}{Fractional Statistics and Anyon Superconductivity}}
  (\bibinfo{publisher}{World Scientific}, \bibinfo{year}{1990}).

\bibitem[{\citenamefont{Wen}(1990)}]{Wentopologicalorder90}
\bibinfo{author}{\bibfnamefont{X.-G.} \bibnamefont{Wen}},
  \bibinfo{journal}{Int. J. Mod. Phys. B} \textbf{\bibinfo{volume}{4}},
  \bibinfo{pages}{239} (\bibinfo{year}{1990}).

\bibitem[{\citenamefont{Wen}(2004)}]{Wenbook}
\bibinfo{author}{\bibfnamefont{X.-G.} \bibnamefont{Wen}},
  \emph{\bibinfo{title}{Quantum Field Theory of Many Body Systems}}
  (\bibinfo{publisher}{Oxford Univ. Press, Oxford}, \bibinfo{year}{2004}).

\bibitem[{\citenamefont{Fradkin}(2013)}]{Fradkinbook}
\bibinfo{author}{\bibfnamefont{E.}~\bibnamefont{Fradkin}},
  \emph{\bibinfo{title}{Field Theories of Condensed Matter Physics}}
  (\bibinfo{publisher}{Cambridge University Press}, \bibinfo{year}{2013}),
  \bibinfo{edition}{2nd} ed.

\bibitem[{\citenamefont{Bonderson and Nayak}(2013)}]{BondersonNayak12}
\bibinfo{author}{\bibfnamefont{P.}~\bibnamefont{Bonderson}} \bibnamefont{and}
  \bibinfo{author}{\bibfnamefont{C.}~\bibnamefont{Nayak}},
  \bibinfo{journal}{Phys. Rev. B} \textbf{\bibinfo{volume}{87}},
  \bibinfo{pages}{\href{http://link.aps.org/doi/10.1103/PhysRevB.87.195451}{195451}}
  (\bibinfo{year}{2013}).

\bibitem[{\citenamefont{Teo and Kane}(2010{\natexlab{a}})}]{TeoKane}
\bibinfo{author}{\bibfnamefont{J.~C.~Y.} \bibnamefont{Teo}} \bibnamefont{and}
  \bibinfo{author}{\bibfnamefont{C.~L.} \bibnamefont{Kane}},
  \bibinfo{journal}{Phys. Rev. B} \textbf{\bibinfo{volume}{82}},
  \bibinfo{pages}{\href{http://link.aps.org/doi/10.1103/PhysRevB.82.115120}{115120}}
  (\bibinfo{year}{2010}{\natexlab{a}}).

\bibitem[{\citenamefont{Doplicher et~al.}(1971)\citenamefont{Doplicher, Haag,
  and Roberts}}]{DoplicherHaagRoberts71}
\bibinfo{author}{\bibfnamefont{S.}~\bibnamefont{Doplicher}},
  \bibinfo{author}{\bibfnamefont{R.}~\bibnamefont{Haag}}, \bibnamefont{and}
  \bibinfo{author}{\bibfnamefont{J.~E.} \bibnamefont{Roberts}},
  \bibinfo{journal}{Comm. Math. Phys.} \textbf{\bibinfo{volume}{23}},
  \bibinfo{pages}{199} (\bibinfo{year}{1971}).

\bibitem[{\citenamefont{Doplicher et~al.}(1974)\citenamefont{Doplicher, Haag,
  and Roberts}}]{DoplicherHaagRoberts74}
\bibinfo{author}{\bibfnamefont{S.}~\bibnamefont{Doplicher}},
  \bibinfo{author}{\bibfnamefont{R.}~\bibnamefont{Haag}}, \bibnamefont{and}
  \bibinfo{author}{\bibfnamefont{J.~E.} \bibnamefont{Roberts}},
  \bibinfo{journal}{Comm. Math. Phys.} \textbf{\bibinfo{volume}{35}},
  \bibinfo{pages}{49} (\bibinfo{year}{1974}).

\bibitem[{\citenamefont{Read}(2003)}]{Read03}
\bibinfo{author}{\bibfnamefont{N.}~\bibnamefont{Read}}, \bibinfo{journal}{J.
  Math. Phys} \textbf{\bibinfo{volume}{44}}, \bibinfo{pages}{558}
  (\bibinfo{year}{2003}).

\bibitem[{\citenamefont{Teo and Kane}(2010{\natexlab{b}})}]{TeoKane09}
\bibinfo{author}{\bibfnamefont{J.~C.~Y.} \bibnamefont{Teo}} \bibnamefont{and}
  \bibinfo{author}{\bibfnamefont{C.~L.} \bibnamefont{Kane}},
  \bibinfo{journal}{Phys. Rev. Lett.} \textbf{\bibinfo{volume}{104}},
  \bibinfo{pages}{\href{http://link.aps.org/doi/10.1103/PhysRevLett.104.046401}{046401}}
  (\bibinfo{year}{2010}{\natexlab{b}}).

\bibitem[{\citenamefont{Freedman
  et~al.}(2011{\natexlab{a}})\citenamefont{Freedman, Hastings, Nayak, Qi,
  Walker, and Wang}}]{FreedmanHastingsNayakQiWalkerWang}
\bibinfo{author}{\bibfnamefont{M.}~\bibnamefont{Freedman}},
  \bibinfo{author}{\bibfnamefont{M.~B.} \bibnamefont{Hastings}},
  \bibinfo{author}{\bibfnamefont{C.}~\bibnamefont{Nayak}},
  \bibinfo{author}{\bibfnamefont{X.-L.} \bibnamefont{Qi}},
  \bibinfo{author}{\bibfnamefont{K.}~\bibnamefont{Walker}}, \bibnamefont{and}
  \bibinfo{author}{\bibfnamefont{Z.}~\bibnamefont{Wang}},
  \bibinfo{journal}{Phys. Rev. B} \textbf{\bibinfo{volume}{83}},
  \bibinfo{pages}{\href{http://link.aps.org/doi/10.1103/PhysRevB.83.115132}{115132}}
  (\bibinfo{year}{2011}{\natexlab{a}}).

\bibitem[{\citenamefont{Freedman
  et~al.}(2011{\natexlab{b}})\citenamefont{Freedman, Hastings, Nayak, and
  Qi}}]{FreedmanHastingsNayakQi}
\bibinfo{author}{\bibfnamefont{M.}~\bibnamefont{Freedman}},
  \bibinfo{author}{\bibfnamefont{M.~B.} \bibnamefont{Hastings}},
  \bibinfo{author}{\bibfnamefont{C.}~\bibnamefont{Nayak}}, \bibnamefont{and}
  \bibinfo{author}{\bibfnamefont{X.-L.} \bibnamefont{Qi}},
  \bibinfo{journal}{Phys. Rev. B} \textbf{\bibinfo{volume}{84}},
  \bibinfo{pages}{\href{http://link.aps.org/doi/10.1103/PhysRevB.84.245119}{245119}}
  (\bibinfo{year}{2011}{\natexlab{b}}).

\bibitem[{\citenamefont{de~Wild~Propitius and Bais}(1996)}]{PropitiusBais96}
\bibinfo{author}{\bibfnamefont{M.}~\bibnamefont{de~Wild~Propitius}}
  \bibnamefont{and} \bibinfo{author}{\bibfnamefont{F.~A.} \bibnamefont{Bais}},
  \bibinfo{journal}{\href{http://arxiv.org/abs/hep-th/9511201}{arXiv:hep-th/9511201}}
   (\bibinfo{year}{1996}).

\bibitem[{\citenamefont{Mochon}(2004)}]{Mochon04}
\bibinfo{author}{\bibfnamefont{C.}~\bibnamefont{Mochon}},
  \bibinfo{journal}{Phys. Rev. A} \textbf{\bibinfo{volume}{69}},
  \bibinfo{pages}{\href{http://link.aps.org/doi/10.1103/PhysRevA.69.032306}{032306}}
  (\bibinfo{year}{2004}).

\bibitem[{spi()}]{spinstatisicsnotes}
\bibinfo{note}{For a more general system, this identification between
  $2\pi\times\mbox{ord}(\lambda)$ rotation and exchange statistics needs to be
  further modified by a constant phase.\cite{TeoRoyXiaoappearsoon}}.

\bibitem[{\citenamefont{Segal}(1988)}]{Segal88}
\bibinfo{author}{\bibfnamefont{G.}~\bibnamefont{Segal}},
  \emph{\bibinfo{title}{Conformal field theory}} (\bibinfo{publisher}{Oxford
  preprint}, \bibinfo{year}{1988}), \bibinfo{note}{and lecture at the IAMP
  Congress, Swansea}.

\bibitem[{\citenamefont{Verlinde}(1988)}]{Verlinde88}
\bibinfo{author}{\bibfnamefont{E.}~\bibnamefont{Verlinde}},
  \bibinfo{journal}{Nucl. Phys. B} \textbf{\bibinfo{volume}{300}},
  \bibinfo{pages}{360} (\bibinfo{year}{1988}).

\bibitem[{\citenamefont{Witten}(1989)}]{WittenJonespolynomials}
\bibinfo{author}{\bibfnamefont{E.}~\bibnamefont{Witten}},
  \bibinfo{journal}{Comm. Math. Phys.} \textbf{\bibinfo{volume}{121}},
  \bibinfo{pages}{351} (\bibinfo{year}{1989}).

\bibitem[{\citenamefont{Keski-Vakkuri and Wen}(1993)}]{KeskiWen93}
\bibinfo{author}{\bibfnamefont{E.}~\bibnamefont{Keski-Vakkuri}}
  \bibnamefont{and} \bibinfo{author}{\bibfnamefont{X.-G.} \bibnamefont{Wen}},
  \bibinfo{journal}{Int. J. Mod. Phys. B} \textbf{\bibinfo{volume}{7}},
  \bibinfo{pages}{4227} (\bibinfo{year}{1993}).

\bibitem[{\citenamefont{Wen}(2012)}]{Wenmodulartransformation12}
\bibinfo{author}{\bibfnamefont{X.-G.} \bibnamefont{Wen}},
  \bibinfo{journal}{\href{http://arxiv.org/abs/1212.5121}{arXiv:1212.5121}}
  (\bibinfo{year}{2012}).

\bibitem[{\citenamefont{Walker}(1991)}]{Walkernotes91}
\bibinfo{author}{\bibfnamefont{K.}~\bibnamefont{Walker}},
  \emph{\bibinfo{title}{On Witten's 3-manifold Invariants}}
  (\bibinfo{year}{1991}), \urlprefix\url{http://canyon23.net/math}.

\bibitem[{\citenamefont{Freedman
  et~al.}(2002{\natexlab{b}})\citenamefont{Freedman, Larsen, and
  Wang}}]{FreedmanLarsenWang00}
\bibinfo{author}{\bibfnamefont{M.}~\bibnamefont{Freedman}},
  \bibinfo{author}{\bibfnamefont{M.}~\bibnamefont{Larsen}}, \bibnamefont{and}
  \bibinfo{author}{\bibfnamefont{Z.}~\bibnamefont{Wang}},
  \bibinfo{journal}{Comm. Math. Phys} \textbf{\bibinfo{volume}{227}},
  \bibinfo{pages}{605} (\bibinfo{year}{2002}{\natexlab{b}}).

\bibitem[{\citenamefont{Wen}(1989)}]{Wentopologicalorder89}
\bibinfo{author}{\bibfnamefont{X.-G.} \bibnamefont{Wen}},
  \bibinfo{journal}{Phys. Rev. B} \textbf{\bibinfo{volume}{40}},
  \bibinfo{pages}{\href{http://link.aps.org/doi/10.1103/PhysRevB.40.7387}{7387}}
  (\bibinfo{year}{1989}).

\bibitem[{\citenamefont{Wen and Niu}(1990)}]{WenNiu90}
\bibinfo{author}{\bibfnamefont{X.-G.} \bibnamefont{Wen}} \bibnamefont{and}
  \bibinfo{author}{\bibfnamefont{Q.}~\bibnamefont{Niu}},
  \bibinfo{journal}{Phys. Rev. B} \textbf{\bibinfo{volume}{41}},
  \bibinfo{pages}{\href{http://link.aps.org/doi/10.1103/PhysRevB.41.9377}{9377}}
  (\bibinfo{year}{1990}).

\bibitem[{\citenamefont{Wilczek and Zee}(1983)}]{WilczekZee}
\bibinfo{author}{\bibfnamefont{F.}~\bibnamefont{Wilczek}} \bibnamefont{and}
  \bibinfo{author}{\bibfnamefont{A.}~\bibnamefont{Zee}},
  \bibinfo{journal}{Phys. Rev. Lett.} \textbf{\bibinfo{volume}{51}},
  \bibinfo{pages}{\href{http://link.aps.org/doi/10.1103/PhysRevLett.51.2250}{2250}}
  (\bibinfo{year}{1983}).

\bibitem[{\citenamefont{Horowitz and Srednicki}(1990)}]{HorowitzSrednicki}
\bibinfo{author}{\bibfnamefont{G.~T.} \bibnamefont{Horowitz}} \bibnamefont{and}
  \bibinfo{author}{\bibfnamefont{M.}~\bibnamefont{Srednicki}},
  \bibinfo{journal}{Commun. Math. Phys.} \textbf{\bibinfo{volume}{130}},
  \bibinfo{pages}{83} (\bibinfo{year}{1990}).

\bibitem[{\citenamefont{Kitaev and Preskill}(2006)}]{KitaevPreskill06}
\bibinfo{author}{\bibfnamefont{A.}~\bibnamefont{Kitaev}} \bibnamefont{and}
  \bibinfo{author}{\bibfnamefont{J.}~\bibnamefont{Preskill}},
  \bibinfo{journal}{Phys. Rev. Lett.} \textbf{\bibinfo{volume}{96}},
  \bibinfo{pages}{\href{http://link.aps.org/doi/10.1103/PhysRevLett.96.110404}{110404}}
  (\bibinfo{year}{2006}).

\bibitem[{\citenamefont{Zhang et~al.}(1989)\citenamefont{Zhang, Hansson, and
  Kivelson}}]{ZhangHanssonKivelson89}
\bibinfo{author}{\bibfnamefont{S.~C.} \bibnamefont{Zhang}},
  \bibinfo{author}{\bibfnamefont{T.~H.} \bibnamefont{Hansson}},
  \bibnamefont{and} \bibinfo{author}{\bibfnamefont{S.}~\bibnamefont{Kivelson}},
  \bibinfo{journal}{Phys. Rev. Lett.} \textbf{\bibinfo{volume}{62}},
  \bibinfo{pages}{\href{http://link.aps.org/doi/10.1103/PhysRevLett.62.82}{82}}
  (\bibinfo{year}{1989}).

\bibitem[{\citenamefont{Blok and Wen}(1990)}]{BlokWen90}
\bibinfo{author}{\bibfnamefont{B.}~\bibnamefont{Blok}} \bibnamefont{and}
  \bibinfo{author}{\bibfnamefont{X.-G.} \bibnamefont{Wen}},
  \bibinfo{journal}{Phys. Rev. B} \textbf{\bibinfo{volume}{42}},
  \bibinfo{pages}{\href{http://link.aps.org/doi/10.1103/PhysRevB.42.8145}{8145}}
  (\bibinfo{year}{1990}).

\bibitem[{\citenamefont{Bullock and Brennen}(2007)}]{Bullock}
\bibinfo{author}{\bibfnamefont{S.~S.} \bibnamefont{Bullock}} \bibnamefont{and}
  \bibinfo{author}{\bibfnamefont{G.~K.} \bibnamefont{Brennen}},
  \bibinfo{journal}{J. Phys. A} \textbf{\bibinfo{volume}{40}},
  \bibinfo{pages}{3481} (\bibinfo{year}{2007}).

\bibitem[{\citenamefont{Schulz et~al.}(2012)\citenamefont{Schulz, Dusuel, Orus,
  Vidal, and Schmidt}}]{Schulz}
\bibinfo{author}{\bibfnamefont{M.~D.} \bibnamefont{Schulz}},
  \bibinfo{author}{\bibfnamefont{S.}~\bibnamefont{Dusuel}},
  \bibinfo{author}{\bibfnamefont{R.}~\bibnamefont{Orus}},
  \bibinfo{author}{\bibfnamefont{J.}~\bibnamefont{Vidal}}, \bibnamefont{and}
  \bibinfo{author}{\bibfnamefont{K.~P.} \bibnamefont{Schmidt}},
  \bibinfo{journal}{New J. Phys.} \textbf{\bibinfo{volume}{14}},
  \bibinfo{pages}{025005} (\bibinfo{year}{2012}).

\bibitem[{\citenamefont{Ran et~al.}(2009)\citenamefont{Ran, Zhang, and
  Vishwanath}}]{RanZhangVishwanath}
\bibinfo{author}{\bibfnamefont{Y.}~\bibnamefont{Ran}},
  \bibinfo{author}{\bibfnamefont{Y.}~\bibnamefont{Zhang}}, \bibnamefont{and}
  \bibinfo{author}{\bibfnamefont{A.}~\bibnamefont{Vishwanath}},
  \bibinfo{journal}{Nat. Phys} \textbf{\bibinfo{volume}{5}},
  \bibinfo{pages}{298} (\bibinfo{year}{2009}).

\bibitem[{\citenamefont{Teo and Hughes}(2013)}]{TeoHughes}
\bibinfo{author}{\bibfnamefont{J.~C.~Y.} \bibnamefont{Teo}} \bibnamefont{and}
  \bibinfo{author}{\bibfnamefont{T.~L.} \bibnamefont{Hughes}},
  \bibinfo{journal}{Phys. Rev. Lett.} \textbf{\bibinfo{volume}{111}},
  \bibinfo{pages}{\href{http://link.aps.org/doi/10.1103/PhysRevLett.111.047006}{047006}}
  (\bibinfo{year}{2013}).

\bibitem[{\citenamefont{Radzihovsky and
  Vishwanath}(2009)}]{RadzihovskyVishwanath08}
\bibinfo{author}{\bibfnamefont{L.}~\bibnamefont{Radzihovsky}} \bibnamefont{and}
  \bibinfo{author}{\bibfnamefont{A.}~\bibnamefont{Vishwanath}},
  \bibinfo{journal}{Phys. Rev. Lett.} \textbf{\bibinfo{volume}{103}},
  \bibinfo{pages}{\href{http://link.aps.org/doi/10.1103/PhysRevLett.103.010404}{010404}}
  (\bibinfo{year}{2009}).

\bibitem[{\citenamefont{Agterberg and
  Tsunetsugu}(2008)}]{AgterbergTsunetsugu08}
\bibinfo{author}{\bibfnamefont{D.~F.} \bibnamefont{Agterberg}}
  \bibnamefont{and}
  \bibinfo{author}{\bibfnamefont{H.}~\bibnamefont{Tsunetsugu}},
  \bibinfo{journal}{Nat. Phys.} \textbf{\bibinfo{volume}{4}},
  \bibinfo{pages}{639} (\bibinfo{year}{2008}).

\bibitem[{\citenamefont{Berg et~al.}(2009)\citenamefont{Berg, Fradkin, and
  Kivelson}}]{BergFradkinKivelson09}
\bibinfo{author}{\bibfnamefont{E.}~\bibnamefont{Berg}},
  \bibinfo{author}{\bibfnamefont{E.}~\bibnamefont{Fradkin}}, \bibnamefont{and}
  \bibinfo{author}{\bibfnamefont{S.~A.} \bibnamefont{Kivelson}},
  \bibinfo{journal}{Nat. Phys.} \textbf{\bibinfo{volume}{5}},
  \bibinfo{pages}{830} (\bibinfo{year}{2009}).

\bibitem[{\citenamefont{Gopalakrishnan
  et~al.}(2013)\citenamefont{Gopalakrishnan, Teo, and
  Hughes}}]{GopalakrishnanTeoHughes13}
\bibinfo{author}{\bibfnamefont{S.}~\bibnamefont{Gopalakrishnan}},
  \bibinfo{author}{\bibfnamefont{J.~C.~Y.} \bibnamefont{Teo}},
  \bibnamefont{and} \bibinfo{author}{\bibfnamefont{T.~L.}
  \bibnamefont{Hughes}}, \bibinfo{journal}{Phys. Rev. Lett.}
  \textbf{\bibinfo{volume}{111}},
  \bibinfo{pages}{\href{http://link.aps.org/doi/10.1103/PhysRevLett.111.025304}{025304}}
  (\bibinfo{year}{2013}).

\bibitem[{\citenamefont{Thouless}(1983)}]{Thouless}
\bibinfo{author}{\bibfnamefont{D.~J.} \bibnamefont{Thouless}},
  \bibinfo{journal}{Phys. Rev. B} \textbf{\bibinfo{volume}{27}},
  \bibinfo{pages}{\href{http://link.aps.org/doi/10.1103/PhysRevB.27.608}{6083}}
  (\bibinfo{year}{1983}).

\bibitem[{\citenamefont{Niu and Thouless}(1984)}]{NiuThouless}
\bibinfo{author}{\bibfnamefont{Q.}~\bibnamefont{Niu}} \bibnamefont{and}
  \bibinfo{author}{\bibfnamefont{D.~J.} \bibnamefont{Thouless}},
  \bibinfo{journal}{J. Phys. A} \textbf{\bibinfo{volume}{17}},
  \bibinfo{pages}{2453} (\bibinfo{year}{1984}).

\bibitem[{\citenamefont{Mermin}(1979)}]{Mermin79}
\bibinfo{author}{\bibfnamefont{N.~D.} \bibnamefont{Mermin}},
  \bibinfo{journal}{Rev. Mod. Phys.} \textbf{\bibinfo{volume}{51}},
  \bibinfo{pages}{\href{http://link.aps.org/doi/10.1103/RevModPhys.51.591}{591}}
  (\bibinfo{year}{1979}).

\bibitem[{\citenamefont{Chaikin and Lubensky}(2000)}]{ChaikinLubensky}
\bibinfo{author}{\bibfnamefont{P.~M.} \bibnamefont{Chaikin}} \bibnamefont{and}
  \bibinfo{author}{\bibfnamefont{T.~C.} \bibnamefont{Lubensky}},
  \emph{\bibinfo{title}{Principles of Condensed Matter Physics}}
  (\bibinfo{publisher}{Cambridge University Press}, \bibinfo{year}{2000}).

\bibitem[{\citenamefont{Nelson}(2002)}]{Nelsonbook}
\bibinfo{author}{\bibfnamefont{D.~R.} \bibnamefont{Nelson}},
  \emph{\bibinfo{title}{Defects and Geometry in Condensed Matter Physics}}
  (\bibinfo{publisher}{Cambridge University Press}, \bibinfo{year}{2002}).

\bibitem[{\citenamefont{Kleman and Friedel}(2008)}]{KlemanFriedel08}
\bibinfo{author}{\bibfnamefont{M.}~\bibnamefont{Kleman}} \bibnamefont{and}
  \bibinfo{author}{\bibfnamefont{J.}~\bibnamefont{Friedel}},
  \bibinfo{journal}{Rev. Mod. Phys.} \textbf{\bibinfo{volume}{80}},
  \bibinfo{pages}{\href{http://link.aps.org/doi/10.1103/RevModPhys.80.61}{61}}
  (\bibinfo{year}{2008}).

\bibitem[{not()}]{notesZ3Wilsonloop}
\bibinfo{note}{We prefer the branched Wilson string on the right for
  convenience in calculating intersection and braiding.}

\bibitem[{\citenamefont{Barkeshli and Wen}(2011)}]{BarkeshliWen11}
\bibinfo{author}{\bibfnamefont{M.}~\bibnamefont{Barkeshli}} \bibnamefont{and}
  \bibinfo{author}{\bibfnamefont{X.-G.} \bibnamefont{Wen}},
  \bibinfo{journal}{Phys. Rev. B} \textbf{\bibinfo{volume}{84}},
  \bibinfo{pages}{\href{http://link.aps.org/doi/10.1103/PhysRevB.84.115121}{115121}}
  (\bibinfo{year}{2011}).

\bibitem[{\citenamefont{MacLane}(1998)}]{MacLanebook}
\bibinfo{author}{\bibfnamefont{S.}~\bibnamefont{MacLane}},
  \emph{\bibinfo{title}{Categories for the Working Mathematician}}
  (\bibinfo{publisher}{Springer}, \bibinfo{year}{1998}), \bibinfo{edition}{2nd}
  ed.

\bibitem[{ord()}]{ord360rotation}
\bibinfo{note}{This is equivalent to a $360^\circ$ rotation on the {\em genon}
  covering space.}

\bibitem[{\citenamefont{L\"utken and Ross}(1992)}]{LutkenRoss92}
\bibinfo{author}{\bibfnamefont{C.~A.} \bibnamefont{L\"utken}} \bibnamefont{and}
  \bibinfo{author}{\bibfnamefont{G.~G.} \bibnamefont{Ross}},
  \bibinfo{journal}{Phys. Rev. B} \textbf{\bibinfo{volume}{45}},
  \bibinfo{pages}{\href{http://link.aps.org/doi/10.1103/PhysRevB.45.11837}{11837}}
  (\bibinfo{year}{1992}).

\bibitem[{\citenamefont{Fradkin and Kivelson}(1996)}]{FradkinKivelson96}
\bibinfo{author}{\bibfnamefont{E.}~\bibnamefont{Fradkin}} \bibnamefont{and}
  \bibinfo{author}{\bibfnamefont{S.}~\bibnamefont{Kivelson}},
  \bibinfo{journal}{Nucl. Phys. B} \textbf{\bibinfo{volume}{474}},
  \bibinfo{pages}{543} (\bibinfo{year}{1996}).

\bibitem[{\citenamefont{Bayntun et~al.}(2011)\citenamefont{Bayntun, Burgess,
  Dolan, and Lee}}]{BayntunBurgessDolanLee11}
\bibinfo{author}{\bibfnamefont{A.}~\bibnamefont{Bayntun}},
  \bibinfo{author}{\bibfnamefont{C.~P.} \bibnamefont{Burgess}},
  \bibinfo{author}{\bibfnamefont{B.~P.} \bibnamefont{Dolan}}, \bibnamefont{and}
  \bibinfo{author}{\bibfnamefont{S.-S.} \bibnamefont{Lee}},
  \bibinfo{journal}{New J. Phys.} \textbf{\bibinfo{volume}{13}},
  \bibinfo{pages}{035012} (\bibinfo{year}{2011}).

\bibitem[{\citenamefont{Witten}(1995)}]{WittenSduality}
\bibinfo{author}{\bibfnamefont{E.}~\bibnamefont{Witten}},
  \bibinfo{journal}{Selecta Math.} \textbf{\bibinfo{volume}{1}},
  \bibinfo{pages}{383} (\bibinfo{year}{1995}),
  \bibinfo{note}{\href{http://arxiv.org/abs/hep-th/9505186}{arXiv:hep-th/9505186}}.

\bibitem[{\citenamefont{Jacak et~al.}(2003)\citenamefont{Jacak, Sitko,
  Wieczorek, and W\'ojs}}]{JacakSitkoWieczorekWojsbook}
\bibinfo{author}{\bibfnamefont{L.}~\bibnamefont{Jacak}},
  \bibinfo{author}{\bibfnamefont{P.}~\bibnamefont{Sitko}},
  \bibinfo{author}{\bibfnamefont{K.}~\bibnamefont{Wieczorek}},
  \bibnamefont{and} \bibinfo{author}{\bibfnamefont{A.}~\bibnamefont{W\'ojs}},
  \emph{\bibinfo{title}{Quantum Hall Systems: Braid Groups, Composite Fermions
  and Fractional Charge}} (\bibinfo{publisher}{Oxford University Press, USA},
  \bibinfo{year}{2003}).

\bibitem[{\citenamefont{Ran et~al.}(2011)\citenamefont{Ran, Hosur, and
  Vishwanath}}]{RanHosurVishwanath11}
\bibinfo{author}{\bibfnamefont{Y.}~\bibnamefont{Ran}},
  \bibinfo{author}{\bibfnamefont{P.}~\bibnamefont{Hosur}}, \bibnamefont{and}
  \bibinfo{author}{\bibfnamefont{A.}~\bibnamefont{Vishwanath}},
  \bibinfo{journal}{Phys. Rev. B} \textbf{\bibinfo{volume}{84}},
  \bibinfo{pages}{\href{http://link.aps.org/doi/10.1103/PhysRevB.84.184501}{184501}}
  (\bibinfo{year}{2011}).

\bibitem[{\citenamefont{Hastings et~al.}(2013)\citenamefont{Hastings, Nayak,
  and Wang}}]{HastingsNayakWang12}
\bibinfo{author}{\bibfnamefont{M.~B.} \bibnamefont{Hastings}},
  \bibinfo{author}{\bibfnamefont{C.}~\bibnamefont{Nayak}}, \bibnamefont{and}
  \bibinfo{author}{\bibfnamefont{Z.}~\bibnamefont{Wang}},
  \bibinfo{journal}{Phys. Rev. B} \textbf{\bibinfo{volume}{87}},
  \bibinfo{pages}{\href{http://link.aps.org/doi/10.1103/PhysRevB.87.165421}{165421}}
  (\bibinfo{year}{2013}).

\bibitem[{\citenamefont{Barkeshli and Wen}(2010)}]{BarkeshliWen10}
\bibinfo{author}{\bibfnamefont{M.}~\bibnamefont{Barkeshli}} \bibnamefont{and}
  \bibinfo{author}{\bibfnamefont{X.-G.} \bibnamefont{Wen}},
  \bibinfo{journal}{Phys. Rev. B} \textbf{\bibinfo{volume}{81}},
  \bibinfo{pages}{\href{http://link.aps.org/doi/10.1103/PhysRevB.81.045323}{045323}}
  (\bibinfo{year}{2010}).

\bibitem[{\citenamefont{Barkeshli and Wen}(2012)}]{BarkeshliWen12}
\bibinfo{author}{\bibfnamefont{M.}~\bibnamefont{Barkeshli}} \bibnamefont{and}
  \bibinfo{author}{\bibfnamefont{X.-G.} \bibnamefont{Wen}},
  \bibinfo{journal}{Phys. Rev. B} \textbf{\bibinfo{volume}{86}},
  \bibinfo{pages}{\href{http://link.aps.org/doi/10.1103/PhysRevB.86.085114}{085114}}
  (\bibinfo{year}{2012}).

\bibitem[{\citenamefont{Lu and Vishwanath}(2013)}]{LuVishwanath13}
\bibinfo{author}{\bibfnamefont{Y.-M.} \bibnamefont{Lu}} \bibnamefont{and}
  \bibinfo{author}{\bibfnamefont{A.}~\bibnamefont{Vishwanath}},
  \bibinfo{journal}{\href{http://arxiv.org/abs/1302.2634}{arXiv:1302.2634}}
  (\bibinfo{year}{2013}).

\bibitem[{\citenamefont{Teo et~al.}()\citenamefont{Teo, Roy, and
  Chen}}]{TeoRoyXiaoappearsoon}
\bibinfo{author}{\bibfnamefont{J.~C.~Y.} \bibnamefont{Teo}},
  \bibinfo{author}{\bibfnamefont{A.}~\bibnamefont{Roy}}, \bibnamefont{and}
  \bibinfo{author}{\bibfnamefont{X.}~\bibnamefont{Chen}}, \bibinfo{note}{to
  appear soon}.

\bibitem[{\citenamefont{Haldane}(1983)}]{Haldanespinchain}
\bibinfo{author}{\bibfnamefont{F.~D.~M.} \bibnamefont{Haldane}},
  \bibinfo{journal}{Phys. Rev. Lett.} \textbf{\bibinfo{volume}{50}},
  \bibinfo{pages}{\href{http://link.aps.org/doi/10.1103/PhysRevLett.50.1153}{1153}}
  (\bibinfo{year}{1983}).

\bibitem[{\citenamefont{Affleck et~al.}(1987)\citenamefont{Affleck, Kennedy,
  Lieb, and Tasaki}}]{AKLT}
\bibinfo{author}{\bibfnamefont{I.}~\bibnamefont{Affleck}},
  \bibinfo{author}{\bibfnamefont{T.}~\bibnamefont{Kennedy}},
  \bibinfo{author}{\bibfnamefont{E.~H.} \bibnamefont{Lieb}}, \bibnamefont{and}
  \bibinfo{author}{\bibfnamefont{H.}~\bibnamefont{Tasaki}},
  \bibinfo{journal}{Phys. Rev. Lett.} \textbf{\bibinfo{volume}{59}},
  \bibinfo{pages}{\href{http://link.aps.org/doi/10.1103/PhysRevLett.59.799}{799}}
  (\bibinfo{year}{1987}).

\bibitem[{\citenamefont{Feiguin et~al.}(2007)\citenamefont{Feiguin, Trebst,
  Ludwig, Troyer, Kitaev, Wang, and
  Freedman}}]{FeiguinTrebstLudwigTroyerKitaevWangFreedman07}
\bibinfo{author}{\bibfnamefont{A.}~\bibnamefont{Feiguin}},
  \bibinfo{author}{\bibfnamefont{S.}~\bibnamefont{Trebst}},
  \bibinfo{author}{\bibfnamefont{A.~W.~W.} \bibnamefont{Ludwig}},
  \bibinfo{author}{\bibfnamefont{M.}~\bibnamefont{Troyer}},
  \bibinfo{author}{\bibfnamefont{A.}~\bibnamefont{Kitaev}},
  \bibinfo{author}{\bibfnamefont{Z.}~\bibnamefont{Wang}}, \bibnamefont{and}
  \bibinfo{author}{\bibfnamefont{M.~H.} \bibnamefont{Freedman}},
  \bibinfo{journal}{Phys. Rev. Lett.} \textbf{\bibinfo{volume}{98}},
  \bibinfo{pages}{\href{http://link.aps.org/doi/10.1103/PhysRevLett.98.160409}{160409}}
  (\bibinfo{year}{2007}).

\bibitem[{\citenamefont{Fendley}(2007)}]{Fendley12}
\bibinfo{author}{\bibfnamefont{P.}~\bibnamefont{Fendley}}, \bibinfo{journal}{J.
  Stat. Mech.} p. \bibinfo{pages}{P11020} (\bibinfo{year}{2007}).

\bibitem[{\citenamefont{Gils et~al.}(2013)\citenamefont{Gils, Ardonne, Trebst,
  Huse, Ludwig, Troyer, and Wang}}]{GilsArdonneTrebstHuseLudwigTroyerWang13}
\bibinfo{author}{\bibfnamefont{C.}~\bibnamefont{Gils}},
  \bibinfo{author}{\bibfnamefont{E.}~\bibnamefont{Ardonne}},
  \bibinfo{author}{\bibfnamefont{S.}~\bibnamefont{Trebst}},
  \bibinfo{author}{\bibfnamefont{D.~A.} \bibnamefont{Huse}},
  \bibinfo{author}{\bibfnamefont{A.~W.~W.} \bibnamefont{Ludwig}},
  \bibinfo{author}{\bibfnamefont{M.}~\bibnamefont{Troyer}}, \bibnamefont{and}
  \bibinfo{author}{\bibfnamefont{Z.}~\bibnamefont{Wang}},
  \bibinfo{journal}{Phys. Rev. B} \textbf{\bibinfo{volume}{87}},
  \bibinfo{pages}{\href{http://link.aps.org/doi/10.1103/PhysRevB.87.235120}{235120}}
  (\bibinfo{year}{2013}).

\bibitem[{\citenamefont{Motruk et~al.}(2013)\citenamefont{Motruk, Berg, Turner,
  and Pollmann}}]{MotrukTurnerBergPollmann13}
\bibinfo{author}{\bibfnamefont{J.}~\bibnamefont{Motruk}},
  \bibinfo{author}{\bibfnamefont{E.}~\bibnamefont{Berg}},
  \bibinfo{author}{\bibfnamefont{A.~M.} \bibnamefont{Turner}},
  \bibnamefont{and} \bibinfo{author}{\bibfnamefont{F.}~\bibnamefont{Pollmann}},
  \bibinfo{journal}{Phys. Rev. B} \textbf{\bibinfo{volume}{88}},
  \bibinfo{pages}{\href{http://link.aps.org/doi/10.1103/PhysRevB.88.085115}{085115}}
  (\bibinfo{year}{2013}).

\bibitem[{\citenamefont{Bondesan and Quella}(2013)}]{BondesanQuella13}
\bibinfo{author}{\bibfnamefont{R.}~\bibnamefont{Bondesan}} \bibnamefont{and}
  \bibinfo{author}{\bibfnamefont{T.}~\bibnamefont{Quella}},
  \bibinfo{journal}{\href{http://arxiv.org/abs/1303.5587}{arXiv:1303.5587}}
  (\bibinfo{year}{2013}).

\bibitem[{\citenamefont{Bonderson et~al.}(2008)\citenamefont{Bonderson,
  Freedman, and Nayak}}]{BondersonFreedmanNayak08}
\bibinfo{author}{\bibfnamefont{P.}~\bibnamefont{Bonderson}},
  \bibinfo{author}{\bibfnamefont{M.}~\bibnamefont{Freedman}}, \bibnamefont{and}
  \bibinfo{author}{\bibfnamefont{C.}~\bibnamefont{Nayak}},
  \bibinfo{journal}{Phys. Rev. Lett.} \textbf{\bibinfo{volume}{101}},
  \bibinfo{pages}{\href{http://link.aps.org/doi/10.1103/PhysRevLett.101.010501}{010501}}
  (\bibinfo{year}{2008}).

\bibitem[{\citenamefont{Alicea et~al.}(2011)\citenamefont{Alicea, Oreg, Refael,
  von Oppen, and Fisher}}]{AliceaOregRefaeOppenFisher11}
\bibinfo{author}{\bibfnamefont{J.}~\bibnamefont{Alicea}},
  \bibinfo{author}{\bibfnamefont{Y.}~\bibnamefont{Oreg}},
  \bibinfo{author}{\bibfnamefont{G.}~\bibnamefont{Refael}},
  \bibinfo{author}{\bibfnamefont{F.}~\bibnamefont{von Oppen}},
  \bibnamefont{and} \bibinfo{author}{\bibfnamefont{M.~P.~A.}
  \bibnamefont{Fisher}}, \bibinfo{journal}{Nat. Phys.}
  \textbf{\bibinfo{volume}{7}}, \bibinfo{pages}{412} (\bibinfo{year}{2011}).

\bibitem[{\citenamefont{Brown et~al.}(2013)\citenamefont{Brown, Bartlett,
  Doherty, and Barrett}}]{BrownBartlettDohertyBarrett13}
\bibinfo{author}{\bibfnamefont{B.~J.} \bibnamefont{Brown}},
  \bibinfo{author}{\bibfnamefont{S.~D.} \bibnamefont{Bartlett}},
  \bibinfo{author}{\bibfnamefont{A.~C.} \bibnamefont{Doherty}},
  \bibnamefont{and} \bibinfo{author}{\bibfnamefont{S.~D.}
  \bibnamefont{Barrett}},
  \bibinfo{journal}{\href{http://arxiv.org/abs/1303.4455}{arXiv:1303.4455}}
  (\bibinfo{year}{2013}).

\bibitem[{\citenamefont{Zhang et~al.}(2012)\citenamefont{Zhang, Grover, Turner,
  Oshikawa, and Vishwanath}}]{ZhangGroverTurnerOshikawaVishwanath12}
\bibinfo{author}{\bibfnamefont{Y.}~\bibnamefont{Zhang}},
  \bibinfo{author}{\bibfnamefont{T.}~\bibnamefont{Grover}},
  \bibinfo{author}{\bibfnamefont{A.}~\bibnamefont{Turner}},
  \bibinfo{author}{\bibfnamefont{M.}~\bibnamefont{Oshikawa}}, \bibnamefont{and}
  \bibinfo{author}{\bibfnamefont{A.}~\bibnamefont{Vishwanath}},
  \bibinfo{journal}{Phys. Rev. B} \textbf{\bibinfo{volume}{85}},
  \bibinfo{pages}{\href{http://link.aps.org/doi/10.1103/PhysRevB.85.235151}{235151}}
  (\bibinfo{year}{2012}).

\bibitem[{\citenamefont{Ocneanu}()}]{Ocneanu}
\bibinfo{author}{\bibfnamefont{A.}~\bibnamefont{Ocneanu}},
  \bibinfo{note}{unpublished}.

\end{thebibliography}
\end{document}